\documentclass[useAMS,usenatbib]{mn2e}

\usepackage{graphicx} 
\usepackage{graphicx}
\usepackage{cite}
\usepackage{verbatim}
\usepackage{array}

\usepackage{graphicx,cite,epsf}  
\usepackage{times}
\usepackage{subfigure}
\usepackage{textcomp}
\usepackage{graphicx,times,multirow,amsmath}
\usepackage{natbib}

\newcommand{\aap}{A\&A}

\newcommand{\apjl}{ApJL}
\newcommand{\apjs}{ApJS}

\def\aap{A\&A}                
\def\apjl{ApJ}
\def\apjl{ApJ}                
\def\apjs{ApJS}               
\def\ao{Appl.~Opt.}           
\def\aap{A\&A}                

\catcode`\@=11
\def\gsim{\ifmmode{\mathrel{\mathpalette\@versim>}}
    \else{$\mathrel{\mathpalette\@versim>}$}\fi}
\def\lsim{\ifmmode{\mathrel{\mathpalette\@versim<}}
    \else{$\mathrel{\mathpalette\@versim<}$}\fi}
\def\@versim#1#2{\lower 2.9truept \vbox{\baselineskip 0pt \lineskip 
    0.5truept \ialign{$\m@th#1\hfil##\hfil$\crcr#2\crcr\sim\crcr}}}
\catcode`\@=12

\def\y1{\hbox{${\rm yr}^{-1}$}}

\title[A method to unveil \textit{blazar}s among UGSs]{A new method to unveil \textit{blazar}s among multi-wavelength counterparts of Unassociated \textit{Fermi} $\gamma$-ray Sources} 

\author[S. Paiano, A. Franceschini, A. Stamerra]
{\parbox{\textwidth}{\raggedright S. Paiano$^{1,2,3}$\thanks{E-mail: \texttt{simona.paiano@gmail.com}}, A. Franceschini$^{1,2,}$, A. Stamerra$^{4,5}$
}\vspace{0.4cm}\\
\parbox{\textwidth}{\raggedright $^{(1)}$Dipartimento di Fisica e Astronomia, Universit\`a di Padova, vicolo dell'Osservatorio 3, I--35122 Padova, Italy.\\
$^{(2)}$INAF - Osservatorio Astronomico di Padova, vicolo dell'Osservatorio 5, I-35122 Padova, Italy.\\
$^{(3)}$INFN - Sezione di Padova, via Marzolo 8, Padova, Italy.\\
$^{(4)}$INAF - Osservatorio Astrofisico di Torino, via P. Giuria 1, 10125 Torino, Italy.\\
$^{(5)}$Scuola Normale Superiore, Piazza dei Cavalieri, 7, 56126 Pisa, Italy.\\
}}

\begin{document}

\date{}

\pagerange{\pageref{firstpage}--\pageref{lastpage}} \pubyear{2015}

\maketitle
\label{firstpage}

\begin{abstract}
We discuss a new method for unveiling the possible \textit{blazar} AGN nature among the numerous population of Unassociated Gamma-ray sources (UGS) in the \textit{Fermi} catalogues. 
Our tool relies on positional correspondence of the \textit{Fermi} object with X-ray sources (mostly from \textit{Swift}-XRT), correlated with other radio, IR and optical data in the field. 
We built a set of Spectral Energy Distributions (SED) templates representative of the various \textit{blazar} classes, and we quantitatively compared them to the observed multi-wavelength flux density data for all \textit{Swift}-XRT sources found within the \textit{Fermi} error-box, by taking advantage of some well-recognised regularities in the broad-band spectral properties of the objects. 
We tested the procedure by comparison with a few well-known \textit{blazar}s, and tested the chance for false positive recognition of UGS sources against known pulsars and other Galactic and extragalactic sources. 
Based on our spectral recognition tool, we find the \textit{blazar} candidate counterparts for 14 2FGL UGSs among 183 selected at high galactic latitudes. Further our tool also allows us rough estimates of the redshift for the candidate blazar. 
In a few cases in which this has been possible (i.e. when the counterpart was a SDSS object), we verified that our estimate is consistent with the measured redshift.  
The estimated redshifts of the proposed UGS counterparts are larger, on average, than those of known Fermi \textit{blazar}s, a fact that might explain the lack of previous association or identification in published catalogues. 
\end{abstract}


\section{Introduction}
\label{intro}

The \textit{Fermi} $\gamma$-ray Observatory \citep{atwood2009} is favouring a dramatic progress in the field of high-energy astrophysics. With its high-throughput, almost instantaneous all-sky vision, the mission has not only offered a tenfold increase in the number of catalogued sources from 100 MeV to 100 GeV with respect to previous $\gamma$-ray missions, but also a continuous monitoring for variability studies and a huge spectral coverage \citep{0fgl, 1fgl, 2fgl, 3FGL}.

As it often happens when such a new revolutionary astronomical facility is put in operation, the \textit{Fermi} mission is also producing unexpected outcomes: of the 1873 sources in the Second Catalogue of the Large Area Telescope (2FGL) \citep{2fgl}, as many as one-third (576 sources) are lacking reliable association with sources detected at other wavelengths, henceforth the Unassociated Gamma-ray sources (UGS). 
Many more are also found in the last release of the \textit{Fermi} catalogue (3FGL) where 1011 sources result to be unassociated among the 3034 objects reported.

The majority (about one thousands) of the 1297 Fermi associated sources in the 2FGL have been classified as active galactic nuclei (AGN), in particular \textit{blazar}s (BL Lac objects and flat spectrum radio quasars). It is thus likely that a large number of the UGSs might hide previously unknown sources of this category.

\textit{Blazar}s are the most extreme engines of nature, producing  the largest amount of radiant energy than any other cosmic source. From  a sub-parsec scale region, they accelerate to relativistic speeds entire plasma clouds transforming fast (even maximally) rotating super-massive black holes and gravitational energy into radiation and mechanical power. 

\textit{Blazar}s are not only unique machines to test extreme physics, but can also be exploited as light-houses to probe the distant universe. 
Their emitted very high energy (VHE; E~$>$~100~GeV) photons are known to interact with the low-energy photon backgrounds \citep[the Extragalactic Background Light, EBL, e.g.][]{franceschini2008} and decay into $e^--e^+$ pairs. 
So observations of the \textit{blazar} VHE spectra and their spectral absorption, e.g. with Cherenkov observatories or \textit{Fermi} satellite itself, are used to test the EBL \citep[e.g.][]{aharonian2006, ackermann2012science}.
On this respect, the identification of distant and high redshift \textit{blazar}s at high energies is particularly relevant, among others, to estimate the earliest EBL components due to the first-light sources (Population III stars, galaxies or quasars) in the universe (see e.g. Franceschini A., \& Rodighiero, G., 2017, to appear in A\&A).

Expanding our knowledge of the \textit{blazar} population at high energies and high redshift is then a priority topic for various reasons. 
Several papers in the literature are dedicated to methods for the identification of the \textit{Fermi} unassociated sources. \citet{ackermann2012x},  \citet{mirabal2012x} and \citet{doert2014x} developed statistical algorithms based on selected UGS $\gamma$-ray features, such as spectral and variability information, able to discriminate between AGNs and pulsars. 
Other works focused on the search for AGN candidates among the 2FGL UGSs by analysing their long-wavelength counterparts: \citet{petrov2013} and \citet{schinzel2015x} compiled a thorough catalogue of ATCA (\textit{Australia Telescope Compact Array)} radio sources lying inside the UGS error-boxes. 
Other radio surveys were published by \citet{nori2014} and \citet{giroletti2016}.  
\citet{massaro2011x, massaro2012x} and \citet{dambrusco2013x} proposed AGN candidates on the basis of the colours of the infrared counterparts in the \textit{Wide-field Infrared Survey Explorer} survey lying within the \textit{Fermi} error ellipses; in \citet{paggi2013x} a complete analysis of the X-ray data provided by the \textit{Swift} satellite has been performed to search for 
X-ray counterparts and in \citet{acero2013x} and \citet{landi2015} a multi-wavelength approach has been adopted. 

The present paper contributes to the effort of exploiting such a unique all-sky $\gamma$-ray survey for a search of high-energy emitting AGN population, with a new approach based on a total-band Spectral-Energy Distribution (SED) analysis.

The \textit{blazar} non-thermal radiation dominates, and often hides, the emission from the host galaxy or from the AGN substructures. 
For most \textit{blazar}s, especially for the BL Lac objects,  this results in a featureless optical spectra, thus hindering redshift measurement.
Our tool for \textit{blazar} recognition among the UGSs has then been tailored to offer at the same time a rough estimate of their redshift. 
This also takes advantage of the known relationship between the frequencies of the synchrotron and IC peaks and the source luminosities (sometimes referred to as the \textit{blazar} sequence, \citealt{fossati1998, ghisellini2017}). 
In any case, our method is completely empirical, model-independent, and not relying on prior assumptions, except for the requirement that the UGSs, proposed as Fermi \textit{blazar} candidates, are detected in X-rays.

Note that in the present paper, we propose a possible physical relationship between the UGS and the \textit{blazar}-like object, to be considered as a candidate for the association. 
In our case, improving from a proposed association to an identification for the source would require, among other, spectroscopic follow-up and confirmation.

With the last \textit{Fermi} catalogue released (3FGL, \citealt{3FGL}), the number of UGSs wherein to exercise our \textit{blazar} recognition tool will be further substantially amplified to as many as 1010 objects out of a total of 3034 sources, with chances to expand the number of $\gamma$-ray selected AGNs. 
While our UGS primary selection relies on the 2FGL catalogue, we make full use of the newer 3FGL to confirm those sources and to improve their error-box and \textit{Fermi} photometry.

The paper is organised as follows.  
In Section 2 we review the UGS selection. 
In Section 3 and Appendix A we discuss our procedures for the counterpart selection of UGSs. In particular Appendix A includes finding charts and the multi-wavelength SED for the UGS counterparts.
\textit{We define as potential association of an UGS a set of sources consistently detected in various bands, all positionally coincident, and within the Fermi error-box.}
In Section 4 we build the library of multi-wavelength SED templates from known \textit{blazar}s, selected from the 3FGL catalogue.
This SED template set is then used in Section 5 to build up our tool for the \textit{blazar} recognition and characterisation. 
In this section the validity of the method is verified on bona-fide \textit{blazar}s, against known Galactic and extragalactic sources, and also the false positive associations are tested. 
We then proceed in Section 6 to present the results of our proposed method for a set of UGSs, and defer to Section 7 some discussion and the conclusions.

Throughout the paper we assume a standard Wilkinson Microwave Anisotropy Probe (WMAP) cosmology with $H_0=70$ km s$^{-1}$, $\Omega_{\Lambda}=0.7$, and $\Omega_{M}=0.3$ \citep{spergel2007}.

\section{The UGS Selection}
\label{sez_ufo_sample}

For as many as 576 sources in the Second \textit{Fermi}-LAT catalogue, no plausible associations or identifications have yet been found. 
This makes an important component of the high-energy sky, and may hide new classes of AGNs, like the \textit{extreme blazars} \citep[]{extreme, extreme2}, the Dark-Matter (DM) candidates \citep{zechlin2012, zechlin2012bis, belikov2012}, or even unexpected high-energy phenomena.

To set up a procedure assisting the recognition of AGN populations among \textit{Fermi} UGSs, we selected the UGS sample from the 2FGL catalogue following this basic selection criteria:

\begin{enumerate}
\item No association in the 2FGL and no association in other $\gamma$-ray catalogues (those from the EGRET and AGILE missions in particular), or catalogues at other wavelengths considered by the \textit{Fermi} collaboration;

\item Sky positions outside the Galactic plane, with a Galactic latitude $|b|>$20$^{\circ}$. 
Many UGSs are in the Galactic plane, but we exclude this region because it is very crowded and confused, and the \textit{Fermi} procedure hardly converges towards correct associations.
Furthermore, this gives us a higher probability to select extragalactic sources.

\end{enumerate}

An additional possible criterion is the variability index on the 2-years baseline of the \textit{Fermi} 2FGL observations. 
This might be used to select DM candidates among UGSs, because they are expected to be stable in time \citep{bertone_review2005}.
In any case, we do not consider flux variability in our primary UGS selected sample, as most of them do not show significant flux variation.

183 UGSs from the 2FGL catalogue survive the selection criteria.
While we referred to the 2FGL for our UGS selection at the time when the present project started, for all subsequent analyses we used data from the 3FGL catalogue, yielding a decisive improvement in the \textit{Fermi}/LAT source position uncertainty and photometry.

\section{Search for UGS Multi-wavelength counterparts}
\label{associations}

In spite of the improvement allowed by the 3FGL, the association and identification of the \textit{Fermi} sources is complicated, or even prevented, by the large \textit{Fermi} LAT error-boxes typically a few arc-minutes radius (for a fraction of UGSs this may even exceed $\sim$10 arcmins). 
Our approach for finding potential counterparts for all of the 183 UGSs was to identify all detected X-ray sources inside the \textit{Fermi} error-box and, if there are X-ray sources, then check for the existence of counterparts at lower energies (radio, IR, optical) to build up a broad-band SED.

Following previous works in the literature, \citep{stephen2010, takahashi2012, takeuchi2013, acero2013x, landi2015}, our UGS recognition procedure is primarily based on the available \textit{Swift}/XRT X-ray imaging data over the \textit{Fermi} source error-box position. 
Without a reliable X-ray counterpart, the method cannot be applied. 

Not all $\gamma$-ray sources have detectable X-ray counterparts inside their error-box. This lack may be due to the intrinsic faintness of the source, to X-ray flux variability, to shallow depth of the X-ray exposure, or the lack of X-ray observations of the field.
Hint about the fraction of X-ray emitting \textit{blazars}, and among them of $\gamma$-ray emitting objects, can be derived from the BZCAT catalogue \citep{BZCAT}.
It represents an exhaustive, although by no means complete, list of sources classified as \textit{blazars}, useful to look for general trends. The sample of 3561 \textit{blazars} of the 5th BZCAT contains 63\% of objects detected in the soft X-ray band and 28\% of \textit{Fermi}/LAT sources. Among the latter, 79\% are X-ray emitters. 
In conclusion, a large fraction of $\gamma$-ray \textit{blazars} contains an X-ray counterpart within the \textit{Fermi}/LAT error-box. 
We then expect that a substantial fraction of \textit{Fermi} UGSs might be within reach of our analysis, in consideration of the sensitivity and extensive coverage by the \textit{Swift}/XRT telescope of the UGS sample.
The counterparts found for the UGSs are proposed associations, to be subsequently verified once new $\gamma$-ray catalogues will be matched to other samples for the next releases.

Thanks to the usually X-ray positional uncertainties of the order of few arc-seconds, we typically have one to a few sources with multi-wavelength photometric data inside the \textit{Swift} source error-box.

Subsequent to the X-ray detection, the radio band is important for our recognition work, since all discovered \textit{blazars} have been identified as radio-loud sources so far. 

Although primarily dedicated to the identification of Gamma-Ray Bursts, the \textit{Swift}/XRT telescope \citep{swift_mission}, thanks to its rapid responsivity and high sensitivity, has been systematically used to obtain X-ray follow-up observations for most of UGSs (e.g. \citet{falcone2011, falcone2014}).
So far, among 183 UGSs selected, $\sim$130 have dedicated \textit{Swift} observations.
In our XRT analysis we only used the PC mode\footnote{Photon-counting mode: PC mode is the more traditional frame transfer operation of an X-ray CCD. It retains full imaging and spectroscopic resolution, but the time resolution is only 2.5 seconds. The instrument is operated in this mode only at very low fluxes (useful below 1 mCrab).} data. 
We analysed them through the UK Swift Science Data Centre XRT tool \footnote{http://www.swift.ac.uk/user\_objects/} 
that provides X-ray images, source positions \citep{evans2009_stdPosition}, spectra \citep{evans2009_spectra} and light curves \citep{evans2009_lc} of any object in the Swift XRT field of view. 

For our purposes, we used the total XRT 0.3-10 keV energy band to generate the X-ray image of the UGS sky field. 
The X-ray sky maps of our UGS sample are reported in the figures of the Appendix A and we checked which X-ray sources (\textit{green circles}) fall inside the 3FGL 95\% confidence error-box (\textit{yellow ellipse}). 
For comparison, we also indicated with magenta ellipses the positional uncertainties of the 2FGL, as white crosses the X-ray sources of the 1SXPS Swift XRT Point Source Catalogue \citep{evans2014_catalog}, and the radio sources of the NRAO VLA Sky Survey (NVSS) and Sydney University Molonglo Sky Survey (SUMSS) catalogues as cyan circles (with radius equal to the semi-major axis of the positional error) or ellipses.

For each of the X-ray sources within the \textit{Fermi} error-box, we provided the position, with the corresponding error radius,   and the X-ray spectrum. 
Two types of position determinations are available for the XRT sources: the \textit{un-enhanced} position, estimated using only a PSF fit, and the \textit{enhanced position} \citep{goad2007_EPosition} where the absolute astrometry is corrected using field stars in the UVOT telescope and the systematic uncertainty is then decreased to 1.4" (90$\%$ confidence), if compared to the 3.5" systematic for the \textit{un-enhanced} positions. 
The X-ray energy spectrum is estimated in the 0.3-10 keV band. 
The output spectra are downloaded and then fitted using the XSPEC software (version 12.8.1g) \citep{xspec1996} of the HEASOFT Ftool package. 
According to the number of total counts, the spectral data are analysed in different ways. 
If the source has less than 25 counts, the total flux is calculated by the Mission Count Rate Simulator WebPIMMS\footnote{http://heasarc.gsfc.nasa.gov/Tools/w3pimms.html}, using a power law model with photon spectral index 2. If the found X-ray counterpart is reported in the 1SXPS catalogue \citep{evans2014_catalog}, we consider the corresponding photometric data points provided by the catalogue and available in the ASI-ASDC database.
With more than 25 total counts, we used an un-binned analysis by applying the Cash statistics. 
For bright sources with at least 150 counts, we binned (with the Ftool \textit{grppha}) the spectra with a minimum of 20 counts per spectral bin.     


Once the list of the X-ray counterparts inside the \textit{Fermi} error-box is defined, our next step is to search for counterparts in radio, infrared and optical bands, around their XRT \textit{enhanced} position (or the \textit{un-enhanced} one if the former is not available), using a search radius corresponding to the 90$\%$ confidence error radius. (\textit{green circle}, as exemplified in Fig. \ref{fig:0102_ass}-\textit{upper panels} of Appendix A). 
The results are displayed in the close-up images (i.e. Fig. \ref{fig:0102_ass}-\textit{upper right panel}) where, on the XRT sky map, we superimpose entries from the radio NVSS catalogue \citep{nvss_catalog} and SUMSS catalogues \citep{sumss_catalog}, from the WISE \citep{allwise} (\textit{blue crosses}) and the 2MASS catalogues \citep{2mass} (\textit{green diamonds}) in the near and mid-infrared bands, and finally from the USNO-B1.0 catalogue \citep{usnob1} or the Sloan Digital Sky Survey (SDSS) catalogue \citep{sdss9, sdss10} (\textit{magenta crosses}) in the optical. 
The error on the optical and IR positions is neglected since it is several times smaller than the uncertainty on the X-ray position.

On the other hand a spurious optical source may fall within the X-ray error box. 
Taking into consideration the number of objects and sky coverage in SDSS and USNO optical catalogues, the source sky density is $\sim$ 38000 deg$^{-2}$; 2MASS and WISE have a lower source density. 
Then the expected number of accidental optical sources in an error box of typical radii from 2" to 5" (the minimum and maximum in our sample) spans from 0.03 to 0.2. 
Therefore in our UGS sample of 14 objects, 5 of them with error box radius of 4", we do expect up to $\sim 1$ spurious optical source in the X-ray error boxes. 
We note that in our sample there is no UGS with more than 1 optical source in the X-ray error box.

As a further check on the goodness of the XRT position estimates, we superimposed on the XRT image the positions of the X-ray sources from the 1SXPS Swift XRT Point Source Catalogue (1SXPS) reported by \citet{evans2014_catalog}. 
As can be seen in the sky maps of the Appendix A, each XRT position found with our procedure is compatible with the 1SXPS positions.  

The multi-wavelength counterpart data set of a given UGS is then used to create the broad-band SEDs (Fig \ref{fig:0102_ass}-\textit{Bottom panel}). 
We combined these data through the SED Builder tool of the ASI ASDC database \footnote{http://tools.asdc.asi.it/SED/}. The \textit{Fermi} flux points are taken from the 3FGL catalogue.
All the X-ray plotted data are corrected for Galactic absorption as available from the XSPEC package. 
If available, we include in the analysis also the X-ray data points reported by \citet{takeuchi2013} (\textit{black points}) and in the 1SXPS catalogue that we can consider a cross-check of our analysis.
 
In Appendix A we report details about our UGS counterpart search procedure for a sample of 14 \textit{Fermi} UGSs among the 183 objects of our primary list.

\section{Definition of a \textit{blazar} SED Template Set}
\label{chap_SEDtemplate}

Since we are interested to recognise \textit{blazars} candidates among the \textit{Fermi} UGS population, we built a tool for the systematic comparison of the broad-band SED of UGS counterparts with spectral templates representing various categories of the \textit{blazar} populations.
One possibility would be to use the so-called \textit{blazar sequence} reported by \citet{fossati1998} and \citet{donato2001}, and updated in \citet{ghisellini2017} that is defined in terms of functional dependencies of the spectral parameters for both the synchrotron and IC components.
However, as explained in Sect. \ref{buil_SED}, we preferred to adopt a different, more empirical, approach.

\subsection{A Sample of Known \textit{Blazars}}
\label{chap_SED}

We defined a reference sample of known \textit{blazars}, for which we collected all the available photometric data, grouping them into four categories defined in the \textit{Fermi} 3LAC catalogue \citep[The third catalogue of AGN detected by the \textit{Fermi} LAT][]{3lac} and characterised by different spectral properties and luminosities: 
the Low-Synchrotron-Peaked sources (LSP, with synchrotron peak frequency $<10^{14}$ Hz), the Intermediate-Synchrotron-Peaked (ISP, with a synchrotron peak frequency between $10^{14}$ and $10^{15}$ Hz), the High-Synchrotron-peaked (HSP, peak frequency $>10^{15}$ Hz), and the extreme High-peaked BL Lacs (EHBL). 
The latter class \citep{Costamante2011} is a new emerging population of BL Lac objects with extreme properties (a large ratio between the X-ray and the radio flux and the hardness of the X-ray continuum locating the synchrotron peak in the medium-hard X-ray band).  

In order to build a \textit{blazar} SED template library, we started our selection from all \textit{blazars} (FSRQ and BL Lac objects) present in the 3LAC catalogue at high Galactic latitude ($|b|>20^{\circ}$), with a certain SED classification and a Likelihood Ratio Reliability between Radio/Gamma bands and X-ray/Gamma greater than 0. 
We cross-matched this preliminary selected sample with the BZCAT and we rejected all objects without X-ray flux and with an uncertain or unknown redshift in the BZCAT.
To ensure a good spectral coverage and a precise SED characterisation, we performed a cross-match with the WISE, 2MASS and Swift (1SWXRT) catalogues. 
We use only the identified LSP of the 3LAC.
Finally, we performed an extensive search in literature to assess the robustness of the published redshifts, also examining the published optical spectra, and we selected only the sources with secure redshift.

We also include the source PG 1553+113, having a very extensive multi-band photometry but a still uncertain distance, for which we adopt a redshift of 0.5 following \citet{danforth2010} and \citet{prandini2010}. 
PG 1553+113 is thought to be among the most distant HSP objects known and was considered as an extragalactic standard candle in the VHE band. Moreover PG 1553+113 shows very moderate variability at all frequencies, which makes it a good candidate to build a robust average SED.

The final list of our adopted \textit{blazar} templates is composed of 50 sources, including 20 LSPs, 12 ISPs, 16 HSPs and 2 EHBLs, and is reported in Table \ref{tab:sample_blazar}, where we indicate the source name, the 3LAC \textit{blazar} SED class, and the redshift from literature.
We have considered this list of objects as sufficiently representative of the various \textit{blazar} categories. 
Further enlarging this template database would be possible and can be done in the future.

\begin{table}
\centering
\begin{tabular}{|c|c|c|}
\hline
\footnotesize{Source Name} & \footnotesize{3LAC SED Class} & \footnotesize{Redshift }\\
\hline
\footnotesize{3C 279} & \footnotesize{LSP} & \footnotesize{0.536}\\
\footnotesize{3C 345} & \footnotesize{LSP} & \footnotesize{0.593}\\
\footnotesize{3C 454} & \footnotesize{LSP} & \footnotesize{0.859}\\
\footnotesize{4C +21.35} & \footnotesize{LSP} & \footnotesize{0.435}\\
\footnotesize{4C +28.07} & \footnotesize{LSP} & \footnotesize{1.213}\\
\footnotesize{4C +38.41} & \footnotesize{LSP} & \footnotesize{1.814}\\
\footnotesize{4C +49.22} & \footnotesize{LSP} & \footnotesize{0.334}\\
\footnotesize{AO 0235+164} & \footnotesize{LSP} & \footnotesize{0.94}\\
\footnotesize{CTA 102} & \footnotesize{LSP} & \footnotesize{1.037}\\
\footnotesize{OJ 287} & \footnotesize{LSP} & \footnotesize{0.306}\\
\footnotesize{PKS 0402-362} & \footnotesize{LSP} & \footnotesize{1.417}\\
\footnotesize{PKS 0420-01} & \footnotesize{LSP} & \footnotesize{0.916}\\
\footnotesize{PKS 0454-234} & \footnotesize{LSP} & \footnotesize{1.003}\\
\footnotesize{PKS 1502+106} & \footnotesize{LSP} & \footnotesize{1.839}\\
\footnotesize{PKS1510-08} & \footnotesize{LSP} & \footnotesize{0.36}\\
\footnotesize{PKS 2142-75} & \footnotesize{LSP} & \footnotesize{1.138}\\
\footnotesize{PKS 2144+092} & \footnotesize{LSP} & \footnotesize{1.113}\\
\footnotesize{PKS 2227-08} & \footnotesize{LSP} & \footnotesize{1.56}\\
\footnotesize{PMN J0017-0512} & \footnotesize{LSP} & \footnotesize{0.227}\\
\footnotesize{S4 1030+61} & \footnotesize{LSP} & \footnotesize{1.401}\\
\hline
\footnotesize{3C 371} & \footnotesize{ISP} & \footnotesize{0.046}\\
\footnotesize{4C +55.17} & \footnotesize{ISP} & \footnotesize{0.899}\\
\footnotesize{87GB 165604.4+601702} & \footnotesize{ISP} & \footnotesize{0.623}\\
\footnotesize{AP Librae} & \footnotesize{ISP} & \footnotesize{0.048}\\
\footnotesize{GB6 J0945+5757} & \footnotesize{ISP} & \footnotesize{0.229}\\
\footnotesize{NRAO 350} & \footnotesize{ISP} & \footnotesize{0.518}\\
\footnotesize{OS 562} & \footnotesize{ISP} & \footnotesize{0.751}\\
\footnotesize{PKS 0403-13} & \footnotesize{ISP} & \footnotesize{0.571}\\
\footnotesize{PKS 1004-217} & \footnotesize{ISP} & \footnotesize{0.33}\\
\footnotesize{PMN J0422-0643} & \footnotesize{ISP} & \footnotesize{0.242}\\
\footnotesize{SBS 1200+608} & \footnotesize{ISP} & \footnotesize{0.0653}\\
\footnotesize{W Comae} & \footnotesize{ISP} & \footnotesize{0.103}\\
\hline
\footnotesize{1ES 0414+009} & \footnotesize{HSP} & \footnotesize{0.287}\\
\footnotesize{1ES 0806+524} & \footnotesize{HSP} & \footnotesize{0.138}\\
\footnotesize{1H 0323+022} & \footnotesize{HSP} & \footnotesize{0.147}\\
\footnotesize{1RXS J023832.6-311658} & \footnotesize{HSP} & \footnotesize{0.232}\\
\footnotesize{Mkn 180} & \footnotesize{HSP} & \footnotesize{0.045}\\
\footnotesize{Mkn 501} & \footnotesize{HSP} & \footnotesize{0.034}\\
\footnotesize{Mkn 421} & \footnotesize{HSP} & \footnotesize{0.031}\\
\footnotesize{PG 1437+398} & \footnotesize{HSP} & \footnotesize{0.349}\\
\footnotesize{PKS 2005-489} & \footnotesize{HSP} & \footnotesize{0.071}\\
\footnotesize{RBS 0334} & \footnotesize{HSP} & \footnotesize{0.411}\\
\footnotesize{RBS 0958} & \footnotesize{HSP} & \footnotesize{0.138}\\
\footnotesize{RX J0847.1+1133} & \footnotesize{HSP} & \footnotesize{0.199}\\
\footnotesize{RX J1136.5+6737 } & \footnotesize{HSP} & \footnotesize{0.136}\\
\footnotesize{TXS 1055+567} & \footnotesize{HSP} & \footnotesize{0.143}\\
\footnotesize{TXS 2106+030} & \footnotesize{HSP} & \footnotesize{0.149}\\
\footnotesize{PG 1553+113} & \footnotesize{HSP} & \footnotesize{0.5*}\\
\hline
\footnotesize{1ES 0229+200} & \footnotesize{EHBL} & \footnotesize{0.140}\\
\footnotesize{1ES 0347-121} & \footnotesize{EHBL} & \footnotesize{0.185}\\
\hline
\end{tabular}
\caption{The \textit{blazars} used for our SED template library. First column is the name of the \textit{blazar}, the second column indicates the 3LAC \textit{blazar} SED classification: LSP denotes the Low-Synchrotron-Peaked sources , ISP and HSP the Intermediate and High-Synchrotron-Peaked respectively. The E-HBL is the emerging class of the extreme High peaked BL Lac objects \citep{Costamante2011}. In the third column is reported the redshift taken from literature. (*The redshift of PG 1553+113 is not spectroscopic.)}
\label{tab:sample_blazar}
\end{table}

The ASI Science Data Centre Database and SED Builder tool have been used to collect the whole set of archived historical observations for every \textit{blazar} of our sample. 
For each source we created a data vector containing the monochromatic luminosities versus emission rest-frame frequencies, computed from the redshift of the object and standard expressions for the luminosity distance.
K-corrections for all the photometric data have been computed assuming a flat frequency-independent spectrum in $\nu F(\nu)$.
The data points range from the radio to the high energy frequencies and the complete list of the data catalogues used is reported in the ASI Science Data Centre SED Builder Tool\footnote{http://tools.asdc.asi.it/SED/docs/SED$\_$catalogs$\_$reference.html}.

\begin{figure}
  \centering
  \begin{minipage}[c]{.45\textwidth}
    \includegraphics[width=1.\textwidth]{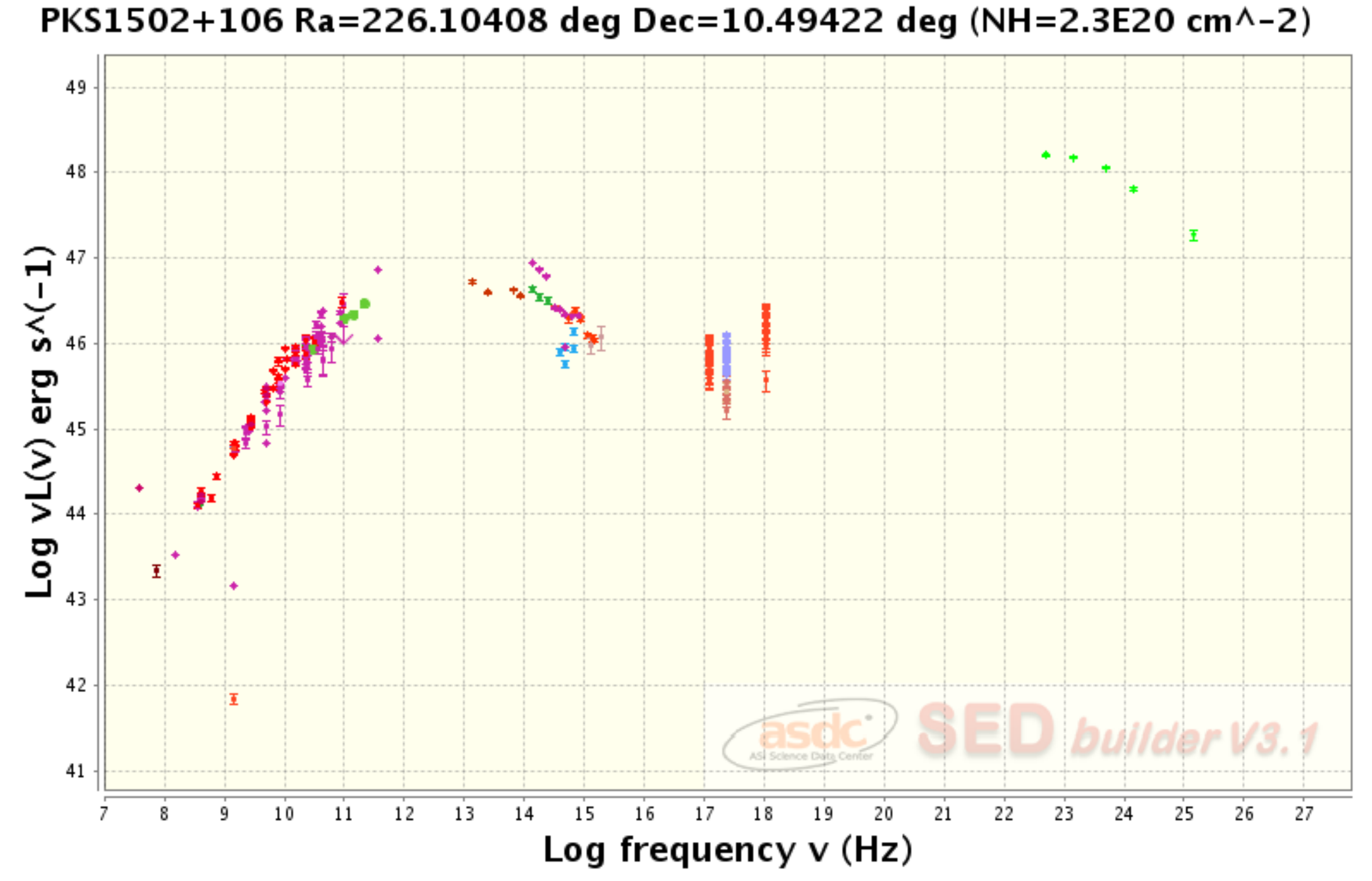}
  \end{minipage}%
  \hspace{3mm}%
  \begin{minipage}[c]{.45\textwidth}
    \includegraphics[width=1.\textwidth]{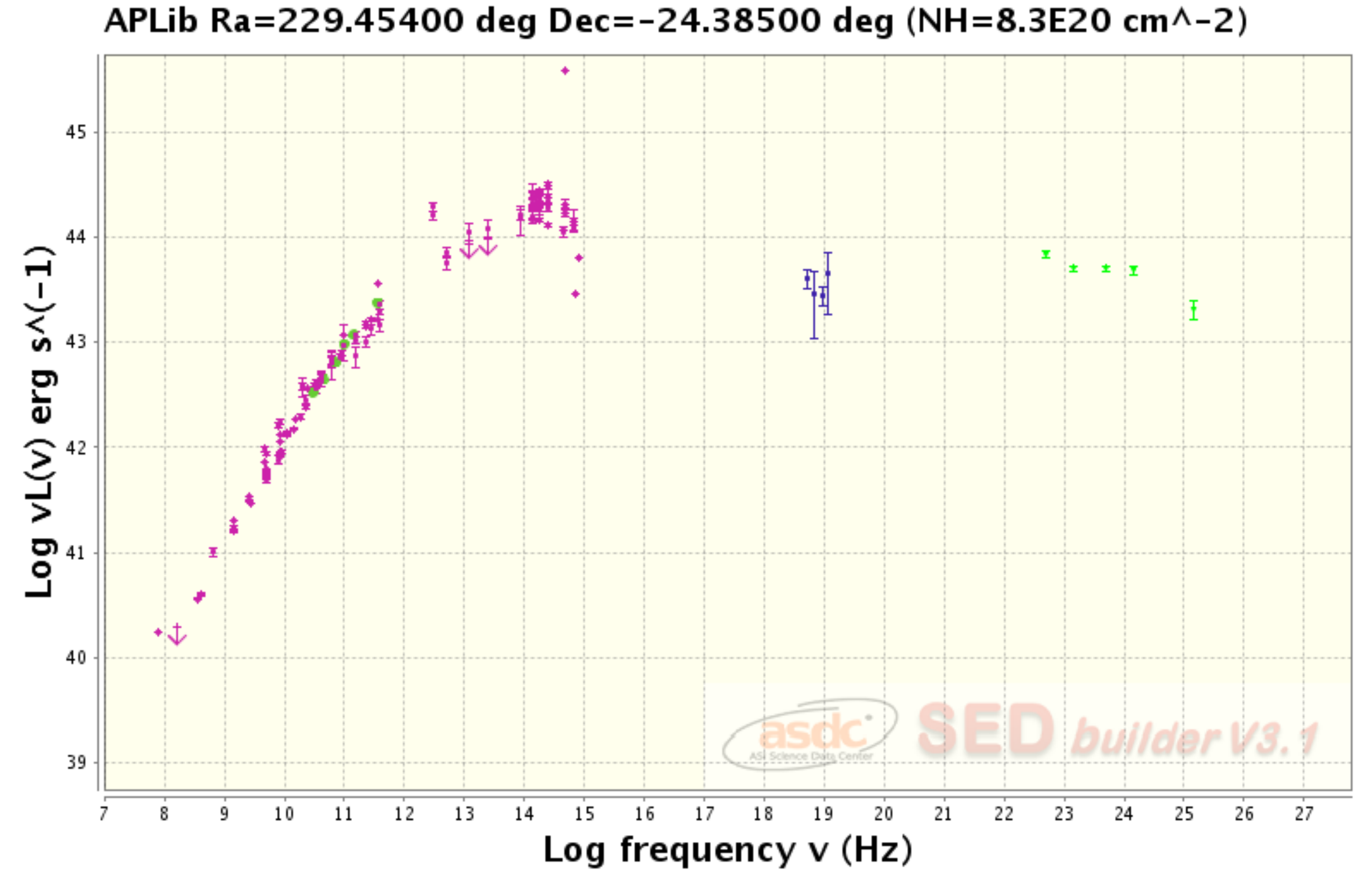}
  \end{minipage}
  \begin{minipage}[c]{.45\textwidth}
    \includegraphics[width=1.\textwidth]{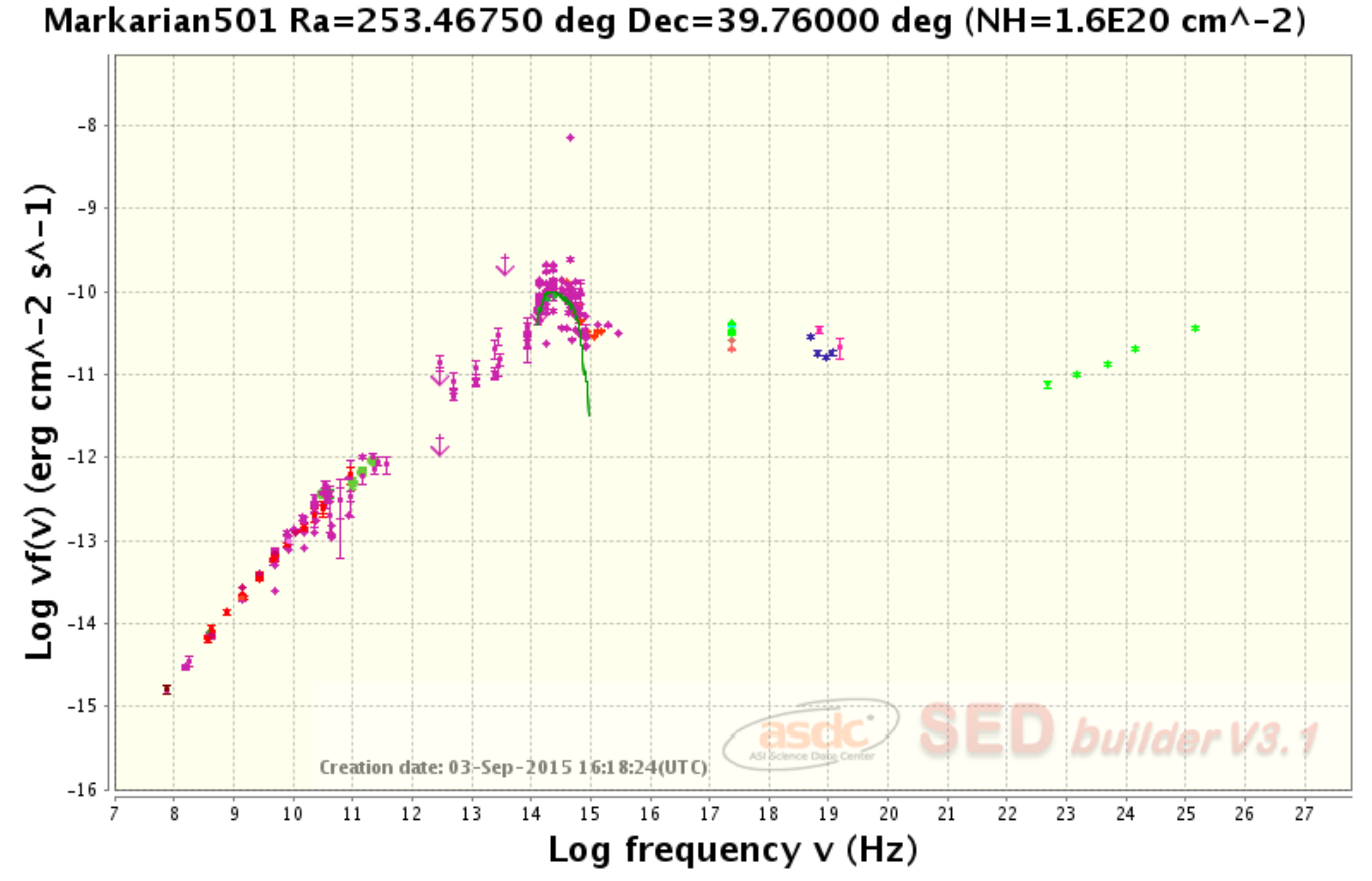}
  \end{minipage}%
  \hspace{3mm}%
  \begin{minipage}[c]{.45\textwidth}
    \includegraphics[width=1.\textwidth]{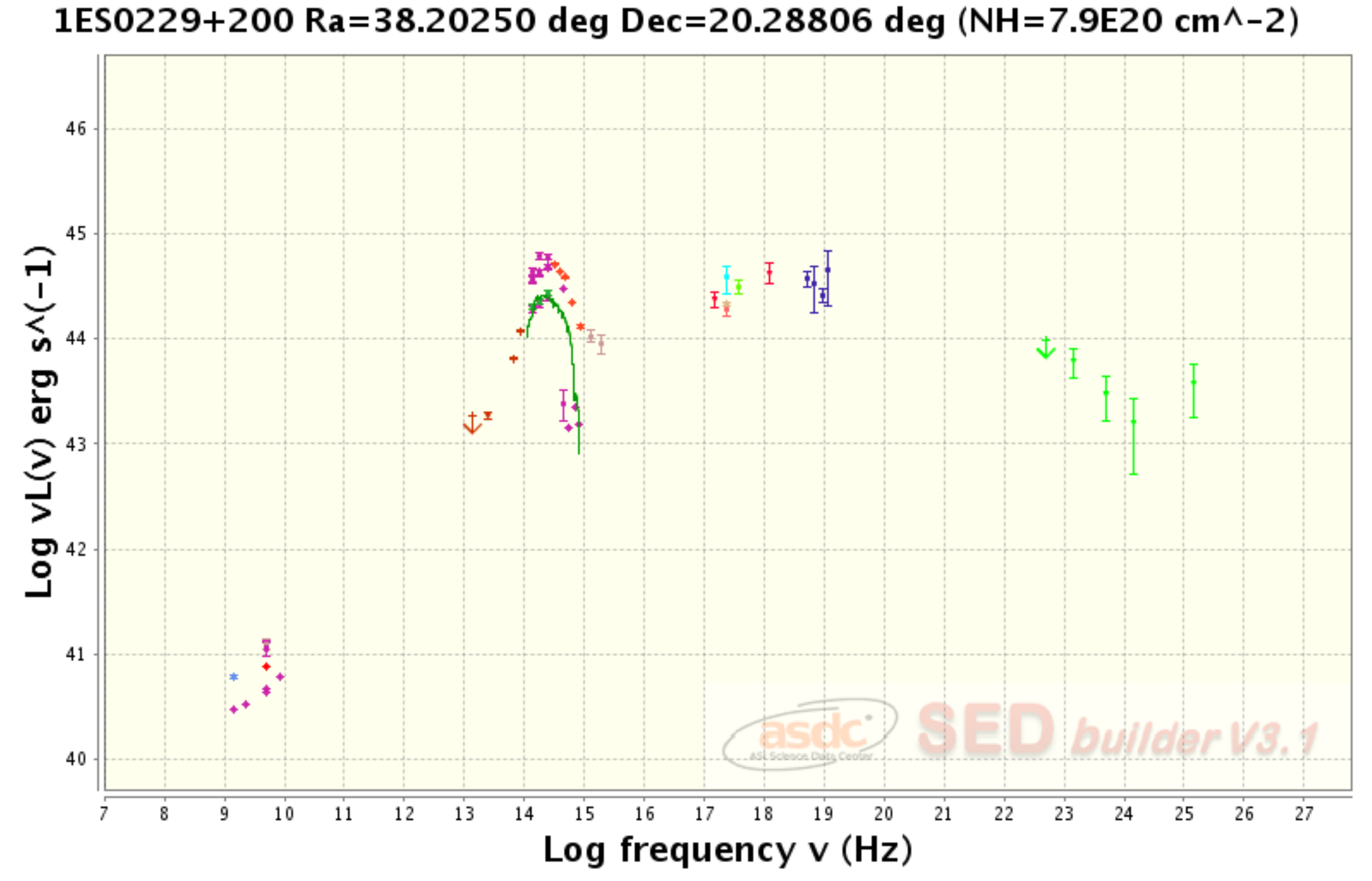}
  \end{minipage}
  \caption{Examples of the SEDs of four \textit{blazars} in our SED template library:  PKS 1502+106 is a LSP \textit{blazar}, ApLib is an ISP object, Markarian 501 is an HSP and 1ES 0229+200 is an Extreme High peaked BL Lac object (EHBL). The SEDs are expressed as $\nu$F$_{\nu}$ versus the frequency $\nu$).}
	\label{fig:sed_asdc}
\end{figure}

Examples of SED data collected for four \textit{blazars} are shown in Figure \ref{fig:sed_asdc}.

\begin{figure}
  \centering
  \begin{minipage}[c]{.45\textwidth}
    \includegraphics[height=.7\textwidth,width=.99\textwidth]{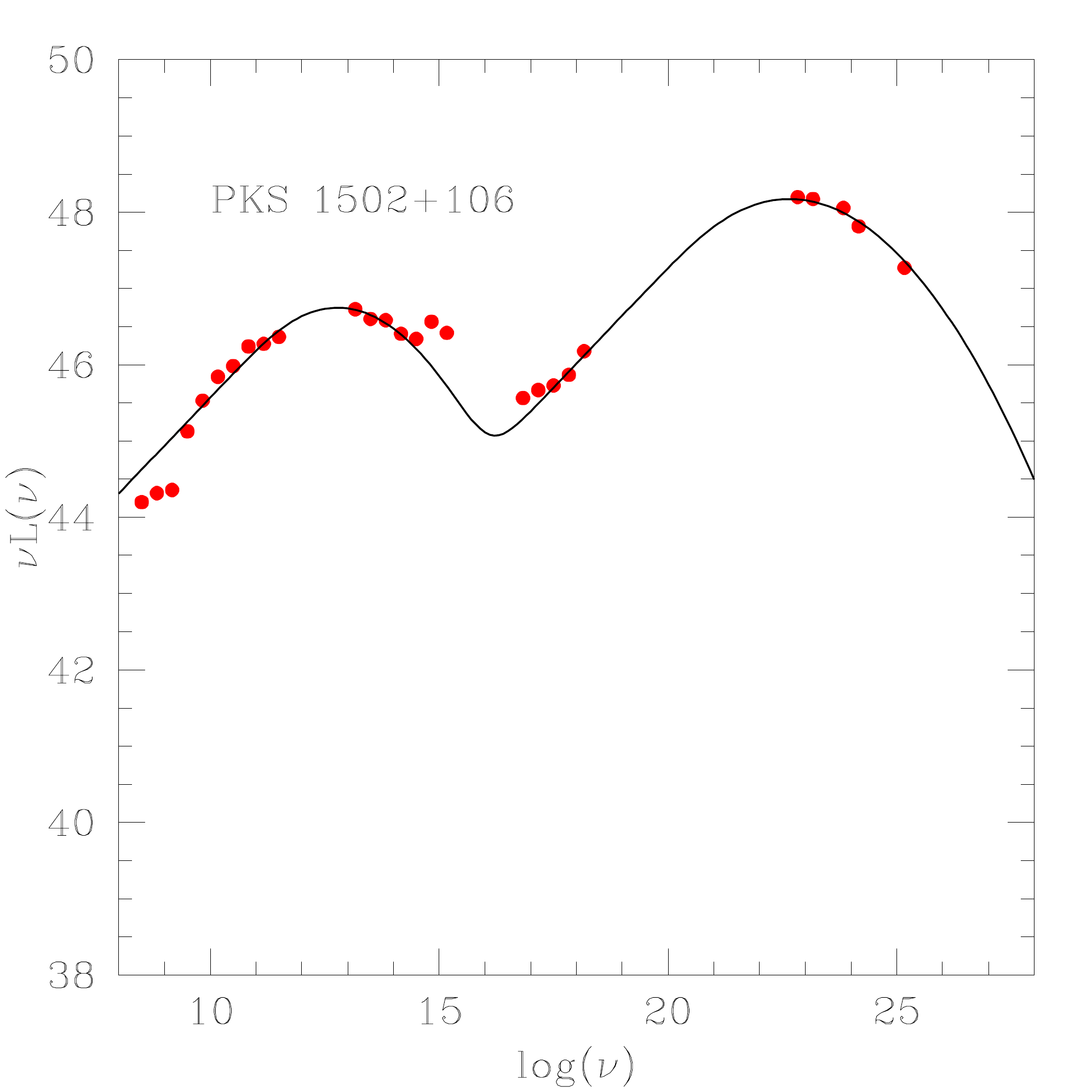}
  \end{minipage}%
  \hspace{3mm}%
  \begin{minipage}[c]{.45\textwidth}
    \includegraphics[height=.7\textwidth,width=.99\textwidth]{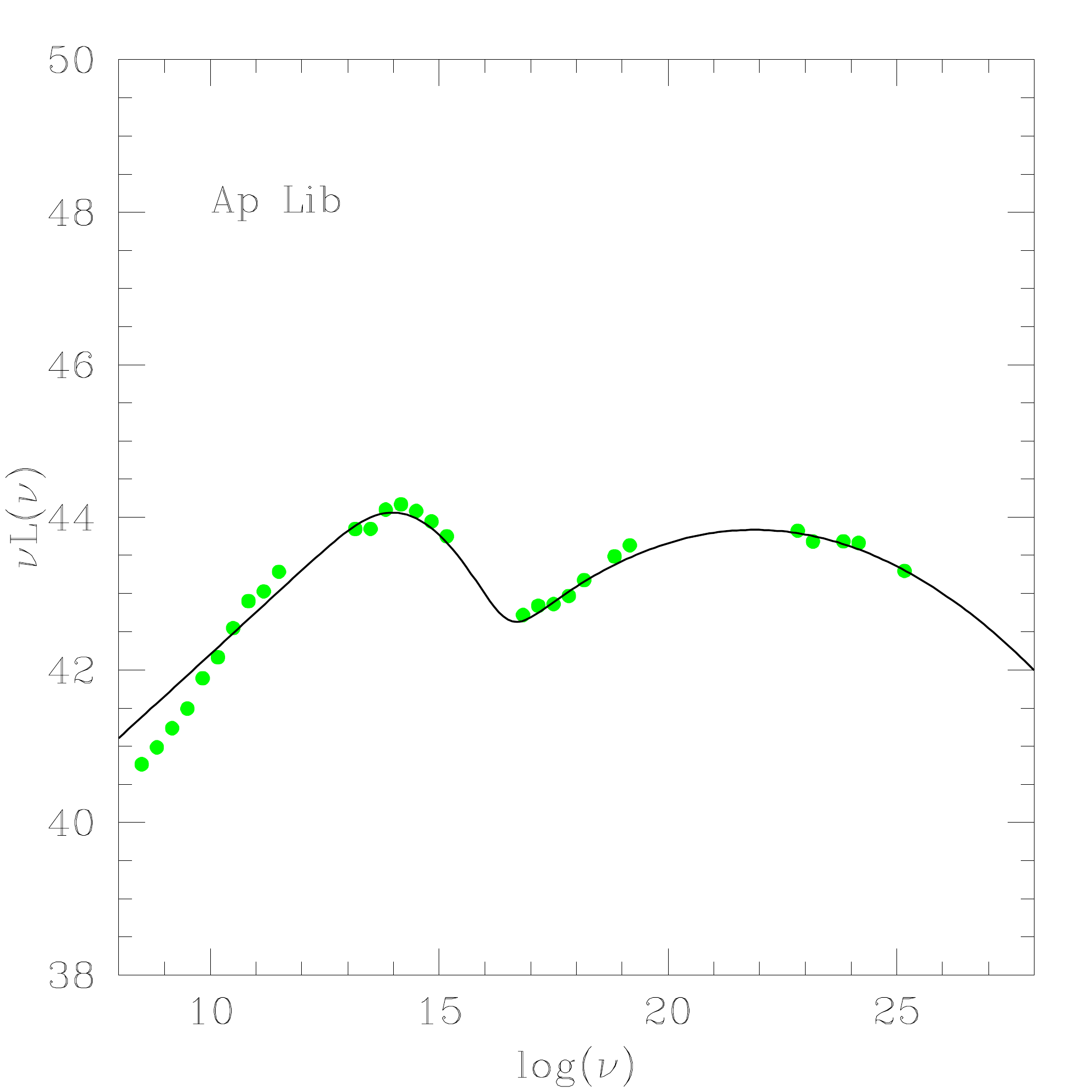}
  \end{minipage}

  \begin{minipage}[c]{.45\textwidth}
     \includegraphics[height=.7\textwidth,width=.99\textwidth]{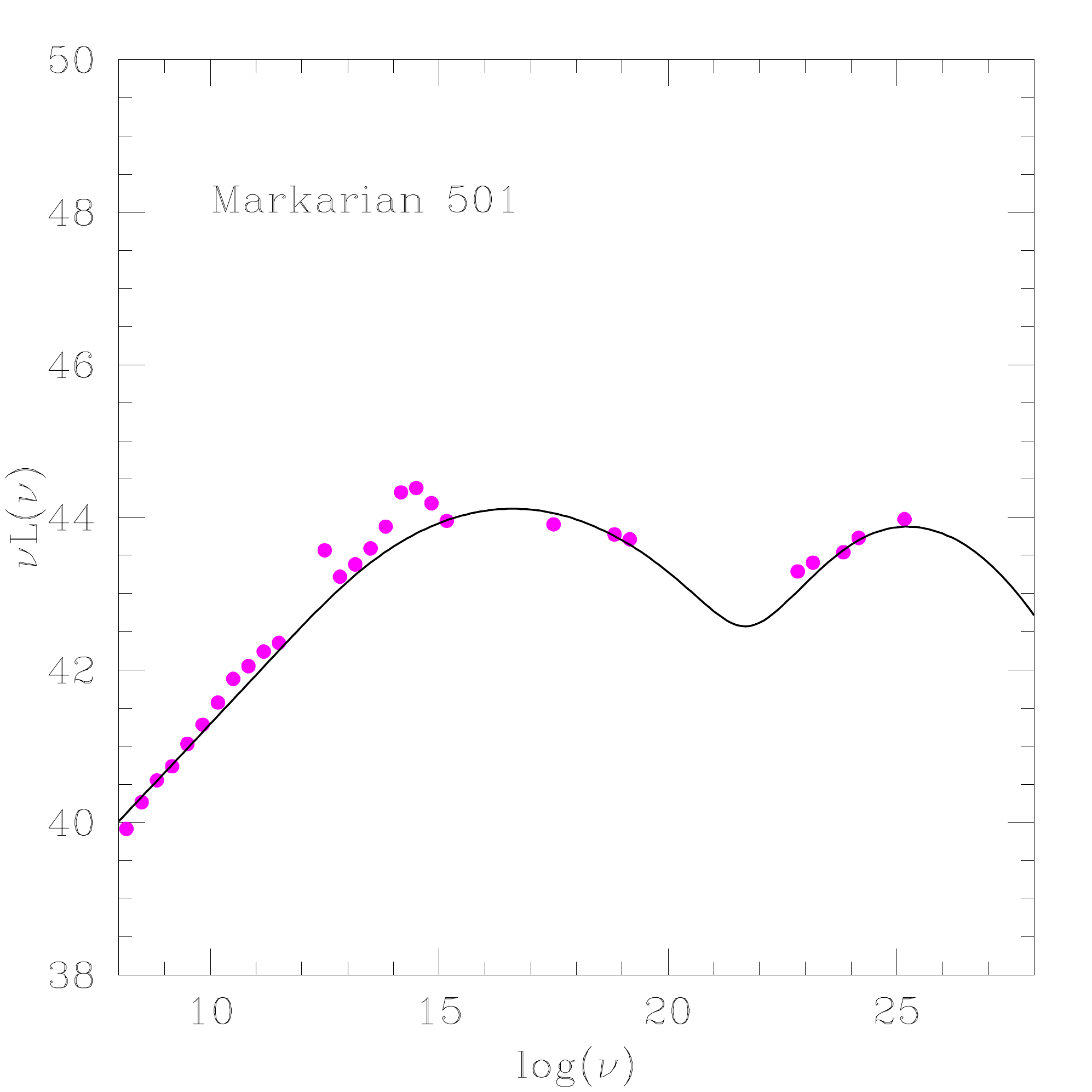}
  \end{minipage}%
  \hspace{3mm}%
  \begin{minipage}[c]{.45\textwidth}
    \includegraphics[height=.7\textwidth,width=.99\textwidth]{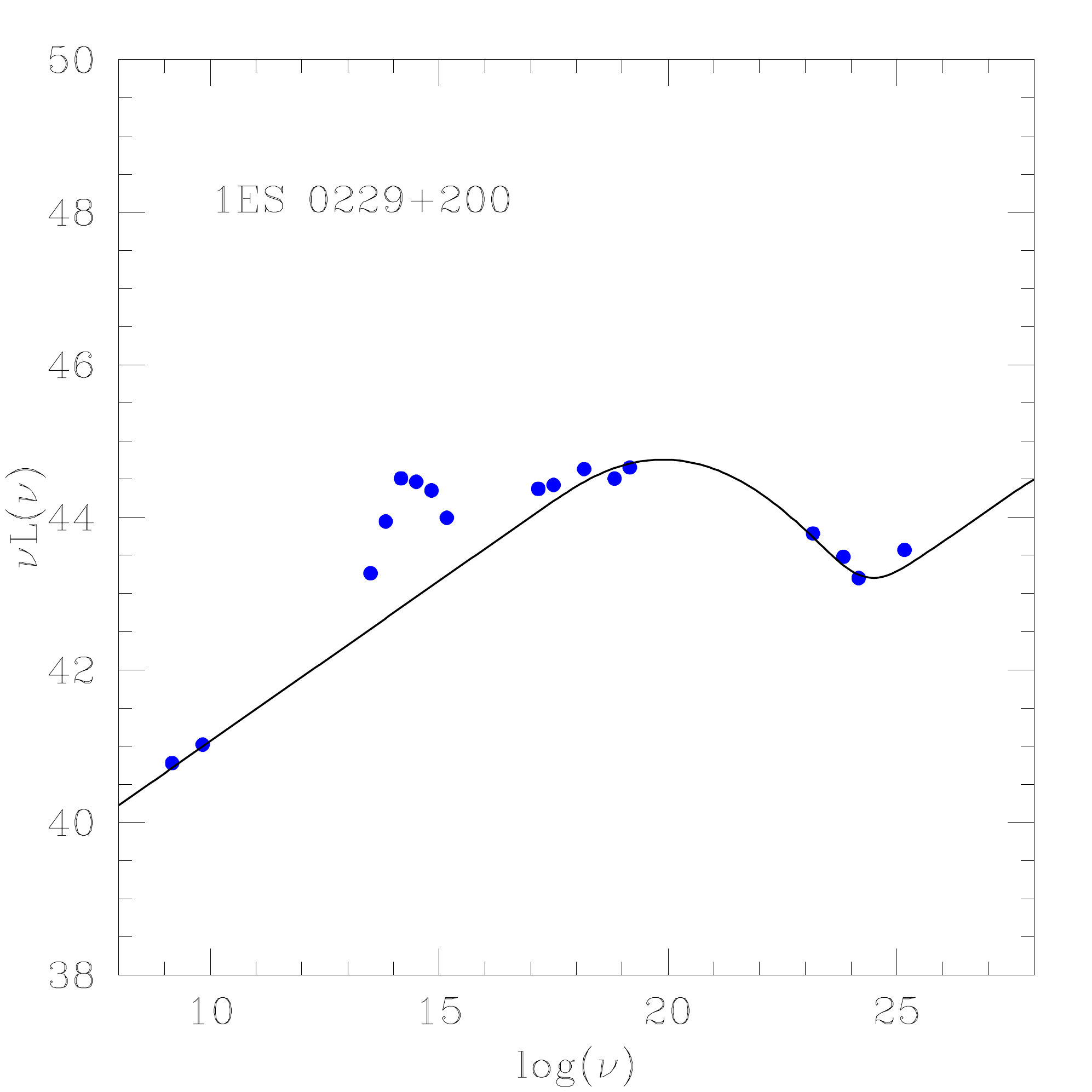}
  \end{minipage}
 \hspace{3mm}%
  \caption{Examples of best-fit for the multi-wavelegth SED data points of the four \textit{blazars} reported in Fig.\ref{fig:sed_asdc}. The fits are based on the analytic form in eq. \ref{eqn:SED_parametrica}.}
	\label{fig:sed_fit}
\end{figure}

\subsection{Building the SED Template Set}
\label{buil_SED}

Once the archive of multi-frequency photometric data for all sample \textit{blazar}s was collected, the next step was to fit these data with simple analytic representations for each one object. 
We first divided the data in equally spaced frequency bins and calculated the average of the logarithms of the luminosity measurements inside each bin.
This allowed us to minimise the effects of flux variability.
Altogether we obtained a library of 50 averaged \textit{blazar} SEDs for the objects in Table \ref{tab:sample_blazar}.

For each one of these sources we fitted the average photometric data using a simply parametrised analytic form with a double power-law with exponential convergence: one component representing the low-frequency synchrotron peak, the other the high-frequency IC peak. 
This function has the following expression:
\begin{equation}
\nu L(\nu)=\nu L_{1}(\nu)+\nu L_{2}(\nu) 
\label{eqn:SED_parametrica}
\end{equation}
with 
\[
\nu L_1(\nu) = A \cdot \left( \frac{\nu}{\nu_1}\right)^{1-\alpha} \cdot \exp \left\{ {-\frac{1}{2\sigma_1^2}     \left[log \left(1+\frac{\nu}{\nu_{1}}\right)\right]^2} \right\} 
\]
\[
\nu L_2(\nu) = B \cdot \left( \frac{\nu}{\nu_2}\right)^{1-\alpha} \cdot \exp \left\{ {-\frac{1}{2\sigma_2^2}     \left[log \left(1+\frac{\nu}{\nu_2}\right)\right]^2} \right\} 
\]
This has 7 free parameters: $A$ and $B$ are the normalisations of the two emission components, $\alpha$ determines the slopes of the two power-law functions, that are assumed to coincide (consistent both with the SSC assumption for the \textit{blazar} modeling, and with the data), $\nu_{1}$ and $\nu_{2}$ are the characteristic frequencies of the two emission bumps, and finally $\sigma_{1}$ and $\sigma_{2}$ determine the two bump widths.

Examples of resulting fitting curves for the average SEDs of the four representative \textit{blazar}s are reported in Fig. \ref{fig:sed_fit} where the SED data are the geometric averages of the luminosity measurements inside each frequency bin. 
Note that two of the sources, the low-redshift Mkn 501 and 1ES 0229+200, show evidence for a narrow peak at $\log \nu \simeq 14$ to be attributed to the host galaxy, that is not fitted by our analytic formula. 
Indeed the formula aims at reproducing only the non-thermal power-law \textit{blazar} emissions. 
When comparing the SED templates to UGS SED data, we check \textit{a-posteriori} if a galactic contribution might show-up (which essentially does not happen in most of our investigated cases, that tend to be sources at high redshifts where the \textit{blazar} emission dominates over the host galaxy).

We adopted this simple analytic representation for the average SEDs instead of using more physical models for \textit{blazar} emission (like the SSC model itself), to be model-independent and to achieve good adherence to the data. 
In particular, SSC models have some difficulties to reproduce data in the radio band.
Unlike the approach of \citet{fossati1998}, \citet{donato2001} and \citet{ghisellini2017}, we did not average out the SEDs of the different sources.

\begin{figure}
  \centering
  \includegraphics[width=90mm]{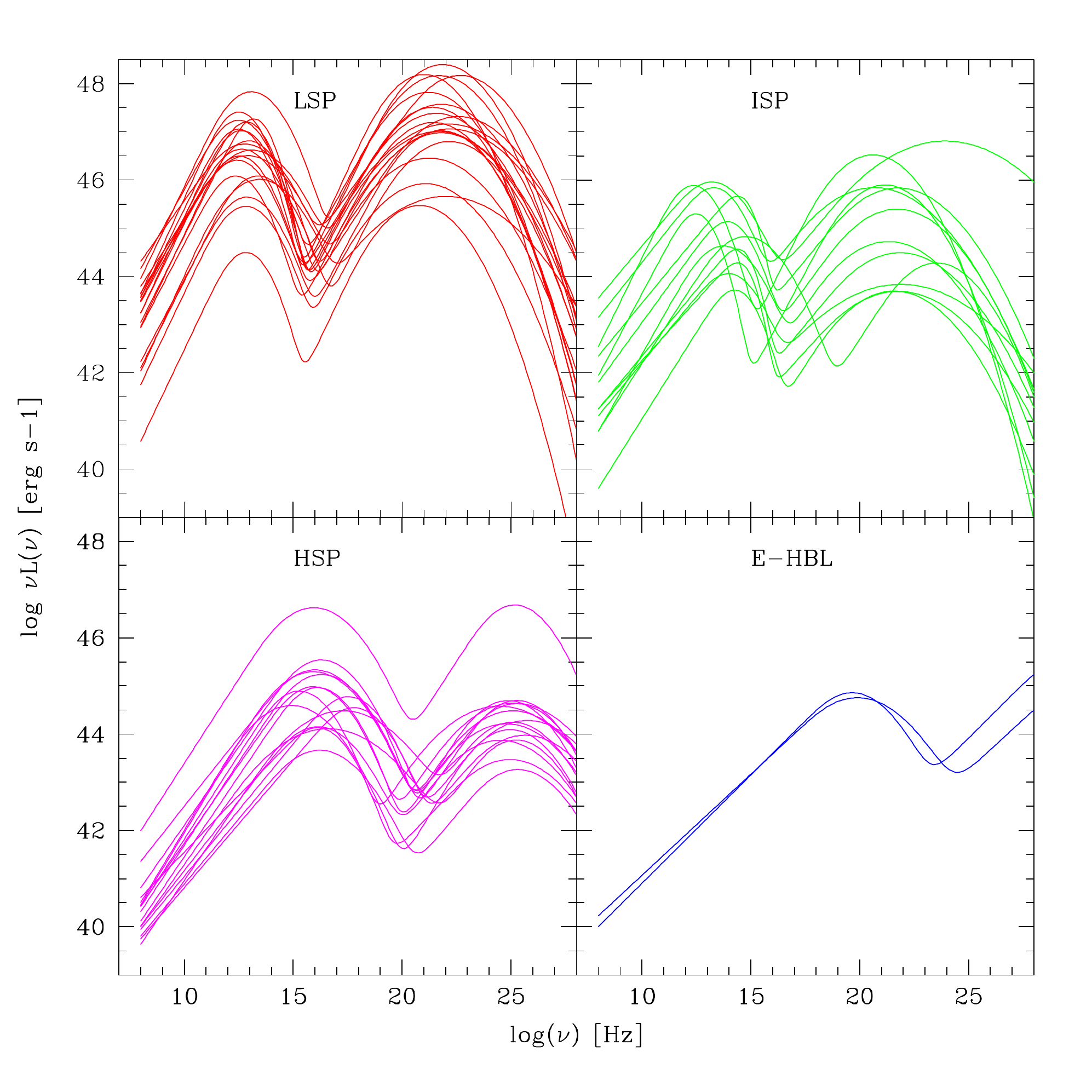}
\caption{Analytic fits with a double power-law and exponential convergence for the 50 average SEDs of the objects of our \textit{\textit{blazar}} library, divided in the four classes: LSP (\textit{red}), ISP (\textit{green}), HSP (\textit{magenta}), and EHBL (\textit{blue}).} 
\label{fig:sed_template_4plot}
\end{figure}

We collect in Fig. \ref{fig:sed_template_4plot} all the SED templates of our spectral library and grouped into the four \textit{blazar} classes: the LSP, ISP, HSP, and E-HBL. 
The plot reveals the general behavior found by \citet{donato2001} and \citet{ghisellini2017}: the LSPs (red curves) occupy the highest luminosity values of the sequence, with the peak emission frequencies falling at lower energies with respect to those of HSP (mainly BL-Lac objects).

\section{Associating UGS to \textit{blazar} classes }
\label{chap_ufo_id}
   
Once the broad-band spectral properties of the \textit{blazar} populations are defined, we proceed to compare them with the SEDs of all our \textit{Fermi} UGSs discussed in Sec. \ref{associations}. 

To this purpose, we developed an algorithm to assess the similarity of the UGS SED with those of \textit{blazar}s of a given class, and to obtain some information on the \textit{blazar} category and the redshift.

\subsection{The algorithm}
\label{blazar_tool}

Our \textit{blazar} recognition tool requires the following steps to be performed \footnote{The numerical code for the \textit{blazar} recognition is written in IDL and SuperMongo, for the ease of graphical comparison between the photometric data and the SED templates}.

\begin{enumerate}

\item
We start by considering the plots of luminosity versus frequency reported in Fig. \ref{fig:sed_template_4plot} for all four \textit{blazar} categories, including the SEDs of all sources in each category. 
The units in these plots are the logarithm of the $\nu L(\nu)$ luminosity in \textit{erg/sec} on the y-axis, and the logarithm of photon frequency $\nu$ in \textit{Hz} on the x-axis.

\item Using the observed multi-wavelength fluxes of a given UGS counterpart, we convert them into $\nu L(\nu)$ luminosities by assuming a suitable grid of redshifts $z$ spanning a range of values from 0.05 to 2.0. 
Here again, we calculate the K-corrections adopting flat spectra to be consistent with Sec. \ref{chap_SED} for the template set.   We then over-plot the luminosity data-points on the SED templates of all \textit{blazar} classes, as illustrated in Fig. \ref{fig:1215_metodo} and followings. 

\item 
For the same UGS counterpart, for every redshift of the grid and with respect to every $j$\textit{-th} SED of the \textit{blazar} template set, the $\chi^{2}_{min}$ statistic is calculated as the minimum of all values of
\begin{equation}
\chi^{2}_{j} = \sum_{i} \frac{ \left\{{\log \left[ \nu_i L_{i}(\nu_{i})\right]-\log \left[\nu_i l_{j}(\nu_{i})\right]}\right\}^{2}} {\left\{0.01 \log \left[\nu_i l_{j}(\nu_{i})\right]\right\}^2} ,  
\label{CHIeq} 
\end{equation}
%
The reduced $\chi^2_{\nu,j}$ is obtained dividing by the number of data-points.
The minimum of this quantity, $\chi^2_{\nu,min}$, offers a measure of how close is that $j$\textit{-th} SED template to the observational UGS SED for a given assumed redshift in the grid. 

\item 
The second quantity that we use to estimate the similarity of the UGS SED with the \textit{blazar} SED templates (and for a given assumed UGS's redshift in the grid) is the \textit{Minimum Average Distance} (MAD), defined as
\begin{equation}
MAD = \frac{1}{N_{SED}} \cdot |\sum_{j}(\chi_{j})| 
\label{MADeq}
\end{equation}
with
\[
  \qquad \chi_{j} = \sum_{i} \frac{\log \left[\nu_i L_{i}(\nu_{i})\right] -\log \left[\nu_i l_{j}(\nu_{i})\right]}{0.01 \log \left[\nu_i l_{j}(\nu_{i})\right]}
\]
where $i$ is running over all photometric data-points for that UGS, $j$ is the index flagging every SED of the \textit{blazar} template set and $\nu_i l_{j}(\nu_{i})$ is the luminosity of the $j$\textit{-th} SED template interpolated at the frequency $\nu_{i}$.  
The normalisation factor $N_{SED}$ is the number of templates of a given class of \textit{blazars}.
MAD is a measure of how far is the UGS SED, for a given assumed redshift of the grid, from the distribution of the SEDs of that \textit{blazar} template category.
Note that while $\chi^2_{\nu,min}$ measures the match of data to a single template SED, MAD refers to the whole distribution of the SEDs in the \textit{blazar} class and how far it is from the object data-points.

\item For a given UGS and its counterparts, the goodness of the recognition, the best-guess redshift and spectral class are found by first considering the $\chi^2_{\nu,min}$ statistics. 
In the cases in which we have more than one counterpart or when there are degeneracies in the $\chi^2$ solutions as for the redshift, the MAD statistics is used to get a qualitative measure of the relative likelihood of these various solutions.

\end{enumerate}

The MAD statistics measures the distance of the assumed UGS luminosity data-points from the whole distribution of the SED templates. 
Consequently it provides us a first hint about the \textit{blazar} class and luminosity, and thus the source redshift by comparison with the observed fluxes. 
Instead the $\chi_{\nu,min}^{2}$ value is more closely related to the spectral shapes of both of the UGS and the individual spectral templates and, in particular, to the slopes of the rising and descending parts of the two spectral components of the UGS SED. 
Hence, it evaluates the degree of similarity in shape of the SED of the UGSs and the \textit{blazar} templates.

\begin{figure}
  \centering
  \includegraphics[width=80mm]{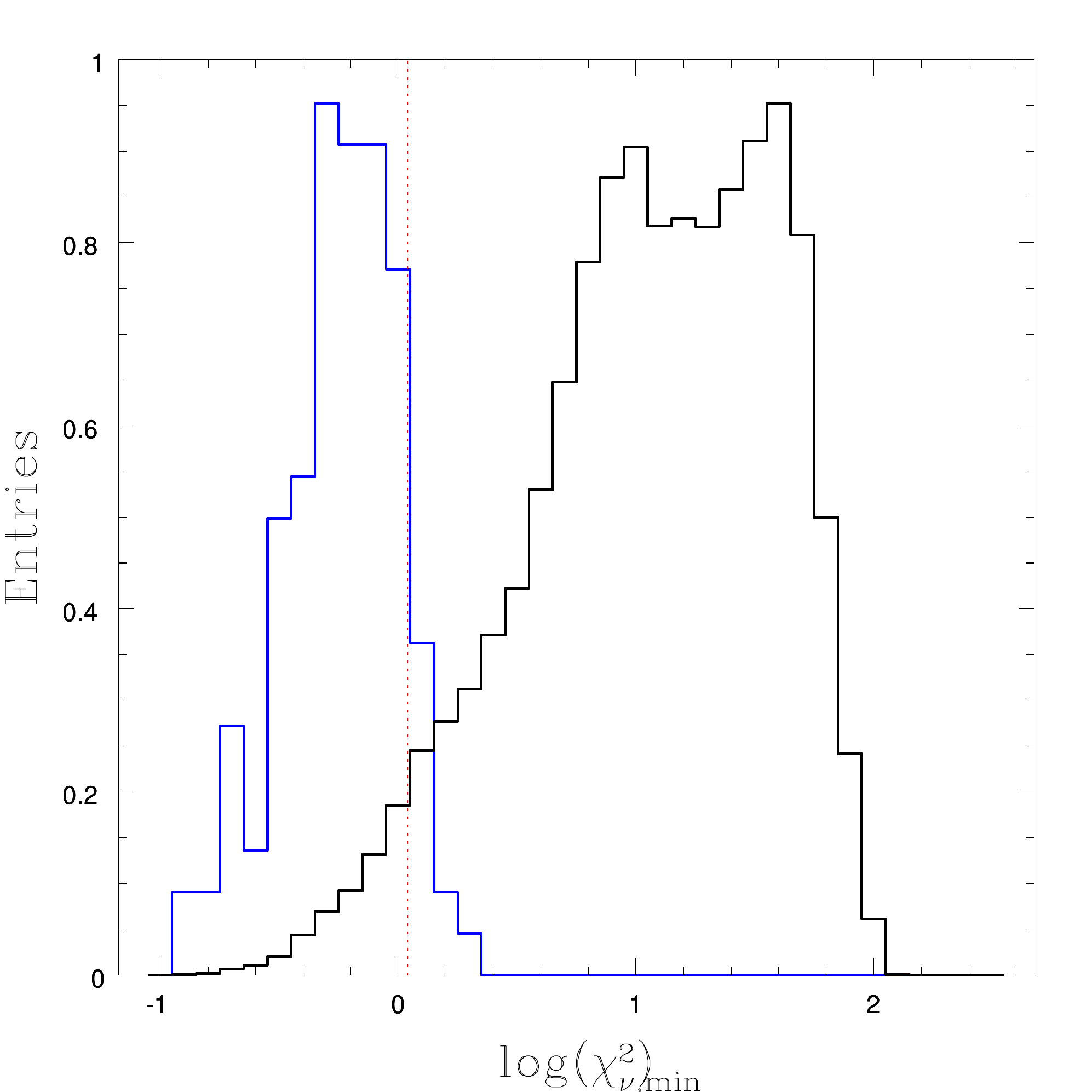}
   \includegraphics[width=80mm]{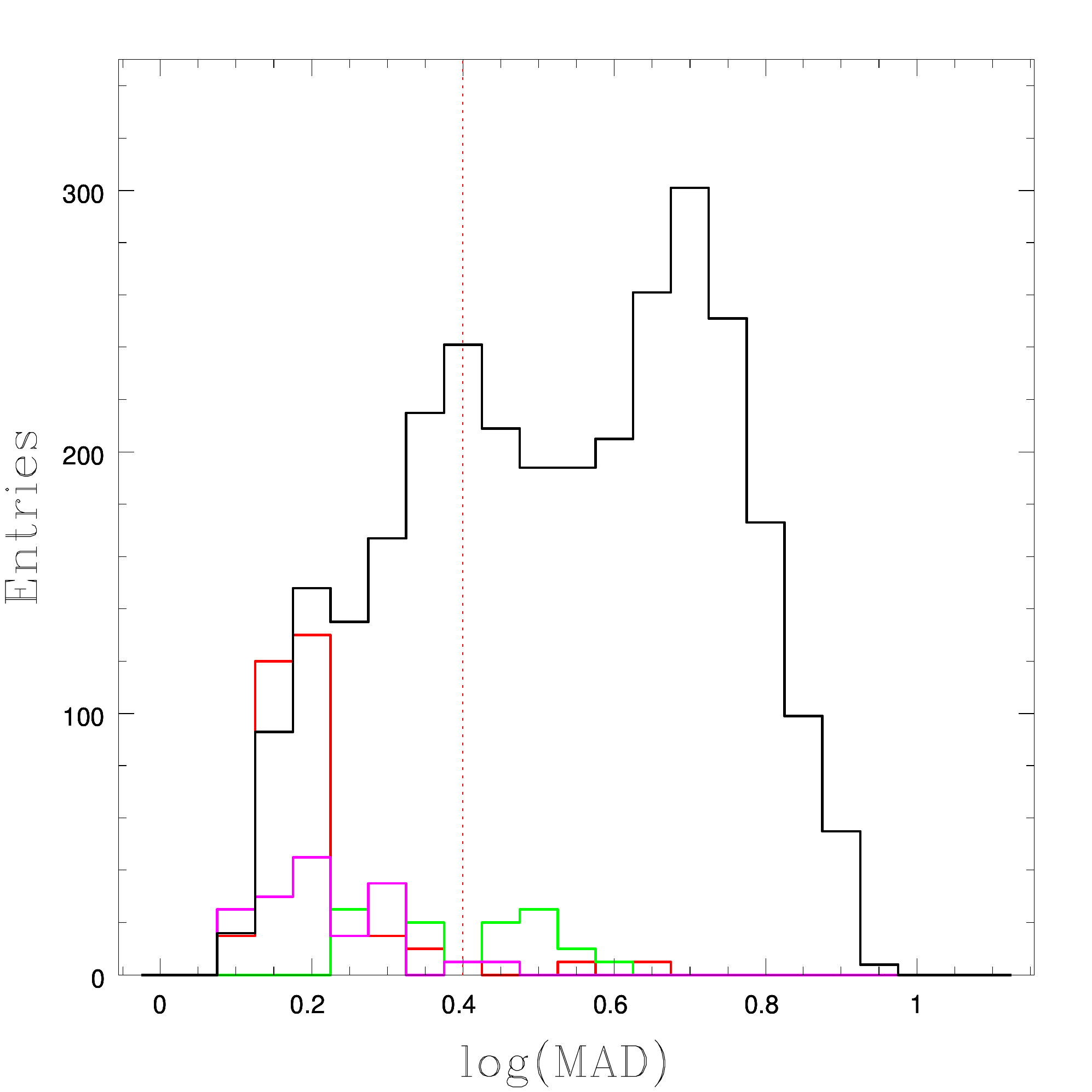}
\caption{The \textit{top panel} illustrates our test statistics for the $\chi_{\nu,min}^{2}$. Black histogram: the distribution of the total $\chi_{\nu,min}^{2}$ values calculated usign the SED of each source of the sample used to build the template and assuming to ignore the real source redshift and the \textit{blazar} class. The blue histogram are the good solutions, as discussed in Sec. \ref{statistics}. The vertical line marks our adopted \textit{confidence} value for $\chi_{\nu,min}^{2}$.
The \textit{bottom panel} illustrates a similar analysis performed for the MAD statistics. Coloured histograms refer to the MAD distribution of "good" solutions for the four different classes of \textit{blazar}s, with the same colour code as in Fig. \ref{fig:sed_template_4plot}: LSP (\textit{red}), ISP (\textit{green}), HSPs (\textit{magenta}). The vertical line marks our adopted \textit{confidence} value for MAD. In this case, this value only indicates wherein preferentially looking for good solutions.
} 
\label{fig:chi_distr}
\end{figure}

So, our \textit{blazar} recognition procedure offers also a method of estimating a tentative redshift for the UGS, in cases in which the agreement between the observational and template SEDs is good. 
We deem this a valuable contribution, considering the difficulty of measuring \textit{blazar} redshifts and the number of objects for which it is unknown.

\subsection{Characterising the $\chi_{min}^{2}$ and MAD statistics }
\label{statistics}

The reduced $\chi_{min}^{2}$ statistics has a well defined $\chi^{2}$ theoretical probability distribution under the assumptions of statistically independent data and Gaussian-distributed errors. 
Unfortunately, this is not our typical case because we are in the presence of flux variability and ill-defined photometric uncertainties. 
Consequently we have adopted arbitrarily fixed errors for all data points (corresponding to 1\% of the luminosity value).
In conclusion, we cannot simply use the $\chi^{2}$ for testing our best-fit solutions.
Nor we have any statistics for our MAD test.
We then proceeded to a rough characterisation of the $\chi_{\nu,min}^{2}$ and MAD statistics in the following way.

For the $\chi_{\nu,min}^{2}$ test, we considered all \textit{blazars} of the template set discussed in the previous section. 
For all these sources we ignored their redshift and we calculated the $\nu L(\nu)$ values by adopting redshifts within our grid of values of z=0.05 to 2. 
Then, we calculated the values of $\chi_{\nu,min}^{2}$ from Eqn. \ref{CHIeq} for all redshifts and all sources by comparing such estimated $\nu L(\nu)$ values with all best-fit SEDs, excluding from the calculation the SED template of the source itself. 
The resulting values are reported as black histograms in Fig. \ref{fig:chi_distr} (top). 
The blue histogram represents instead the $\chi_{\nu,min}^{2}$ distribution for the \textit{a-priori} known good solutions, that are the solutions for which the \textit{blazar} class and redshift are consistent with the real source properties (we considered a good solution if the found redshift value is within $\delta z/z \leq 0.1$ of the real redshift). 

So, the black histogram details the distribution of the $\chi_{\nu,min}^{2}$ values that would be obtained from a blind application of the test to \textit{blazar}s of unknown class and redshift. 
The fraction of random solutions with $\chi_{\nu,min}^{2}\leq 1.1$ is $4\%$, that can be considered as our approximate \textit{confidence} figure.
Note that we expect that the $\chi_{\nu,min}^{2}$ test would obtain higher values on average when applied to \textit{Fermi} objects other than \textit{blazar}s. This indeed will be checked against non-blazar sources in Sec. \ref{pulsars}. 
So the figure of $4\%$ can be considered as a conservative one.
Our good solutions will be obtained for values of our $\chi_{\nu,min}^{2}$ statistics $\chi_{\nu,min}^{2}\leq 1.1$.

We have performed a similar characterisation of the MAD statistics, whose results are reported in the bottom panel of Fig. \ref{fig:chi_distr}. Here the black histogram is calculated from eq. \ref{MADeq} for all sources in the template set, assuming that we do not know \textit{a-priori} the source redshift and class. The coloured histograms show the corresponding histograms for "good" redshift solutions ($\delta z/z \leq 0.1$ of the real value) for the three main \textit{blazar} classes. 
We see that the MAD test performs well in identifying good solutions for the LSP and HSP, less well for the ISP, whose MAD distribution for the good solutions has substantial overlap with that of the random population.
As a guideline, we will consider as potentially good \textit{blazar} recognitions those with MAD$<2.5$, however without excluding solutions with higher MAD values.
It is clear that, as anticipated, MAD offers a rather complementary test potentially useful for disentangling among degenerate solutions.

In all our later analyses, when applying our test to \textit{Fermi} UGS and other sources, we offer a graphical summary of the test performance in the form of plots of $\chi_{\nu,min}^{2}$ versus MAD statistics for the four \textit{blazar} classes and various redshifts (see e.g. Fig. \ref{fig:1215_metodo} below). 
The quadrant at $\chi_{\nu,min}^{2}<1.1$ and MAD$<2.5$ indicates regions where to look for potentially good solutions.

In general, our analysis can be effectively applied in cases in which there is a sufficient sampling of the synchrotron component of the \textit{blazar} SED. This means having at least three reliable data-points over the synchrotron part, and assuming that the Fermi and X-ray data are sufficient to sample the IC component. 


\begin{figure*}
\centering
\mbox{%
\begin{minipage}{.65\textwidth}
\includegraphics[height=0.9\textwidth,width=\textwidth]{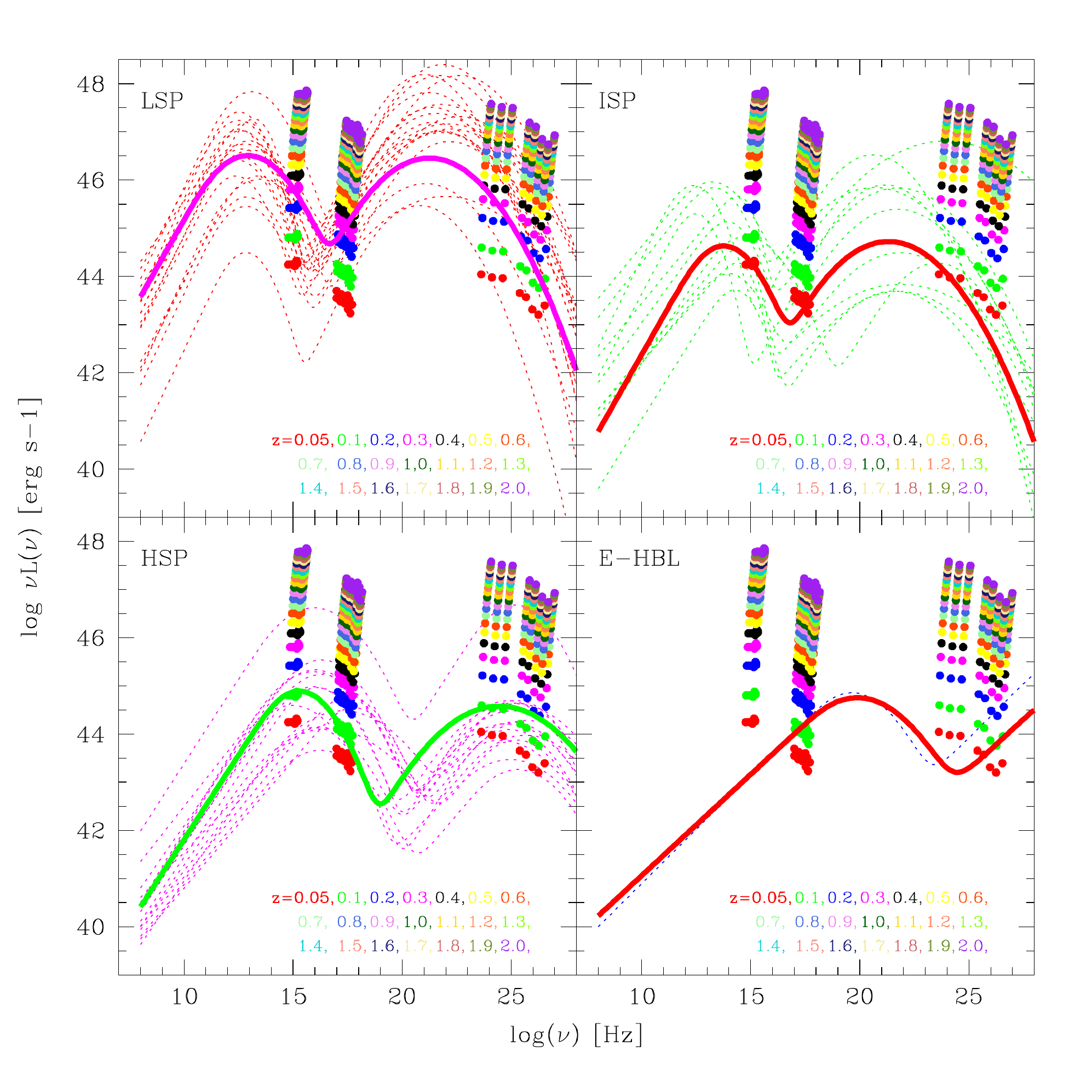}
\end{minipage}%
\begin{minipage}[c]{.40\textwidth}
\quad \quad \quad \quad
 \begin{tabular}{cccc}
\hline
\footnotesize{Class}  & \footnotesize{$\chi_{\nu,min}^{2}$} &  \footnotesize{MAD} &\footnotesize{$z$}\\
\hline
\footnotesize{LSP}  & \footnotesize{0.44} & \footnotesize{1.73} & \footnotesize{0.2}\\
\footnotesize{ISP}  & \footnotesize{0.37} & \footnotesize{1.79} & \footnotesize{0.05}\\
\footnotesize{\bf{HSP}}  & \footnotesize{\bf{0.09}} & \footnotesize{\bf{1.48}} & \footnotesize{\bf{0.1}}\\
\footnotesize{EHBL} & \footnotesize{3.42} & \footnotesize{1.78} & \footnotesize{0.05}\\
\hline
\end{tabular}
  \\
\includegraphics[height=.9\textwidth,width=.9\textwidth]{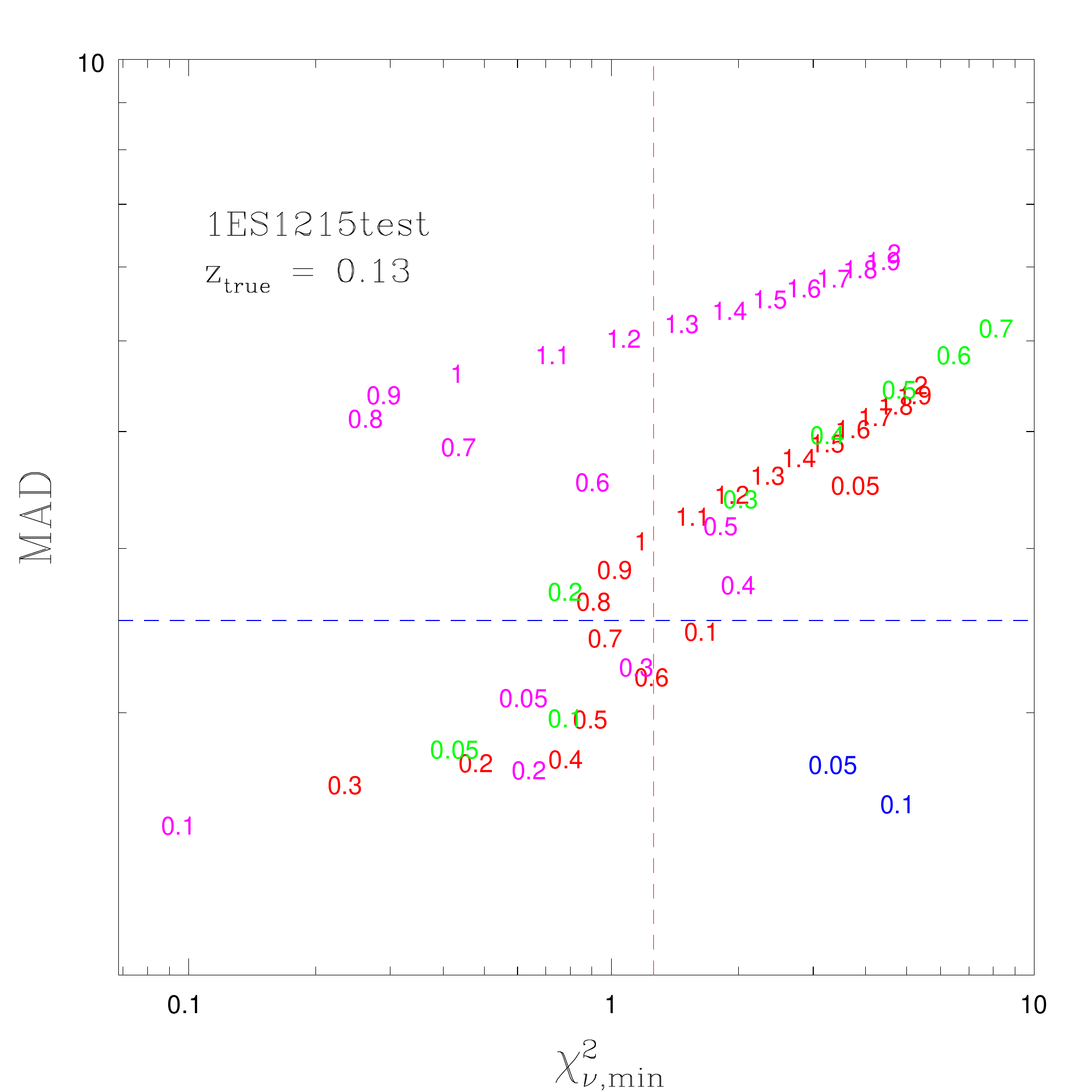}
\end{minipage}
}
\caption{Spectral luminosity points of 1ES 1215+303 for different assumed redshifts (from 0.05 to 2.0), compared to our \textit{blazar} SED template sets. The latter are built from archive data of known \textit{blazars} for the four classes of LSP, ISP, HSP and EHBL. 
The coloured SEDs indicate the best-fits redshift solutions for each class.
The  values of MAD and $\chi_{min}^{2}$ in the table are referred to the best-fit SED template of each class (and shown as \textit{bold coloured lines}). 
The plot on the right shows the values of MAD and $\chi_{min}^{2}$ obtained by comparing photometric data for the source and the SED template set for different assumed redshifts.
For 1ES 1215+303 our best-guess recognition is an HSP at z$\sim$0.1 (\textit{green line}). }
\label{fig:1215_metodo}
\end{figure*}


\begin{figure*}
\centering
\mbox{%
\begin{minipage}{.65\textwidth}
\includegraphics[height=0.9\textwidth,width=\textwidth]{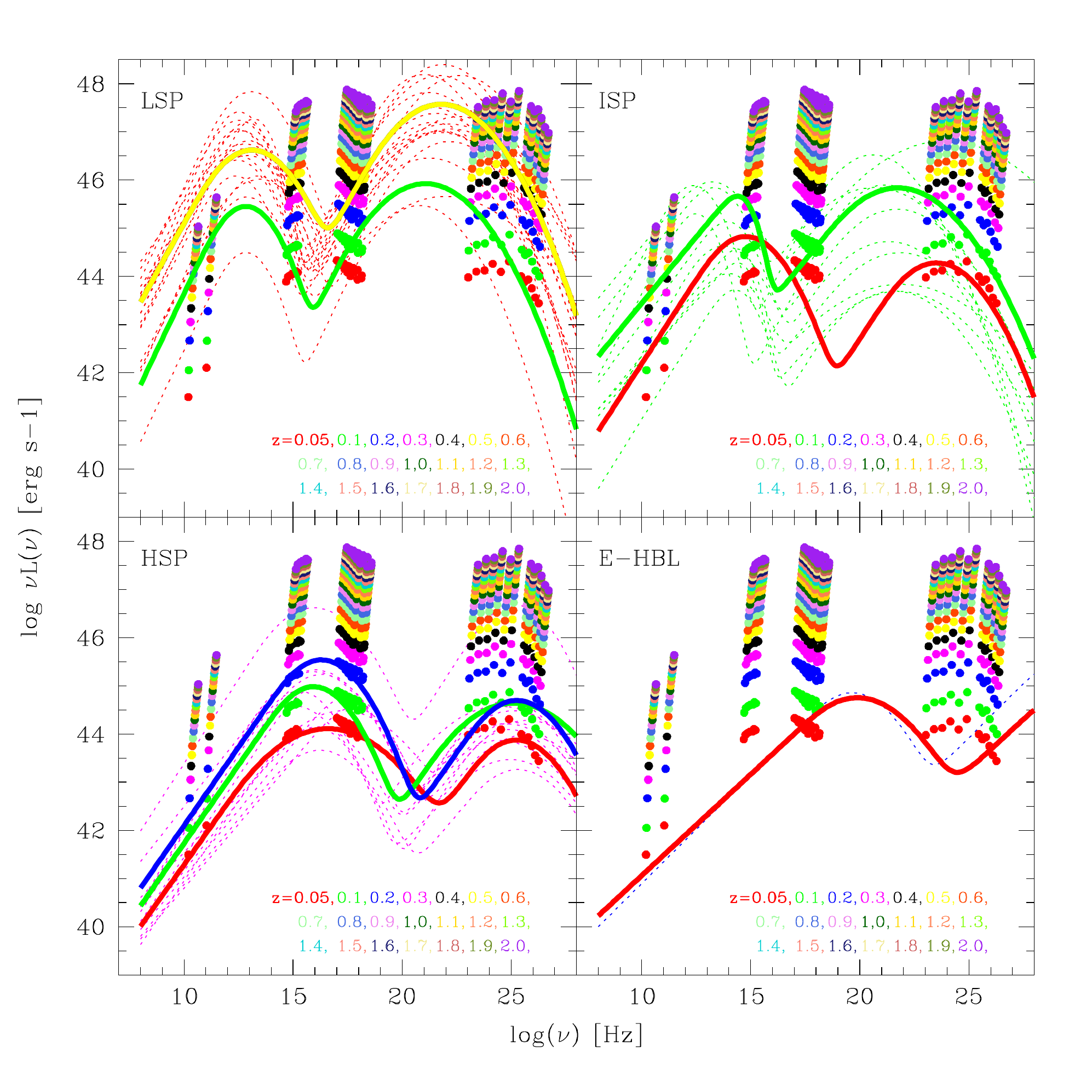}
\end{minipage}%
\begin{minipage}[c]{.40\textwidth}
\quad \quad \quad \quad
 \begin{tabular}{cccc}
\hline
\footnotesize{Class}  & \footnotesize{$\chi_{\nu,min}^{2}$} &  \footnotesize{MAD} &\footnotesize{$z$}\\
\hline
\footnotesize{LSP}  & \footnotesize{2.32} & \footnotesize{2.50} & \footnotesize{0.1}\\
\footnotesize{LSP}  & \footnotesize{2.78} & \footnotesize{2.23} & \footnotesize{0.5}\\
\footnotesize{ISP}  & \footnotesize{2.35} & \footnotesize{2.61} & \footnotesize{0.1}\\
\footnotesize{ISP}  & \footnotesize{2.14} & \footnotesize{2.63} & \footnotesize{0.05}\\
\footnotesize{\bf{HSP}}  & \footnotesize{\bf{0.29}} & \footnotesize{\bf{1.43}} & \footnotesize{\bf{0.1}}\\
\footnotesize{HSP}  & \footnotesize{0.37} & \footnotesize{1.42} & \footnotesize{0.05}\\
\footnotesize{HSP}  & \footnotesize{0.77} & \footnotesize{2.23} & \footnotesize{0.2}\\
\footnotesize{EHBL} & \footnotesize{1.39} & \footnotesize{0.95} & \footnotesize{0.05}\\
\hline
\end{tabular}
  \\
\includegraphics[height=.9\textwidth,width=.9\textwidth]{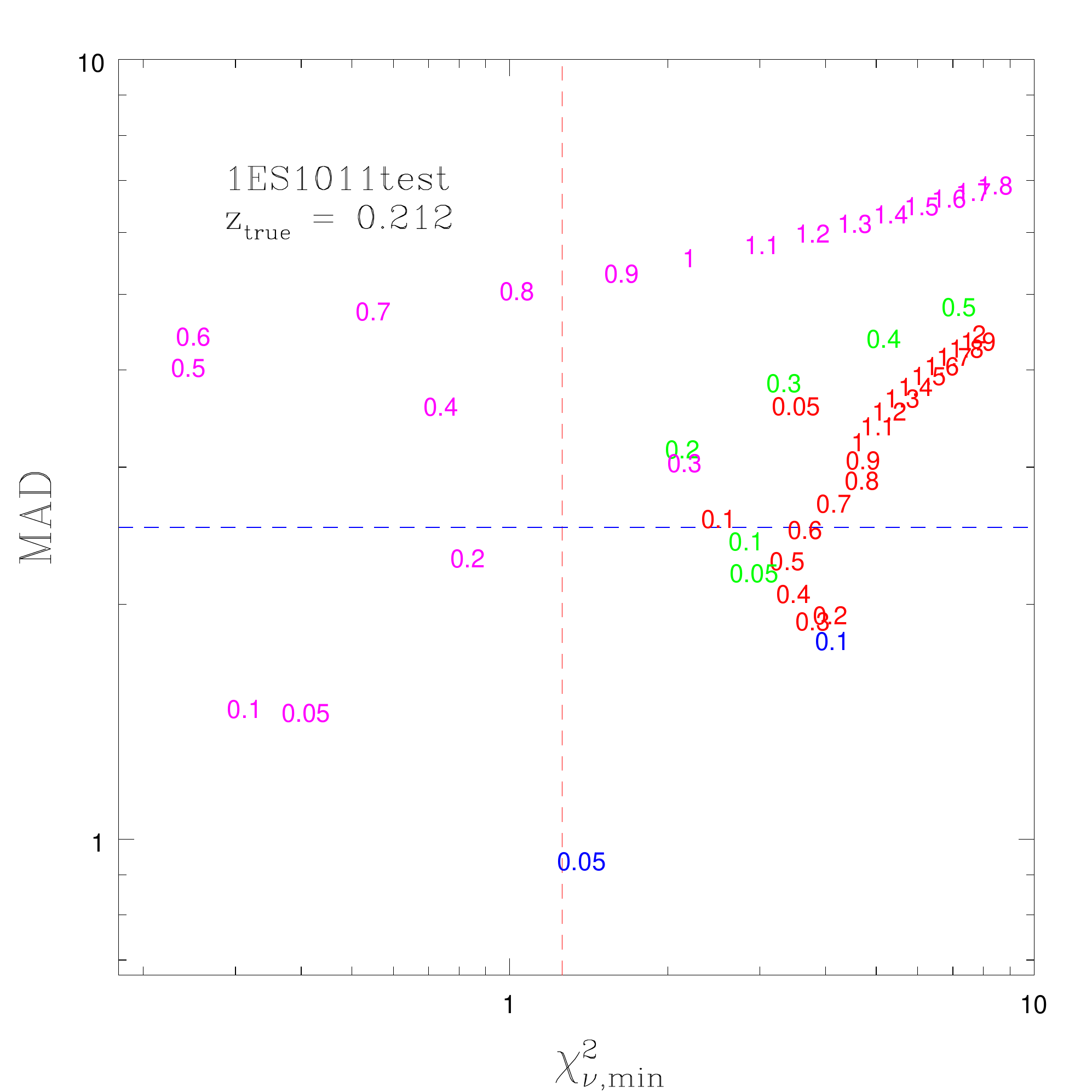}
\end{minipage}
}
\caption{The same diagnostic plot as in Fig.\ref{fig:1215_metodo} for 1ES 1011+496. Here the best-guess recognition is a HSP at z$\sim$0.1 (\textit{green line}).}
\label{fig:1011_metodo}
\end{figure*}


\begin{figure*}
\centering
\mbox{%
\begin{minipage}{.65\textwidth}
\includegraphics[height=0.9\textwidth,width=\textwidth]{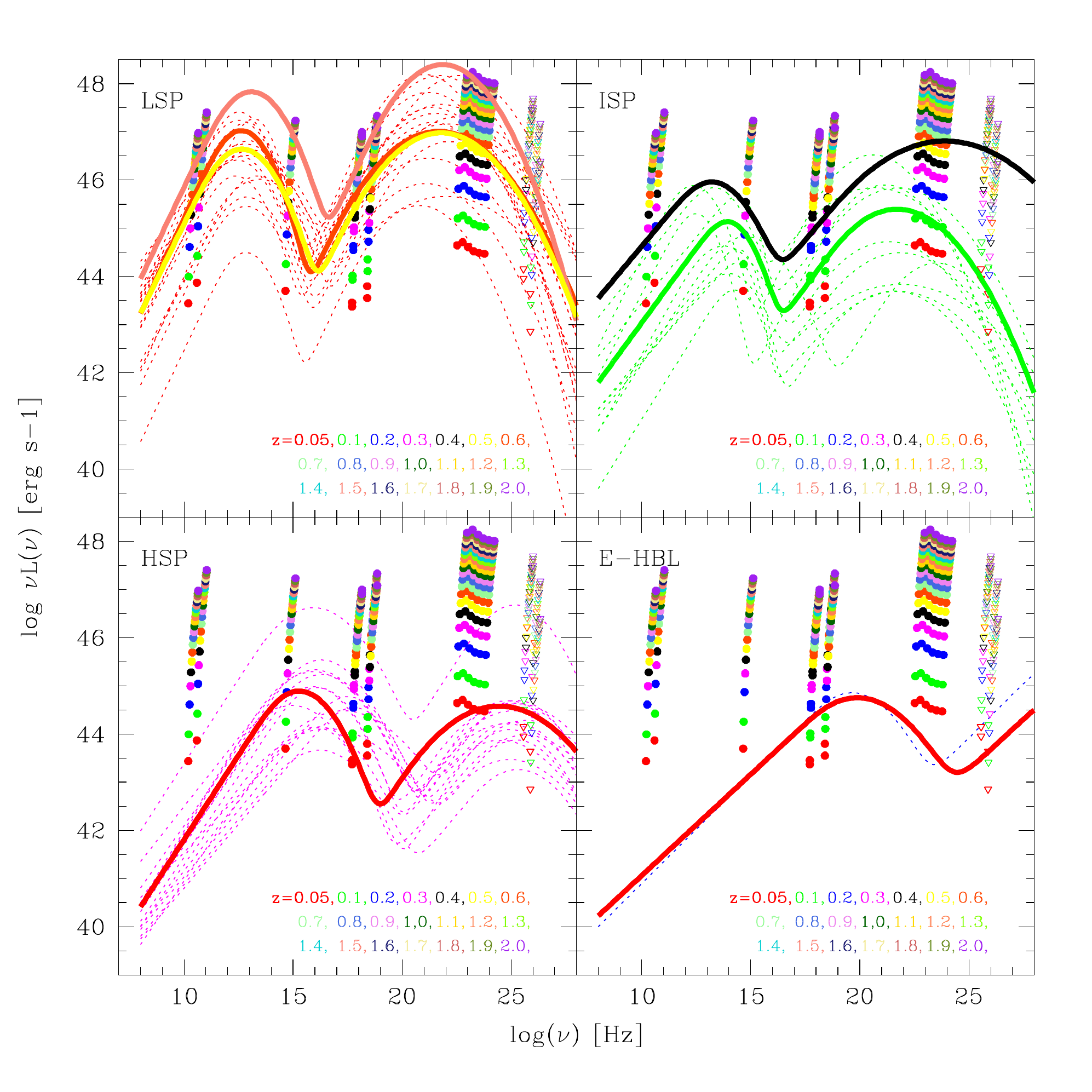}
\end{minipage}%
\begin{minipage}[c]{.40\textwidth}
\quad \quad \quad \quad
 \begin{tabular}{cccc}
\hline
\footnotesize{Class}  & \footnotesize{$\chi_{\nu,min}^{2}$} &  \footnotesize{MAD} &\footnotesize{$z$}\\
\hline
\footnotesize{\bf{LSP}}  & \footnotesize{\bf{0.10}} & \footnotesize{\bf{1.34}} & \footnotesize{\bf{0.5}}\\
\footnotesize{LSP}  & \footnotesize{0.19} & \footnotesize{1.31} & \footnotesize{0.6}\\
\footnotesize{LSP}  & \footnotesize{0.24} & \footnotesize{2.42} & \footnotesize{1.5}\\
\footnotesize{ISP}  & \footnotesize{0.71} & \footnotesize{3.58} & \footnotesize{0.4}\\
\footnotesize{ISP}  & \footnotesize{0.82} & \footnotesize{2.04} & \footnotesize{0.1}\\
\footnotesize{HSP}  & \footnotesize{2.91} & \footnotesize{2.29} & \footnotesize{0.05}\\
\footnotesize{EHBL} & \footnotesize{7.56} & \footnotesize{2.60} & \footnotesize{0.05}\\
\hline
\end{tabular}
  \\
\includegraphics[height=.9\textwidth,width=.9\textwidth]{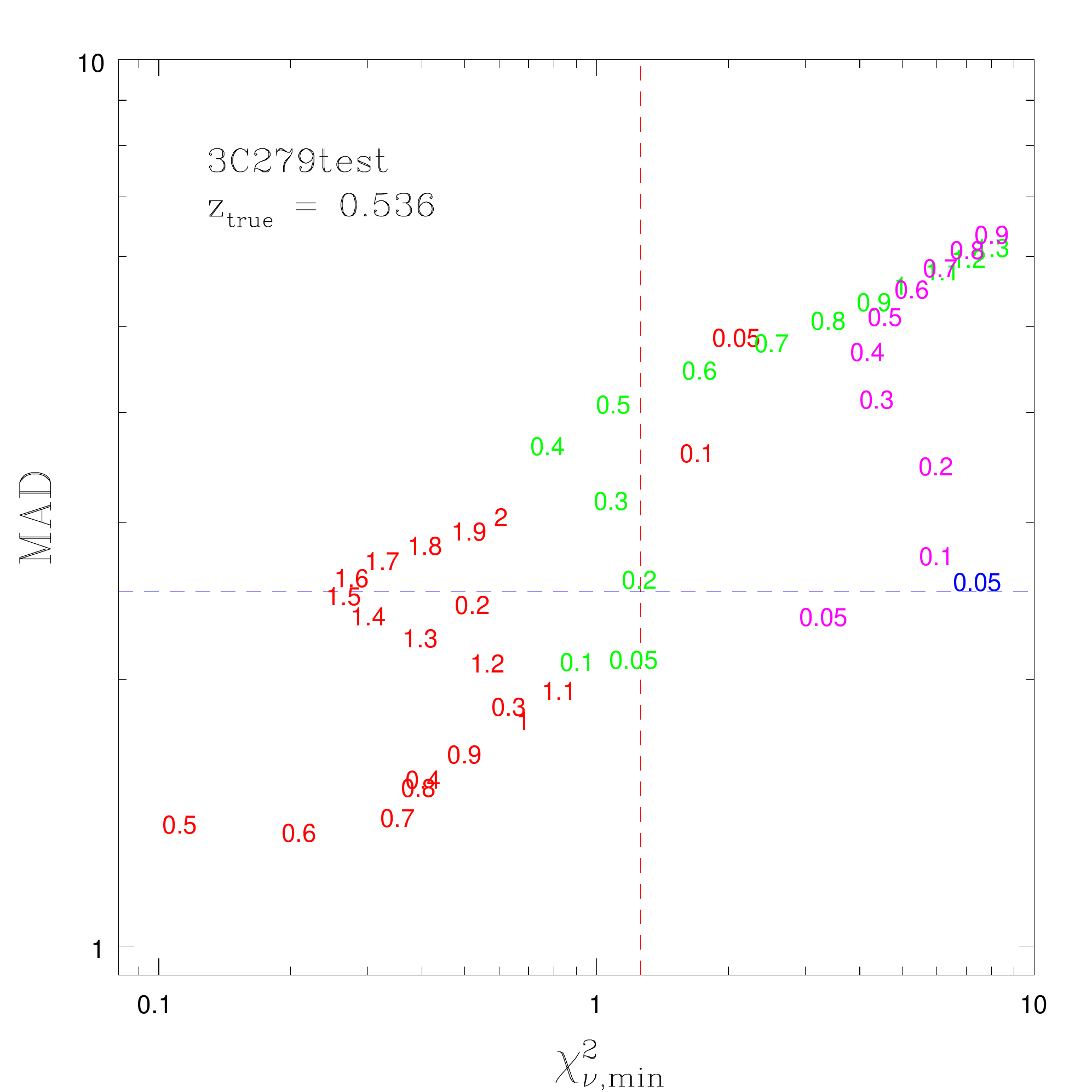}
\end{minipage}
}
\caption{The same diagnostic plot as Fig.\ref{fig:1215_metodo} for 3C 279. Here the best-guess recognition is a LSP at $z\sim$0.5 (\textit{yellow line}). }
\label{fig:3c279_metodo}
\end{figure*}


\begin{figure*}
\centering
\mbox{%
\begin{minipage}{.65\textwidth}
\includegraphics[height=0.9\textwidth,width=\textwidth]{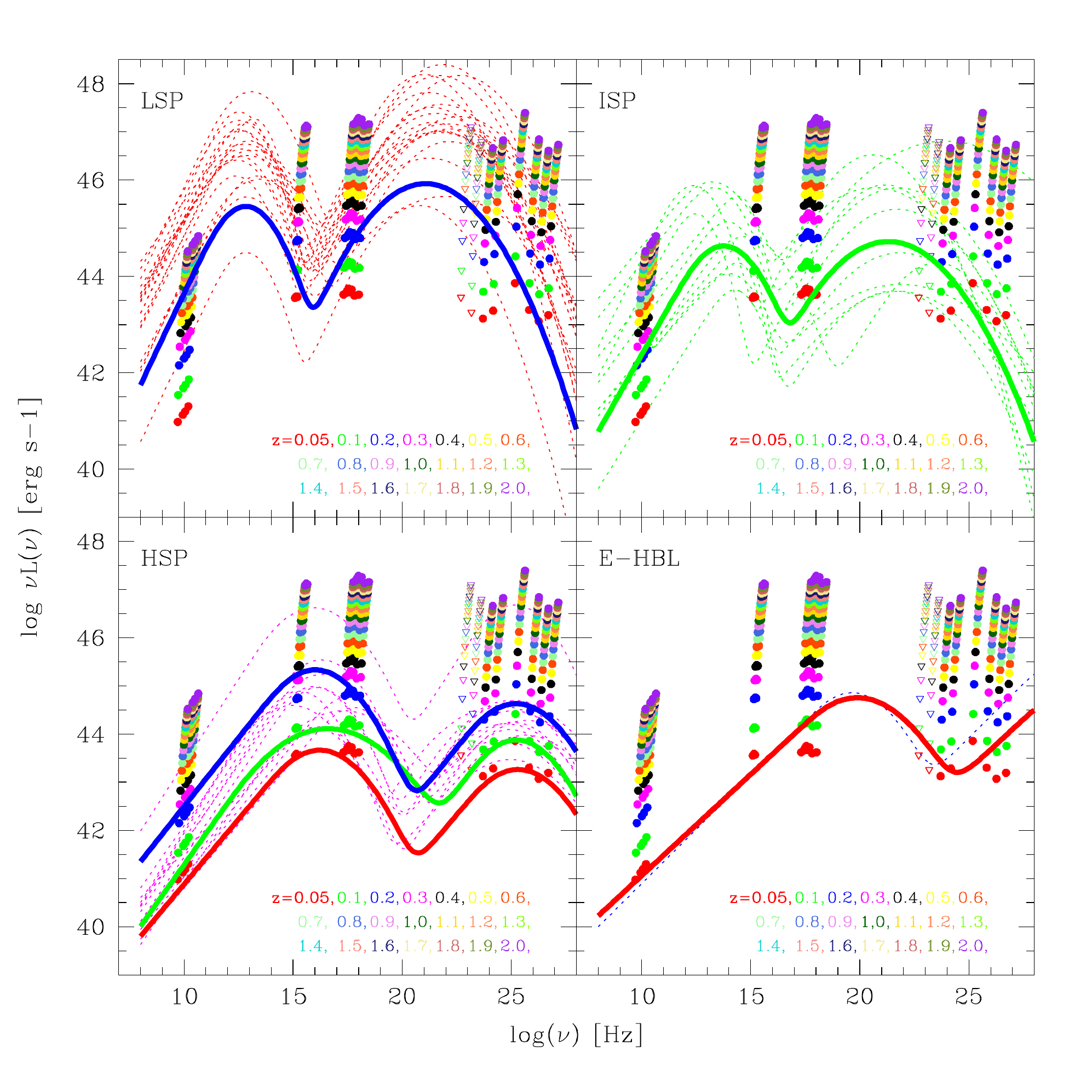}
\end{minipage}%
\begin{minipage}[c]{.40\textwidth}
\quad \quad \quad \quad
 \begin{tabular}{cccc}
\hline
\footnotesize{Class}  & \footnotesize{$\chi_{\nu,min}^{2}$} &  \footnotesize{MAD} &\footnotesize{$z$}\\
\hline
\footnotesize{LSP}  & \footnotesize{3.61} & \footnotesize{2.49} & \footnotesize{0.3}\\
\footnotesize{ISP}  & \footnotesize{2.39} & \footnotesize{2.29} & \footnotesize{0.1}\\
\footnotesize{ISP}  & \footnotesize{2.79} & \footnotesize{2.35} & \footnotesize{0.2}\\
\footnotesize{\bf{HSP}}  & \footnotesize{\bf{0.34}} & \footnotesize{\bf{1.57}} & \footnotesize{\bf{0.2}}\\
\footnotesize{\bf{HSP}}  & \footnotesize{\bf{0.40}} & \footnotesize{\bf{1.91}} & \footnotesize{\bf{0.05}}\\
\footnotesize{HSP}  & \footnotesize{0.44} & \footnotesize{1.37} & \footnotesize{0.1}\\
\footnotesize{EHBL} & \footnotesize{1.05} & \footnotesize{0.96} & \footnotesize{0.05}\\
\hline
\end{tabular}
  \\
\includegraphics[height=.9\textwidth,width=.9\textwidth]{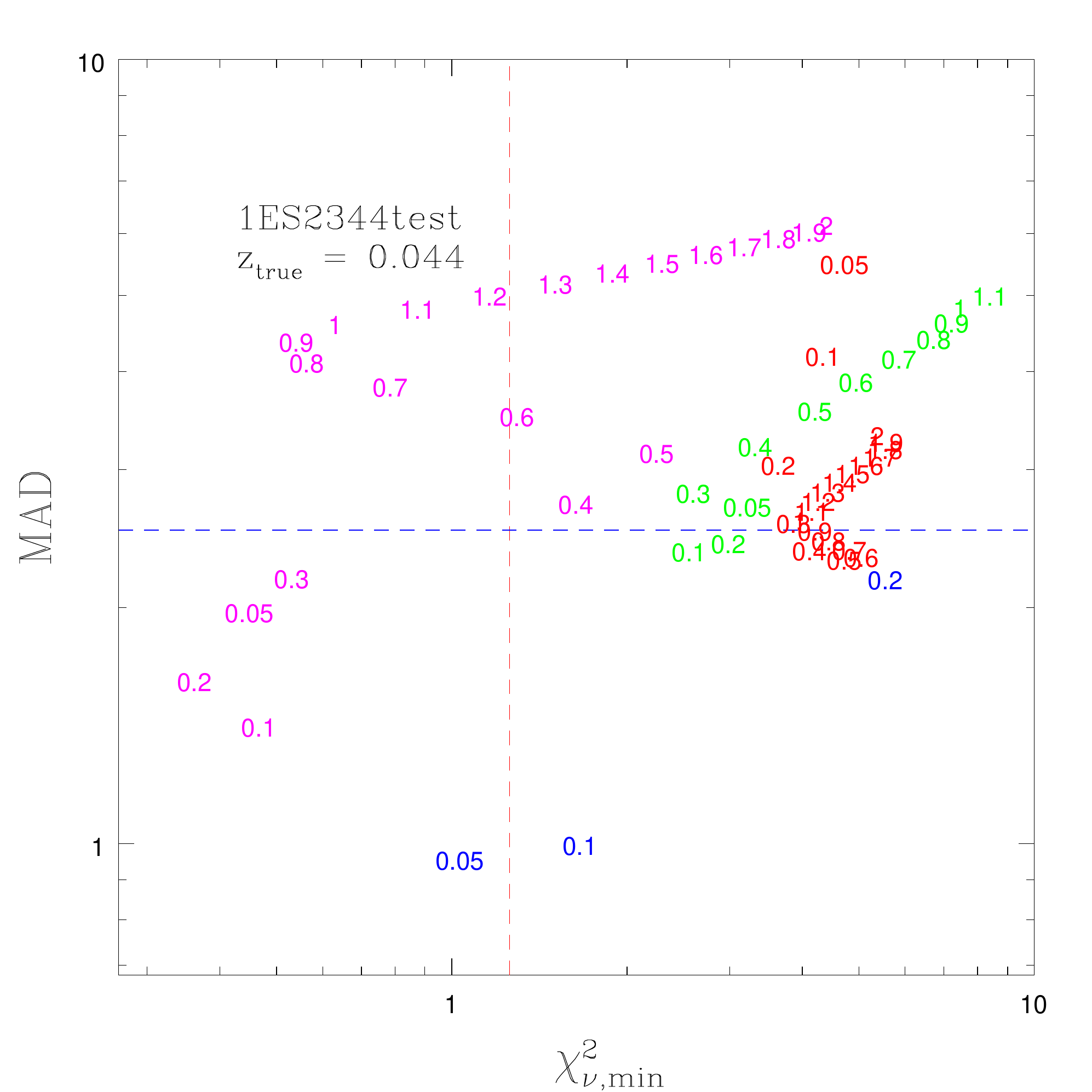}
\end{minipage}
}
\caption{The same diagnostic plot as in Fig.\ref{fig:1215_metodo} for 1ES 2344+514. Here the best-guess recognition is a HSP at $z\sim$0.2 (\textit{blue line}). }
\label{fig:2344_metodo}
\end{figure*}


\begin{figure*}
\centering
\mbox{%
\begin{minipage}{.65\textwidth}
\includegraphics[height=0.9\textwidth,width=\textwidth]{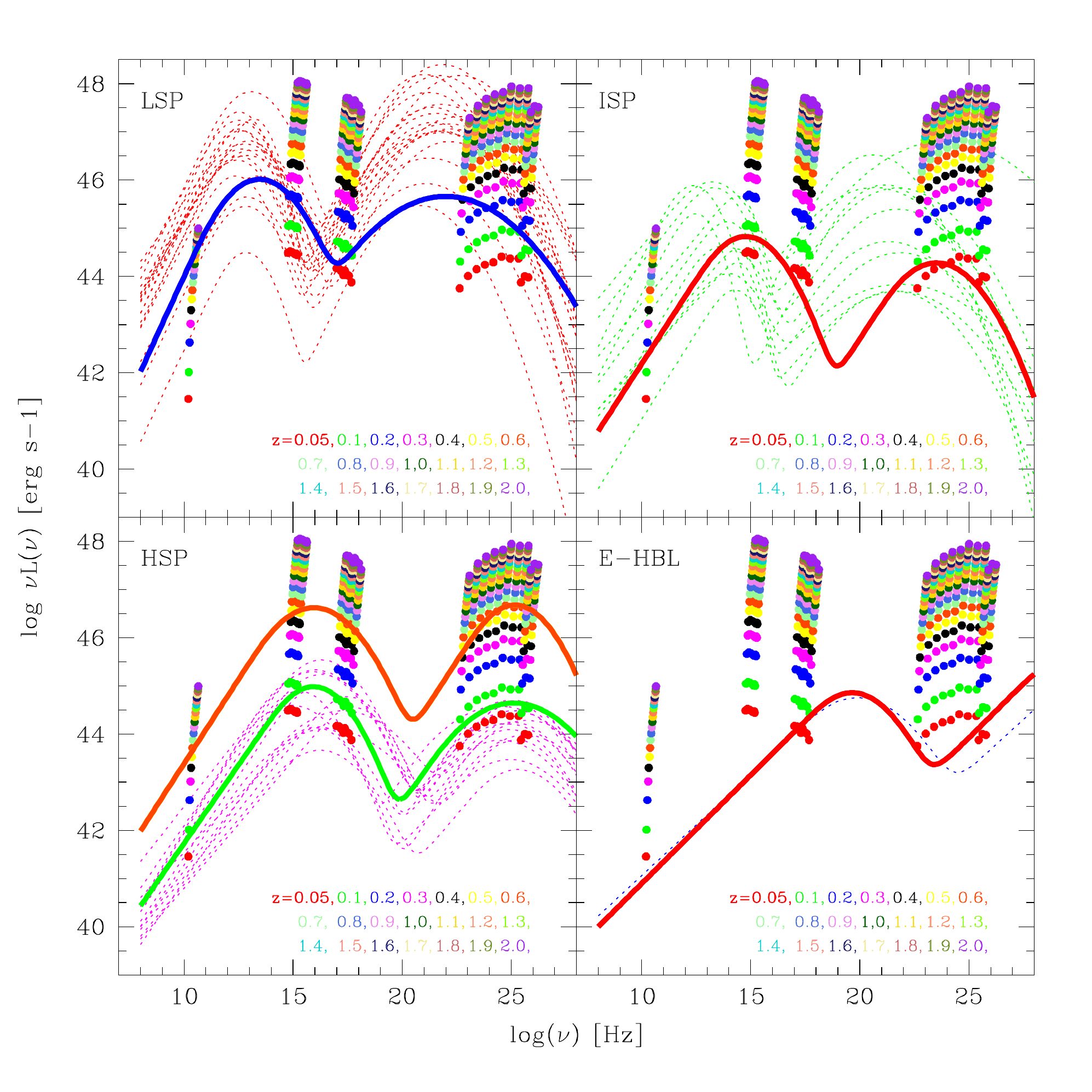}
\end{minipage}%
\begin{minipage}[c]{.40\textwidth}
\quad \quad \quad \quad
 \begin{tabular}{cccc}
\hline
\footnotesize{Class}  & \footnotesize{$\chi_{\nu,min}^{2}$} &  \footnotesize{MAD} &\footnotesize{$z$}\\
\hline
\footnotesize{LSP}  & \footnotesize{2.29} & \footnotesize{2.06} & \footnotesize{0.2}\\
\footnotesize{ISP}  & \footnotesize{0.91} & \footnotesize{2.01} & \footnotesize{0.05}\\
\footnotesize{\bf{HSP}}  & \footnotesize{\bf{0.09}} & \footnotesize{\bf{4.44}} & \footnotesize{\bf{0.6}}\\
\footnotesize{HSP}  & \footnotesize{0.21} & \footnotesize{1.56} & \footnotesize{0.1}\\
\footnotesize{EHBL} & \footnotesize{3.02} & \footnotesize{1.43} & \footnotesize{0.05}\\
\hline
\end{tabular}
  \\
\includegraphics[height=.9\textwidth,width=.9\textwidth]{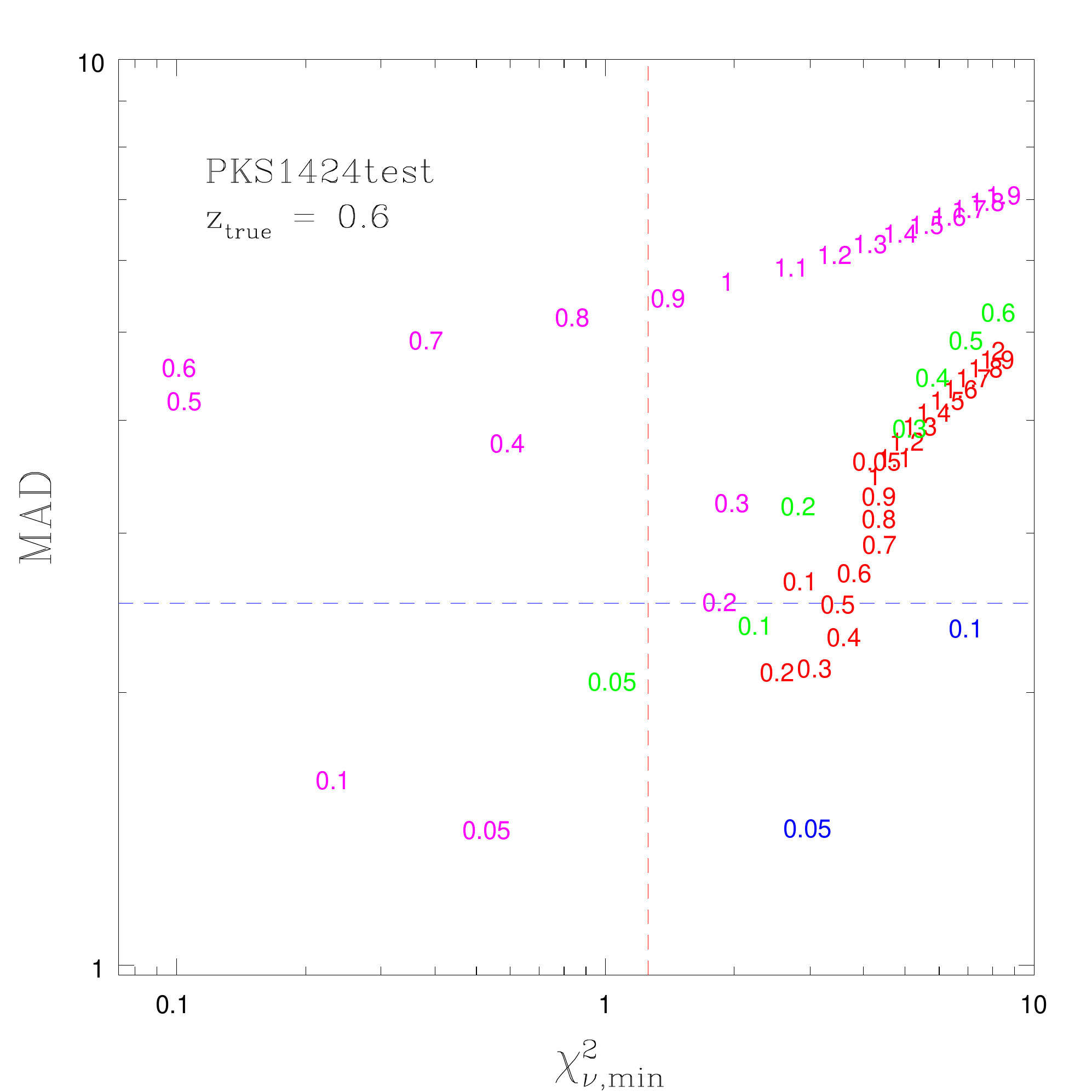}
\end{minipage}
}
\caption{The same diagnostic plot as in Fig.\ref{fig:1215_metodo} for 1ES 1424+240. Here the best-guess recognition is a HSP at $z\sim$0.6 (\textit{orange line}).}
\label{fig:1424_metodo}
\end{figure*}


\begin{figure*}
\centering
\mbox{%
\begin{minipage}{.65\textwidth}
\includegraphics[height=0.9\textwidth,width=\textwidth]{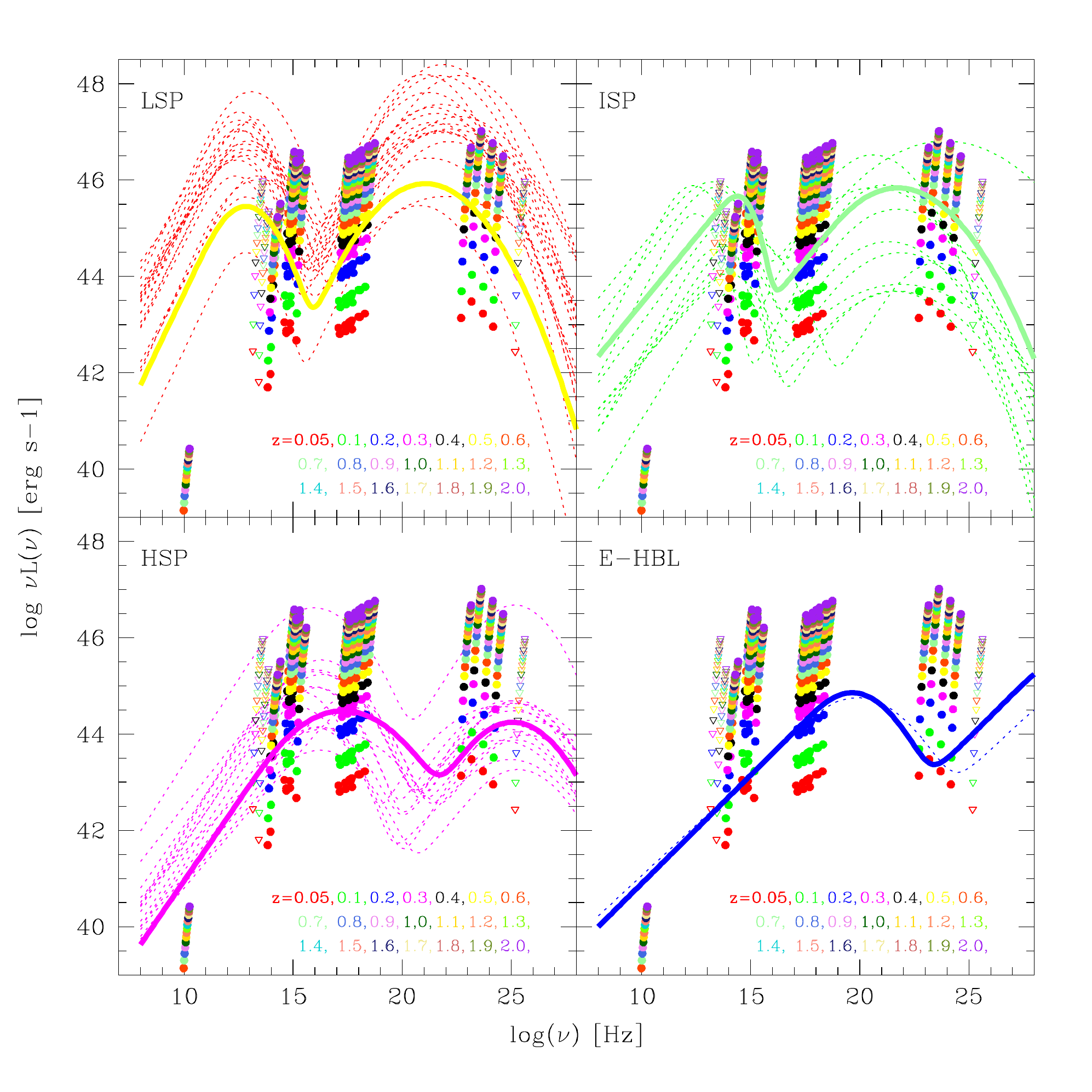}
\end{minipage}%
\begin{minipage}[c]{.40\textwidth}
\quad \quad \quad \quad
 \begin{tabular}{cccc}
\hline
\footnotesize{Class}  & \footnotesize{$\chi_{\nu,min}^{2}$} &  \footnotesize{MAD} &\footnotesize{$z$}\\
\hline
\footnotesize{LSP}  & \footnotesize{7.22} & \footnotesize{2.58} & \footnotesize{0.5}\\
\footnotesize{ISP}  & \footnotesize{7.19} & \footnotesize{3.17} & \footnotesize{0.7}\\
\footnotesize{HSP}  & \footnotesize{2.70} & \footnotesize{1.85} & \footnotesize{0.3}\\
\footnotesize{EHBL} & \footnotesize{4.15} & \footnotesize{1.33} & \footnotesize{0.2}\\
\hline
\end{tabular}
  \\
\includegraphics[height=.9\textwidth,width=.9\textwidth]{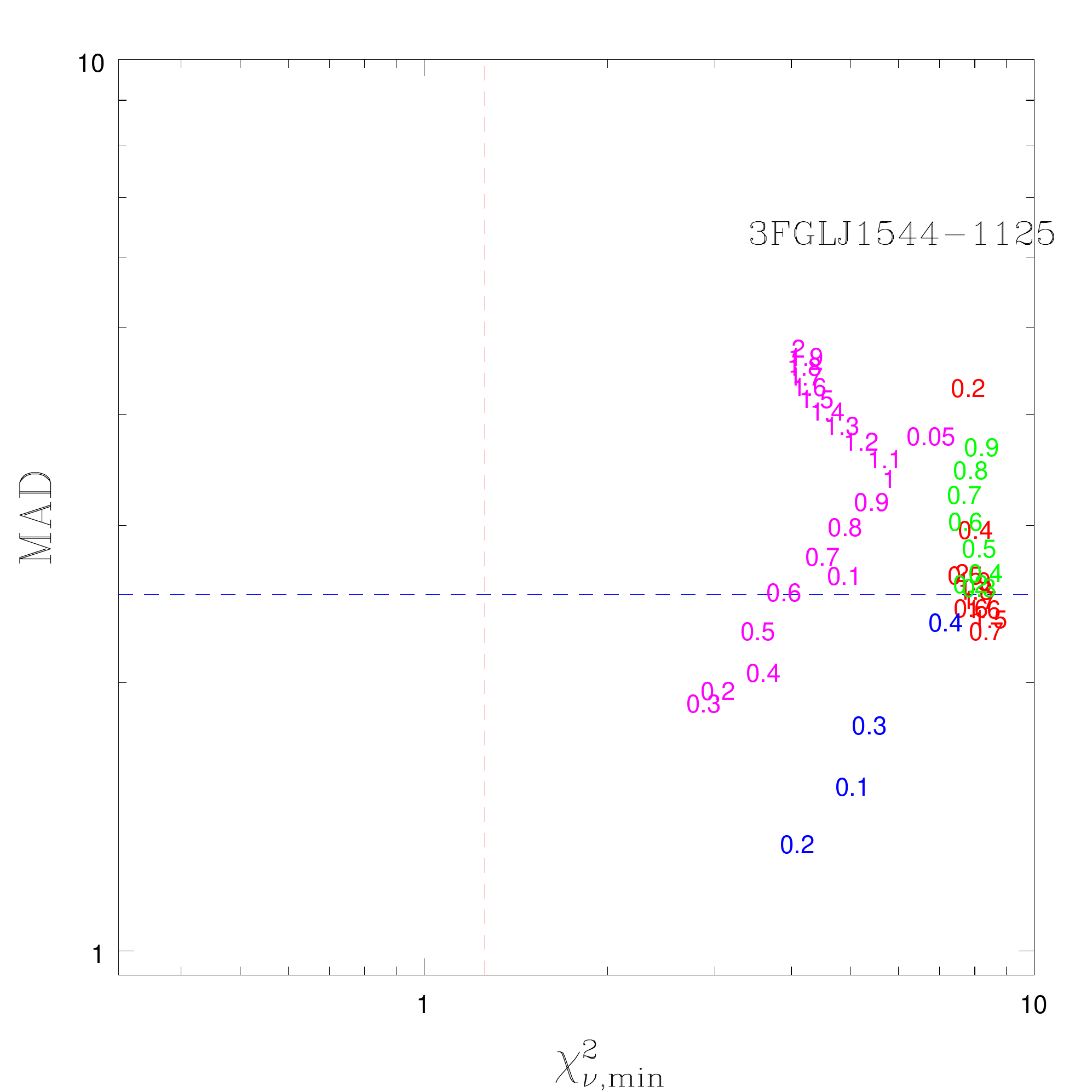}
\end{minipage}
}
\caption{Spectral luminosity points of 2FGL J1544.5-1126 for different assumed redshift (from 0.05 to 2.0), same as in Fig. \ref{fig:1215_metodo}. For 2FGL J1544.5-1126 there is not a clear solution within our considered \textit{blazar} categories.}
\label{fig:1544_metodo}
\end{figure*}


\subsection{Testing of the \textit{blazar} recognition tool on known objects}

As a sanity check and to test the effectiveness of our method in recognising \textit{blazar}-like sources among UGSs, we applied it on a few well-known \textit{blazars}: 1ES 1215+303, 1ES 1011+496, 1ES 2344+514, and 3C 279 (the latter also present in the sample used to build the SED templates). 
We have also made a test on the well-known high-redshift HSP 1ES 1424+240 with redshift z$=$0.604 \citep{paiano2017}.
We assumed these to be sources with unknown class and redshift, and run blindly the algorithm on the simultaneous photometric data collected during dedicated campaigns for each \textit{blazar}. 
Note that, to make it a meaningful test for the source 3C 279, we use here a different flux dataset to that used for building the SED template set in Sec. \ref{chap_SED}: in that case the whole set of historical observations in the ASDC archive was used, while here the test is done instead on completely independent sets of simultaneous observations, as detailed below for the 5 test sources.

Following the above described recognition procedure, from the flux data of the five sources we determined the corresponding luminosity values as a function of the redshift values of our grid. 
Then the luminosity data were over-plotted with different colours in the four panels for the \textit{blazar} classes, as shown in Figs. \ref{fig:1215_metodo}, \ref{fig:1011_metodo}, \ref{fig:3c279_metodo}, \ref{fig:2344_metodo}, and \ref{fig:1424_metodo}. 
The MAD and $\chi^2_\nu$ statistics were computed for every redshift and every SED template.
The results for the five well-known \textit{blazar} sources are hereby briefly discussed.

\begin{itemize}

\item \textbf{Results for 1ES 1215+303}: 
This is an HSP \textit{blazar} \citep{abdo2010_1es1215} with redshift z$=$0.129 \citep{paiano2017}.
For this source we used the flux data collected during a multi-wavelength campaign performed in 2011 \citep{aleksic2012_1es1215} and triggered by an optical outburst of the source. 
The data were taken simultaneously from radio to VHE band. 
The minima of the MAD and  $\chi_{\nu,min}^{2}$ statistics are reported in Fig. \ref{fig:1215_metodo} and are used to select the best-fit SED template. 
The minimum value of $\chi_{\nu,min}^{2}$ is associated with a SED template of the HSP class (\textit{green curve}), suggesting that 1ES 1215+303 is an HSP object with a redshift of about 0.1. 
\newline

\item \textbf{Results for 1ES 1011+496}: 
For the test \textit{blazar} 1ES 1011+496 at z=0.212, we used the simultaneous multi-wavelength data obtained by the observational campaign in 2012 \citep{magic_1011_2016}.   
We constructed the diagnostic plots, shown in Fig. \ref{fig:1011_metodo}, with the corresponding table including the MAD and $\chi_{\nu,min}^{2}$ values. 
The minimum $\chi_{\nu,min}^{2}$ corresponds to the HSP class with a significant level of degeneracy. We found three best-fitting SED template candidates with the value of MAD and $\chi_{\nu,min}^{2}$ in the \textit{good solution} quadrant defined in Sec. \ref{statistics}. 
The first solution assumes for 1ES 1011+496 a redshift of $z=0.1$, providing a $\chi_{\nu,min}^{2}\sim 0.29$, while the second one, with $\chi_{\nu,min}^{2}\sim 0.37$, corresponds to a redshift of 0.05. The third solution provides a $\chi_{\nu,min}^{2}\sim 0.77$, with a higher MAD value.

Although this level of degeneracy, our tool is pretty in agreement with the real \textit{blazar} classification and the real redshift of $z=0.212$ \citep{albert2007_z1011}. 
\newline

\item \textbf{Results for 3C 279}: 
Another test \textit{blazar} studied is 3C 279, the first LSP discovered to emit VHE $\gamma$-rays in 2006 \citep{albert2008_3c279} and with a redshift of 0.536 \citep{hewitt1993}, making it one of most distant VHE emitting sources discovered so far.
The multi-wavelength data, used in the \textit{blazar} diagnostic plots, are taken from \citet{aleksic2014_3c279}, obtained during the 2011 observational campaign performed from February 8 to April 11, when the source was in a low state.
We see from Fig. \ref{fig:3c279_metodo} that we get an excellent match only for an LSP class and a redshift which is very consistent, within the uncertainties of our method, with the observed redshift.
Such a good match with this well known high-redshift source makes us confident about the validity of the test even for distant high-luminosity objects.
\newline

\item \textbf{Results for 1ES 2344+514}: 
This BL Lac object is classified as an HSP with a redshift of z~$=$~0.044 \citep{perlman1996_z2344}. It was targeted in 2008 by a simultaneous broad-band observational campaign from radio to VHE energies, during which this source was found in a low flux state \citep{aleksic2013_1es2344}. 
We used these simultaneous multi-wavelength data to test our \textit{blazar} recognition tool. 
Concerning the HE $\gamma$-ray flux, we decided to use the 1FGL catalogue flux points because we can consider them quasi-simultaneous with the data collected during the 2008 season. 
The diagnostic plots for this source are shown in Fig. \ref{fig:2344_metodo} with the values of the MAD and $\chi_{\nu,min}^{2}$ statistics. 
Our procedure indicates a good solution with an HSP at z~$\simeq$~0.2, but few other solutions at lower redshift are within the confidence limits of $\chi_{\nu,min}^{2}$ and MAD defined in Sec. \ref{statistics}.
On one side for this source, our test fully confirms the classification as an HSP \textit{blazar}, on the other it clearly indicates for it a low redshift, in spite of some degeneracy.
\newline

\item \textbf{Results for PKS 1424+240}: 
The last \textit{blazar} used as test for our tool is the BL Lac object PKS 1424+240 belonging to the class of the HSPs.
Recently the redshift of the source, z$=$0.604, was determined \citep{paiano2017}, which makes it among the most distant TeV BL Lac objects. 
PKS 1424+240 was observed in the framework of a multi-wavelength campaign during 2009 and 2010, allowing us to build a well covered simultaneous broad-band SED from radio to VHE regime \citep{aleksic2014_pks1424}.  
The diagnostic plots, resulting by our \textit{blazar} tool, are displayed in Fig. \ref{fig:1424_metodo}. 
We found excellent match with an HSP with z~$\simeq$~0.6 assuming a very high luminosity, although a correspondingly bad value of MAD.

\end{itemize}


\subsection{Test on non-blazar \textit{Fermi} extragalactic sources}

We have tested our method on extragalactic sources of the 3LAC catalogue that are classified as non-\textit{blazars}.
The results are summarised in Table \ref{tab:NOBZ_sample}. 
Double values correspond to multiple solutions with acceptable fit.

The only type-1 Seyfert galaxy, Circinus is clearly classified by our tool as a non-\textit{blazar} object.

In the 3LAC catalogue there are 5 steep spectrum radio quasars, but only 2 have a good X-ray coverage. For these we find that the global SEDs are not evidently distinguishable from classical \textit{blazars}.

We have considered 14 radio galaxies among the 16 present in the 3LAC catalogue (for the excluded sources there is no X-ray coverage): all objects are rejected as \textit{blazars} by our algorithm because of a global misfit, with the exception for PKS 0625-35, the only radio galaxy with a HSP SED reported in the 3LAC catalogue, and NGC 1218 that shows a marginal MAD value.

Finally about the Narrow-line Seyfert-1 galaxy (NLSG1) class, there are 5 objects reported in the 3LAC catalogue and we analysed them with our recognition tool. 
Note that the most marginal source among these is 1H 0323+342, having a marginally acceptable $\chi_{min}^{2}$. 
We classify this as an ISP with a consistent tentative redshift estimate by our method with respect to the correct one. 
All other four NLSG1s reveal fairly acceptable fits in our test as blazar objects. 
We do not consider in this paper the reason for this similarity between apparently different classes of sources, that will be discussed instead in a future paper.

\begin{table*}
\centering
\begin{tabular}{|c|c|c|c|c|c|c|c|c|}
\hline
\footnotesize{Name} & \footnotesize{3FGL Name } & \footnotesize{Opt Class} & \footnotesize{3FGL SED Class} & \footnotesize{Redshift} & \footnotesize{$\chi_{min}^{2}$} & \footnotesize{MAD} & \footnotesize{AGN Class} & \footnotesize{Redshift} \\ 
\footnotesize{ } & \footnotesize{ } & \footnotesize{ } & \footnotesize{ } & \footnotesize{ }  & \footnotesize{ } & \footnotesize{ } & \footnotesize{proposed} & \footnotesize{proposed } \\ 
\hline\hline
\footnotesize{Circinus galaxy } & \footnotesize{3FGL J1413.2-6518} & \footnotesize{sy} & \footnotesize{HSP}& \footnotesize{0.0015}  & \footnotesize{ 6.62}& \footnotesize{ 3.22} & \footnotesize{LSP} & \footnotesize{0.1} \\ 
\hline
\footnotesize{3C 207} & \footnotesize{3FGL J0840.8+1315 } & \footnotesize{ssrq } & \footnotesize{LSP } & \footnotesize{0.681}  & \footnotesize{0.53}& \footnotesize{1.54 } & \footnotesize{LSP} & \footnotesize{1.2} \\ 
\hline
\footnotesize{3C 380 } & \footnotesize{3FGL J1829.6+4844 } & \footnotesize{ssrq } & \footnotesize{LSP } & \footnotesize{0.695 }  & \footnotesize{0.91} & \footnotesize{1.81 } & \footnotesize{LSP} & \footnotesize{0.7 } \\ 
\hline
\footnotesize{ PKS 2004-447} & \footnotesize{3FGL J2007.8-4429 } & \footnotesize{nlsy1 } & \footnotesize{LSP } & \footnotesize{0.24 }  & \footnotesize{0.28 - 0.26 }& \footnotesize{1.83 - 2.34} & \footnotesize{ISP - ISP} & \footnotesize{0.4 - 0.7 } \\ 
\hline
\footnotesize{PMN J0948+0022 } & \footnotesize{3FGL J0948.8+0021 } & \footnotesize{nlsy1 } & \footnotesize{LSP } & \footnotesize{0.585 }  & \footnotesize{0.83 - 0.98 }& \footnotesize{2.62 - 1.81 } & \footnotesize{LSP - LSP} & \footnotesize{0.4 - 1.6 } \\ 
\hline
\footnotesize{SBS 0846+513 } & \footnotesize{3FGL J0849.9+5108 } & \footnotesize{nlsy1 } & \footnotesize{LSP } & \footnotesize{0.584 }  & \footnotesize{0.42 - 0.43 }& \footnotesize{1.87 -  1.92 } & \footnotesize{LSP - ISP} & \footnotesize{0.9 - 0.4 } \\ 
\hline
\footnotesize{ 1H 0323+342} & \footnotesize{3FGL J0325.2+3410 } & \footnotesize{nlsy1 } & \footnotesize{ HSP} & \footnotesize{ 0.061}  & \footnotesize{1.08}& \footnotesize{ 2.11 } & \footnotesize{ISP} & \footnotesize{ 0.1 } \\ 
\hline
\footnotesize{PKS 1502+036 } & \footnotesize{3FGL J1505.1+0326 } & \footnotesize{nlsy1 } & \footnotesize{LSP} & \footnotesize{0.409 }  & \footnotesize{0.40 - 0.45 }& \footnotesize{ 1.95 -  1.78 } & \footnotesize{LSP - ISP} & \footnotesize{ 0.9 - 0.4} \\ 
\hline
\footnotesize{NGC 1275 } & \footnotesize{3FGL J0319.8+4130 } & \footnotesize{rdg } & \footnotesize{LSP } & \footnotesize{0.0175 }  & \footnotesize{1.41 }& \footnotesize{2.45} & \footnotesize{LSP} & \footnotesize{0.1 } \\ 
\hline
\footnotesize{ IC 310} & \footnotesize{3FGL J0316.6+4119 } & \footnotesize{rdg } & \footnotesize{ - } & \footnotesize{0.019 }  & \footnotesize{2.27 }& \footnotesize{2.31 } & \footnotesize{ISP} & \footnotesize{0.1 } \\ 
\hline
\footnotesize{M 87 } & \footnotesize{3FGL J1230.9+1224 } & \footnotesize{ rdg} & \footnotesize{- } & \footnotesize{0.0043 }  & \footnotesize{1.51 }& \footnotesize{2.61} & \footnotesize{ISP} & \footnotesize{0.1 } \\ 
\hline
\footnotesize{ 3C 303} & \footnotesize{3FGL J1442.6+5156} & \footnotesize{rdg} & \footnotesize{-} & \footnotesize{ 0.014}  & \footnotesize{ 1.33}& \footnotesize{2.11 } & \footnotesize{ LSP} & \footnotesize{ 0.7} \\ 
\hline
\footnotesize{ 4C +39.12} & \footnotesize{3FGL J0334.2+3915} & \footnotesize{rdg} & \footnotesize{-} & \footnotesize{ 0.021}  & \footnotesize{ 5.19 } & \footnotesize{ 2.87} & \footnotesize{ ISP} & \footnotesize{0.1 } \\ 
\hline
\footnotesize{ NGC 6251} & \footnotesize{3FGL J1630.6+8232} & \footnotesize{rdg} & \footnotesize{LSP} & \footnotesize{ 0.02}  & \footnotesize{ 1.50}& \footnotesize{ 2.58 } & \footnotesize{ ISP } & \footnotesize{ 0.3} \\ 
\hline
\footnotesize{ Pic A} & \footnotesize{3FGL J0519.2-4542} & \footnotesize{rdg} & \footnotesize{LSP} & \footnotesize{ 0.035}  & \footnotesize{ 0.59 } & \footnotesize{ 2.96 } & \footnotesize{ ISP} & \footnotesize{0.3 } \\ 
\hline
\footnotesize{ PKS 0625-35} & \footnotesize{3FGL J0627.0-3529} & \footnotesize{rdg} & \footnotesize{HSP} & \footnotesize{0.055}  & \footnotesize{1.01 }& \footnotesize{ 1.88 } & \footnotesize{ HSP} & \footnotesize{ 0.05} \\ 
\hline
\footnotesize{ NGC 2484} & \footnotesize{3FGL J0758.7+3747} & \footnotesize{rdg} & \footnotesize{ISP} & \footnotesize{ 0.042}  & \footnotesize{  3.38  }& \footnotesize{  2.62 } & \footnotesize{ ISP} & \footnotesize{ 0.1} \\ 
\hline
\footnotesize{ Cen B} & \footnotesize{3FGL J1346.6-6027} & \footnotesize{rdg} & \footnotesize{ISP} & \footnotesize{ 0.013}  & \footnotesize{ 1.99 }& \footnotesize{ 2.012} & \footnotesize{ LSP} & \footnotesize{ 0.3} \\ 
\hline
\footnotesize{ Cen A Core} & \footnotesize{3FGL J1325.4-4301} & \footnotesize{rdg} & \footnotesize{LSP} & \footnotesize{ 0.0018}  & \footnotesize{ 2.59  }& \footnotesize{ 2.09 } & \footnotesize{ LSP} & \footnotesize{ 0.3} \\ 
\hline
\footnotesize{ 3C 111} & \footnotesize{3FGL J0418.5+3813c} & \footnotesize{rdg} & \footnotesize{ - } & \footnotesize{0.049 }  & \footnotesize{ 1.57   }& \footnotesize{2.11} & \footnotesize{ ISP} & \footnotesize{ 0.05} \\ 
\hline
\footnotesize{ NGC 1218 } & \footnotesize{3FGL J0308.6+0408} & \footnotesize{rdg} & \footnotesize{ISP} & \footnotesize{ 0.029}  & \footnotesize{  0.58}& \footnotesize{  2.30  } & \footnotesize{ ISP } & \footnotesize{ 0.05} \\ 
\hline
\footnotesize{ 3C 264} & \footnotesize{3FGL J1145.1+1935} & \footnotesize{rdg} & \footnotesize{-} & \footnotesize{ 0.022}  & \footnotesize{  1.55  }& \footnotesize{ 2.46  } & \footnotesize{ ISP} & \footnotesize{ 0.05} \\ 
\hline
\end{tabular}
\caption{Summary of the results of our test on non-\textit{blazar} sources.   Col.1: Object Name, Col.2: 3FGL Name, Col.3: Optical classification taken from 3LAC, Col.4: 3FGL SED Classification, Col.5: Redshift taken from NED, Col.6: $\chi_{min}^{2}$, Col.7: MAD, Col.8: Tool Classification proposed, Col.9: Redshift proposed. \newline   We remark that the solutions with $\chi_{min}^{2} >$1.1 or MAD $>$ 2.5 cannot be considered acceptable and the object is not recognised as a \textit{blazar}-like source.
}
\label{tab:NOBZ_sample}
\end{table*}


\subsection{A counter-example:  2FGL J1544.5-1126  (3FGL J1544.6-1125)} 
\label{1544_id}

Our tool is also suited to exclude a \textit{blazar} recognition: we consider, for example,  the source 2FGL J1544.5-1126 in our UGS catalogue, a rather complicated case.

\begin{figure*}
  \centering
  \begin{minipage}[c]{.45\textwidth}
    \includegraphics[width=1.\textwidth, angle=0]{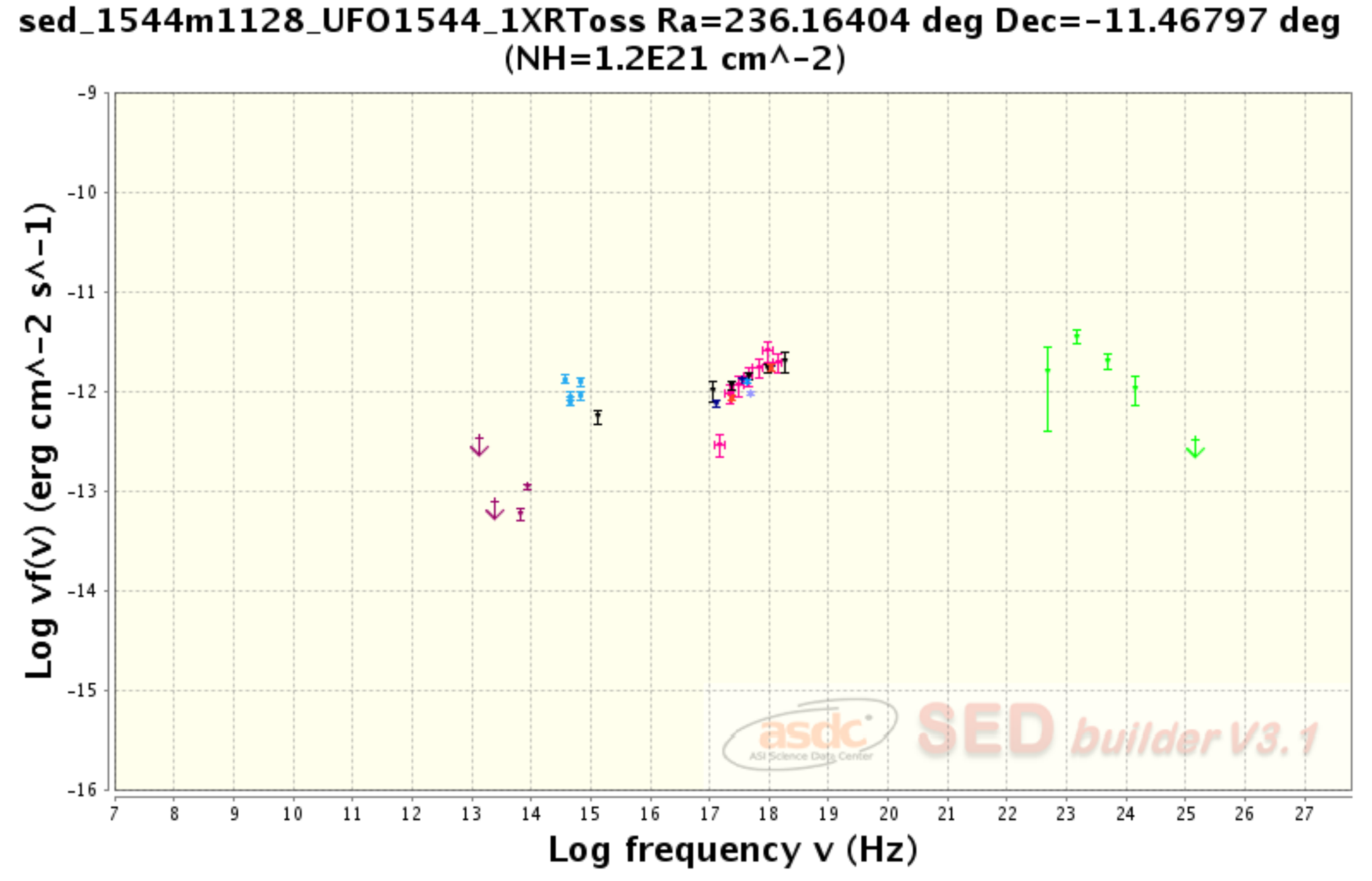}
  \end{minipage}%
  \hspace{3mm}%
  \begin{minipage}[c]{.45\textwidth}
    \includegraphics[width=1.\textwidth, angle=0]{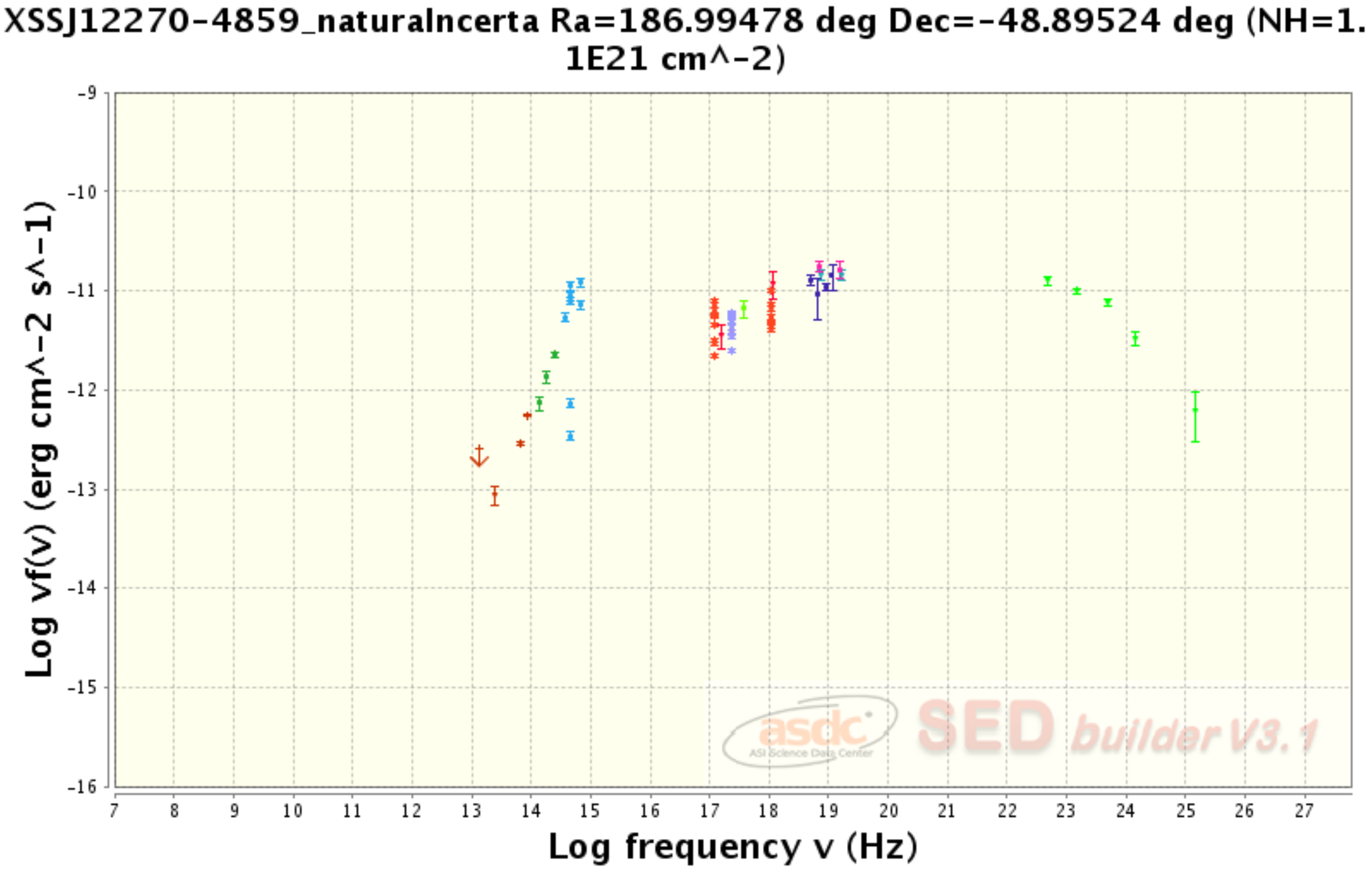}
  \end{minipage}
  \caption{Comparison of the SEDs of 2FGL J1544.5-1126 and the low-galactic latitude transitional millisecond pulsar binaries XSS J1227-4859 (3FGL J1227.9-4854), two sources for which a galactic origin is suggested. The two SED behaviour appear virtually identical.}
	\label{fig:1544_sed_comparison}
\end{figure*}

From the \textit{Swift}/XRT observations of this UGS error-box (see Fig. \ref{fig:1544_ass} in Appendix A), we proposed the X-ray source 1RXS J154439.4-112820, the brightest X-ray source in the field, as the likely X-ray counterpart of the \textit{Fermi} source. 
This association is also proposed by \citet{bogdanov2015}, who identify 2FGL J1544.5-1126 as a transitional millisecond pulsar binary in an accretion state. 
They also note that the \textit{Fermi} source 3FGL J1227.9-4854 (2FGL J1227.7-4853 in the 2FGL catalogue), associated with the transitional millisecond pulsar binaries XSS J12270-4859 \citep{bonnet2012_xss}, has a radio-to-$\gamma$ SED very similar to the 2FGL J1544.5-1126 SED as illustrated in Fig. \ref{fig:1544_sed_comparison}.

We applied our \textit{blazar} recognition procedure (see Fig. \ref{fig:1544_metodo}), assuming 1RXS J154439.4-112820 as the most likely counterpart.
From the $\chi^2_{\nu,min}$~/~MAD plot, we note that actually no SED template can suitably match the observational data, independently of the assumed redshift (all $\chi^2_{\nu,min}$ values are too high).
Hence our results suggest that a standard \textit{blazar} classification for this source is quite unlikely.

The nature of the brighter X-ray source 1RXS J154439.4-112820 has been studied with optical spectroscopy by \citet{masetti2013_rosat}. These data show a Galactic source characterised by broad emission lines (Balmer series and helium transitions, EW$\sim$20~$\AA$, FWHM $>$ 800 km s$^{-1}$).

\subsection{\textbf{Testing the rate of false positive recognitions with known pulsars and other Galactic sources}}
\label{pulsars}

We have further tested the validity of our \textit{blazar} recognition tool against the observed SEDs of 15 Galactic sources of different HE classes included in the 3FGL catalogue, in order to verify the chance of false recognition.
The objects have a spectral coverage similar to that of our \textit{blazar} SED template set and the multi-wavelength fluxes are retrieved from the ASDC archive.
The selected sample of Galactic objects includes the following.

\begin{itemize}
\item Seven pulsars detected in the \textit{Fermi} surveys with good multi-wavelength coverage and with a galactic latitude higher than 20 degrees. 
They are PSR J0437-4715, PSR J0614-3329, PSR J1024-0719, PSR J1614-2230, PSR J2124-3358, plus the well-known Vela pulsar and the brightest HE millisecond pulsar PSR J2339-0533. 
We excluded the Geminga pulsar because lacking sufficient spectral coverage.

\item The Crab Nebula, whose pulsar wind nebula is assumed as the standard candle in high-energy astrophysics.

\item V407 Cyg, the only nova known in the 3FGL catalogue.

\item The HE binary Eta Carinae and the three high-mass binaries LS 5039, 1FGL J1018.6-5856, and LS I+61 303.

\item The Supernova Remnants Cas A and SNR G349+002, also present in the 3FGL catalogue.
\end{itemize}

\begin{figure*}
\includegraphics[height=.3\textwidth,width=0.32\textwidth]{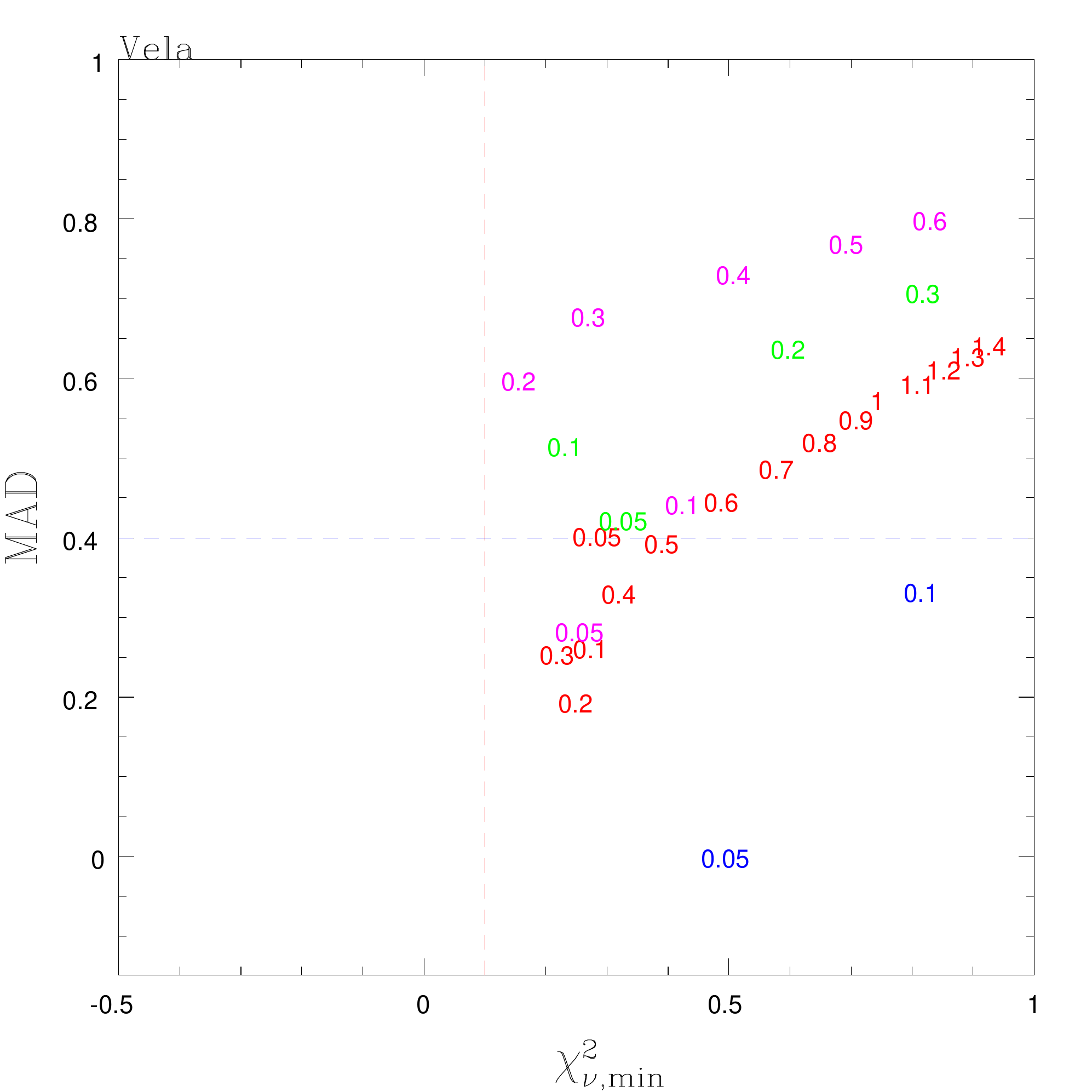}
\includegraphics[height=.3\textwidth,width=0.32\textwidth]{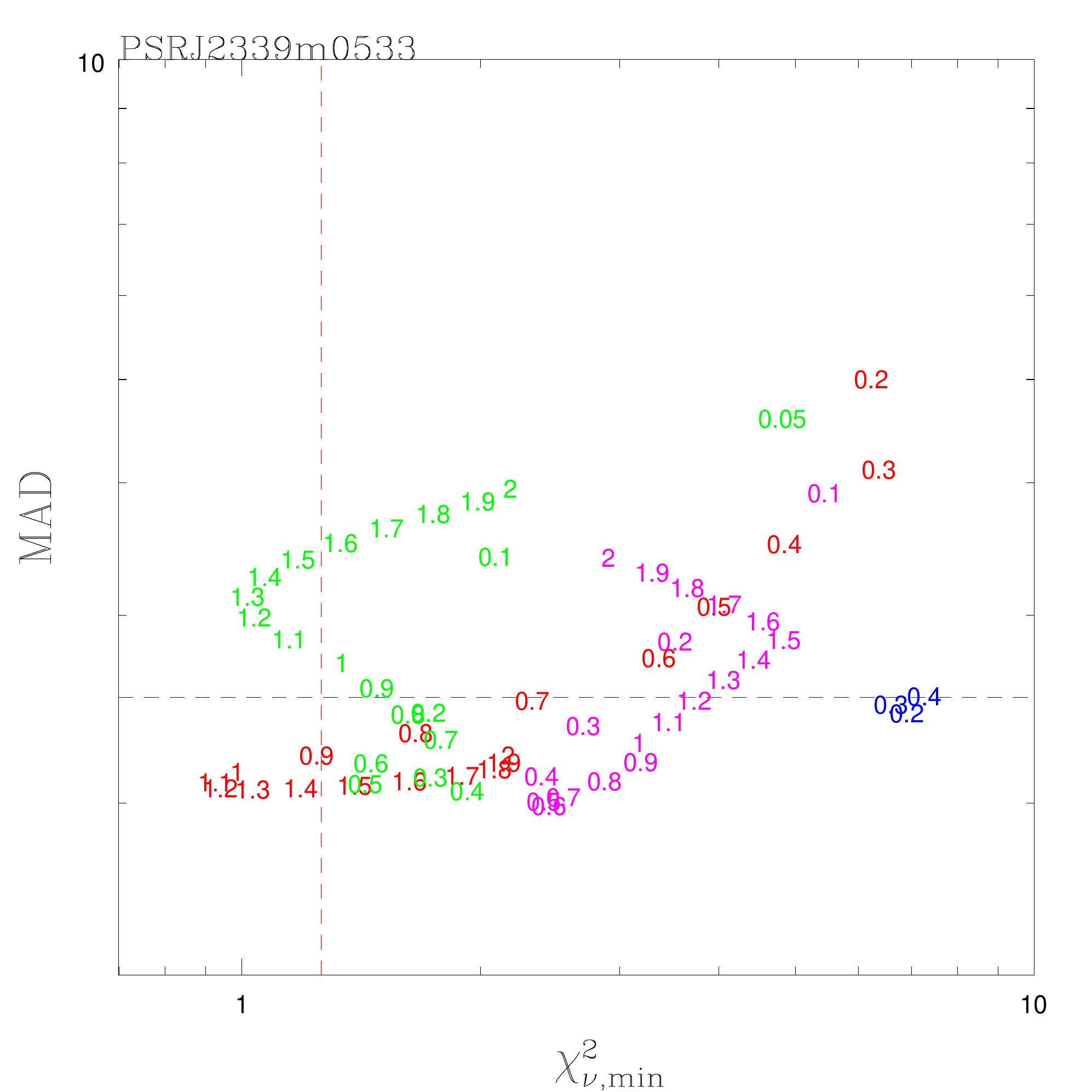}
\includegraphics[height=.3\textwidth,width=.32\textwidth]{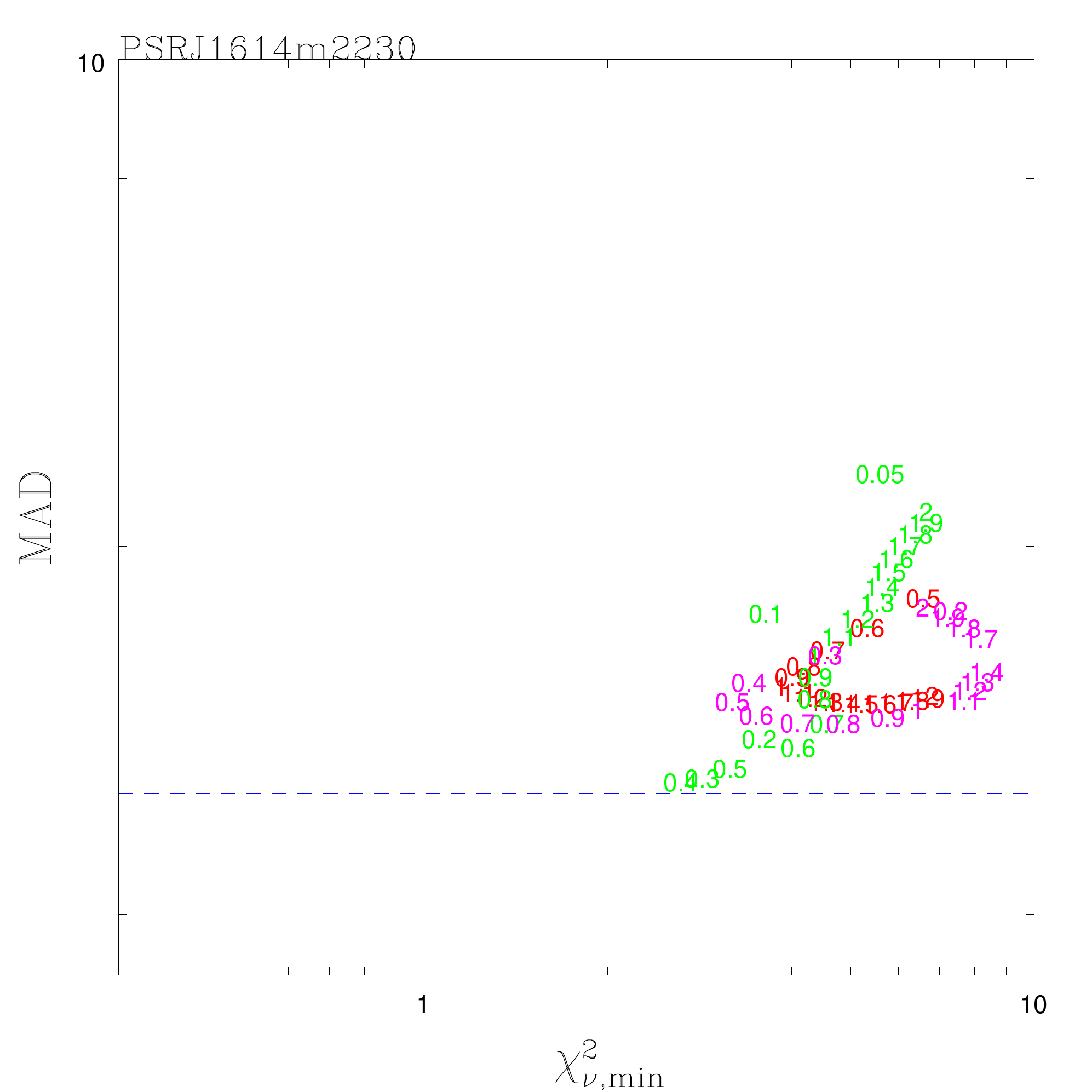}
\includegraphics[height=.3\textwidth,width=.32\textwidth]{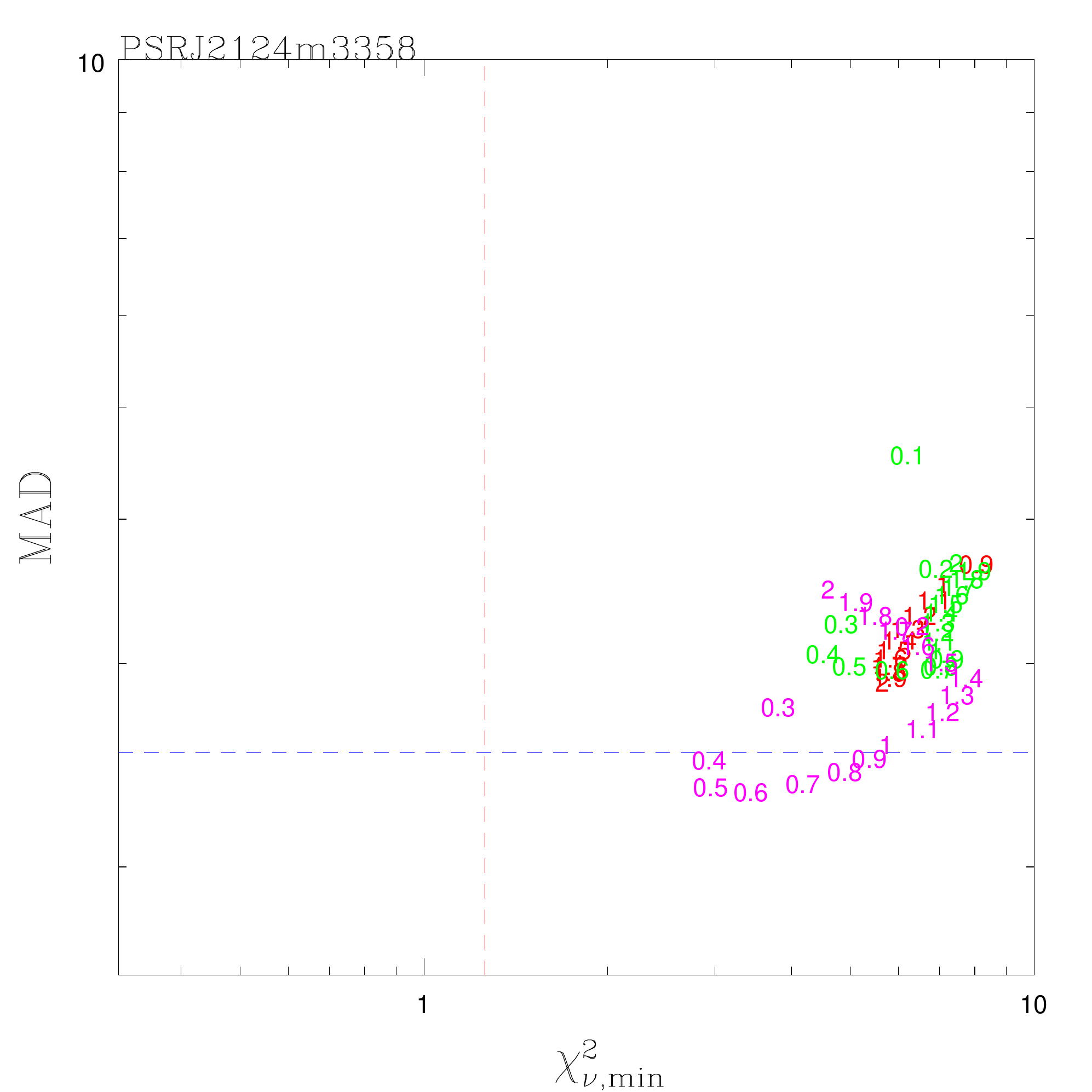}
\includegraphics[height=.3\textwidth,width=.32\textwidth]{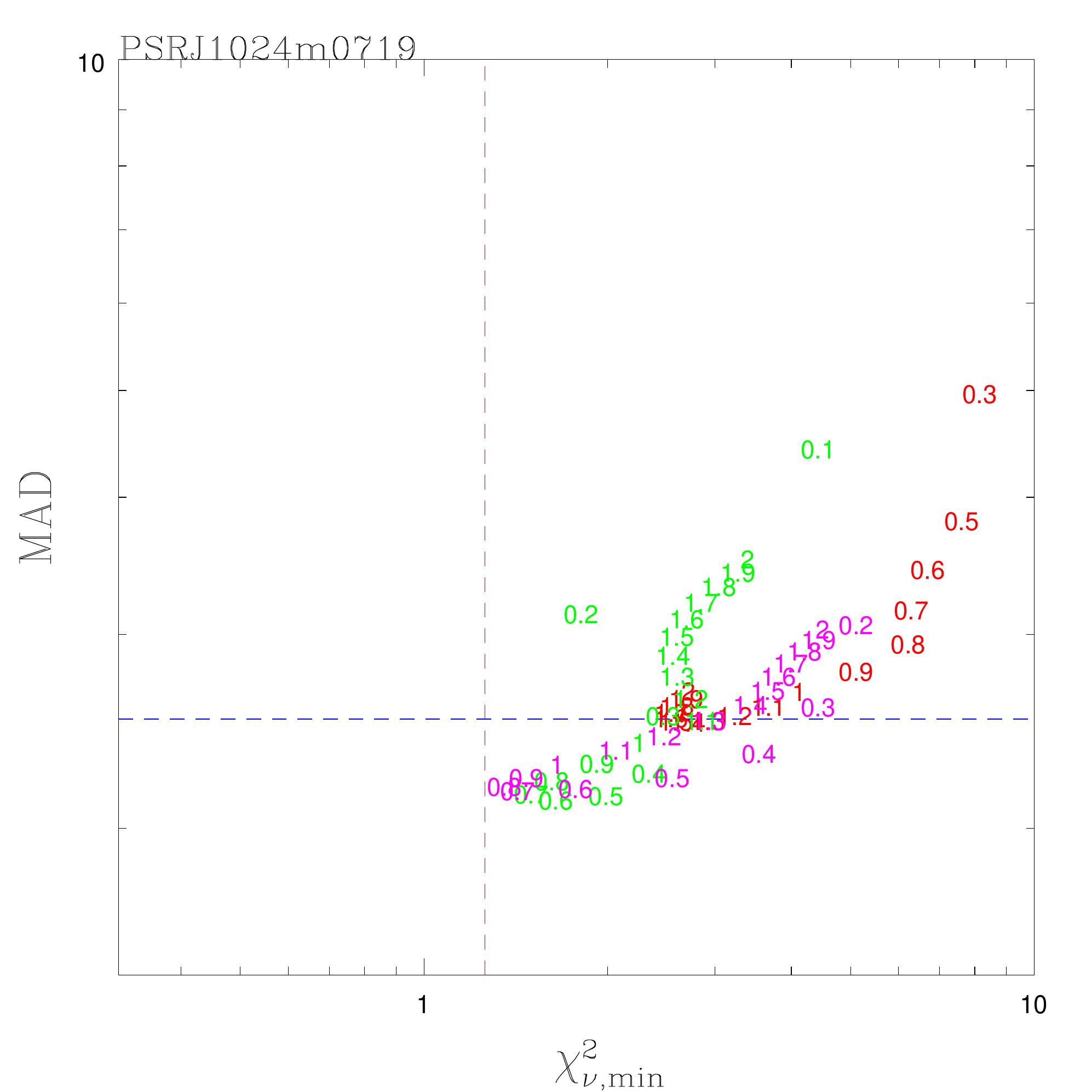}
\includegraphics[height=.3\textwidth,width=.32\textwidth]{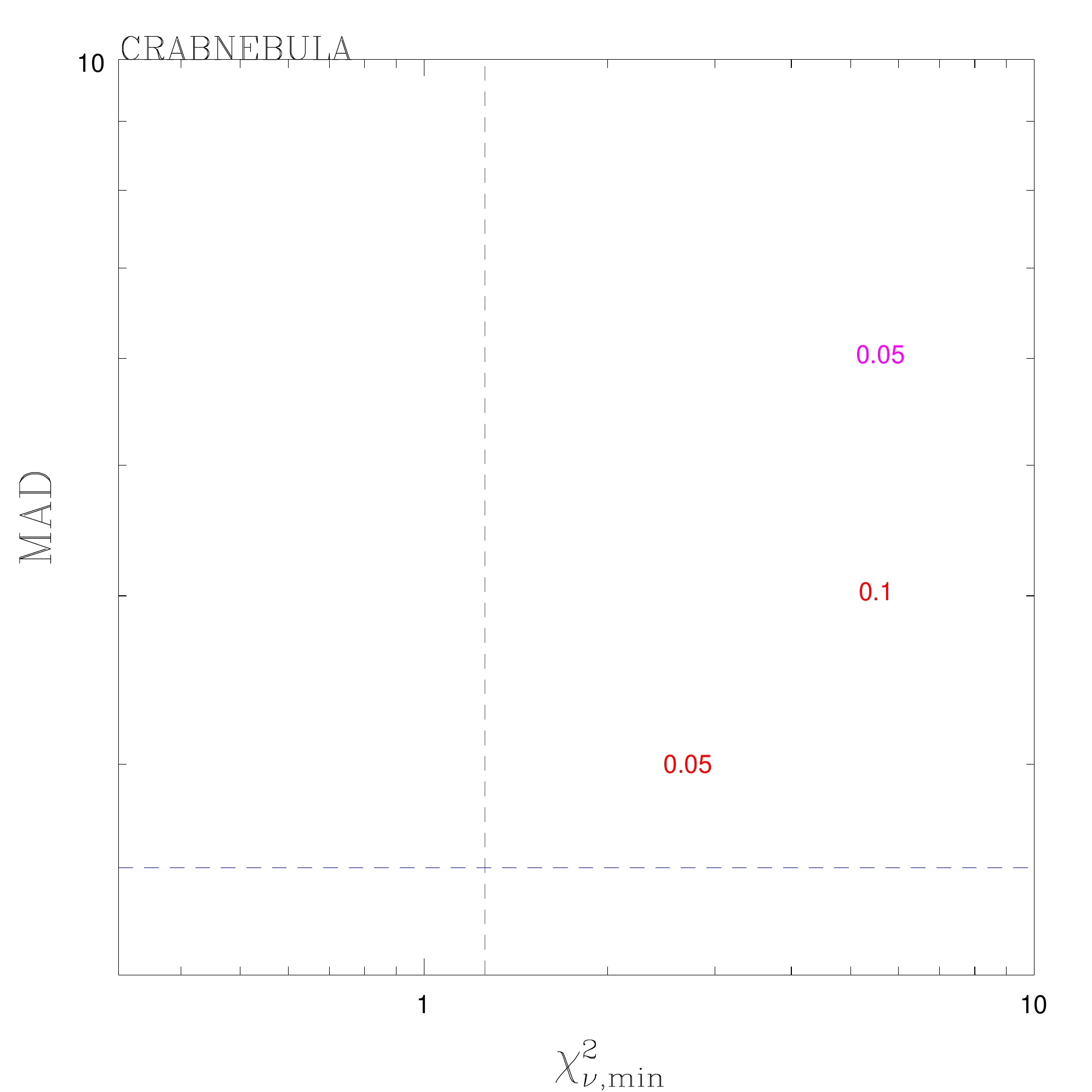}
\includegraphics[height=.3\textwidth,width=.32\textwidth]{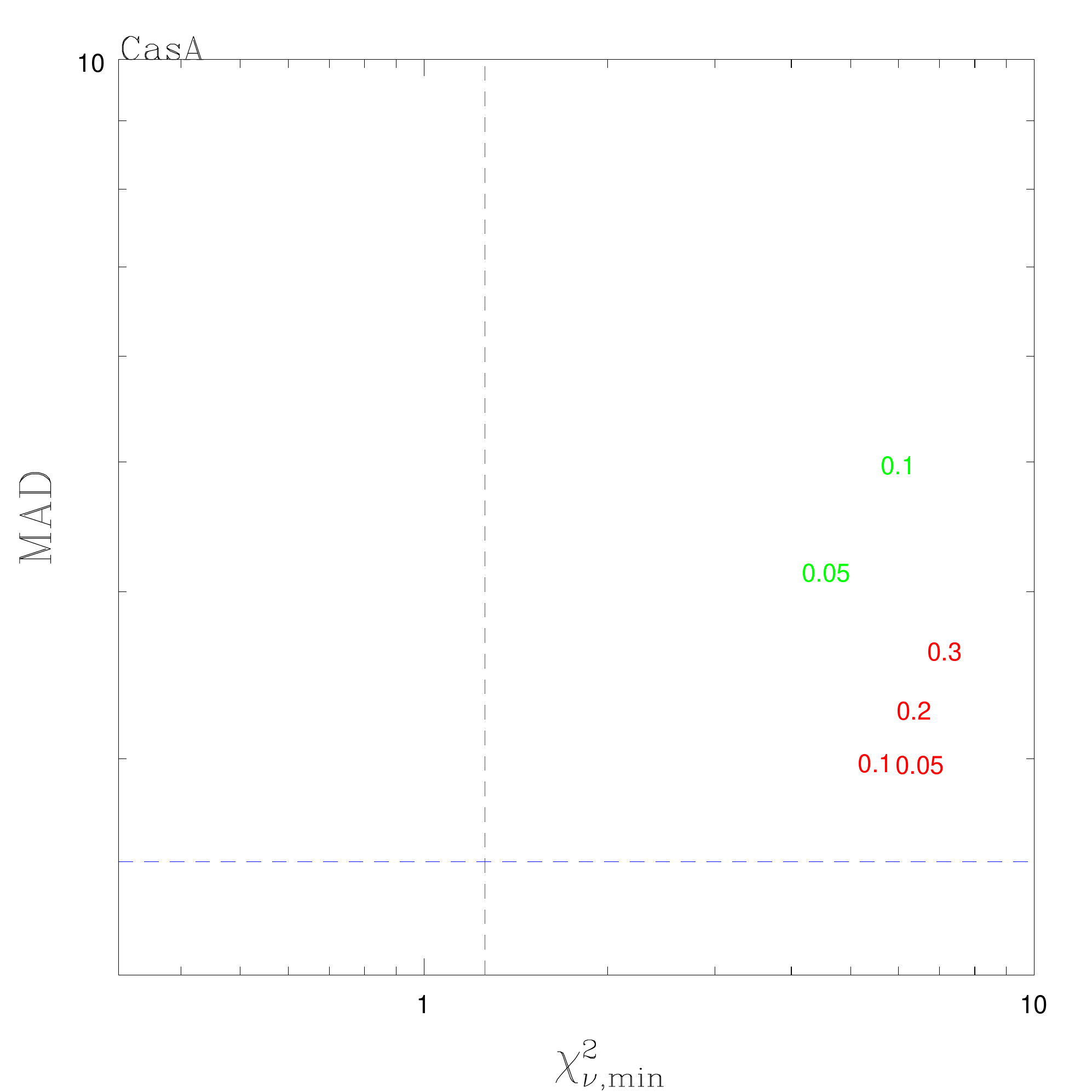}
\includegraphics[height=.3\textwidth,width=.32\textwidth]{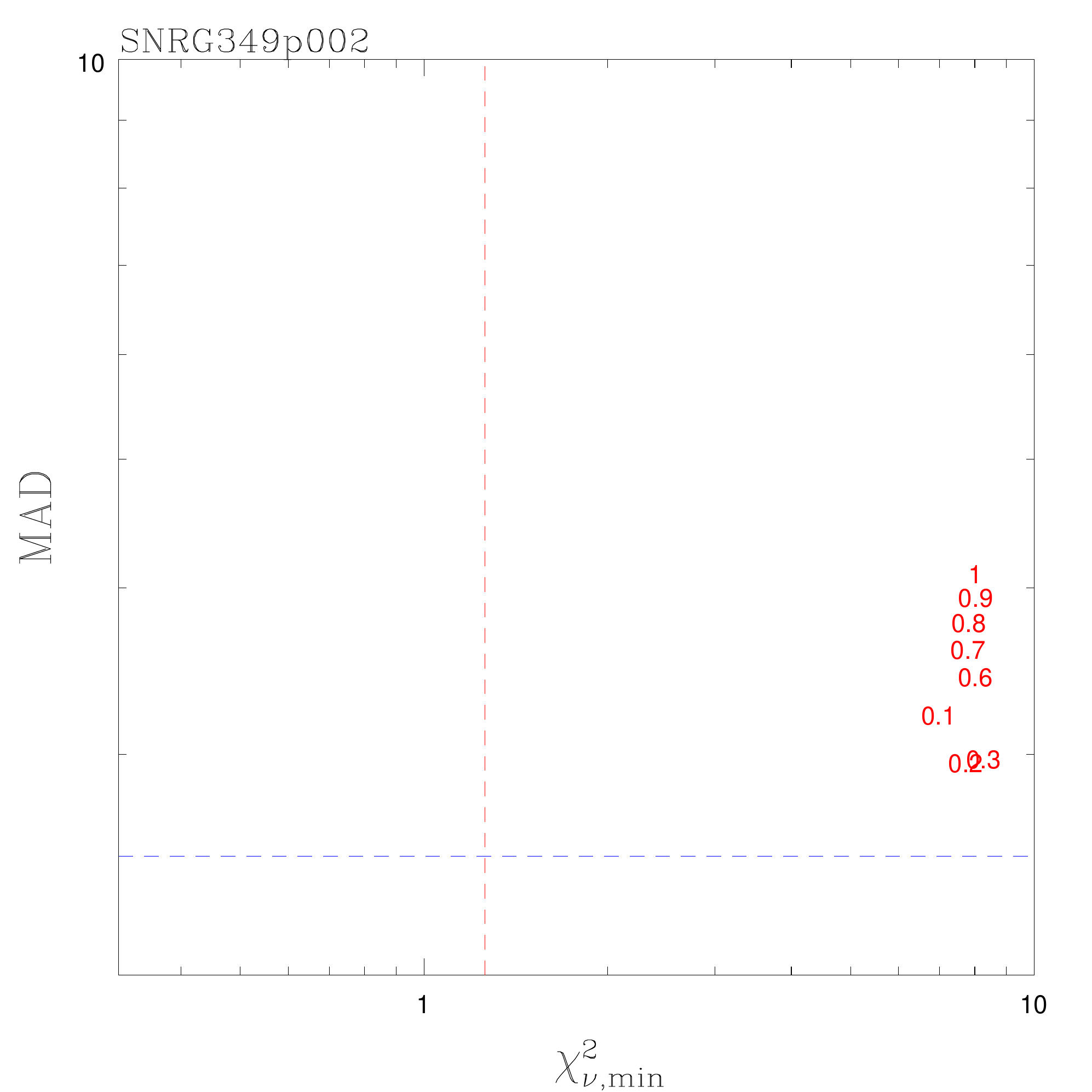}
\includegraphics[height=.3\textwidth,width=.32\textwidth]{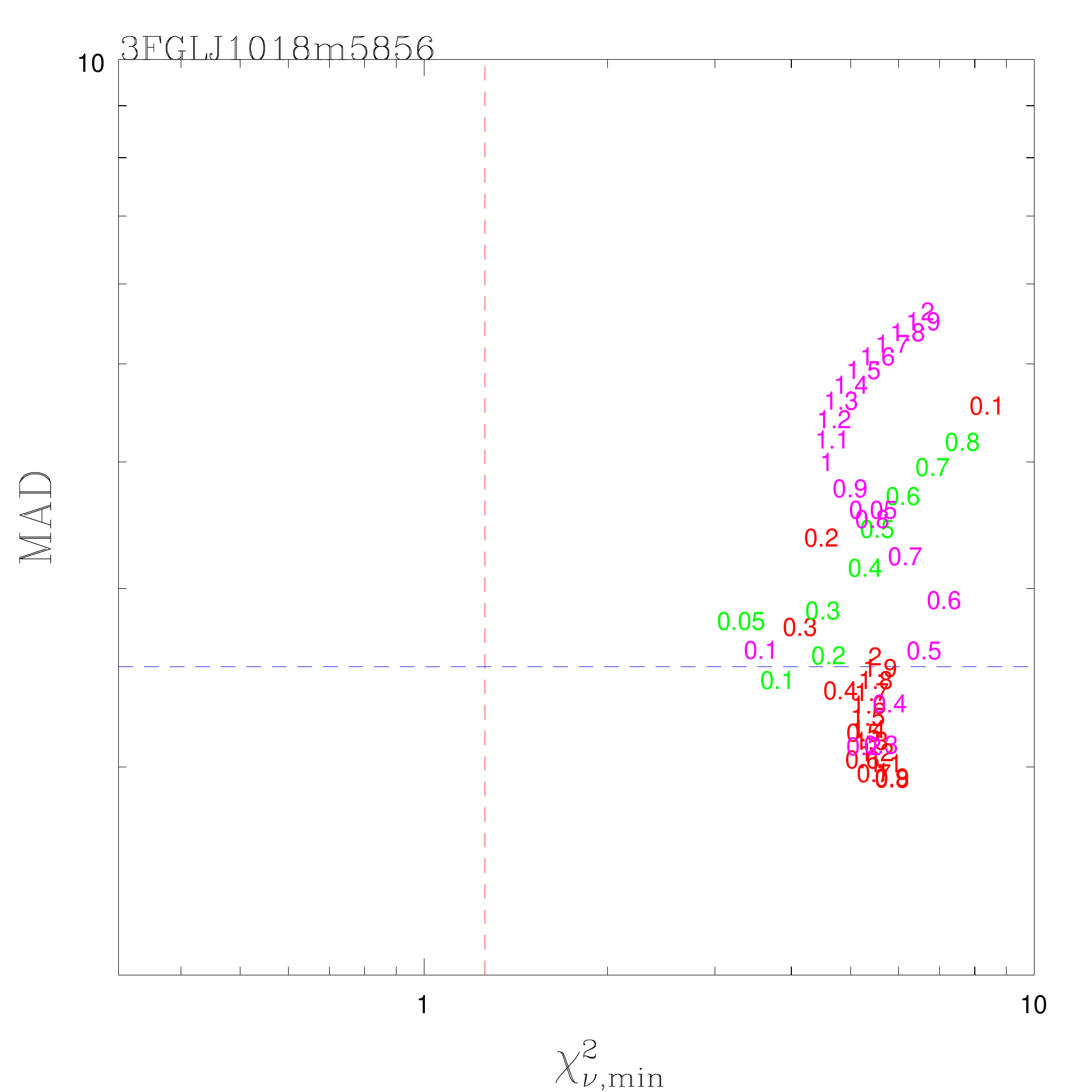}
\includegraphics[height=.3\textwidth,width=.32\textwidth]{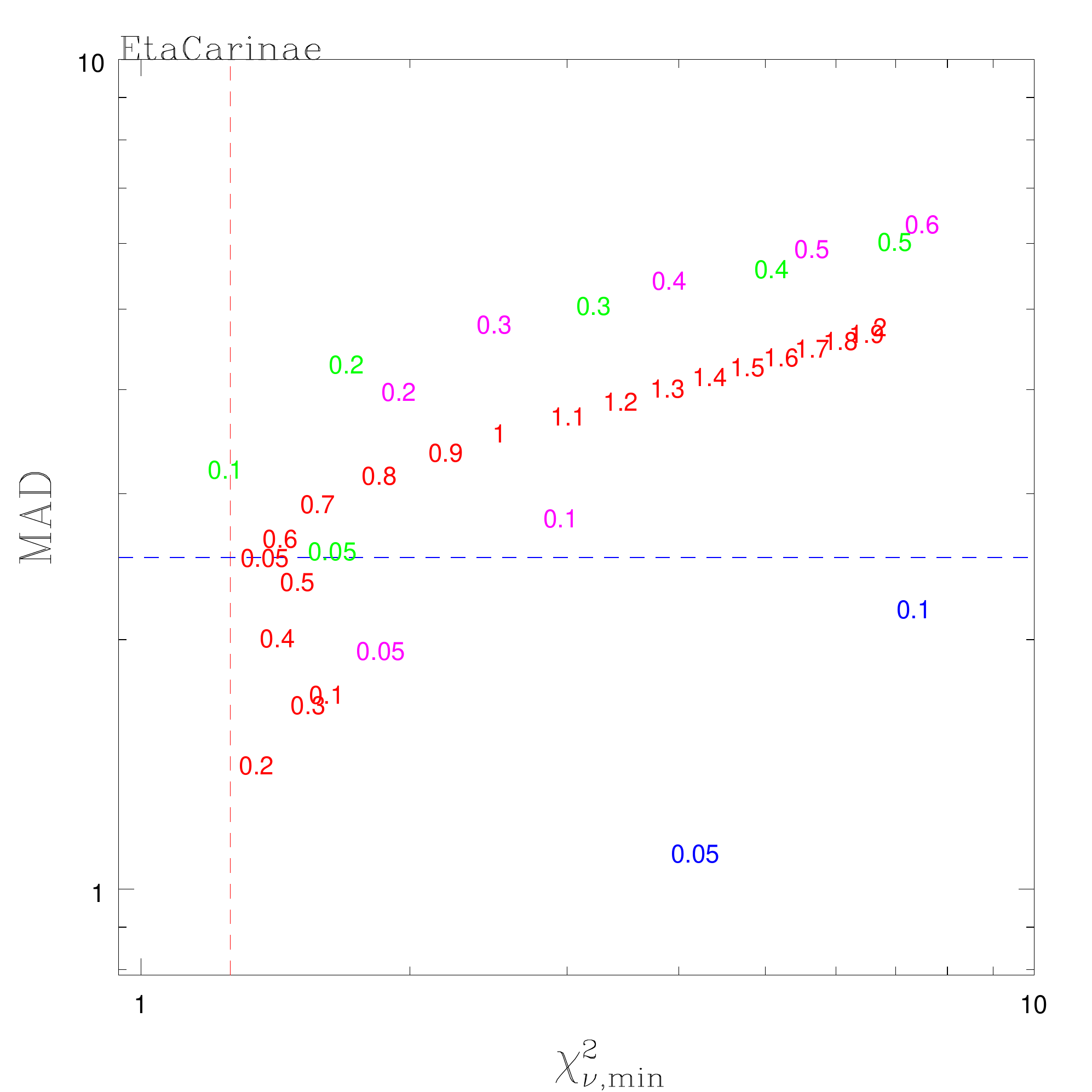}
\includegraphics[height=.3\textwidth,width=.32\textwidth]{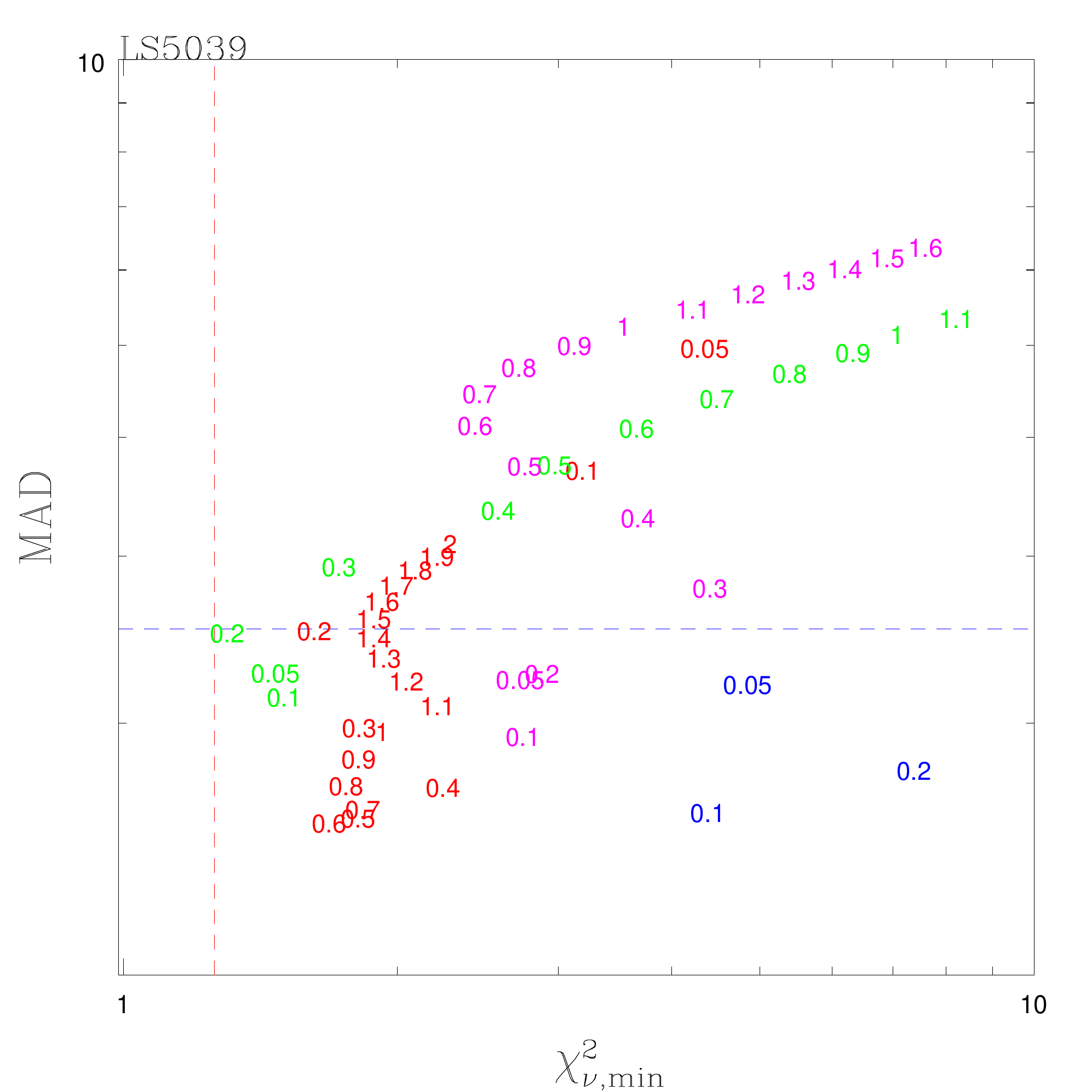}
\caption{A graphical summary of the results of our \textit{blazar} recognition method for Galactic sources.
The plots show the values of MAD and $\chi_{min}^{2}$ obtained by comparing photometric data for the sources and the SED template set for a wide range of redshifts. The high values of the statistics indicate these sources as non-\textit{blazar}s.
}
  \label{fig:pulsar_results}
\end{figure*}

In Fig. \ref{fig:pulsar_results}  we report a graphical summary of the results obtained applying our method to 11 of the test Galactic sources. 
With no exception, these show high or very high values of $\chi^2_{\nu,min}>1$, and also typically large values for MAD. 
For the remaining four cases,  the inferred values of $\chi^2_{\nu,min}$ and MAD exceed our considered boundaries of 10 in both statistics and therefore the algorithm plots are not shown. 
It is worth to note that for the pulsar PSR J2339-0533, our method would find acceptable fits for the LSP class at high redshift (z=1.1-1.3). 
Indeed we know that the source is located at 450 pc \citep{rayewass2015}, such that our inferred radio luminosity 
($\sim$10$^{44}$ erg s$^{-1}$) from the best-fit spectrum would imply an enormous continuum flux if located at that distance, that is excluded by the (unpublished) observational radio upper limits reported in \citet{rayewass2015}.
A radio detection or an upper limit would immediately rule out a \textit{blazar} classification for this source. 
This illustrates that the radio data may result significant constraints on the UGS classification.   
\newline
In summary, our \textit{blazar} recognition tool appears rather robust against mis-interpreting pulsars and Galactic sources as AGNs.

\begin{figure*}
\centering
\mbox{%
\begin{minipage}{.65\textwidth}
\includegraphics[height=0.9\textwidth,width=\textwidth]{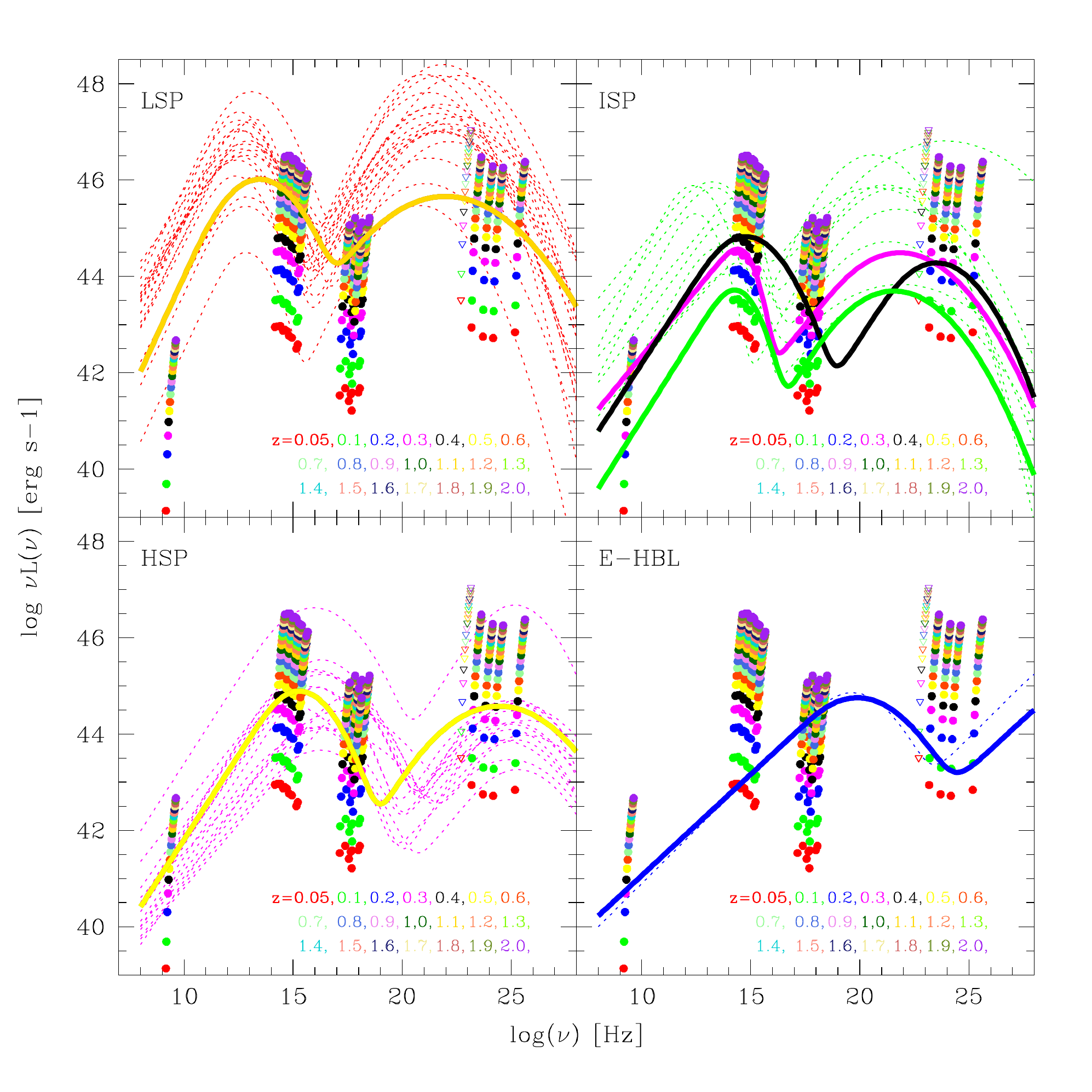}
\end{minipage}%
\begin{minipage}[c]{.40\textwidth}
\quad \quad \quad \quad
 \begin{tabular}{cccc}
\hline
\footnotesize{Class}  & \footnotesize{$\chi_{\nu,min}^{2}$} &  \footnotesize{MAD} &\footnotesize{$z$}\\
\hline
\footnotesize{LSP}  & \footnotesize{0.88} & \footnotesize{2.29} & \footnotesize{1.1}\\
\footnotesize{ISP}  & \footnotesize{0.64} & \footnotesize{3.33} & \footnotesize{0.1}\\
\footnotesize{ISP}  & \footnotesize{0.71} & \footnotesize{1.96} & \footnotesize{0.3}\\
\footnotesize{ISP}  & \footnotesize{0.78} & \footnotesize{1.92} & \footnotesize{0.4}\\
\footnotesize{\bf{HSP}}  & \footnotesize{\bf{0.43}} & \footnotesize{\bf{1.81}} & \footnotesize{\bf{0.5}}\\
\footnotesize{EHBL} & \footnotesize{7.31} & \footnotesize{2.47} & \footnotesize{0.2}\\
\hline
\end{tabular}
  \\
\includegraphics[height=.9\textwidth,width=.9\textwidth]{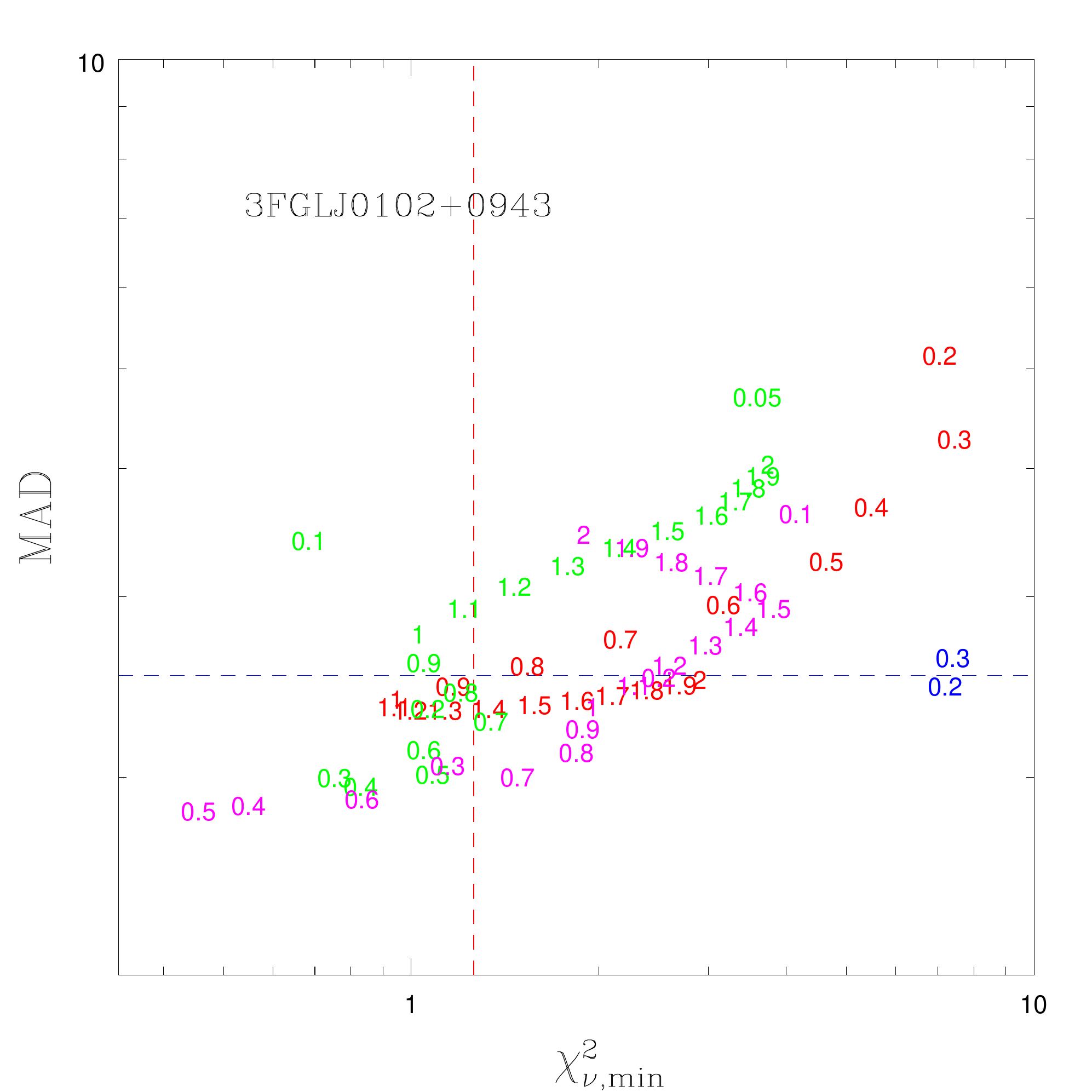}
\end{minipage}
}
\caption{Spectral luminosity points of 2FGL J0102.2+0943 for different assumed redshift (from 0.05 to 2.0), compared to the \textit{blazar} SED templates. The latter are built from archive data of known \textit{blazars} for the four classes: LSP, ISP, HSP and EHBL. The values of MAD and $\chi_{\nu,min}^{2}$ reported in the table are referred to the best fit SED template of each class (\textit{bold coloured line}). The colour of the selected SED template refers to the best-fit redshift indicated by our method. For 2FGL J0102.2+0943 the best-guess recognition is an HSP at $z\sim$0.5 (\textit{yellow line}).}
\label{fig:0102_metodo}
\end{figure*}

\begin{figure*}
\centering
\mbox{%
\begin{minipage}{.65\textwidth}
\includegraphics[height=0.9\textwidth,width=\textwidth]{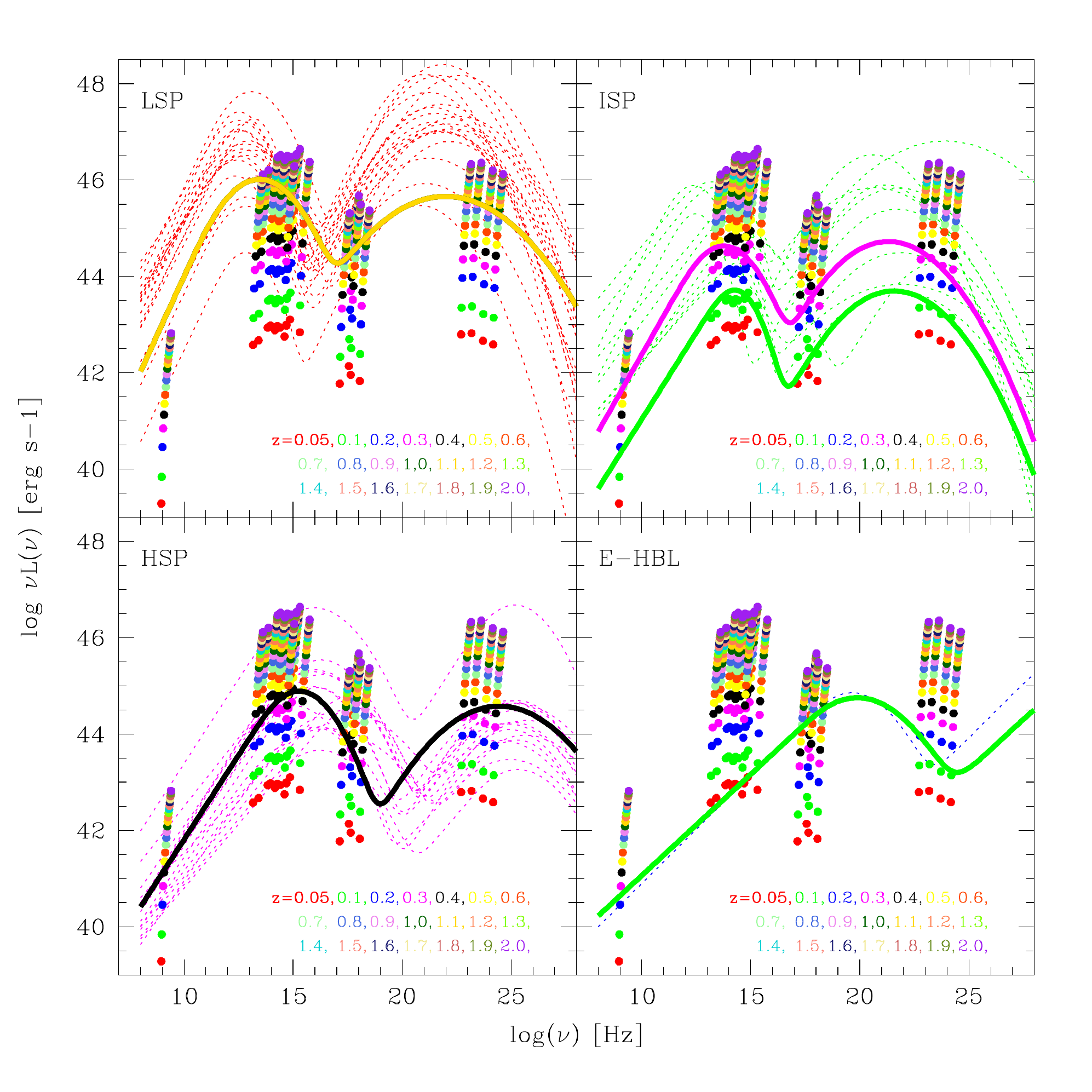}
\end{minipage}%
\begin{minipage}[c]{.40\textwidth}
\quad \quad \quad \quad
 \begin{tabular}{cccc}
\hline
\footnotesize{Class}  & \footnotesize{$\chi_{\nu,min}^{2}$} &  \footnotesize{MAD} &\footnotesize{$z$}\\
\hline
\footnotesize{LSP}  & \footnotesize{0.67} & \footnotesize{2.11} & \footnotesize{1.1}\\
\footnotesize{ISP}  & \footnotesize{0.33} & \footnotesize{3.08} & \footnotesize{0.1}\\
\footnotesize{ISP}  & \footnotesize{0.41} & \footnotesize{1.78} & \footnotesize{0.3}\\
\footnotesize{\bf{HSP}}  & \footnotesize{\bf{0.27}} & \footnotesize{\bf{1.62}} & \footnotesize{\bf{0.4}}\\
\footnotesize{EHBL} & \footnotesize{4.80} & \footnotesize{1.78} & \footnotesize{0.1}\\
\hline
\end{tabular}
  \\
\includegraphics[height=.9\textwidth,width=.9\textwidth]{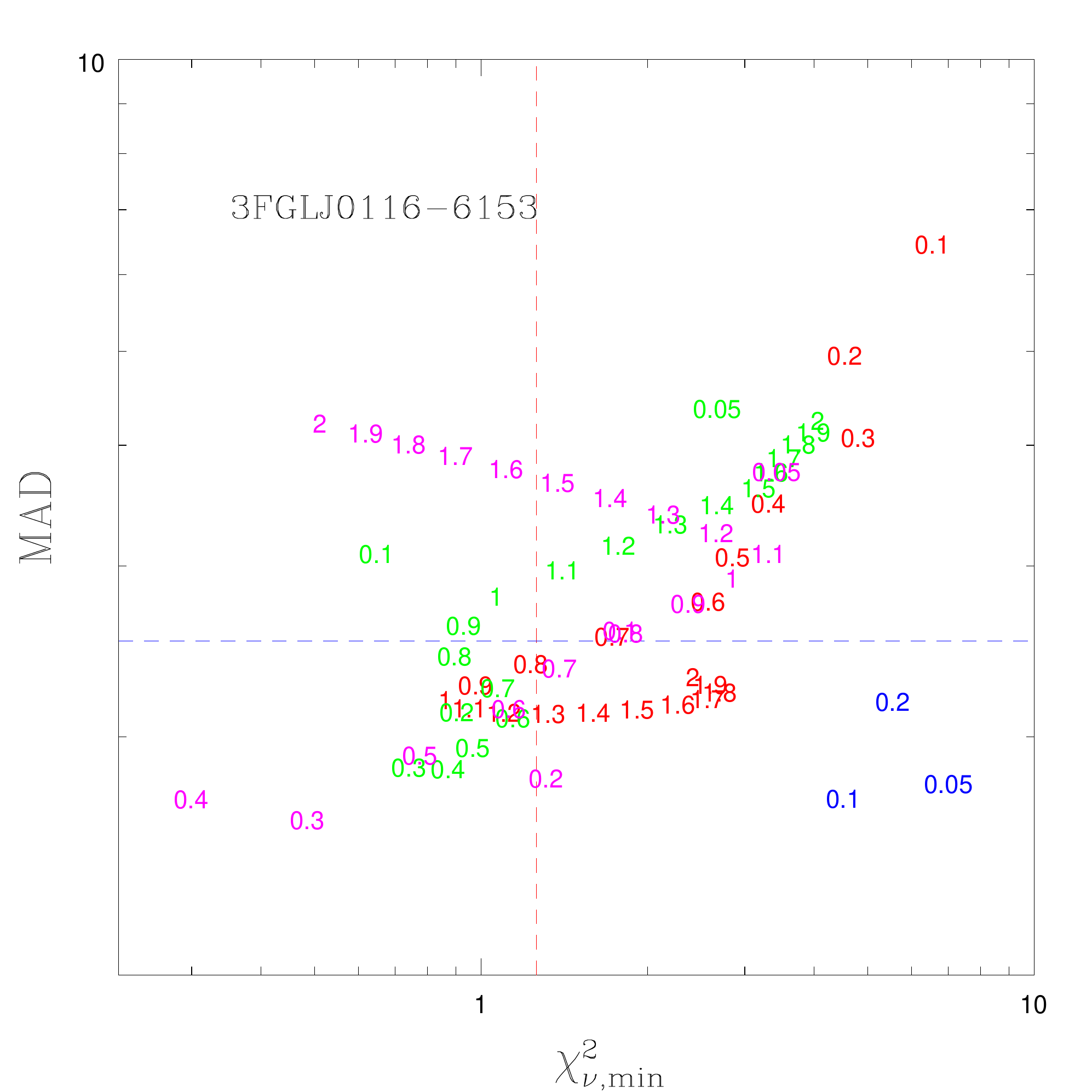}
\end{minipage}
}
\caption{The same diagnostic plots as in Fig. \ref{fig:0102_metodo} for the source 2FGL J0116.6-6153. The best-guess recognition is an HSP at $z\sim$0.4 (\textit{black line}). }
\label{fig:0116_metodo}
\end{figure*}


\begin{figure*}
\centering
\mbox{%
\begin{minipage}{.65\textwidth}
\includegraphics[height=0.9\textwidth,width=\textwidth]{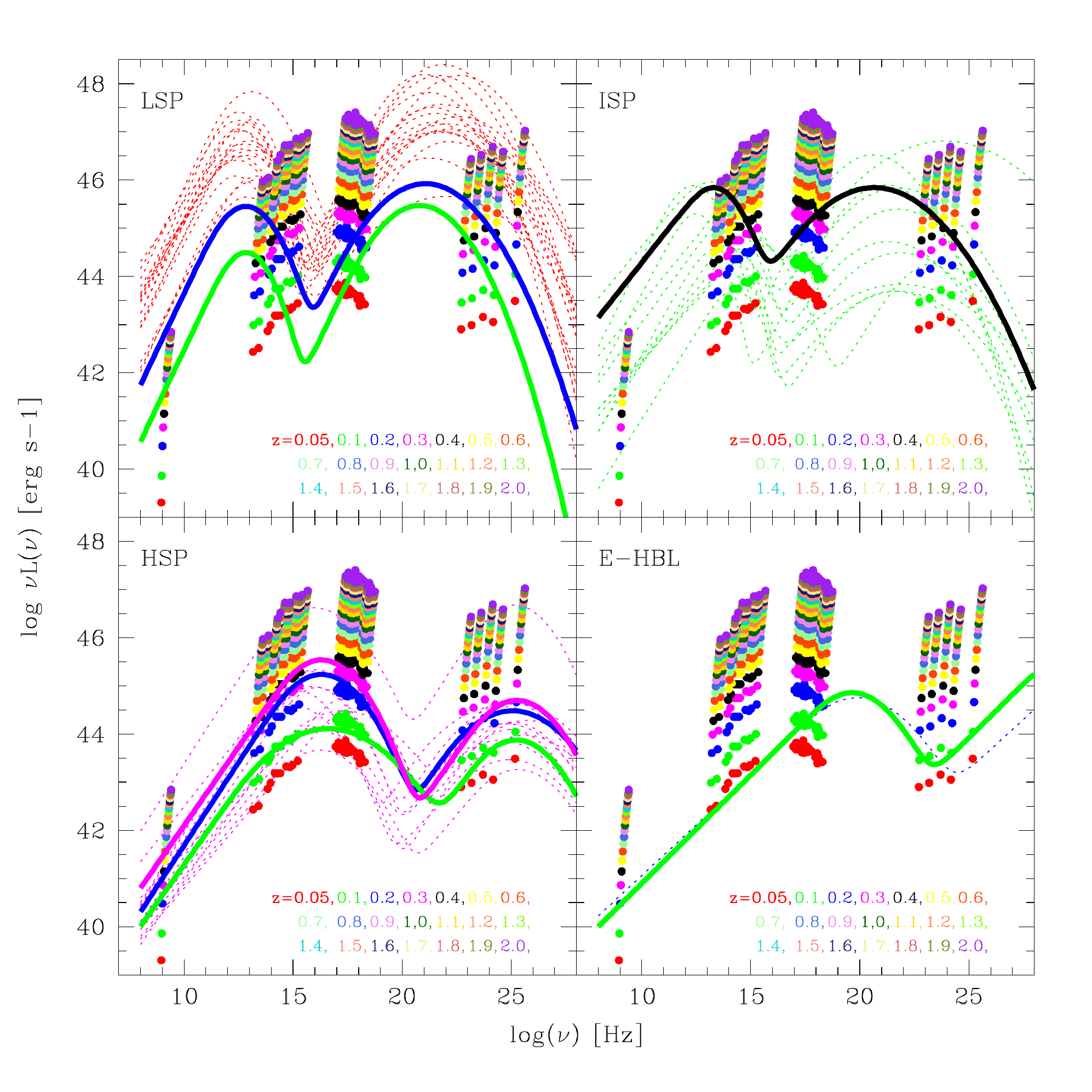}
\end{minipage}%
\begin{minipage}[c]{.40\textwidth}
\quad \quad \quad \quad
 \begin{tabular}{cccc}
\hline
\footnotesize{Class}  & \footnotesize{$\chi_{\nu,min}^{2}$} &  \footnotesize{MAD} &\footnotesize{$z$}\\
\hline
\footnotesize{LSP}  & \footnotesize{2.75} & \footnotesize{3.75} & \footnotesize{0.1}\\
\footnotesize{LSP}  & \footnotesize{2.76} & \footnotesize{2.63} & \footnotesize{0.2}\\
\footnotesize{ISP}  & \footnotesize{2.45} & \footnotesize{3.14} & \footnotesize{0.4}\\
\footnotesize{\bf{HSP}}  & \footnotesize{\bf{0.27}} & \footnotesize{\bf{1.51}} & \footnotesize{\bf{0.1}}\\
\footnotesize{\bf{HSP}}  & \footnotesize{\bf{0.28}} & \footnotesize{\bf{1.45}} & \footnotesize{\bf{0.2}}\\
\footnotesize{\bf{HSP}}  & \footnotesize{\bf{0.26}} & \footnotesize{\bf{1.87}} & \footnotesize{\bf{0.3}}\\
\footnotesize{EHBL} & \footnotesize{1.09} & \footnotesize{0.81} & \footnotesize{0.1}\\
\hline
\end{tabular}
  \\
\includegraphics[height=.9\textwidth,width=.9\textwidth]{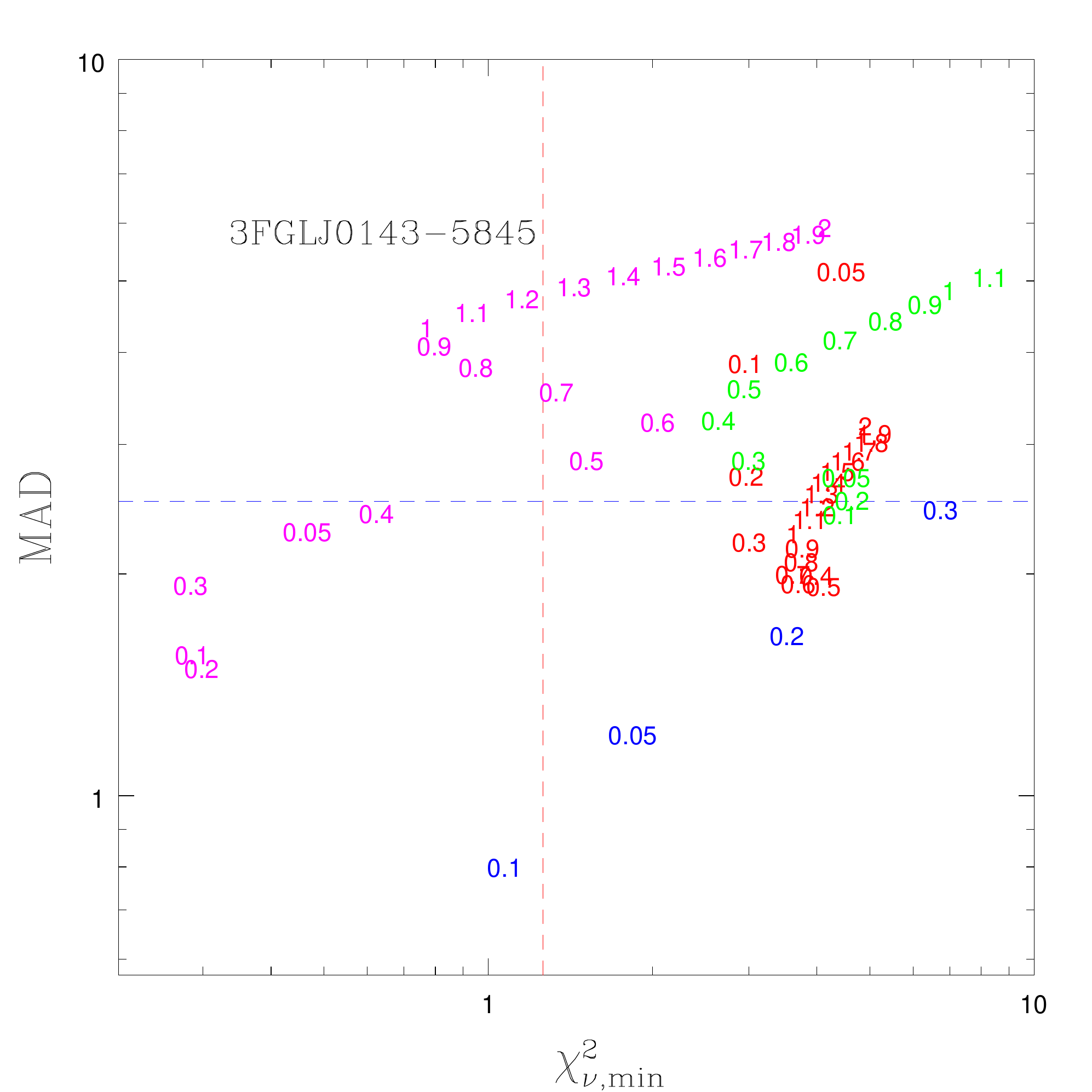}
\end{minipage}
}
\caption{The same diagnostic plot as in Fig. \ref{fig:0102_metodo} for the source 2FGL J0143.6-5844. Here the best-guess recognition is a HSP at $z~\sim$~0.1- 0.3 (\textit{green, blue and magenta line}).}
\label{fig:0143_metodo}
\end{figure*}


\begin{figure*}
\centering
\mbox{%
\begin{minipage}{.65\textwidth}
\includegraphics[height=0.9\textwidth,width=\textwidth]{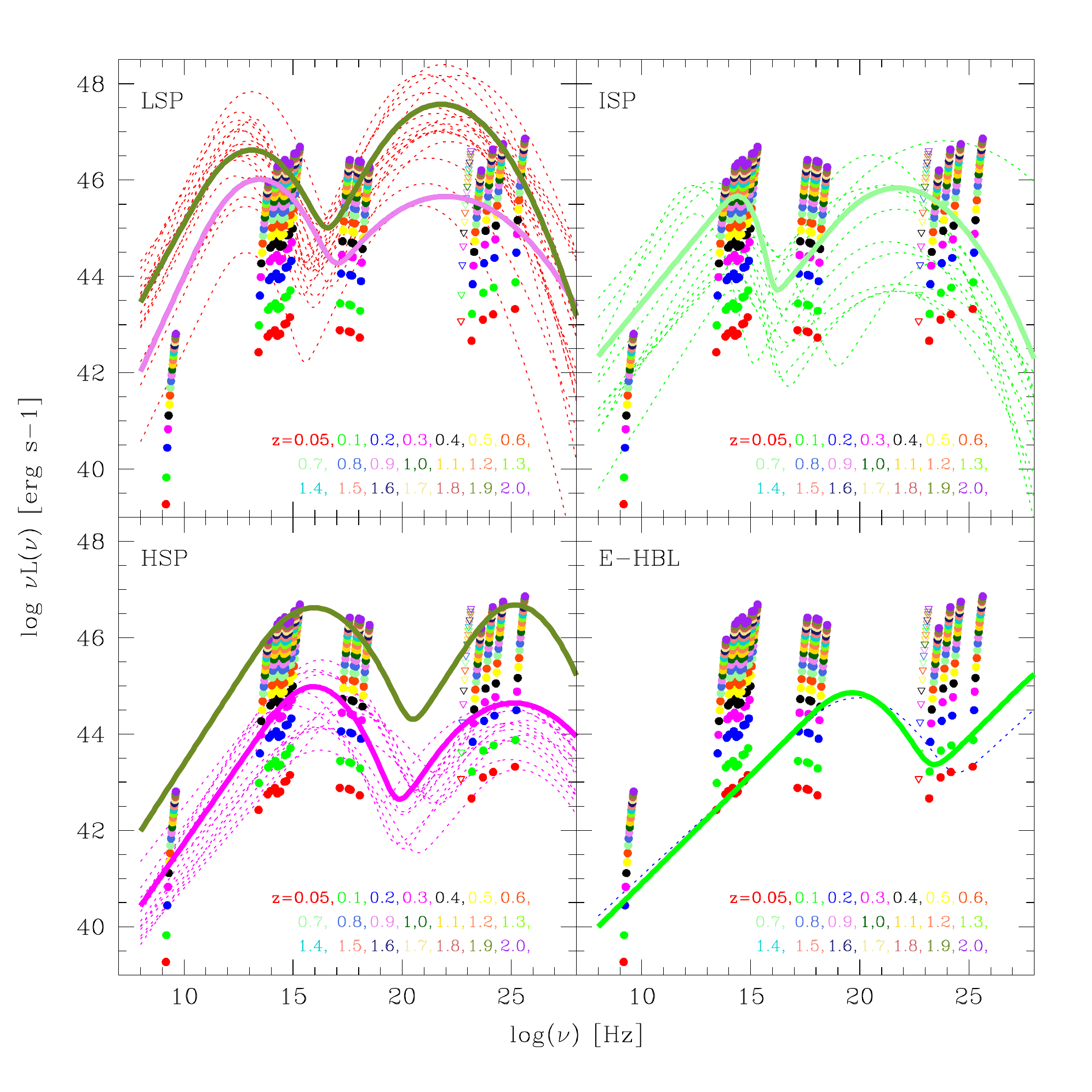}
\end{minipage}%
\begin{minipage}[c]{.40\textwidth}
\quad \quad \quad \quad
 \begin{tabular}{cccc}
\hline
\footnotesize{Class}  & \footnotesize{$\chi_{\nu,min}^{2}$} &  \footnotesize{MAD} &\footnotesize{$z$}\\
\hline
\footnotesize{LSP}  & \footnotesize{2.57} & \footnotesize{1.96} & \footnotesize{0.9}\\
\footnotesize{LSP}  & \footnotesize{2.44} & \footnotesize{2.22} & \footnotesize{1.9}\\
\footnotesize{ISP}  & \footnotesize{2.09} & \footnotesize{2.53} & \footnotesize{0.7}\\
\footnotesize{\bf{HSP}}  & \footnotesize{\bf{0.17}} & \footnotesize{\bf{1.46}} & \footnotesize{\bf{0.3}}\\
\footnotesize{HSP}  & \footnotesize{0.17} & \footnotesize{4.37} & \footnotesize{1.9}\\
\footnotesize{EHBL} & \footnotesize{2.23} & \footnotesize{1.39} & \footnotesize{0.1}\\
\hline
\end{tabular}
  \\
\includegraphics[height=.9\textwidth,width=.9\textwidth]{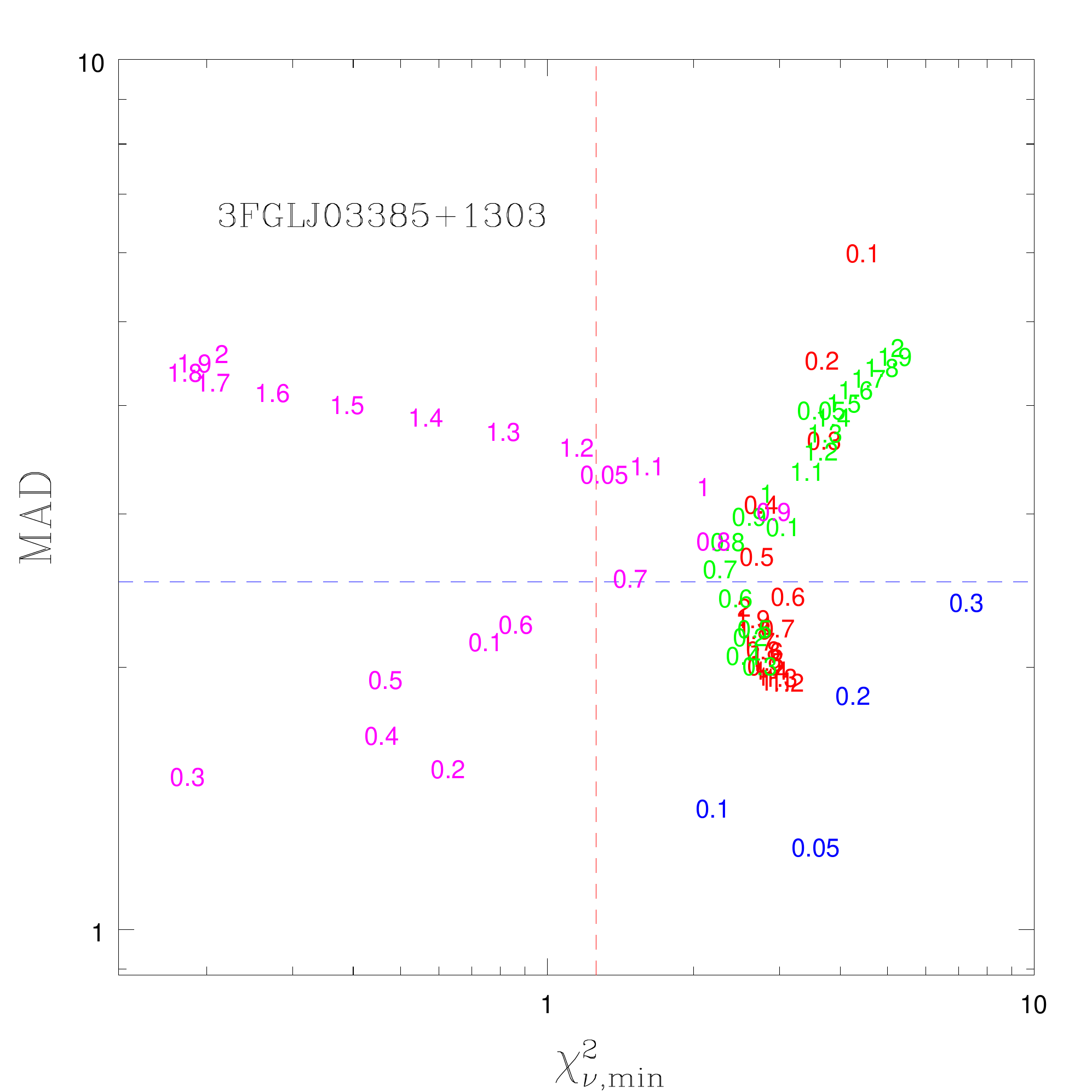}
\end{minipage}
}
\caption{The same diagnostic plot as in Fig. \ref{fig:0102_metodo} for the source 2FGL J0338.2+1306. Here the best-guess recognition is a HSP at $z\sim$0.3 (\textit{magenta line}). }
\label{fig:0338_metodo}
\end{figure*}


\begin{figure*}
\centering
\mbox{%
\begin{minipage}{.65\textwidth}
\includegraphics[height=0.9\textwidth,width=\textwidth]{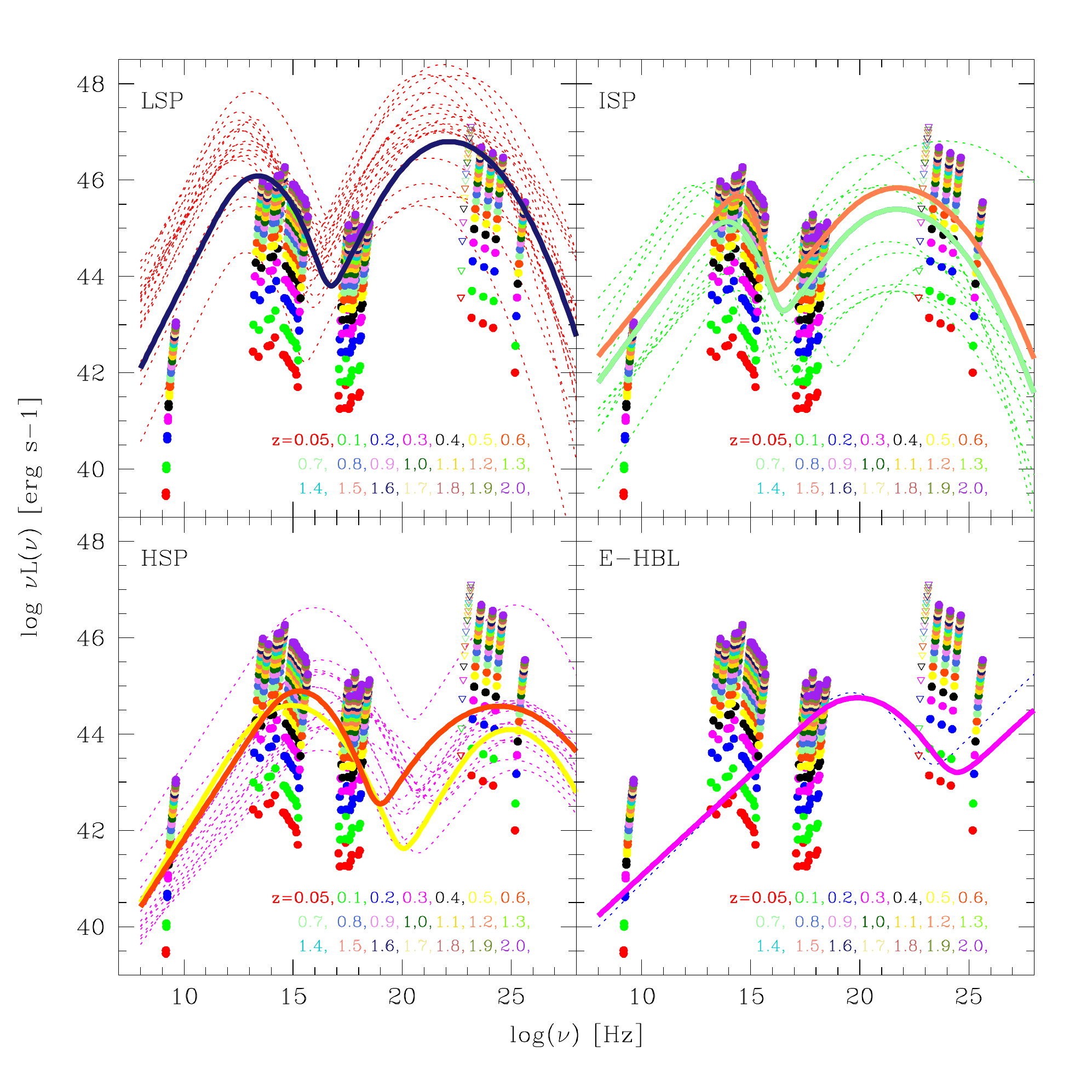}
\end{minipage}%
\begin{minipage}[c]{.40\textwidth}
\quad \quad \quad \quad
 \begin{tabular}{cccc}
\hline
\footnotesize{Class}  & \footnotesize{$\chi_{\nu,min}^{2}$} &  \footnotesize{MAD} &\footnotesize{$z$}\\
\hline
\footnotesize{\bf{LSP}}  & \footnotesize{\bf{0.49}} & \footnotesize{\bf{1.97}} & \footnotesize{\bf{1.6}}\\
\footnotesize{ISP}  & \footnotesize{0.54} & \footnotesize{1.88} & \footnotesize{0.7}\\
\footnotesize{ISP}  & \footnotesize{0.57} & \footnotesize{2.40} & \footnotesize{1.2}\\
\footnotesize{HSP}  & \footnotesize{1.28} & \footnotesize{1.90} & \footnotesize{0.5}\\
\footnotesize{HSP}  & \footnotesize{1.15} & \footnotesize{1.88} & \footnotesize{0.6}\\
\footnotesize{EHBL} & \footnotesize{6.04} & \footnotesize{2.28} & \footnotesize{0.3}\\
\hline
\end{tabular}
  \\
\includegraphics[height=.9\textwidth,width=.9\textwidth]{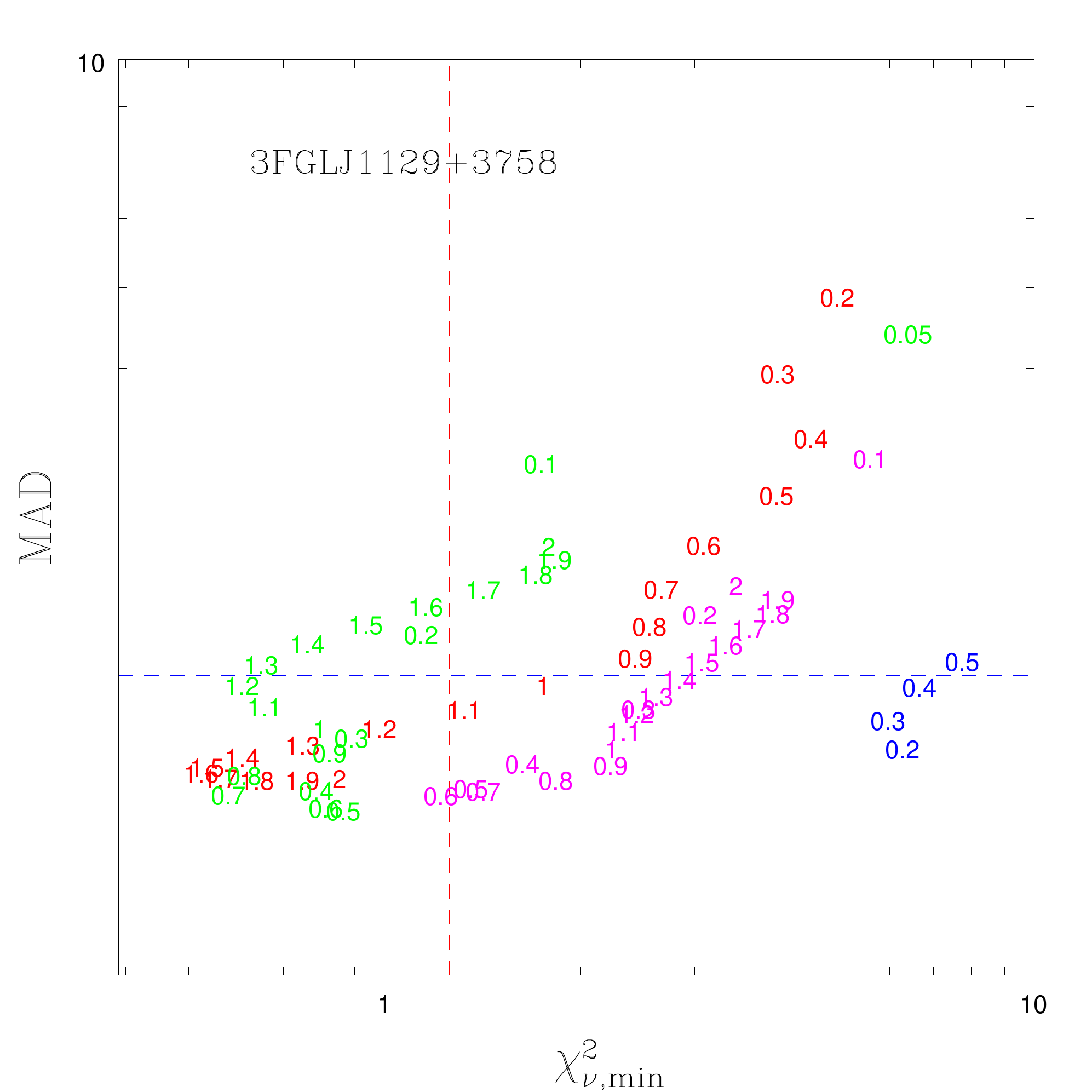}
\end{minipage}
}
\caption{The same diagnostic plot as in Fig.\ref{fig:0102_metodo} for the source 2FGL J1129.5+3758. Here the best-guess recognition is a LSP at $z\sim$1.6 (\textit{dark blue line}).}
\label{fig:1129_metodo}
\end{figure*}


\begin{figure*}
\centering
\mbox{%
\begin{minipage}{.65\textwidth}
\includegraphics[height=0.9\textwidth,width=\textwidth]{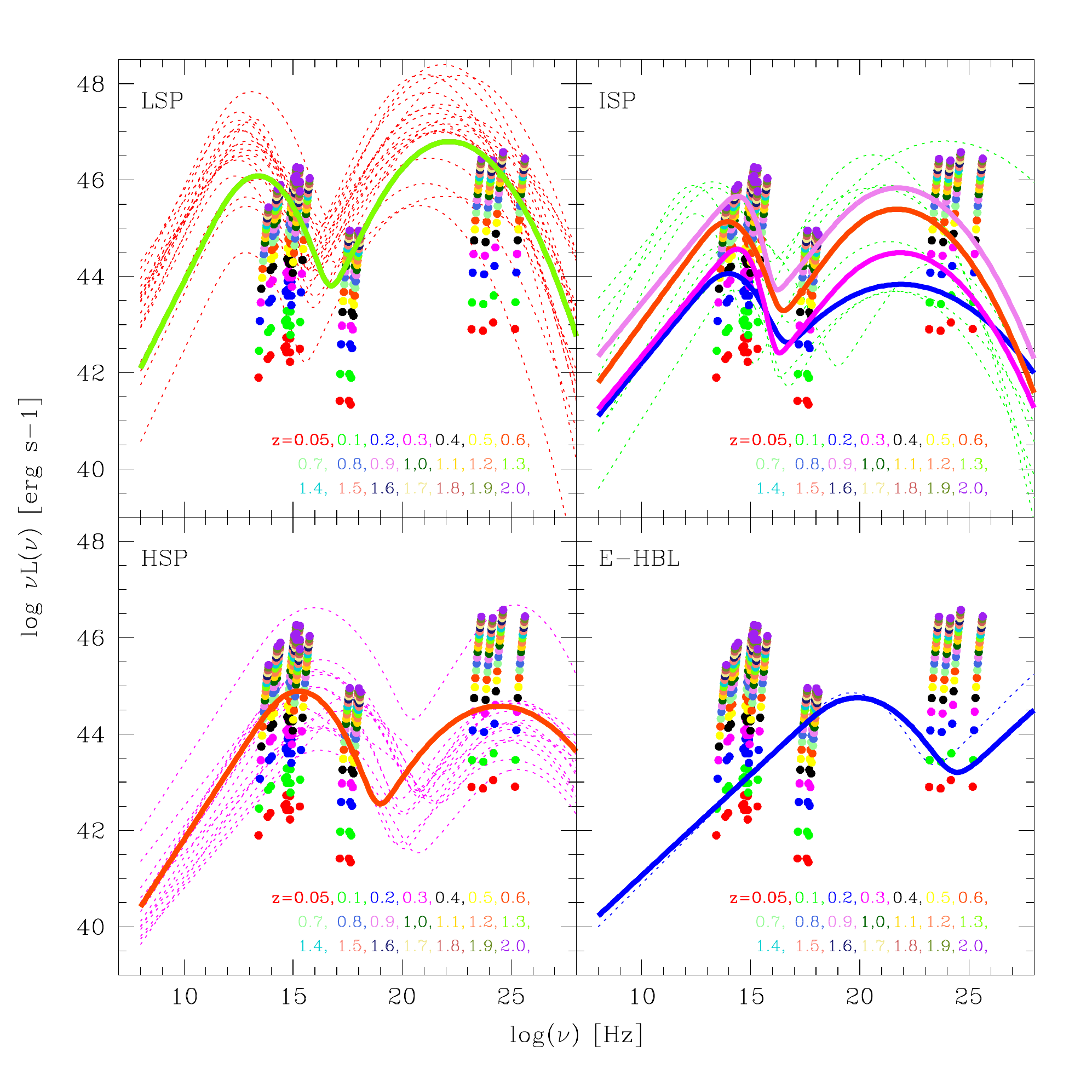}
\end{minipage}%
\begin{minipage}[c]{.40\textwidth}
\quad \quad \quad \quad
 \begin{tabular}{cccc}
\hline
\footnotesize{Class}  & \footnotesize{$\chi_{\nu,min}^{2}$} &  \footnotesize{MAD} &\footnotesize{$z$}\\
\hline
\footnotesize{LSP}  & \footnotesize{0.99} & \footnotesize{2.00} & \footnotesize{1.3}\\
\footnotesize{ISP}  & \footnotesize{1.05} & \footnotesize{2.36} & \footnotesize{0.2}\\
\footnotesize{ISP}  & \footnotesize{1.14} & \footnotesize{1.95} & \footnotesize{0.3}\\
\footnotesize{ISP}  & \footnotesize{1.01} & \footnotesize{1.90} & \footnotesize{0.6}\\
\footnotesize{ISP}  & \footnotesize{1.13} & \footnotesize{2.34} & \footnotesize{0.9}\\
\footnotesize{\bf{HSP}}  & \footnotesize{\bf{0.63}} & \footnotesize{\bf{1.82}} & \footnotesize{\bf{0.5}}\\
\footnotesize{\bf{HSP}}  & \footnotesize{\bf{0.59}} & \footnotesize{\bf{1.83}} & \footnotesize{\bf{0.6}}\\
\footnotesize{EHBL} & \footnotesize{4.24} & \footnotesize{1.82} & \footnotesize{0.2}\\
\hline
\end{tabular}
  \\
\includegraphics[height=.9\textwidth,width=.9\textwidth]{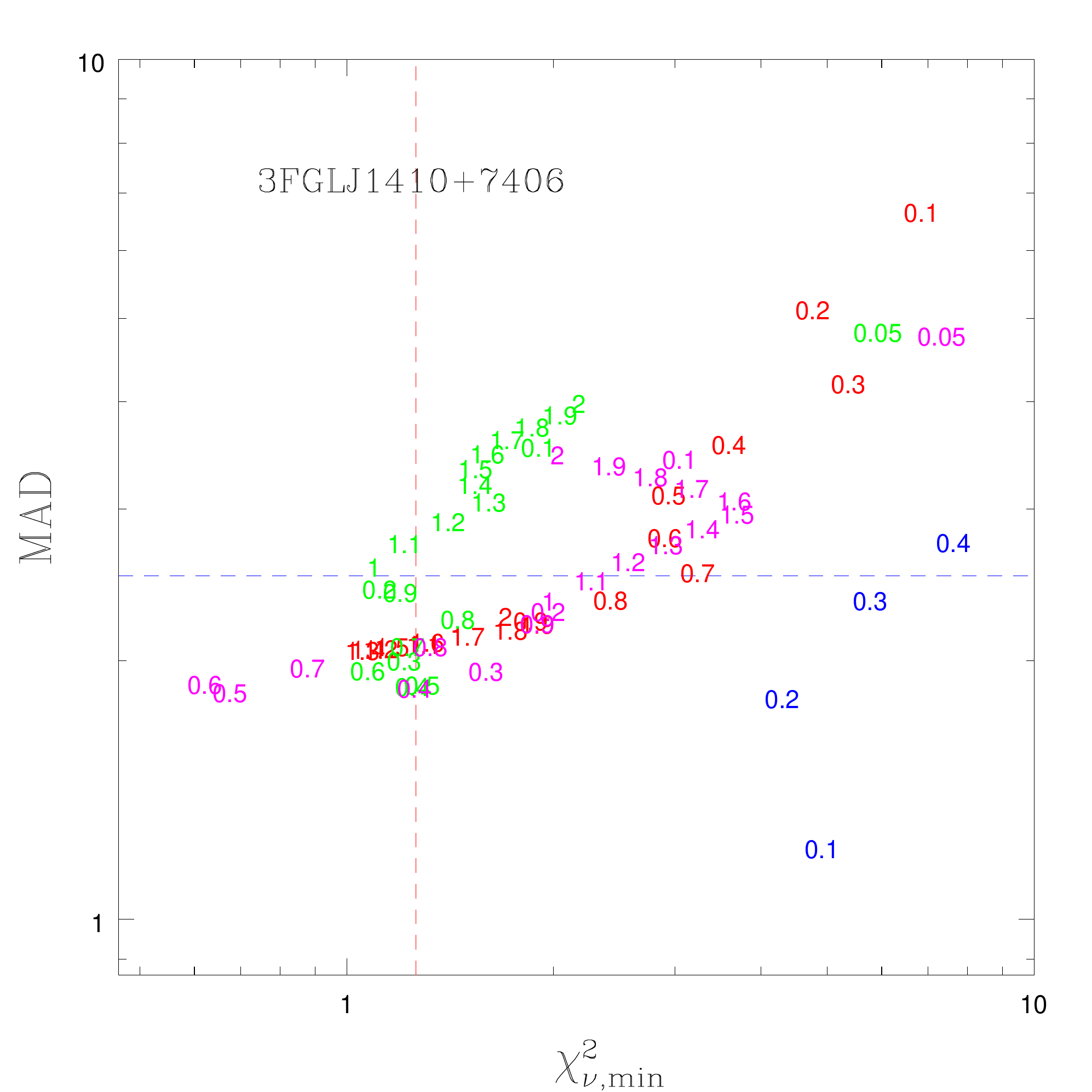}
\end{minipage}
}
\caption{The same diagnostic plot as in Fig.\ref{fig:0102_metodo} for the source XRT J141045+740509. Here the best-guess recognition is a HSP at $z\sim$0.5-0.6 (\textit{yellow and orange lines} that result overlapped).}
\label{fig:1410_metodo}
\end{figure*}


\begin{figure*}
\centering
\mbox{%
\begin{minipage}{.65\textwidth}
\includegraphics[height=0.9\textwidth,width=\textwidth]{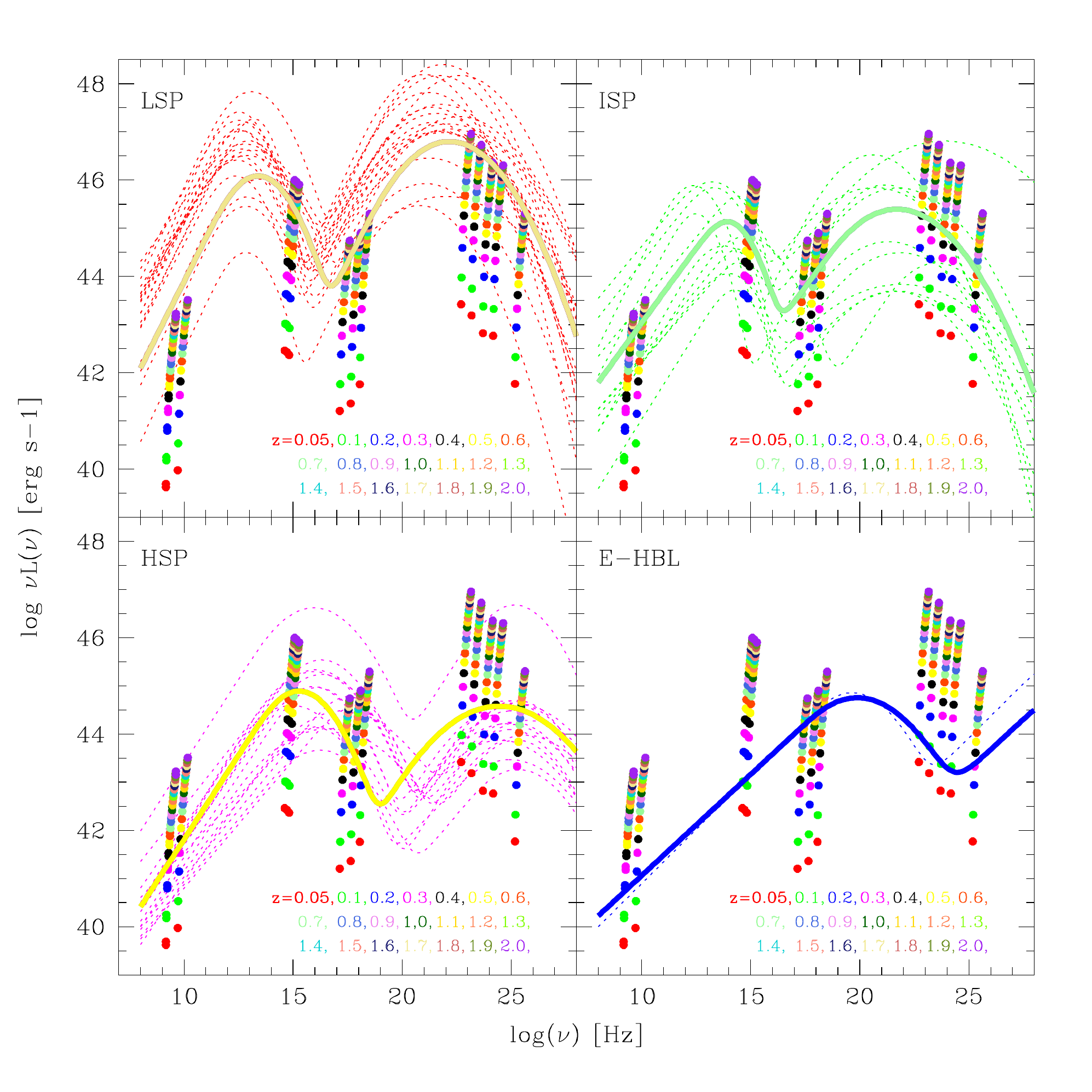}
\end{minipage}%
\begin{minipage}[c]{.40\textwidth}
\quad \quad \quad \quad
 \begin{tabular}{cccc}
\hline
\footnotesize{Class}  & \footnotesize{$\chi_{\nu,min}^{2}$} &  \footnotesize{MAD} &\footnotesize{$z$}\\
\hline
\footnotesize{\bf{LSP}}  & \footnotesize{\bf{0.61}} & \footnotesize{\bf{2.20}} & \footnotesize{\bf{1.7}}\\
\footnotesize{ISP}  & \footnotesize{0.75} & \footnotesize{2.08} & \footnotesize{0.7}\\
\footnotesize{HSP}  & \footnotesize{1.59} & \footnotesize{2.04} & \footnotesize{0.5}\\
\footnotesize{EHBL} & \footnotesize{4.29} & \footnotesize{1.80} & \footnotesize{0.2}\\
\hline
\end{tabular}
  \\
\includegraphics[height=.9\textwidth,width=.9\textwidth]{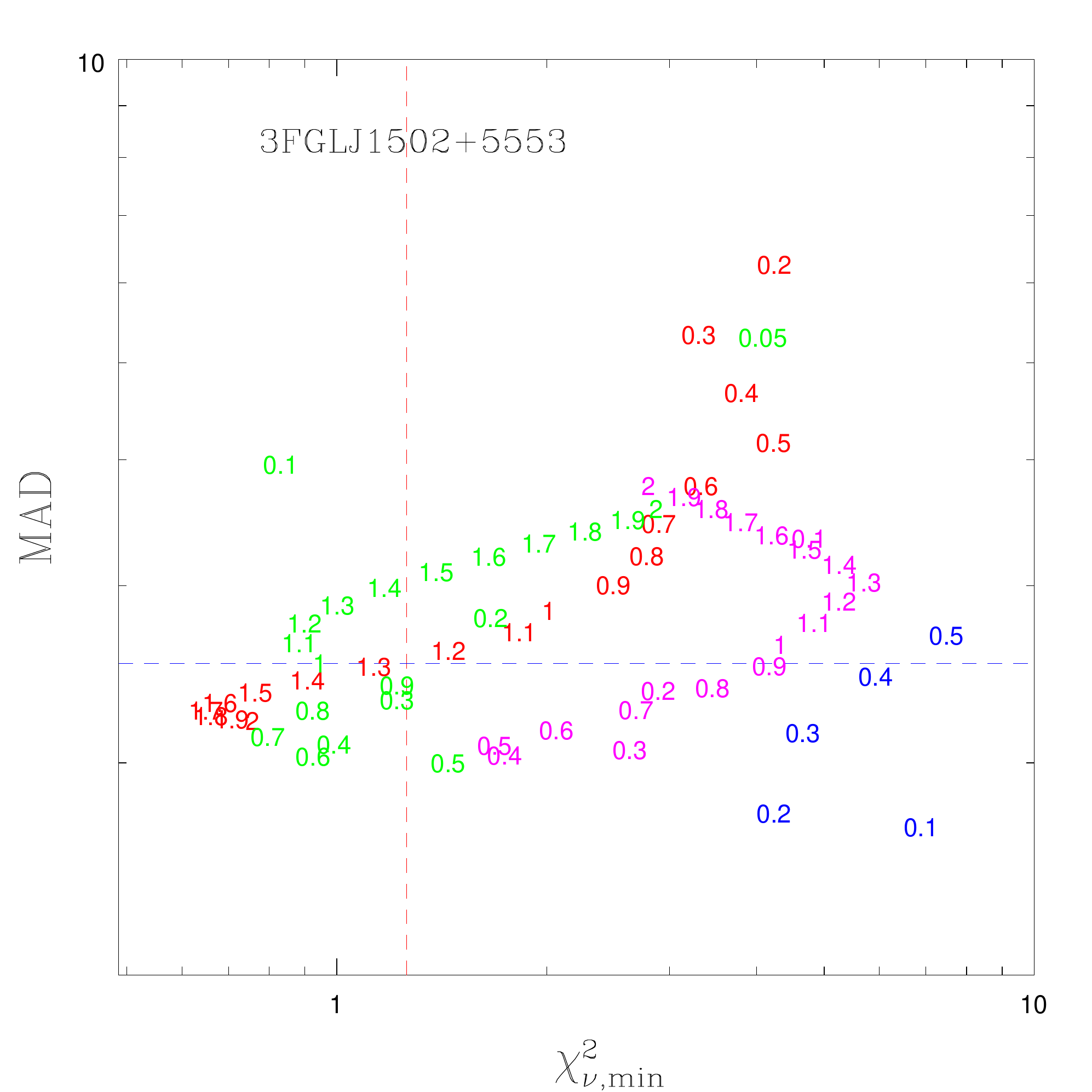}
\end{minipage}
}
\caption{The same diagnostic plot as in Fig.\ref{fig:0102_metodo} for the source 1SXPS J150229.0+555204. Here the best-guess recognition is a LSP at $z\sim$1.7 (\textit{beige line}). }
\label{fig:1502_metodo}
\end{figure*}


\begin{figure*}
\centering
\mbox{%
\begin{minipage}{.65\textwidth}
\includegraphics[height=0.9\textwidth,width=\textwidth]{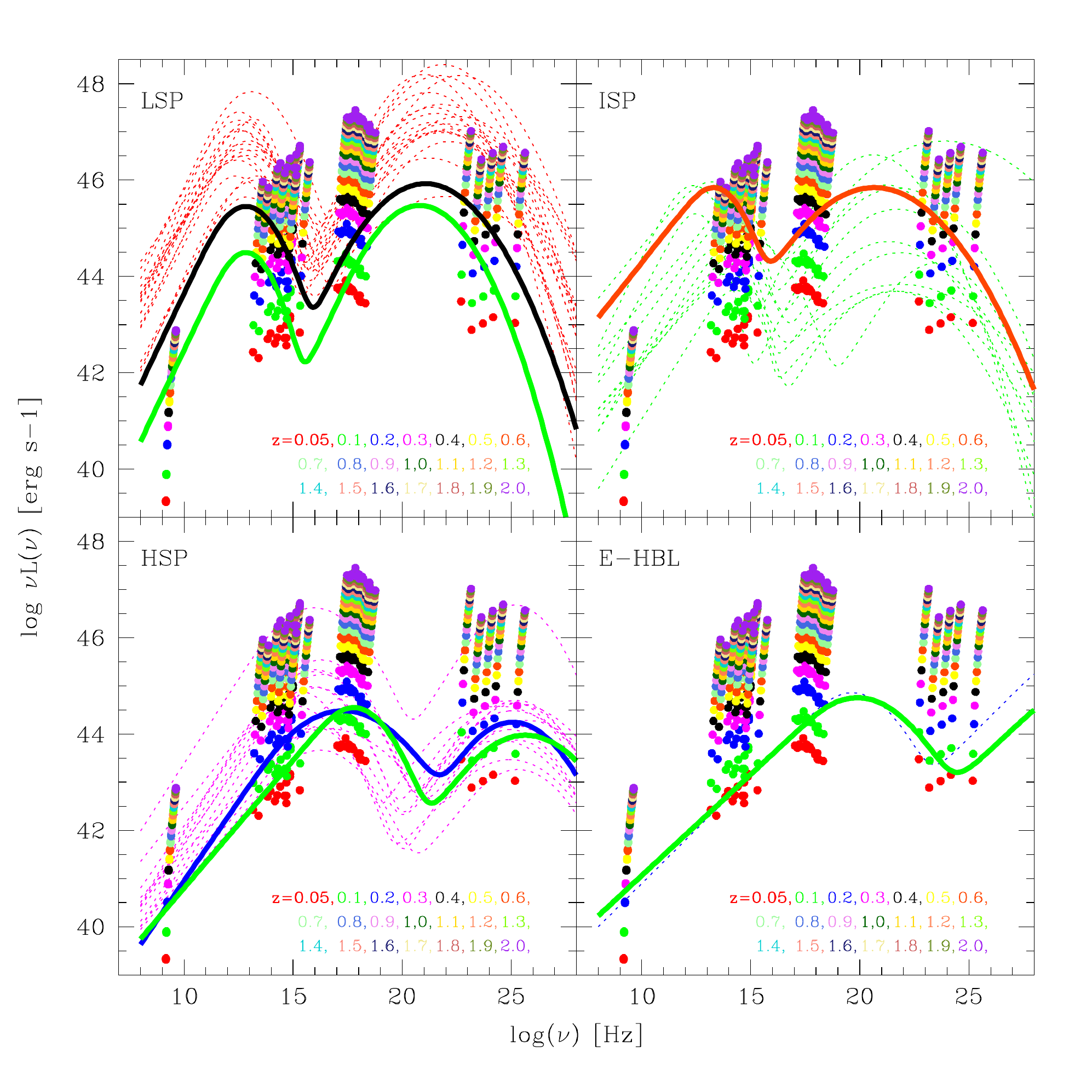}
\end{minipage}%
\begin{minipage}[c]{.40\textwidth}
\quad \quad \quad \quad
 \begin{tabular}{cccc}
\hline
\footnotesize{Class}  & \footnotesize{$\chi_{\nu,min}^{2}$} &  \footnotesize{MAD} &\footnotesize{$z$}\\
\hline
\footnotesize{LSP}  & \footnotesize{3.40} & \footnotesize{4.76} & \footnotesize{0.1}\\
\footnotesize{LSP}  & \footnotesize{3.47} & \footnotesize{2.49} & \footnotesize{0.4}\\
\footnotesize{ISP}  & \footnotesize{3.85} & \footnotesize{3.10} & \footnotesize{0.6}\\
\footnotesize{\bf{HSP}}  & \footnotesize{\bf{0.60}} & \footnotesize{\bf{1.86}} & \footnotesize{\bf{0.1}}\\
\footnotesize{\bf{HSP}}  & \footnotesize{\bf{0.68}} & \footnotesize{\bf{1.55}} & \footnotesize{\bf{0.2}}\\
\footnotesize{EHBL} & \footnotesize{0.90} & \footnotesize{0.80} & \footnotesize{0.1}\\
\hline
\end{tabular}
  \\
\includegraphics[height=.9\textwidth,width=.9\textwidth]{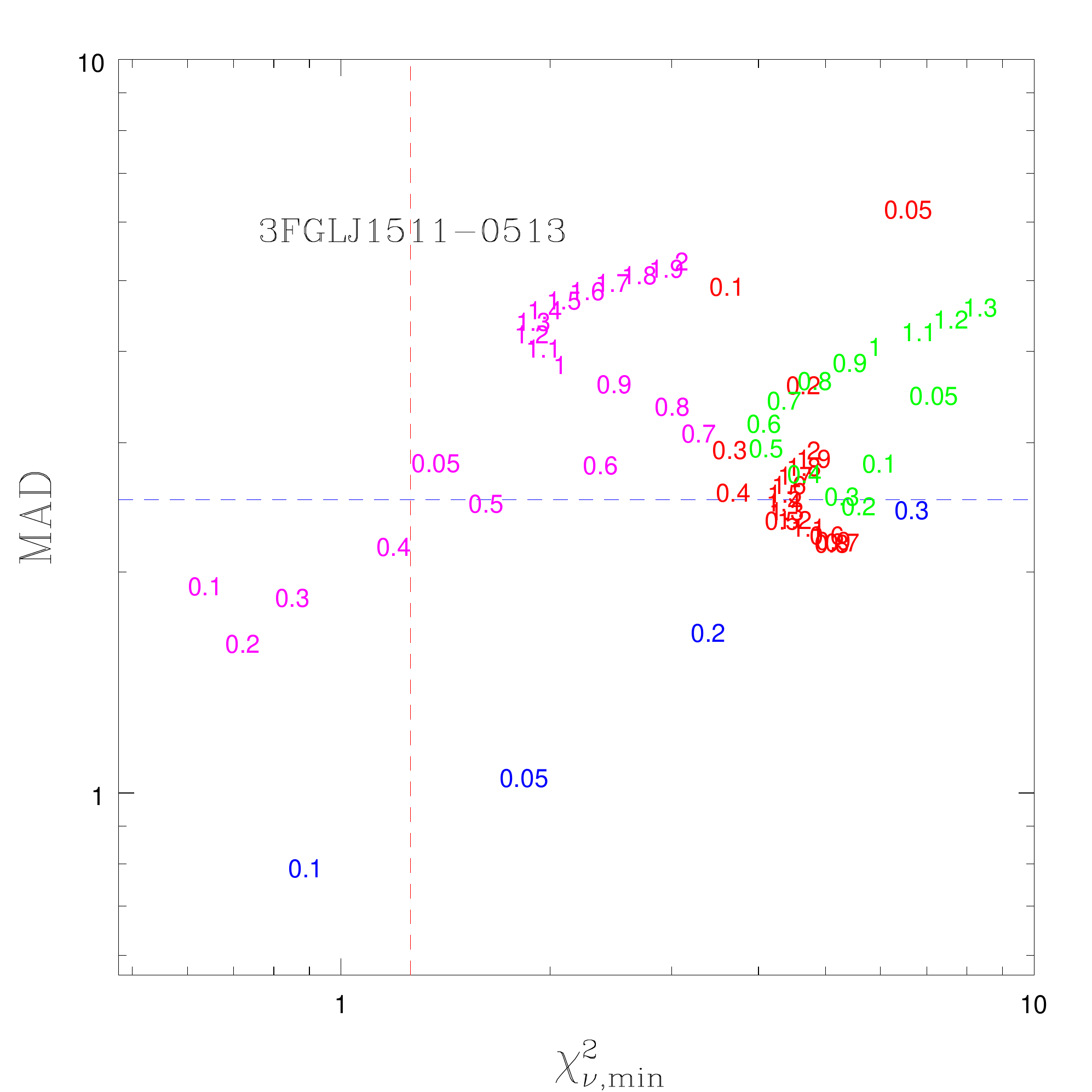}
\end{minipage}
}
\caption{The same diagnostic plot of Fig.\ref{fig:0102_metodo} for the source XRT J151148-051348, the X-ray counterpart candidate for 2FGL J1511.8-0513 for different assumed redshift (from 0.05 to 2.0). Here the best-guess recognition is a HSP at $z\sim$0.1-0.2 (\textit{green and blue lines} respectively).}
\label{fig:1511_metodo}
\end{figure*}


\begin{figure*}
\centering
\mbox{%
\begin{minipage}{.65\textwidth}
\includegraphics[height=0.9\textwidth,width=\textwidth]{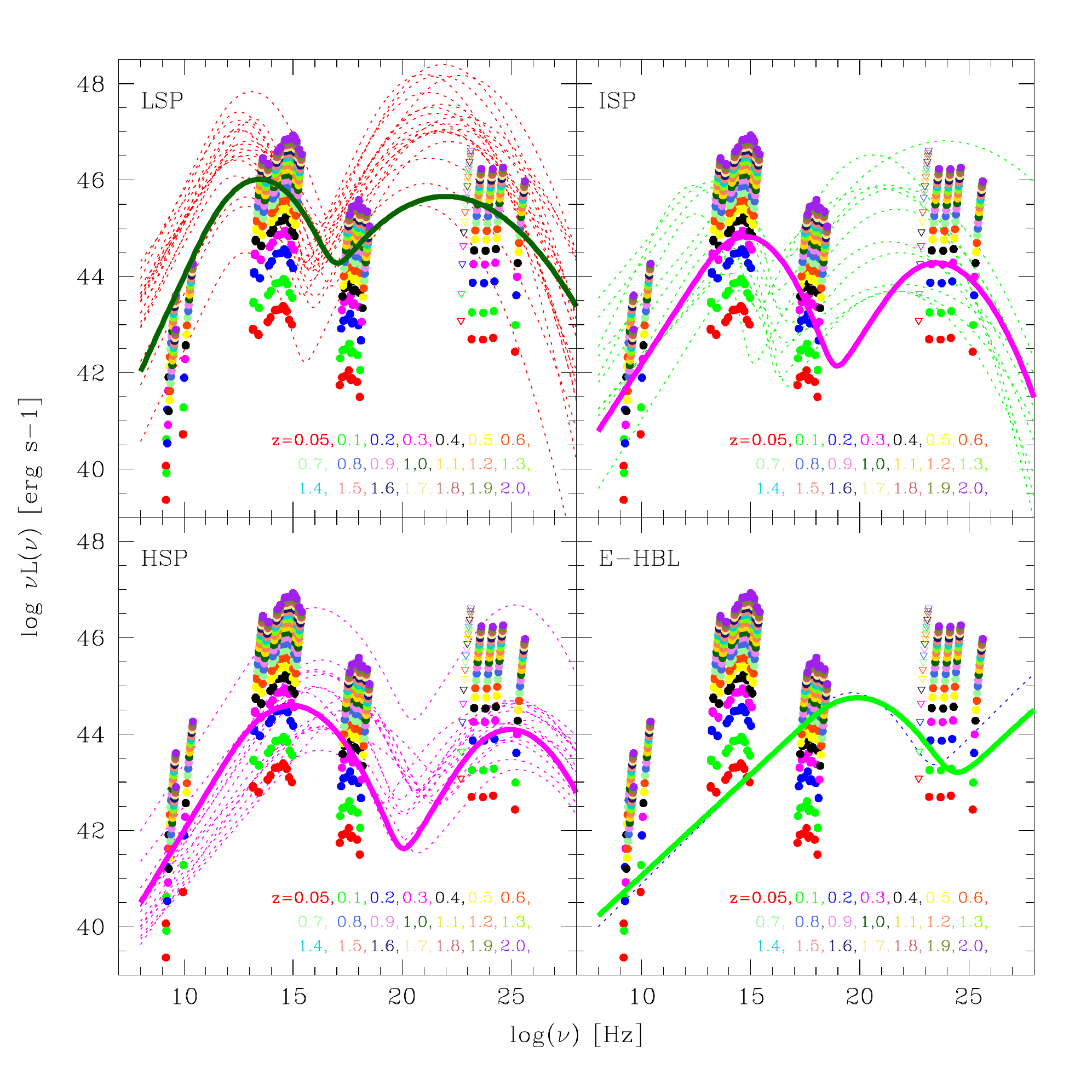}
\end{minipage}%
\begin{minipage}[c]{.40\textwidth}
\quad \quad \quad \quad
 \begin{tabular}{cccc}
\hline
\footnotesize{Class}  & \footnotesize{$\chi_{\nu,min}^{2}$} &  \footnotesize{MAD} &\footnotesize{$z$}\\
\hline
\footnotesize{LSP}  & \footnotesize{0.84} & \footnotesize{2.19} & \footnotesize{1.0}\\
\footnotesize{\bf{ISP}}  & \footnotesize{\bf{0.27}} & \footnotesize{\bf{1.82}} & \footnotesize{\bf{0.3}}\\
\footnotesize{HSP}  & \footnotesize{0.39} & \footnotesize{1.82} & \footnotesize{0.3}\\
\footnotesize{EHBL} & \footnotesize{7.45} & \footnotesize{2.41} & \footnotesize{0.1}\\
\hline
\end{tabular}
  \\
\includegraphics[height=.9\textwidth,width=.9\textwidth]{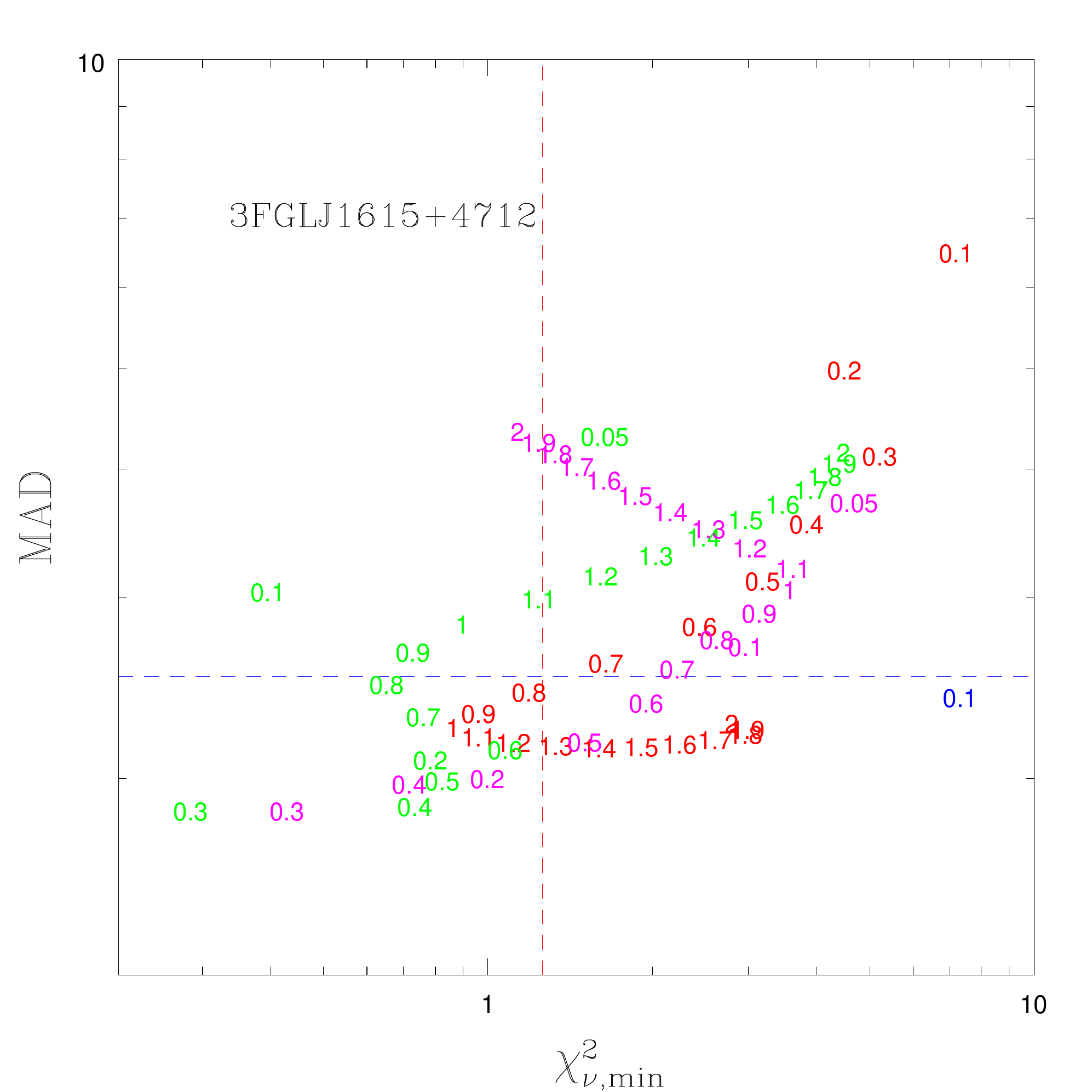}
\end{minipage}
}
\caption{The same diagnostic plot as in Fig. \ref{fig:0102_metodo} for the source 2FGL J1614.8+4703. Here the best-guess recognition is an ISP at $z\sim$0.3 (\textit{magenta line}).}
\label{fig:1615_metodo}
\end{figure*}


\begin{figure*}
\centering
\mbox{%
\begin{minipage}{.65\textwidth}
\includegraphics[height=0.9\textwidth,width=\textwidth]{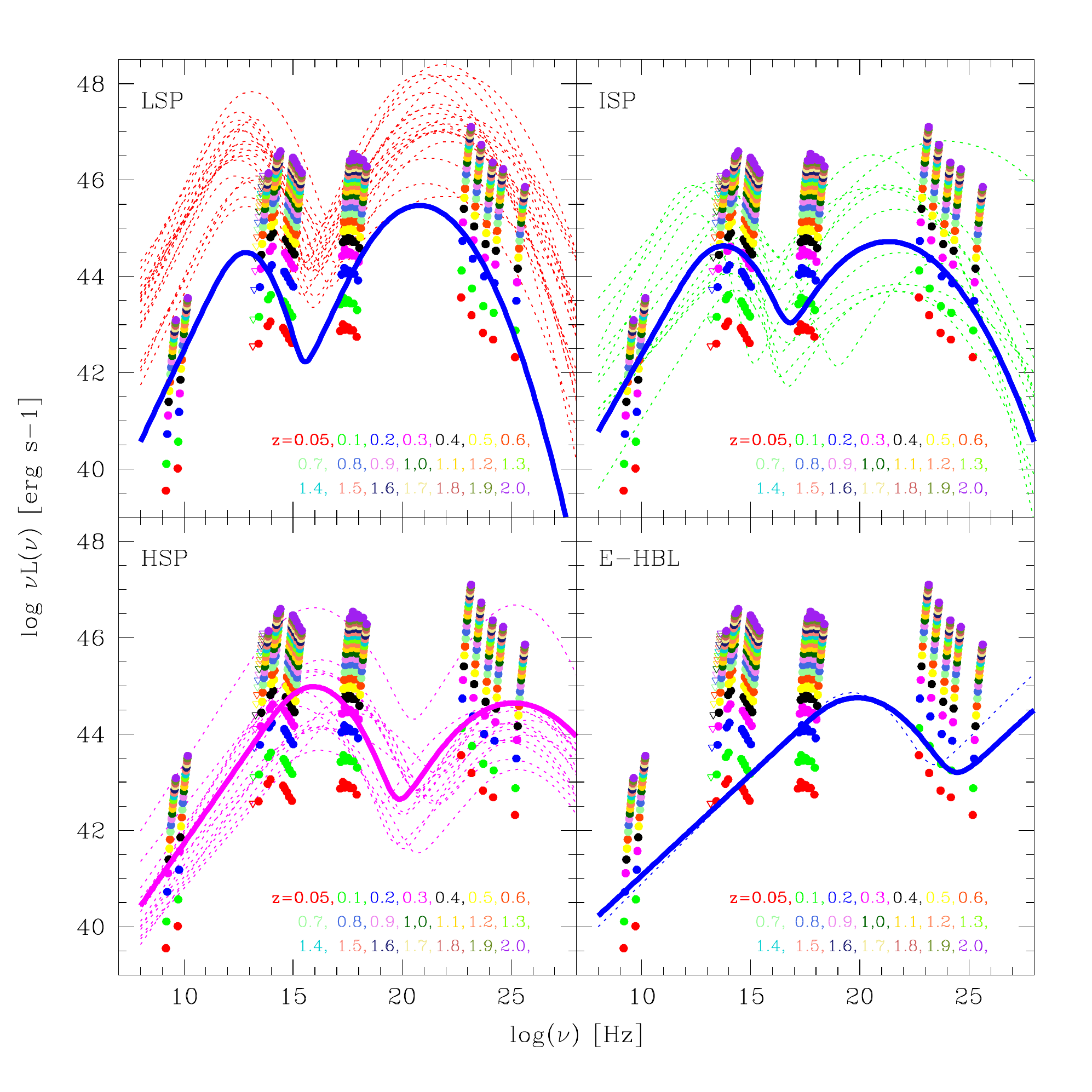}
\end{minipage}%
\begin{minipage}[c]{.40\textwidth}
\quad \quad \quad \quad
 \begin{tabular}{cccc}
\hline
\footnotesize{Class}  & \footnotesize{$\chi_{\nu,min}^{2}$} &  \footnotesize{MAD} &\footnotesize{$z$}\\
\hline
\footnotesize{LSP}  & \footnotesize{1.39} & \footnotesize{4.46} & \footnotesize{0.2}\\
\footnotesize{ISP}  & \footnotesize{2.15} & \footnotesize{2.27} & \footnotesize{0.2}\\
\footnotesize{\bf{HSP}}  & \footnotesize{\bf{0.63}} & \footnotesize{\bf{1.45}} & \footnotesize{\bf{0.3}}\\
\footnotesize{\bf{EHBL}}  & \footnotesize{\bf{0.78}} & \footnotesize{\bf{0.79}} & \footnotesize{\bf{0.2}}\\
\hline
\end{tabular}
  \\
\includegraphics[height=.9\textwidth,width=.9\textwidth]{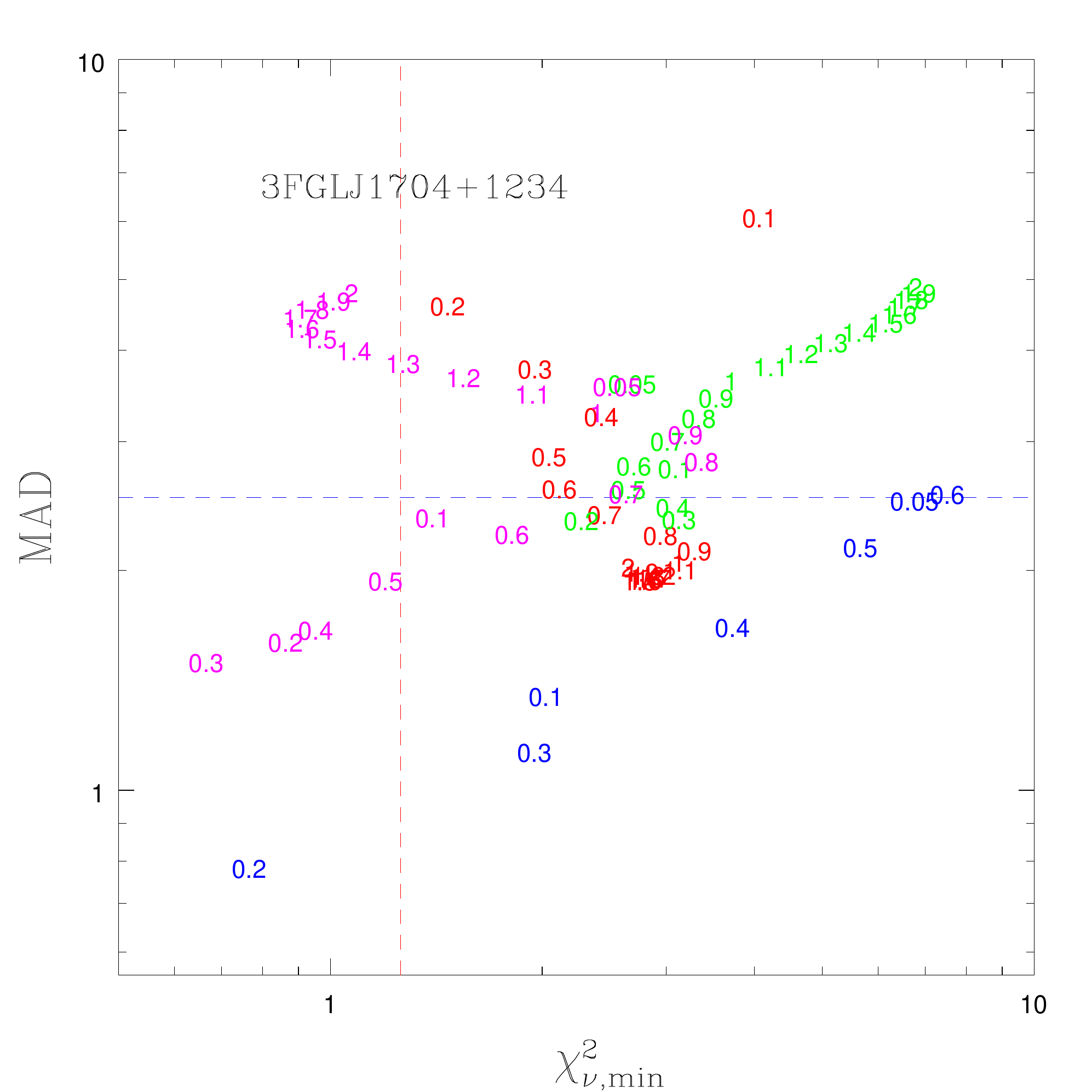}
\end{minipage}
}
\caption{The same diagnostic plot as in Fig. \ref{fig:0102_metodo} for the source 2FGL J1704.3+1235.  Here we propose two best-guess solutions in terms of an HSP object at $z\sim$0.3 (\textit{magenta line}) and EHBL at  $z\sim$0.2 .}
\label{fig:1704_12_metodo}
\end{figure*}


\begin{figure*}
\centering
\mbox{%
\begin{minipage}{.65\textwidth}
\includegraphics[height=0.9\textwidth,width=\textwidth]{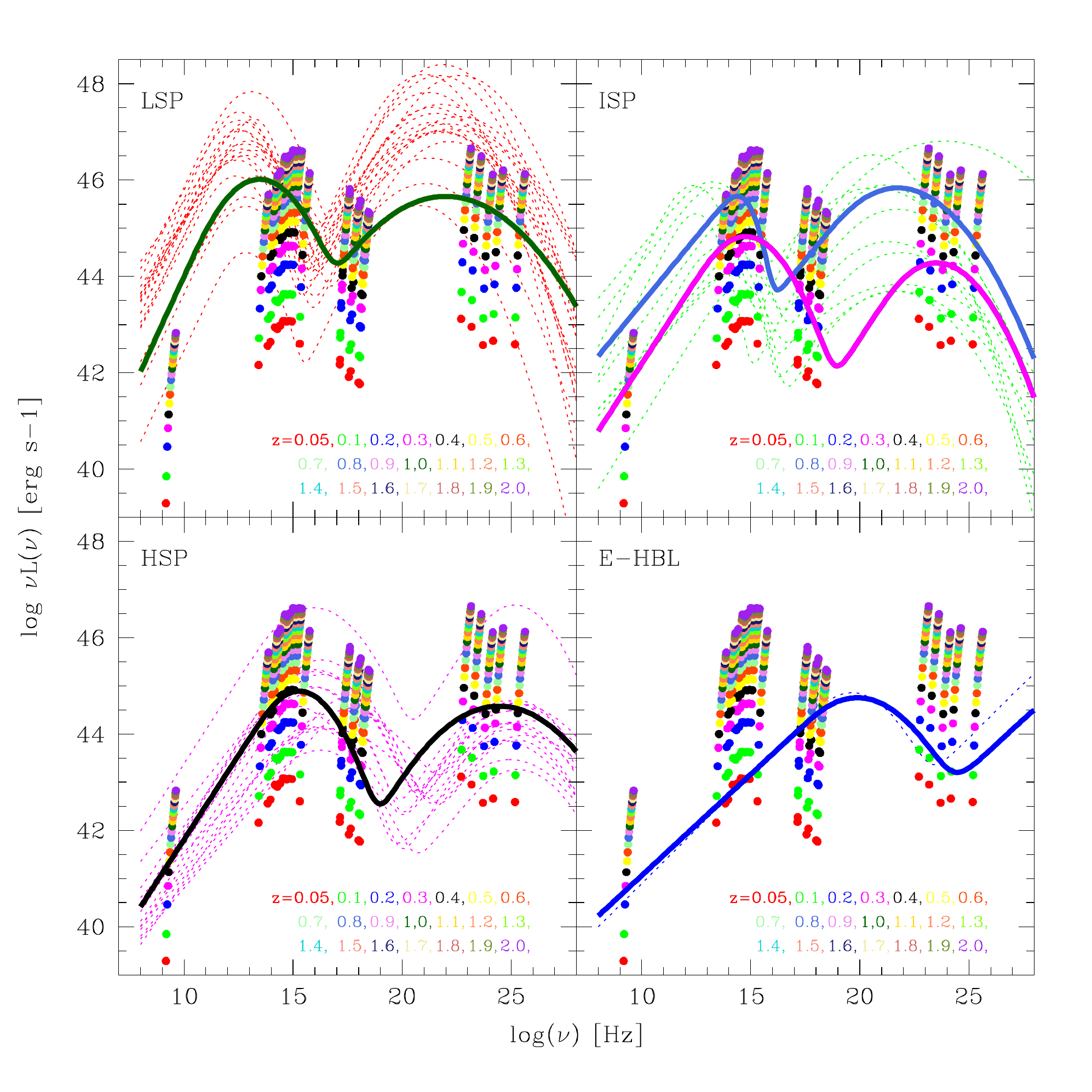}
\end{minipage}%
\begin{minipage}[c]{.40\textwidth}
\quad \quad \quad \quad
 \begin{tabular}{cccc}
\hline
\footnotesize{Class}  & \footnotesize{$\chi_{\nu,min}^{2}$} &  \footnotesize{MAD} &\footnotesize{$z$}\\
\hline
\footnotesize{LSP}  & \footnotesize{1.16} & \footnotesize{2.17} & \footnotesize{1.0}\\
\footnotesize{ISP}  & \footnotesize{0.86} & \footnotesize{1.86} & \footnotesize{0.3}\\
\footnotesize{ISP}  & \footnotesize{0.96} & \footnotesize{2.42} & \footnotesize{0.8}\\
\footnotesize{\bf{HSP}}  & \footnotesize{\bf{0.27}} & \footnotesize{\bf{1.62}} & \footnotesize{\bf{0.4}}\\
\footnotesize{EHBL} & \footnotesize{5.08} & \footnotesize{2.12} & \footnotesize{0.2}\\
\hline
\end{tabular}
  \\
\includegraphics[height=.9\textwidth,width=.9\textwidth]{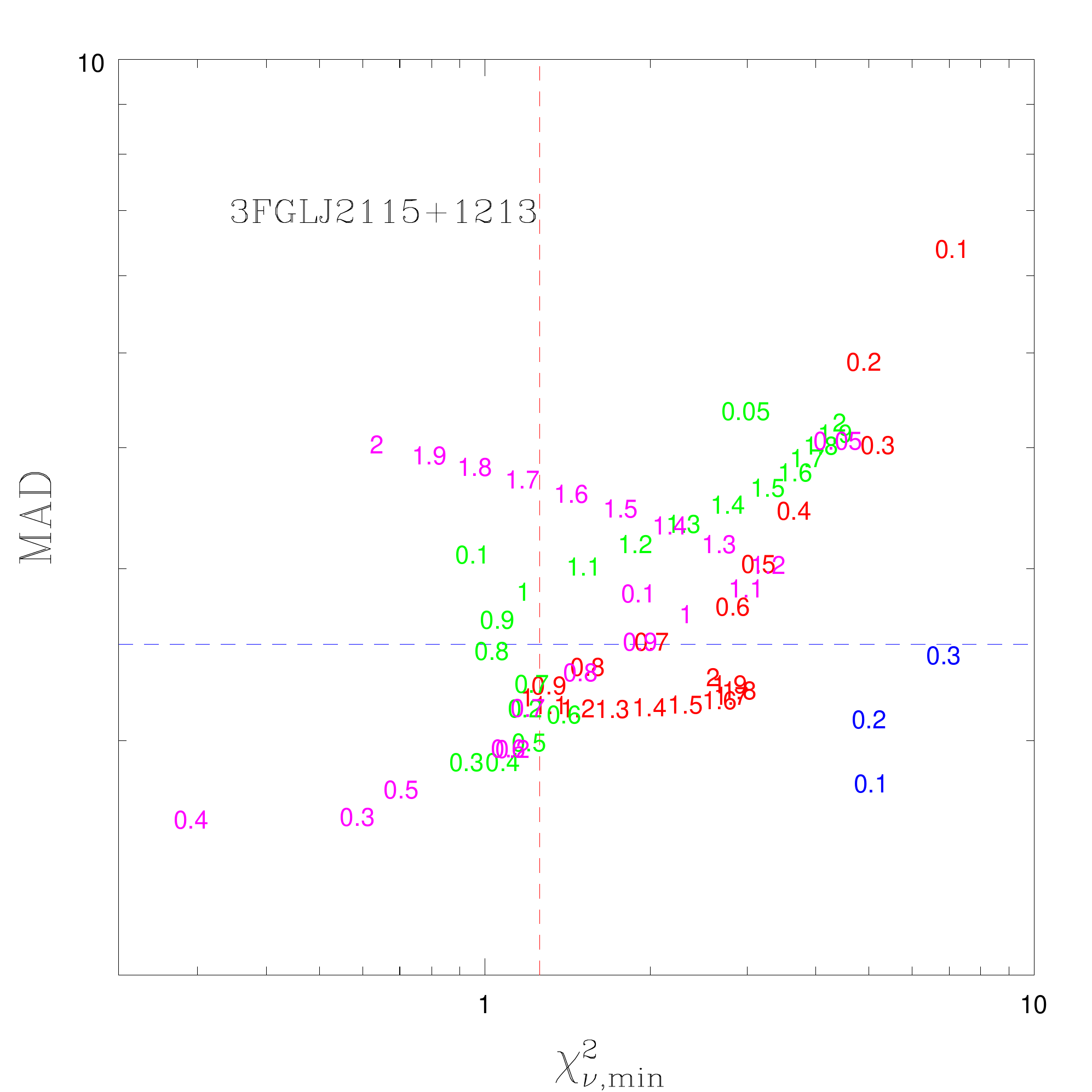}
\end{minipage}
}
\caption{The same diagnostic plot as in Fig. \ref{fig:0102_metodo} for the source 2FGL J2115.3+1549. For this source,  we propose the best-guess solution in terms of an HSP at $z\sim$0.4 (\textit{black line}).}
\label{fig:2115_metodo}
\end{figure*}


\begin{figure*}
\centering
\mbox{%
\begin{minipage}{.65\textwidth}
\includegraphics[height=0.9\textwidth,width=\textwidth]{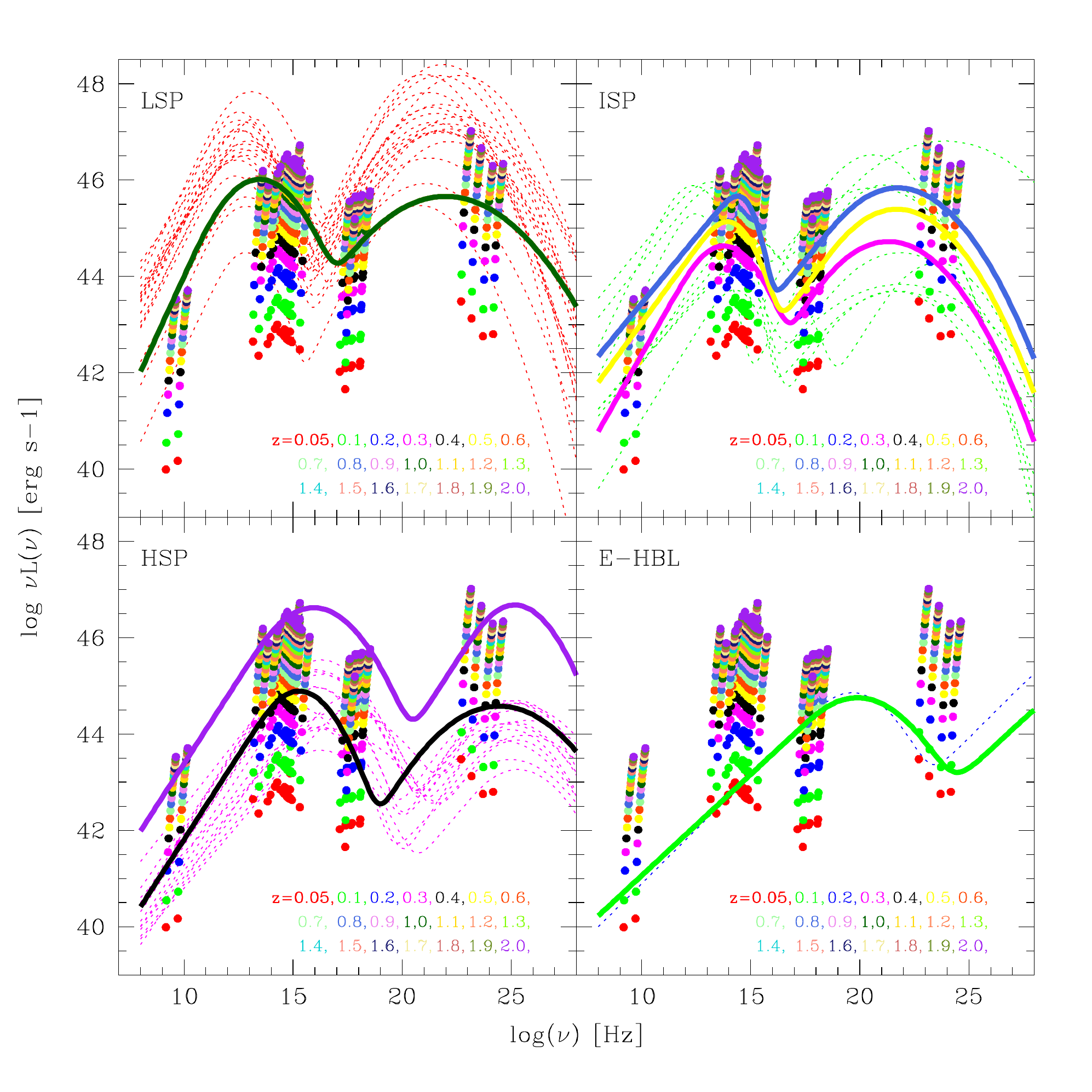}
\end{minipage}%
\begin{minipage}[c]{.40\textwidth}
\quad \quad \quad \quad
 \begin{tabular}{cccc}
\hline
\footnotesize{Class}  & \footnotesize{$\chi_{\nu,min}^{2}$} &  \footnotesize{MAD} &\footnotesize{$z$}\\
\hline
\footnotesize{LSP}  & \footnotesize{0.71} & \footnotesize{1.98} & \footnotesize{1.0}\\
\footnotesize{\bf{ISP}}  & \footnotesize{\bf{0.38}} & \footnotesize{\bf{1.76}} & \footnotesize{\bf{0.3}}\\
\footnotesize{ISP}  & \footnotesize{0.43} & \footnotesize{1.81} & \footnotesize{0.5}\\
\footnotesize{ISP}  & \footnotesize{0.37} & \footnotesize{2.31} & \footnotesize{0.8}\\
\footnotesize{HSP}  & \footnotesize{0.71} & \footnotesize{1.67} & \footnotesize{0.4}\\
\footnotesize{HSP}  & \footnotesize{0.70} & \footnotesize{3.93} & \footnotesize{2.0}\\
\footnotesize{EHBL} & \footnotesize{4.37} & \footnotesize{1.60} & \footnotesize{0.1}\\
\hline
\end{tabular}
  \\
\includegraphics[height=.9\textwidth,width=.9\textwidth]{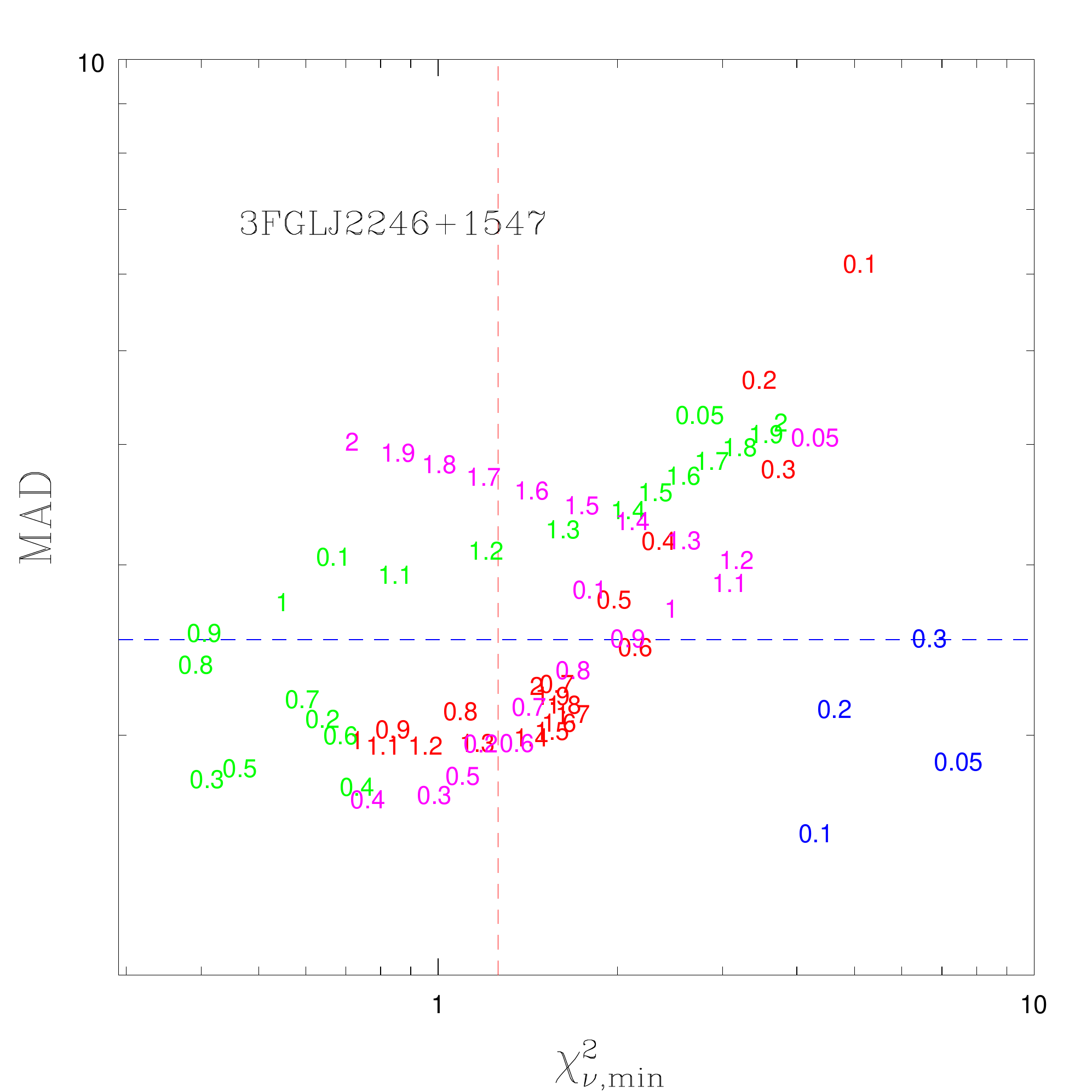}
\end{minipage}
}
\caption{The same diagnostic plot as in Fig. \ref{fig:0102_metodo} for the source 2FGL J2246.3+1549. For 2FGL J2246.3+1549 we propose two best-guess solutions in term of an ISP at either $z\sim$0.3 (\textit{magenta line}) or at $z\sim$0.8 (\textit{blue line}).}
\label{fig:2246_metodo}
\end{figure*}


\begin{figure*}
\centering
\mbox{%
\begin{minipage}{.65\textwidth}
\includegraphics[height=0.9\textwidth,width=\textwidth]{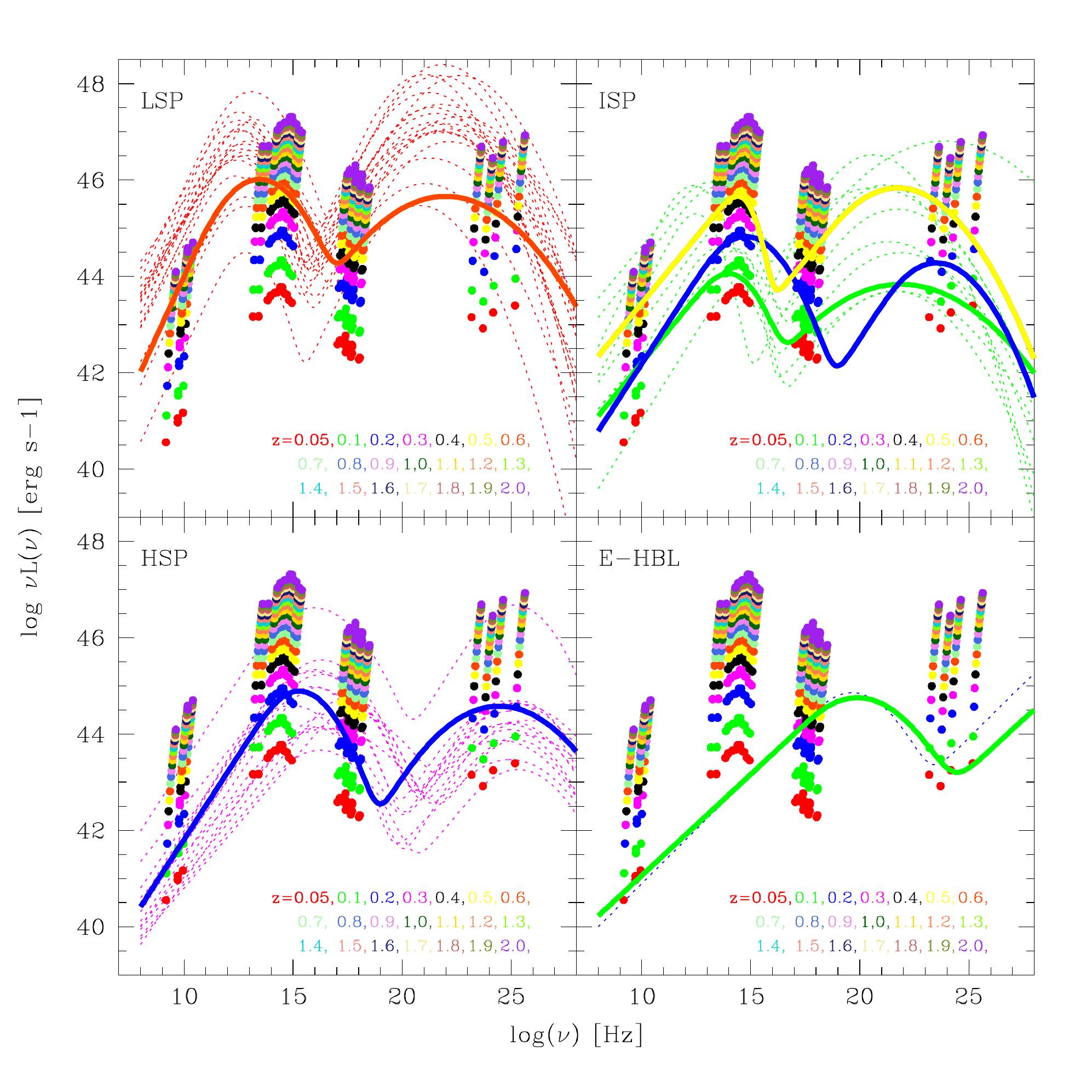}
\end{minipage}%
\begin{minipage}[c]{.40\textwidth}
\quad \quad \quad \quad
 \begin{tabular}{cccc}
\hline
\footnotesize{Class}  & \footnotesize{$\chi_{\nu,min}^{2}$} &  \footnotesize{MAD} &\footnotesize{$z$}\\
\hline
\footnotesize{LSP}  & \footnotesize{0.76} & \footnotesize{1.99} & \footnotesize{0.6}\\
\footnotesize{ISP}  & \footnotesize{0.50} & \footnotesize{2.14} & \footnotesize{0.1}\\
\footnotesize{\bf{ISP}}  & \footnotesize{\bf{0.37}} & \footnotesize{\bf{1.77}} & \footnotesize{\bf{0.2}}\\
\footnotesize{ISP}  & \footnotesize{0.37} & \footnotesize{2.37} & \footnotesize{0.5}\\
\footnotesize{\bf{HSP}}  & \footnotesize{\bf{0.34}} & \footnotesize{\bf{1.77}} & \footnotesize{\bf{0.2}}\\
\footnotesize{EHBL} & \footnotesize{6.77} & \footnotesize{2.52} & \footnotesize{0.1}\\
\hline
\end{tabular}
  \\
\includegraphics[height=.9\textwidth,width=.9\textwidth]{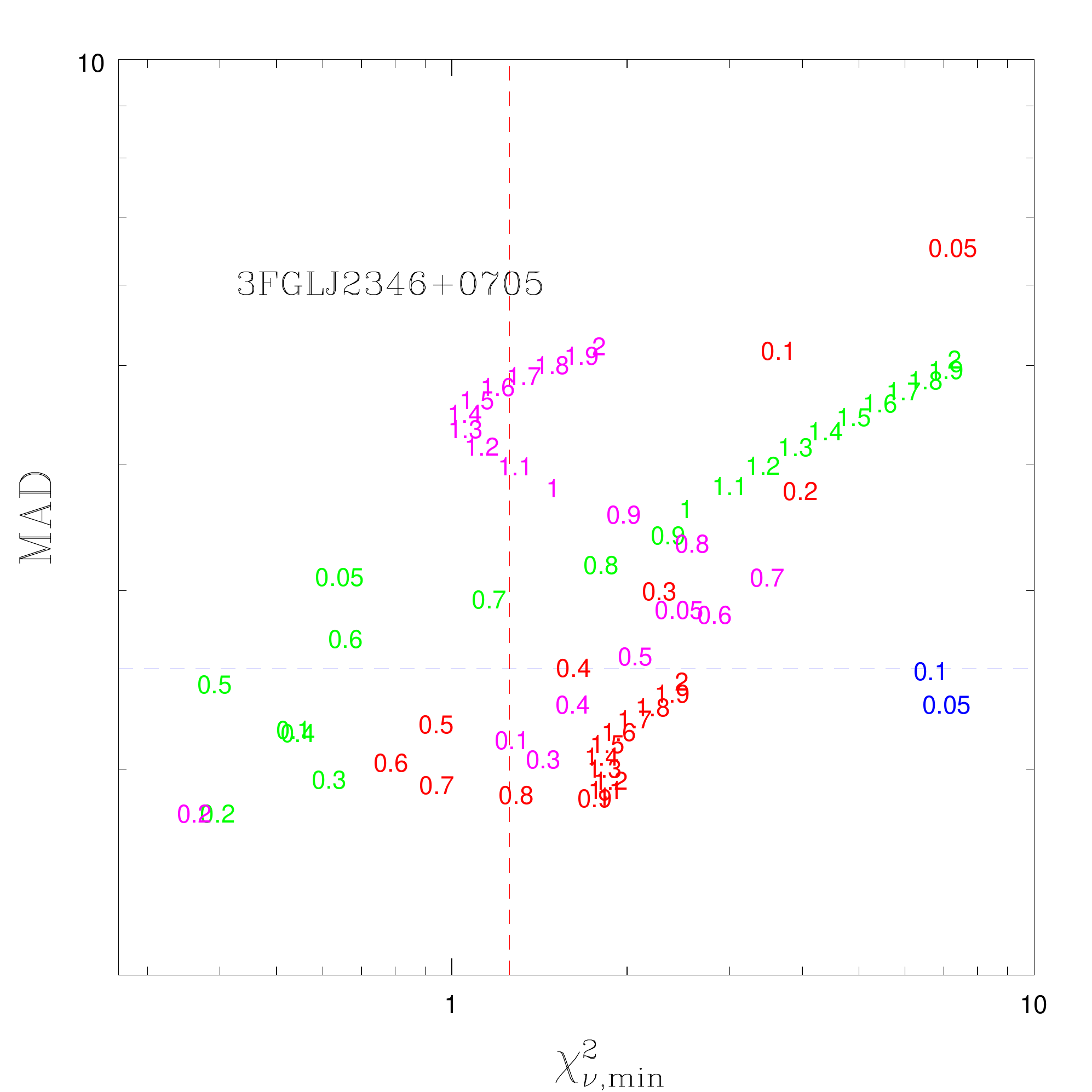}
\end{minipage}
}
\caption{The same diagnostic plot as in Fig. \ref{fig:0102_metodo} for the source 2FGL J2347.2+0707. Here the best-guess recognition is a HSP-ISP at $z\sim$0.2 (\textit{yellow line}).}
\label{fig:2346_metodo}
\end{figure*}


\section{UGS recognition results }
\label{ufo_id_result}

We then proceeded to exploit our \textit{blazar} recognition tool for the analysis of 14 \textit{Fermi} UGSs of the 2FGL catalogue, whose multi-wavelength counterparts have been previously defined and discussed. 
These make a sub-set of our complete, flux-limited UGS sample described in Sec. 2 for which we will discuss a clear evidence in favour of a \textit{blazar} recognition and provide a tentative estimate about the redshift.
Results of our recognition procedure for them are detailed in the following.

\subsection{2FGL J0102.2+0943 (3FGL J0102.1+0943)}
The error-box area of this S/N=7.09 \textit{Fermi} detection was observed for a total of about 4 ksec by \textit{Swift} XRT (details are reported in Appendix A). 
We found only one counterpart, for which we got data in X-ray (\textit{Swift}/XRT), optical (SDSS),  infrared (2MASS) and radio (NVSS) bands.
The diagnostic plots obtained by our \textit{blazar} recognition code, with the MAD and $\chi_{\nu,min}^{2}$ values corresponding to the best-fitting SED templates, are reported in Fig. \ref{fig:0102_metodo}.
The template with the minimum $\chi_{\nu,min}^{2}$ corresponds to an HSP SED, for a best-fit redshift of $z\sim 0.5$.
An HSP blazar at about such redshift is our proposed classification for 2FGL J0102.2+0943.


\subsection{2FGL J0116.6-6153 (3FGL J0116.3-6153)} 
 This $\gamma$-ray source is reported with a 9.9$\sigma$ significance in the 3FGL. 
 It was unassociated in the 2FGL catalogue, but in the 3FGL and 3LAC catalogues it is classified as an ISP BL Lac object with unknown redshift.
In the 3.3 ksec \textit{Swift}/XRT image, we found only one X-ray source as a possible counterpart (Fig. \ref{0116_ass}). 
The broad-band SED, obtained combining the multi-wavelength fluxes of this counterpart, was analysed with our method and the resulted plots are displayed in Fig. \ref{fig:0116_metodo}.
These indicate, as best-guess classification, an HSP \textit{blazar} with a tentative redshift of $\sim 0.4$. 

This result is in agreement with the association and the classification of the 3FGL catalogue and with the optical spectroscopic classification as a BL Lac object reported in \citet{landoni2015} for the IR counterpart WISE J011619.62-615343.4. 
No spectroscopic redshift estimates are provided by them due to the lack of optical emission or absorption features in the optical spectrum.


\subsection{2FGL J0143.6-5844 (3FGL J0143.7-5845)} 
This bright $\gamma$-ray source is classified as an UGS in the 2FGL, but as a BL Lac object with unknown redshift in the 3FGL and 3LAC catalogues 
The source was observed by \textit{Swift}/XRT for about 4.5 ksec.
As discussed in Sec. \ref{0143_ass}, within the 3FGL \textit{Fermi} error-box a very bright X-ray source has been found with multi-wavelength counterparts and the resulting multi-wavelength SED (Fig. \ref{fig:0143_ass}-bottom panel) presents a good spectral coverage. 
Our \textit{blazar} recognition tool (Fig. \ref{fig:0143_metodo}) indicates clearly a minimum $\chi_{min}^{2}$ corresponding to an HSP template that fits the source luminosity data assuming a redshift of 0.1-0.3. 
Our proposed \textit{association} and classification is in agreement with the 3LAC classification and with the optical spectrum reported in \citet{landoni2015} where the source is classified as a BL Lac object with unknown redshift for lack of optical features. 

Since the source shows a hard \textit{Fermi} spectrum\footnote{According to the 3FGL catalogue the hard-spectrum sources have a spectral index $\Gamma < 2.2$. } 
of  $\Gamma < $1.84  and is reported in the Second Fermi LAT catalogue of High energy sources \citep[2FHL catalog]{2fhl}, it could be an interesting target for TeV observations, once account is taken of the EBL absorption.


\subsection{2FGL J0338.2+1306 (3FGL J0338.5+1303)}
This source is reported in the 3FGL and 3LAC catalogue as a \textit{blazar} candidates of uncertain type of the second sub-type (BCU-II)\footnote{The 3LAC sources classified as \textit{blazar} candidates of uncertain type are divided in three sub-types: the BCU-I sources where the counterpart has a published optical spectrum but is not sensitive enough for a classification as an FSRQ or a BL Lac; the BCU-II objects whit the counterpart lacking of an optical spectrum but a reliable evaluation of the SED synchrotron-peak position is possible; the BCU-III sources for which the counterpart is lacking both an optical spectrum and an estimated synchrotron-peak position but shows \textit{blazar}-like broadband emission and a flat radio spectrum.} and with a detection significance of 11.90 $\sigma$ (in the 2FGL it was classified as unassociated). 
The error-box field is analysed in Sec. \ref{0338_ass}, where only one candidate counterpart is found.  
The broad-band SED of this object is reported in Fig. \ref{fig:0338_ass}, including the \textit{Swift}/XRT imaging photometry. 
The output plots of our \textit{blazar} recognition tool are shown in Fig. \ref{fig:0338_metodo}. 
We have two best-fit solutions for the minimum $\chi_{\nu,min}^{2}$ belonging to the HSP class with z=0.3 and z=1.9, the latter having a very large MAD value. 
For such a high redshift value even the \textit{Fermi} fluxes would be strongly damped because of the pair-production by the extragalactic background light (EBL) \citep{franceschini2008}: the last \textit{Fermi} point at $\sim$e$^{25}$ Hz by about a factor 10, which is not seen in the data.  

In conclusion, we consider the HSP solution with redshift z$\sim 0.3$ as our preferred solution and it is worth to note that our proposal is confirmed by the recent work of \citet{marchesini2016}, where the optical spectrum of the counterpart reveals a BL Lac object nature with an unknown redshift due to the lack of emission and absorption lines.


\subsection{2FGL J1129.5+3758 (3FGL J1129.0+3758)}
The error-box area of this S/N=10.25 $\gamma$-ray emitter was observed for a total of about 4.7 ksec by \textit{Swift}/XRT and the X-ray sky map is reported in Fig. \ref{fig:1129_ass}.
We proposed the object XRT J1129-375857 as the likely X-ray counterpart and we were able to build its multi-frequency SED spanning from radio to HE band.
Our \textit{blazar}-like recognition code results in the diagnostic plots reported in Fig. \ref{fig:1129_metodo} and the SED template with the minimum $\chi_{\nu,min}^{2}$ corresponds to an LSP SED, for a best-fit redshift of z~$\sim$1.6. 
However we can see a significant degeneracy with other solutions belong to the same class of LSP at z~$\sim$1.2 to 1.5 and to the ISP class at z$\sim$ 0.5 to 1.2. 
In either case, a high value of the redshift is indicated.


\subsection{2FGL J1410.4+7411 (3FGL J1410.9+7406)}
Thanks to the new reduced 3FGL error box of this \textit{Fermi} UGS, we can find an X-ray source that can be proposed as likely counterpart for the source (see details in \ref{1410_ass}). 
Despite the lack of a radio counterpart, that could help the tool to constrain the classification and the redshift for this object, we have a good spectral coverage at the other frequencies and we can build a multi-wavelength SED for the counterpart XRT J141045+740509.  
Based on our \textit{blazar}-like SED recognition tool (Fig. \ref{fig:1410_metodo}), we suggest that the nature of 2FGL J1410.4+7411 is an HSP object with a high tentative redshift of $z$=0.5-0.6. 
An optical classification of our proposed counterpart is provided by \citet{marchesini2016}. The optical spectrum shows emission lines allowing to classify the source as a new NLSY1 with a z$=$0.429.


\subsection{2FGL J1502.1+5548 (3FGL J1502.2+5553)}
The source is still an UGS in the 3FGL catalogue with a detection significance of 12.6$\sigma$. 
In the 3FGL error-box region of the source, the only X-ray source found is 1SXPS J150229.0+555204, which is spatially coincident with a radio source. 
We propose it as the likely X-ray counterpart for 2FGL J1502.1+5548 (the broad-band SED and details in Sect. \ref{1502_ass}).
The resulting plots by our \textit{blazar} SED-recognition tool are shown in Fig. \ref{fig:1502_metodo}. 
The best-fitting SED template, with minimum $\chi_{\nu,min}^{2}$, belongs to the LSP class at redshift $\sim$1.6-1.9, but a similarly good solution is found with a template of the ISP class at lower redshift (z$\sim$0.4-0.7). 
Hence  for this source the \textit{blazar} classification and redshift are uncertain (but a high redshift is indicated), probably due to the limited spectral coverage of the synchrotron peak for this source.
Photon-photon absorption by the EBL is not expected to seriously affect the \textit{Fermi} fluxes even for the high-redshift solution.


\subsection{2FGL J1511.8-0513 (3FGL J1511.8-0513)}	
This object is present in the 3FGL and 3LAC catalogues with a significance of 10.59 $\sigma$ and its new classification is a \textit{blazar} candidate of uncertain type with unknown redshift. 
Two X-ray sources are found in the source region observed by \textit{Swift}-XRT (App. \ref{1511_ass}), but only the brightest, XRT J151148-051348, is in the reduced 3FGL error ellipse and also proposed as counterpart in the \textit{Fermi} catalogue. 
The diagnostic plots for the X-ray counterpart are shown in Fig. \ref{fig:1511_metodo} and the best-fitting SED template corresponds to an HSP with a tentative redshift $z=$0.1-0.2. 
Our classification is in agreement with the result reported in \citet{alvarez2016}, where the source is classified as a BL Lac object with unknown redshift, on the basis of its featureless optical spectrum.


\subsection{2FGL J1614.8+4703 (3FGL J1615.8+4712)} 
The multi-wavelength counterpart set for this source is discussed in Sec. \ref{1614_ass} and we propose the \textit{Swift} source XRT J161541+471110 as the likely X-ray counterpart, in agreement with the association reported in the 3FGL and 3LAC catalogues. 
In Fig. \ref{fig:1614_ass}, we show its multi-wavelength SED built from its counterpart set. 
Based on our \textit{blazar} recognition tool (Fig. \ref{fig:1615_metodo}) and on the minimum value of $\chi_{min}^{2}$, we suggest that 2FGL J1614.8+4703 is an ISP object at redshift $z$=0.3. 
For this object, the SDSS survey has reported the presence of an early-type spiral (Sa) (\ref{fig:1614_ass}) in the position of our optical counterpart, with a measured spectroscopic redshift of z=0.19, that may represent the host galaxy of a very faint low-z \textit{blazar}.


\subsection{2FGL J1704.3+1235 (3FGL J1704.1+1234) } 
Inside the error-box of this 3FGL S/N=9.43 \textit{Fermi} source (details in Sect. \ref{1704_12_ass}), we found only one bright X-ray counterpart with data in the radio, optical and IR. This appears  to be as a robust and unique counterpart for 2FGL J1704.3+1235.
For this source, the SDSS survey reports the presence of an un-resolved reddish object, in the source position, classified as a \textit{star}. 
Based on our \textit{blazar}-like SED recognition tool (Fig. \ref{fig:1704_12_metodo}), we find two possible solutions. 
One is in terms of an HSP object with tentative redshift $z$=0.3.  A fit at this redshift appears to be confirmed by some evidence of an host galaxy contribution in the optical, as illustrated in Fig. \ref{fig:1704_12_ass}. 
Despite our resulting fit to the \textit{Fermi} data turns out to be quite poor, our result is in broad agreement with the classification provided by \citet{alvarezcrespo2016}, where the optical spectrum of the proposed potential counterpart suggests a BL Lac object nature with a redshift of z = 0.45 .
The other solution is instead for an EHBL classification with z=0.2. 
This source certainly requires further scrutiny, given the robustness and the uniqueness of the \textit{association}.
 

\subsection{2FGL J2115.4+1213 (3FGL J2115.2+1213)} 
Of the two X-ray sources found in the 3FGL error-box of 2FGL J2115.4+1213, as discussed in App. \ref{2115_ass}, the fainter one has essentially no counterparts in other bands. 
Instead for the brighter X-ray source, XRT J211522+121801, we find counterparts in all bands and for this reason it is proposed as our likely X-ray counterpart. 
Based on our \textit{blazar} recognition algorithm (see Fig. \ref{fig:2115_metodo}), we suggest that 2FGL J2115.4+1213 is a \textit{blazar} of the HSP class at redshift $z=$0.4.
About the optical counterpart, the SDSS survey reports the presence of an un-resolved object that is classified as a \textit{star}. 


\subsection{2FGL J2246.3+1549 (3FGL J2246.2+1547)} 
This $\gamma$-ray emitter is reported in the 3FGL and 3LAC catalogue with a detection significance of 9.5$\sigma$ and it is classified as a \textit{blazar} candidate of unknown type (BCU-II) with an ISP SED classification and an unknown redshift. 
From the analysis of the XRT data covering the error-box field (discussed in Sect. \ref{2246_ass}), we found only one faint X-ray source with positional counterparts in various bands.
Despite this source is not within the 3FGL, we suggest this as X-ray counterpart because it is the only X-ray source detected around the 2FGL J2246.3+1549 sky region and moreover our proposal is in agreement with the 3FGL association. 
The plots based on our tool are shown in Fig. \ref{fig:2246_metodo}. 
The best-fitting SED template indicates a classification as ISP object with a tentative redshift from $z\sim 0.3$ to $\sim 0.8$, although the upper value corresponds to an only marginally acceptable MAD.


\subsection{2FGL J2347.2+0707 (3FGL J2346.7+0705)} 
Inside the 3FGL error-box of this S/N=13.83 \textit{Fermi} source of 3 arc-mins (see Sec. \ref{2347_ass}), we found a bright X-ray source with good counterparts in the radio, optical and IR. This counterpart set is in agreement with the 3FGL and 3LAC association where the source is classified as a BCU-II with a ISP SED and with unknown redshift.
Based on our \textit{blazar} recognition code (see Fig. \ref{fig:2346_metodo}), we suggest that the source is a \textit{blazar} of the ISP-HSP class with a best-fit redshift of $z=$0.2.
The SDSS survey (dr12) reports the presence of a r$=$16.62 BL Lac object in our proposed counterpart position, for which a spectroscopic redshift of z$\sim$0.17 is provided by the SDSS automatic analysis procedure and that is in a good agreement and supportive of our result. 
Further dedicated optical observations are needed to confirm this result.


\begin{table*}
\centering
\begin{tabular}{|c|c|c|c|c|c|c|}
\hline
\footnotesize{2FGL Name} &  \footnotesize{3FGL Name } & \footnotesize{3FGL (SED)} & \footnotesize{Optical coordinates        } & \footnotesize{AGN Class}  & \footnotesize{Redshift} & \footnotesize{Classification and Redshift}\\
\footnotesize{Counterpart name} & \footnotesize{  } & \footnotesize{Classification} & \footnotesize{                   (RA,DEC)} & \footnotesize{proposed }  & \footnotesize{from tool}& \footnotesize{from spectroscopy}\\
\hline\hline
\footnotesize{2FGL J0102.2+0943} &\footnotesize{3FGL J0102.1+0943 } & \footnotesize{UGS (-)} & \footnotesize{} & \footnotesize{} & \footnotesize{} & \footnotesize{}\\
\footnotesize{XRT J010217+094411} & \footnotesize{  } & \footnotesize{  } & \footnotesize{(01 02 17.11, +09 44 9.53)*} & \footnotesize{HSP} & \footnotesize{0.5}& \footnotesize{-}\\
\hline
\footnotesize{2FGL J0116.6-6153} & \footnotesize{3FGL J0116.3-6153} & \footnotesize{BLLac (ISP)} & \footnotesize{} & \footnotesize{} & \footnotesize{}& \footnotesize{}\\
\footnotesize{XRT J011619-615340} & \footnotesize{  } & \footnotesize{  } & \footnotesize{(01 16 19.58, -61 53 43.08)} & \footnotesize{HSP} & \footnotesize{0.4}& \footnotesize{BLL, z= ?}\\
\hline
\footnotesize{2FGL J0143.6-5844} & \footnotesize{3FGL J0143.7-5845} & \footnotesize{BLLac (HSP)} & \footnotesize{} & \footnotesize{} & \footnotesize{} & \footnotesize{}\\
\footnotesize{XRT J014347-584551} & \footnotesize{  } & \footnotesize{  } & \footnotesize{(01 43 47.40,-58 45 51.48)} & \footnotesize{HSP} & \footnotesize{0.3} & \footnotesize{-}\\
\hline
\footnotesize{2FGL J0338.2+1306} & \footnotesize{3FGL J0338.5+1303} & \footnotesize{BCU-II (HSP)} & \footnotesize{} & \footnotesize{} & \footnotesize{} & \footnotesize{}\\
\footnotesize{XRT J033829+130216} & \footnotesize{  } & \footnotesize{  } & \footnotesize{(03 38 29.26, +13 02 15.72)} & \footnotesize{HSP} & \footnotesize{0.3}& \footnotesize{BLL, z= ?}\\
\hline
\footnotesize{2FGL J1129.5+3758} & \footnotesize{3FGL J1129.0+3758} & \footnotesize{UGS (-)} & \footnotesize{} & \footnotesize{} & \footnotesize{}& \footnotesize{}\\
\footnotesize{XRT J112903-375857} & \footnotesize{  } & \footnotesize{  } & \footnotesize{(11 29 03.36, +37 56 56.07)} & \footnotesize{LSP} & \footnotesize{1.6}& \footnotesize{-}\\
\hline
\footnotesize{2FGL J1410.9+7406} & \footnotesize{3FGL J1410.9+7406} & \footnotesize{UGS (-)} & \footnotesize{} & \footnotesize{} & \footnotesize{}& \footnotesize{}\\
\footnotesize{XRT J141045+740509} & \footnotesize{  } & \footnotesize{  } & \footnotesize{(14 10 45.84, +74 05 11.04)} & \footnotesize{HSP} & \footnotesize{0.5-0.6}& \footnotesize{NSLY1, z$=$0.429}\\
\hline
\footnotesize{2FGL J1502.1+5548} & \footnotesize{3FGL J1502.2+5553} & \footnotesize{UGS (-)} & \footnotesize{} & \footnotesize{} & \footnotesize{}& \footnotesize{}\\
\footnotesize{1SXPS J150229.0+555204} & \footnotesize{  } & \footnotesize{  } & \footnotesize{(15 02 29.07, 55 52 04.09)} & \footnotesize{LSP-ISP} & \footnotesize{0.7-1.7}& \footnotesize{-}\\
\hline
\footnotesize{2FGL J1511.8-0513} & \footnotesize{3FGL J1511.8-0513} & \footnotesize{BCU-III (-)} & \footnotesize{} & \footnotesize{} & \footnotesize{}& \footnotesize{}\\
\footnotesize{XRT J151148-051348} & \footnotesize{  } & \footnotesize{  } & \footnotesize{(15 11 48.48, -05 13 46.74)} & \footnotesize{HSP} & \footnotesize{0.1-0.2}& \footnotesize{BLL, z= ?}\\
\hline
\footnotesize{2FGL J1544.5-1126} & \footnotesize{3FGL J1544.6-1125} & \footnotesize{UGS (-)} & \footnotesize{} & \footnotesize{} & \footnotesize{}& \footnotesize{}\\
\footnotesize{XRT J154439-112804} & \footnotesize{  } & \footnotesize{  } & \footnotesize{(15 44 39.36, -11 28 04.44)} & \footnotesize{counter-example} & \footnotesize{-}& \footnotesize{-}\\
\hline
\footnotesize{2FGL J1614.8+4703} & \footnotesize{3FGL J1615.8+4712} & \footnotesize{FSRQ (LSP)} & \footnotesize{} & \footnotesize{} & \footnotesize{}& \footnotesize{}\\
\footnotesize{XRT J161541+471110} & \footnotesize{  } & \footnotesize{  } & \footnotesize{(16 15 41.28, +47 11 11.76)*} & \footnotesize{ISP} & \footnotesize{0.3}& \footnotesize{G , z$=$0.19}\\
\hline
\footnotesize{3FGL J1704.3+1235} & \footnotesize{3FGL J1704.1+1234} & \footnotesize{UGS (-)} & \footnotesize{} & \footnotesize{} & \footnotesize{}& \footnotesize{}\\
\footnotesize{XRT J170409+123421} & \footnotesize{  } & \footnotesize{  } & \footnotesize{(17 04 09.60, +12 34 21.36)*} & \footnotesize{HSP} & \footnotesize{0.3}& \footnotesize{BLL, z$=$0.45}\\
\hline
\footnotesize{2FGL J2115.4+1213} & \footnotesize{3FGL J2115.2+1213} & \footnotesize{UGS (-)} & \footnotesize{} & \footnotesize{} & \footnotesize{}& \footnotesize{}\\
\footnotesize{XRT J211522+121801} & \footnotesize{  } & \footnotesize{  } & \footnotesize{(21 15 22.08, +12 18 02.88)} & \footnotesize{HSP} & \footnotesize{0.4}& \footnotesize{-}\\
\hline
\footnotesize{2FGL J2246.3+1549} & \footnotesize{3FGL J2246.2+1547} & \footnotesize{BCU-II (ISP)} & \footnotesize{} & \footnotesize{} & \footnotesize{}& \footnotesize{}\\
\footnotesize{XRT J224005+154434} & \footnotesize{  } & \footnotesize{  } & \footnotesize{(22 46 05.04, +15 44 35.52)*} & \footnotesize{ISP} & \footnotesize{0.3-0.8}& \footnotesize{-}\\
\hline
\footnotesize{3FGL J2347.2+0707} & \footnotesize{3FGL J2346.7+0705} & \footnotesize{BCU-II (ISP)} & \footnotesize{} & \footnotesize{} & \footnotesize{}& \footnotesize{}\\
\footnotesize{XRT J234640+070507} & \footnotesize{  } & \footnotesize{  } & \footnotesize{(23 46 39.84, +07 05 06.86)*} & \footnotesize{ISP-HSP} & \footnotesize{0.2} & \footnotesize{BLL, z$=$0.17}\\
\hline
\end{tabular}
\caption{Summary of our proposed \textit{blazar} classification and SED analysis for the UGS sample considered in this work.  In the first column we report the 2FGL name and the X-ray source name proposed as likely counterpart, in the second and third column the 3FGL name, and the classification (UGS= Unassociated Gamma-ray Source, BCU= active galaxy of uncertain type, BLLac= BL Lac object type, FSRQ= Flat Spectrum Radio Quasar)and the SED class, fourth column are the coordinates of the optical source of the proposed X-ray counterpart (the $\ast$ symbol indicates that the proposed counterpart is a SDSS source), the fifth and sixth column are the classification type and the estimated redshift found running our \textit{blazar} recognition tool, and finally the spectroscopic classification (BLL = BL Lac object, NLSY1= Narrow Line Seyfer 1 galaxy, G= elliptical galaxy) and the redshift reported in literature (see for details the text of the notes for each source) for the proposed optical source counterpart.}
\label{tab:UGS_sample}
\end{table*}

\begin{figure}
  \centering
  \includegraphics[width=0.5\textwidth,height=0.5\textwidth]{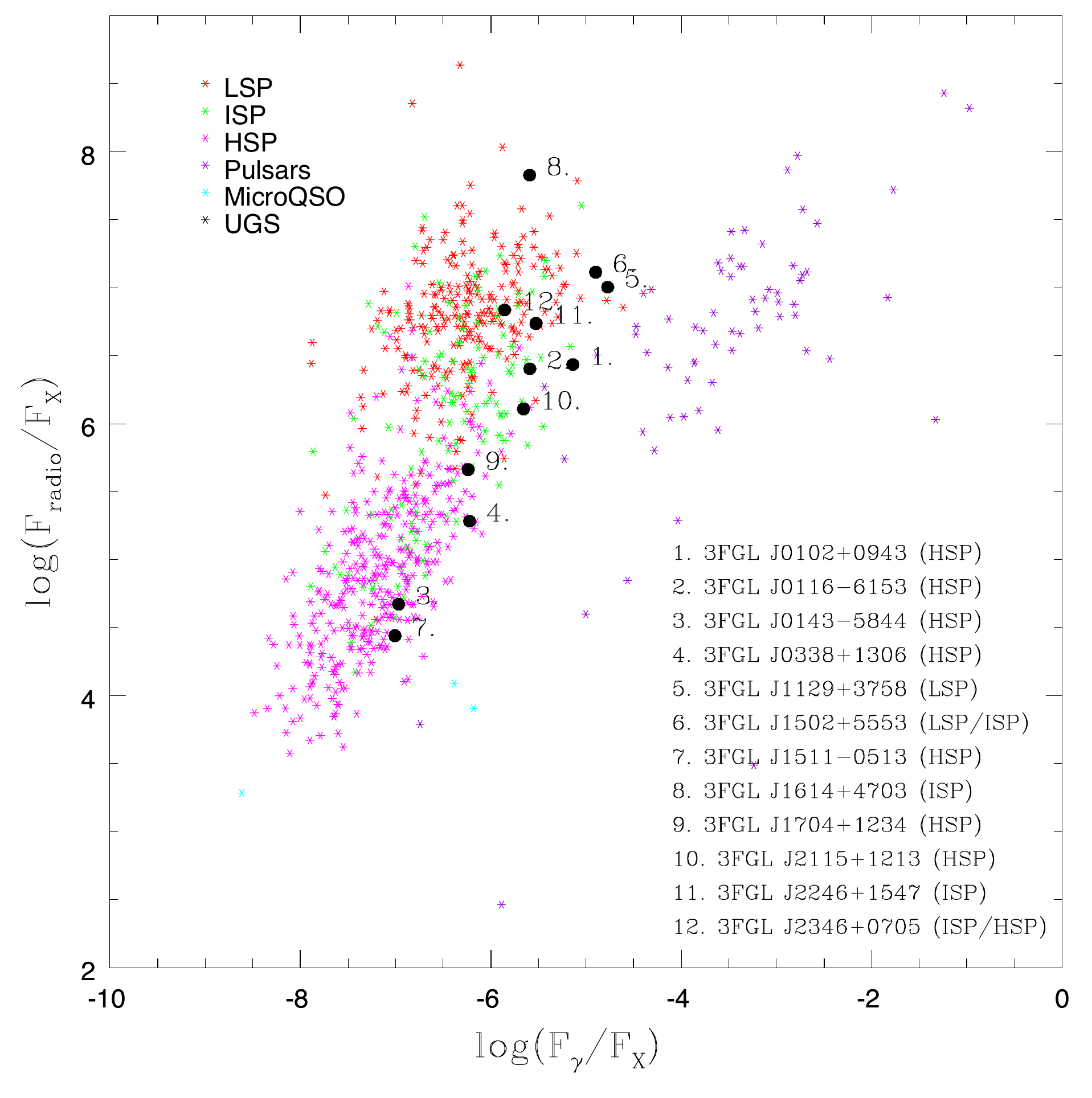}
  \includegraphics[width=0.5\textwidth,height=0.4\textwidth]{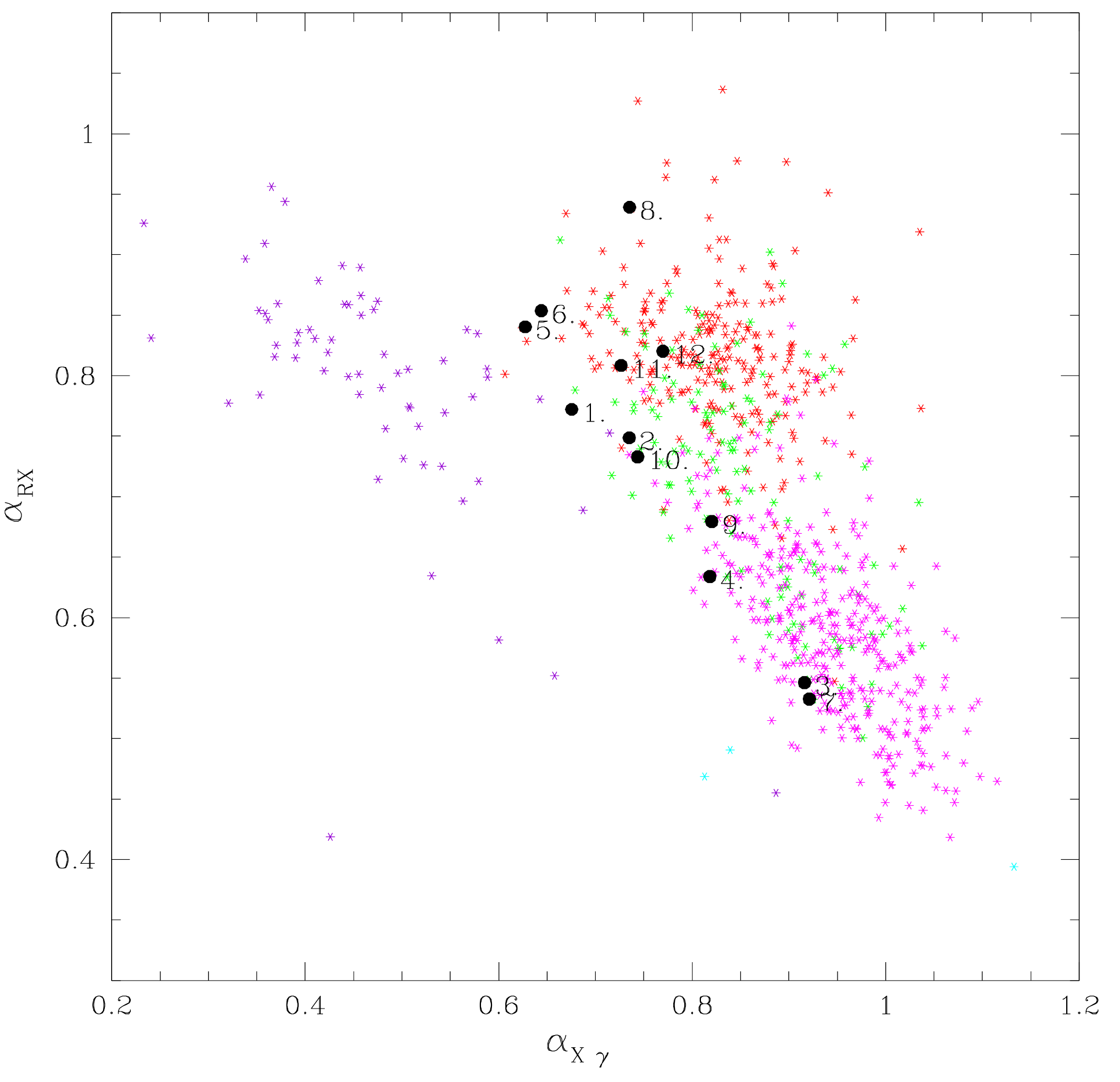}
  \caption{\textit{Top panel:} the $\gamma$-ray to X-ray versus radio to X-ray flux ratios for 3FGL objects of different classes. 
\textit{Bottom panel:} the $\gamma$-ray to X-ray spectral indices versus radio to X-ray spectral indices for the same 3FGL objects. The broad-band spectral index is defined as $\alpha_{ij}\equiv -log(F_j/F_i)/log(\nu_j/\nu_i)$. Red, magenta and green stars indicate the \textit{blazar} sources divided in Low-synchrotron-peaked, Intermediate-synchrotron-peaked and High-synchrotron-peaked objects. Violet stars represent the data points of sources that are associated with pulsars in the 2PC and 2FGL catalogue.
The black points show 12 UGSs studied in this paper and for which we provide a possible recognition as \textit{blazars}. 
The insert details the UGS names and our proposed \textit{blazar} classification.} \label{fig:multiplot_ufo}
\end{figure}

\section{Discussion and Conclusions} 

The \textit{Fermi} mission unveiled a mine of information about the high-energy Universe, which is far from having been completely exploited yet.
In particular, a large fraction of the sources of the 2FGL catalogue and a comparable fraction in 3FGL, are still waiting for reliable identification. 
As many as 576 of those high-confidence UGSs may be either pulsars, other kinds of Galactic objects, or more likely high-energy emitting AGN, mainly BL Lac objects or flat spectrum radio quasars. 
There is also a non-negligible chance that these signals might hide entirely new classes of sources, and even the electromagnetic signatures of (either decaying or annihilating) non-baryonic massive particles that are expected to constitute the dark matter in the Universe. 
In the recently released 3FGL catalogue, there are 1010 unidentified sources, exactly one-third of the 3034 detected sources. 

As a further step towards a more complete characterisation of the UGS population, we discussed in this paper a new method for recognising sources with \textit{blazar}-like SED among the UGSs.
This tool is based on the observed multi-wavelength flux density data, and takes advantage of some well-recognised regularities in the spectral properties of the \textit{blazar} population, like the dependence of the peak frequencies of the synchrotron and IC on source luminosity, and the spectral slopes.
The procedure is tested by comparison with a few well-known \textit{blazar}s, pulsars, and other Galactic sources, and then used for proposing the recognition of 14 UGSs selected in the 2FGL catalogue at high galactic latitudes. 
The 3FGL classification for these 14 sources includes 7 unassociated $\gamma$ sources (UGS), 3 \textit{blazars} (two BL Lac objects and one FSRQ), and 4 active galaxies of unknown type (BCU). 

A summary of our results is reported in Table \ref{tab:UGS_sample}, and for all sources of our UGS sample we report our proposed \textit{blazar} typology and a rough estimate of the redshift.
We find \textit{blazar}-like counterparts for 13 of these UGSs (the remaining one is 2FGL J1544.5-1126, our counter-example for which we disfavour an AGN classification): the majority of them belongs to the HSP class, a couple are of the LSP class and two to ISP class.
This is in agreement with the results of previous works as \citet{mirabal2012x} and \citet{doert2014x}, and with the \textit{Fermi} 3LAC classification when a given UGS is classified as AGN in the 3FGL catalogue.
Identification works based on optical spectroscopic observations, \citep[i.e.][]{masetti2013_rosat, landoni2015, alvarez2016, marchesini2016},  show the typical power-law optical spectrum for 7 sources of our UGS sample and therefore they confirm our classification and redshift (in case of presence of emission and absorption lines).
For our proposed counterparts, we suggest substantial values of redshift, from about $z\sim 0.2$ upwards. These relatively high redshift may partly explain their lack of previous association or identification in published catalogues, although other explanations are possible.

To better understand the general properties of these new counterparts, and to further test the reliability of our method, we have built colour-colour diagrams for \textit{Fermi} sources of various nature, based on existing multi-wavelength data. 
These sources include pulsars, micro-quasars and AGNs, these latter classified by the 3LAC catalogue into high-, intermediate- and low-synchrotron-peaked objects (referred to as HSPs, ISPs and LSPs), corresponding to our \textit{blazar} classification scheme. 

In Fig. \ref{fig:multiplot_ufo} (\textit{upper panel}), we plot the $\gamma$-ray to X-ray flux ratios versus radio to X-ray ratios. For ease of comparison with previous works, we also plot the corresponding broad-band spectral indices (\textit{bottom panel}). The radio and X-ray fluxes of the counterparts associated with the 3FGL sources of different astronomical classes have been derived from the 3LAC and 2PC (the Second Pulsar \textit{Fermi} catalogue, \citealt{2pc_catalog}) catalogues.
As we see, there is a clear separation between the HSP and LSP classes of sources, while the ISP objects span the whole range of properties from HSP to LSP.
Twelve of the 2FGL UGSs that we recognise as \textit{blazar}s in this paper (and for which we could calculate the radio-to-$\gamma$-ray spectral index) are shown as black points in Fig. \ref{fig:multiplot_ufo}. 
All of them are situated in the \textit{blazar} region, in agreement with our proposed classification. 
There is also a good agreement between the colour region and our estimate of the UGS \textit{blazar} classes, perhaps with the exception of sources 1 and 2 in the plot, that we classify as HSP, but that fall in the ISP colour region. The rest of our associated objects fall instead in the expected regions.
We also note a tendency for our UGSs to fall closer to the right-side border of the multi-wavelength colour distributions (to the left of the spectral index region), while the radio to X-ray colour appear consistent. This is likely explained as a selection effect due to the higher than average $\gamma$-ray and lower than average X-ray fluxes for our UGSs compared to standard luminous \textit{blazar}s in the \textit{Fermi} catalogues.

The obvious next step will be to obtain spectroscopic observations of our proposed UGS counterparts lacking of a optical spectrum. 
Given the brightness of the sources and their characteristic featureless spectrum, confirming a BL Lac nature of the candidates, or the LSP nature from the strong emission lines, should be a relatively easy task. Much more difficult, or even impossible, might instead be the redshift measurement, for which however our analysis offers at least a guideline.

From tests carried out in the present paper, our new method to study unassociated \textit{Fermi} objects, based on the analysis of radio-to-$\gamma$ total-band spectral energy distributions, appears to offer a valuable tool to assist in the investigation of the large number of $\gamma$-ray sources still missing a physical interpretation.

\section*{Acknowledgments}
We acknowledge helpful discussions and suggestions by Stefano Vercellone and Patrizia Romano. 
This work has benefited by an extensive collaboration with the MAGIC project, in particular the Padova MAGIC team.  
Edoardo Iani helped us to draw some of the figures.

This work made use of data supplied by the UK Swift Science Data Centre at the University of Leicester.
Part of this work is based on archival data, software and online services provided by the ASI science data centre (ASDC).

We are grateful to an anonymous referee for his careful reading and numerous suggestions that helped improving the paper.

The financial contributions by the contracts INAF ASTRI (PI E. Giro) and INAF E-ELT (PI R. Falomo), and by the Padova University, are also acknowledged.

\addcontentsline{toc}{section}{bibliography}
\bibliographystyle{mn2e}
\bibliography{biblio}

\appendix

\section[]{Multi-Wavelength counterparts for a selected sample of \textit{Fermi} UGSs}

\subsection{2FGL J0102.2+0943} 
\label{0102}

\begin{figure*}
\raisebox{-.5\height}  {\includegraphics[width=.7\textwidth]{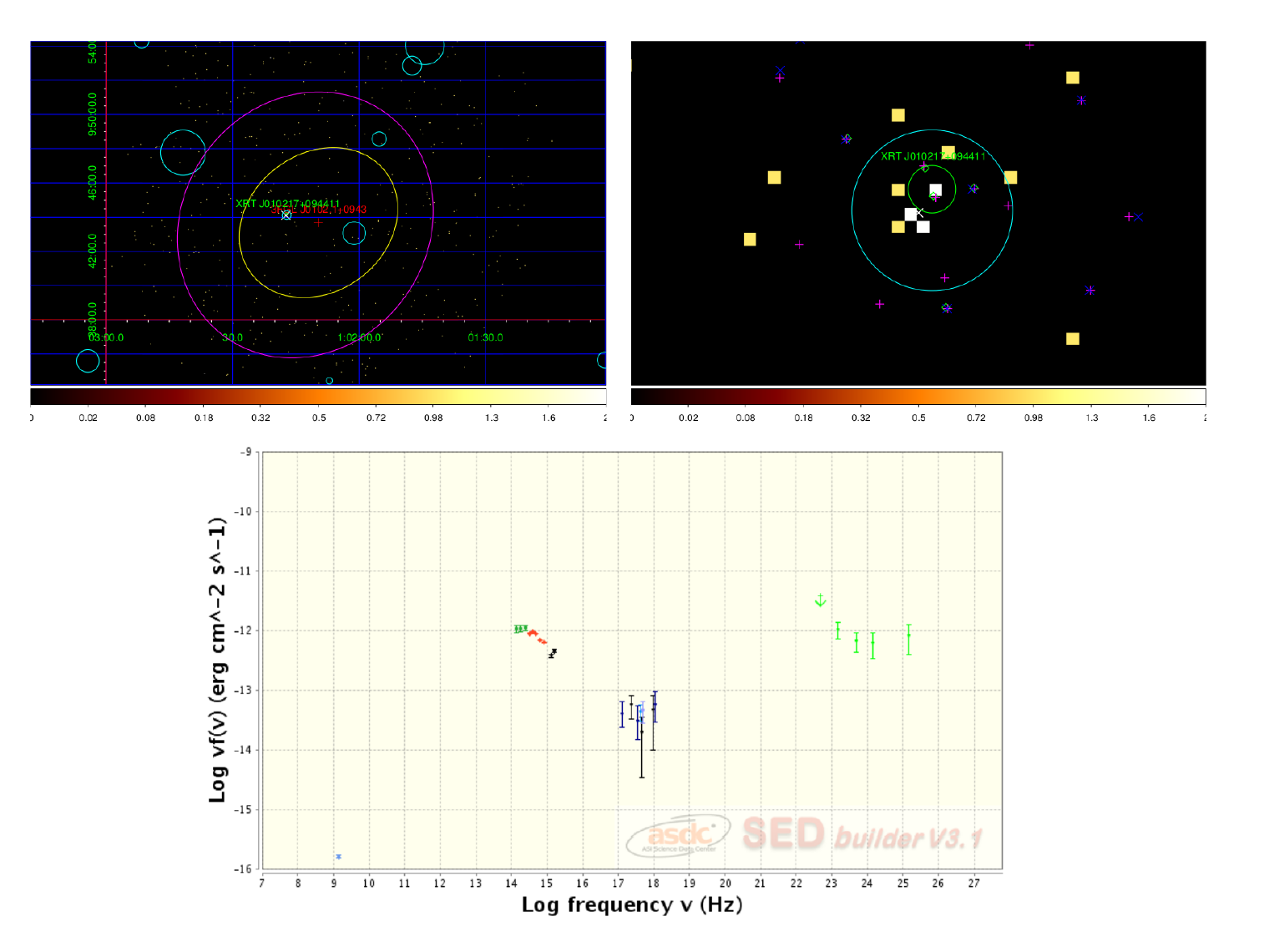} } 
  \caption{\texttt{Upper panel left}:   \textit{Swift}/XRT images of 2FGL J0102.2+0943 created using the online data analysis tool of UK Swift Science Data Centre. The red cross is the position of 2FGL J0102.2+0943 as reported in the 3FGL catalogue and the yellow (magenta) ellipse the 95$\%$ error region of 3FGL (2FGL) catalogue. The XRT source detected in this work, XRT 010217+094411, is displayed as a green circle. The cyan circles show the error circles of the NVSS radio sources and the white crosses are \textit{Swift}/XRT objects of the 1SXPS Swift XRT Point Source catalogue \citep{evans2014_catalog}. 
\texttt{Upper panel right}:   Close-up of XRT J010217+094411 sky map. The white cross is the position of the XRT source in the 1SXPS catalogue. The blue and magenta crosses are the positions of WISE and SDSS objects and the green diamonds corresponds to 2MASS sources.\newline
\texttt{Bottom panel}: Broadband SED of 2FGL J0102.2+0943 created using the SED Builder tool of the ASI ASDC Data Centre. We combine NVSS radio data (light blue point), IR 2MASS data (dark green points), SDSS optical data (red points) with the HE $\gamma$-ray data (green points) from the 3FGL catalogue. The X-ray flux is taken from the 1SXPS catalogue  (blue points) and from \citep{takeuchi2013} (black points).  
}
\label{fig:0102_ass}
\end{figure*}

This UGS shows a detection significance of  7.09 (5.5) $\sigma$ and an error-box of 4.8' (7.8') in the 3FGL (2FGL) catalogue. Two observations are performed by \textit{Swift}/XRT for a total exposure time of about 4000 sec.
Using the XRT imaging analysis tool of the UK Swift Science Data Centre, in the X-ray sky map (Fig. \ref{fig:0102_ass}, upper-left panel), for this \textit{Fermi} source, only one faint X-ray source is detected within the 3FGL error-box (\textit{yellow ellipse}), with (RA,DEC) = (01 02 17.15, 09 44 11.16) and a 90\% positional error radius of 4.5". 
The estimated count rate is (2.624$\times$10$^{-3}$ $\pm$ 8.314$\times$10$^{-4}$) cts/s for a total of 11 counts.
Superimposing the catalogues of the other wavelengths with the DS9 plotting tool (Fig. \ref{fig:0102_ass} upper-right panel), we find a positional coincidence with the radio source NVSS J010217+094407, the IR source 2MASS 01021713+0944098 and the optical SDSS10 1237678833220911130. 
Another IR source appears positionally coincident within the XRT error-box, but we do not consider it because the coincidence is very marginal and the corresponding optical source is outside the region. 
Through the SED builder tool of the ASI ASDC Data Centre we build the multifrequency SED (Fig. \ref{fig:0102_ass}, bottom panel) combining the fluxes of the proposed set of counterparts and including also the XRT flux data from \citep{takeuchi2013} and the X-ray data points taken from 1SXPS catalogue \citep{evans2014_catalog} .

\subsection{2FGL J0116.6-6153} 
\label{0116_ass}

\begin{figure*}
\raisebox{-.5\height}  {\includegraphics[width=.7\textwidth]{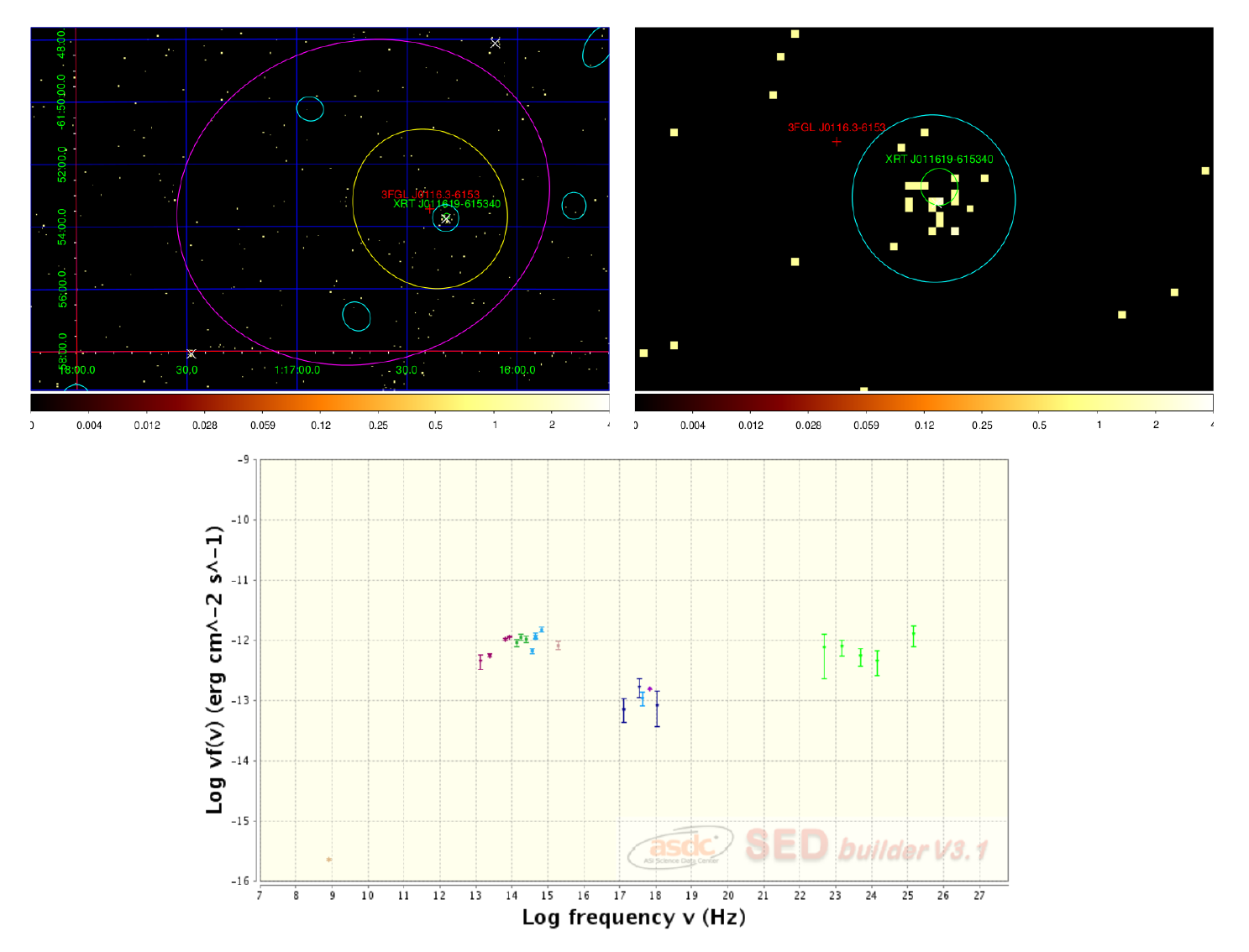} } 
 \caption{\texttt{Upper panel left}: \textit{Swift}/XRT images of 2FGL J0116.6-6153 created using the online data analysis tool of UK Swift Science Data Centre. The red cross is the 2FGL J0116.6-6153 position as reported in the 3FGL, and the yellow (magenta) ellipse the 95$\%$ error region of 3FGL (2FGL) catalogue. The X-ray source XRT J011619-615340, detected in this work, is displayed as a green circle. The cyan ellipses show the error ellipses of the SUMSS radio sources and the white crosses are \textit{Swift}/XRT objects of the 1SXPS Swift XRT Point Source catalogue \citep{evans2014_catalog}. 
\texttt{Upper panel right}: Close-up sky-map of XRT J011619-615340. The blue and magenta crosses are the positions of WISE and USNO objects and the green diamond corresponds to 2MASS source. \newline
\texttt{Bottom panel}: Broadband SED of 2FGL J0116.6-6153 created using the SED Builder tool of the ASI ASDC Data Centre. We combine SUMSS radio data (beige point), WISE and 2MASS IR data (violet and green points), USNOB1.0 optical data (blue points) with the HE $\gamma$-ray data (green points) from the 3FGL catalogue. The X-ray data points (blue points) are taken from the 1SXPS catalogue.}
  \label{fig:0116_ass}
\end{figure*}

In the 3FGL (2FGL) catalogue this object is reported with a detection significance of 9.90 (5.5) $\sigma$ and the 95\% \textit{semi major axis} is 0.04$^{\circ}$ (6'). 
In the 3FGL Its new classification is a \textit{blazar} of BL Lac type. 
Two \textit{Swift}-XRT observations are available for a total of 3276 sec. 
Through the UK \textit{Swift} data analysis, we obtain the X-ray image shown in Fig. \ref{fig:0116_ass} (upper-left panel).  
Within the 3FGL error-box (yellow ellipse), we detect only one X-ray source with (RA,DEC) = (01 16 19.24, -61 53 40.2) with a 90$\%$ error radius of 5.7''. 
The estimated net count rate is (6.424$\times$10$^{-3}$ $\pm$ 1.432$\times$10$^{-3}$) cts/s. 
Hence we propose this object as the most likely counterpart of 2FGL J0116.6-6153.
From the close-up image in Fig. \ref{fig:0116_ass} (upper-right panel), the radio source SUMSS J011619-615343, the IR sources WISE J011619-615343 and 2MASS 01161959-6153434, and the optical source USNOB U0281-0014602 are spatially coincident with the X-ray position of XRT J011619-615340. 
The multi-wavelength SED (\ref{fig:0116_ass} bottom panel) is built by combining all available flux data of this set of counterparts. 

\subsection{2FGL J0143.6-5844} 
\label{0143_ass}
\begin{figure*}
\raisebox{-.5\height}  {\includegraphics[width=.7\textwidth]{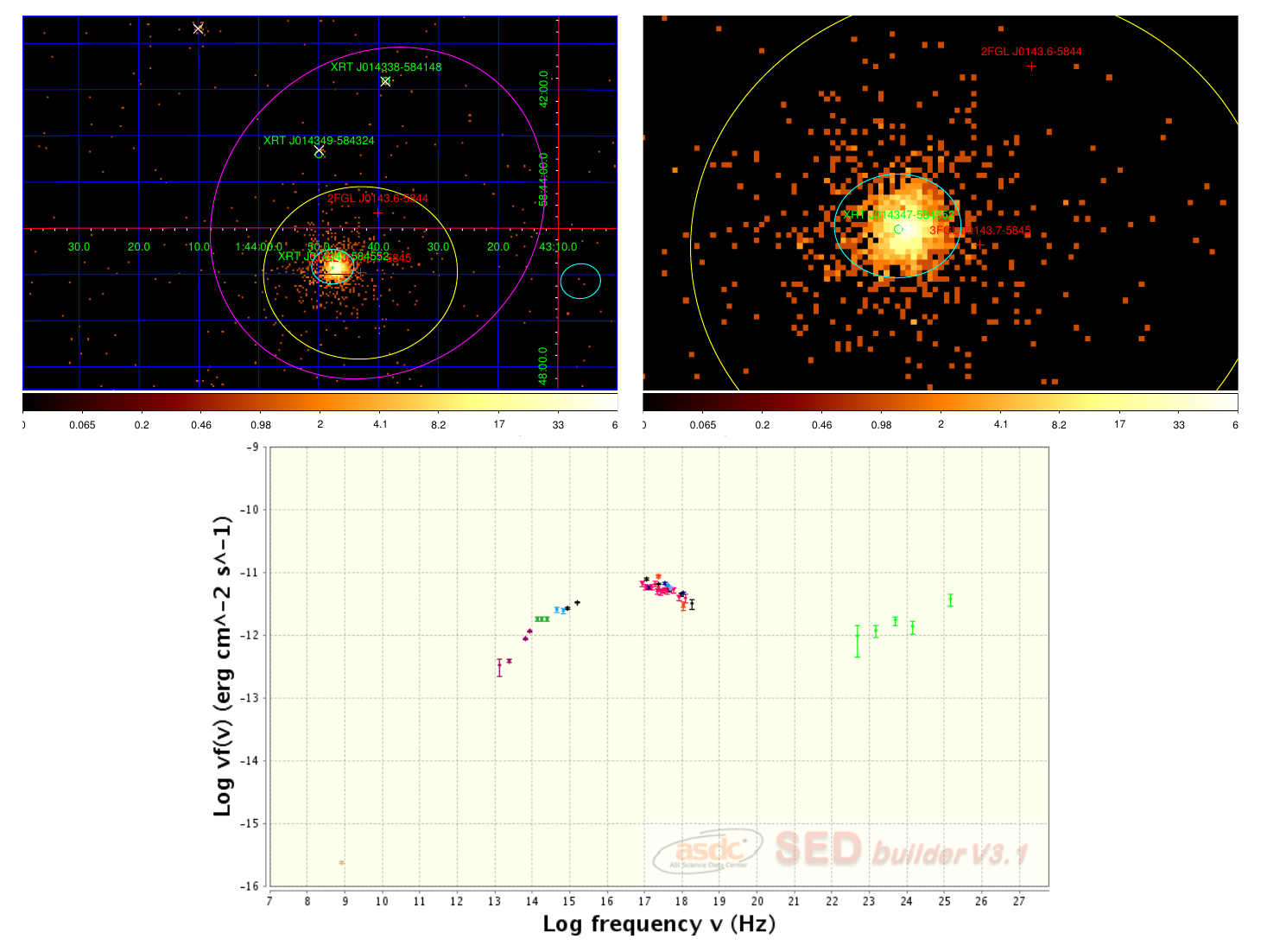} } 
  \caption{\texttt{Upper panel left}:  \textit{Swift}/XRT images of 2FGL J0143.6-5844 created using the online data analysis tool of UK Swift Science Data Centre. The red crosses are the 2FGL J0143.6-5844 and 3FGL J0143.6-5845 position, and the yellow (magenta) ellipse the 95$\%$ error region of 3FGL catalogue. The XRT sources detected in this work are displayed as green circles with radius equal to the XRT error radius. The cyan ellipses show the error ellipses of the SUMSS radio sources and the with crosses are \textit{Swift}/XRT objects of the 1SXPS catalogue. \texttt{Upper panel right}:  Close-up of XRT 014347-584551 sky map. The white cross is the position of the XRT source in the 1SXPS catalogue. The blue and magenta crosses are the positions of WISE and USNO objects and the green diamonds correspond to 2MASS sources. \newline
\texttt{Bottom panel}: Broadband SED of 2FGL J0143.6-5844 created using the SED Builder tool of the ASI ASDC Data Centre. We combine SUMSS radio data (beige point), WISE and 2MASS IR data (violet and green points), USNOA2.0 optical data (light blue points) with the HE $\gamma$-ray data (green points) from the 3FGL catalogue. The X-ray data points are given from our \textit{Swift}/XRT analysis (magenta points), from \citep{takeuchi2013} (black points) and from the 1SXPS catalogue (blue points).}
  \label{fig:0143_ass}
\end{figure*}

In the 3FGL (2FGL) catalogue, this source is reported with a detection significance of 18.98 (14.2) $\sigma$ and a 95$\%$ \textit{semi major axis} of 2.4' (3.6'). 
The new 3FGL classification for the source is \textit{blazar} of BL Lac type. 
2FGL J0143.6-5844 has been observed by \textit{Swift}/XRT, that was pointing at the coordinates of the 1FGL J0143.9-5845 and collecting 4348 seconds of good exposure time.
The XRT sky map is shown in Fig. \ref{fig:0143_ass}-(upper-left panel), with only one X-ray source detected within the 3FGL error-box.  
We suggest XRT J014347-584551 as the likely X-ray counterpart for 2FGL J0143.6-5844. 
It is a very bright X-ray source with a count rate of (3.765$\times$10$^{-1}$ $\pm$ 9.337$\times$10$^{-1}$) cts/s. 
The XRT enhanced position is (RA,DEC)=(01 43 47.57, -58 45 51.6) with an error radius of 1.9''. 
In the close-up image (Fig. \ref{fig:0143_ass} upper-right panel), we find that the radio source SUMSS J014347-584550, together with the infrared sources WISE J014347-584551 and 2MASS 01434742-5845514, and the optical object USNOA2.0 U0300\_00524092 are spatially coincident with the error region of the X-ray counterpart.
In the bottom panel of the same figure, the corresponding broad-band SED built is shown. 
The magenta points are the X-ray data calculated from our dedicated XRT analysis, the black points are the X-ray spectrum taken from \citep{takeuchi2013} and the blue points from 1SXPS catalogue.

\subsection{2FGL J0338.2+1306} 
\label{0338_ass}

\begin{figure*}
\raisebox{-.5\height}  {\includegraphics[width=.7\textwidth]{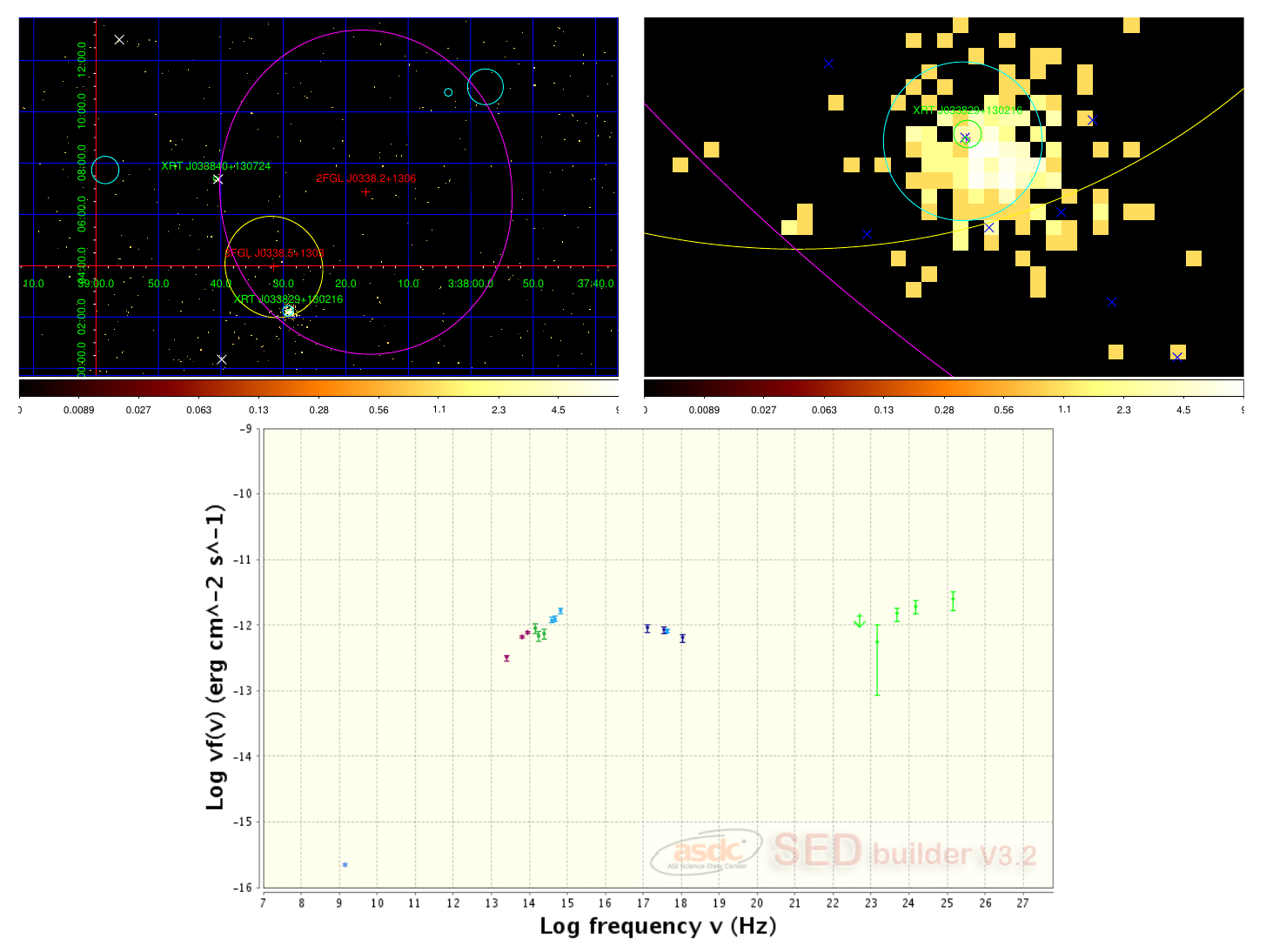} } 
  \caption{\texttt{Upper panel left}:  \textit{Swift}/XRT images of the 2FGL J0338.2+1306 created using the online data analysis tool of UK Swift Science Data Centre. The red crosses are the 2FGL and 3FGL position and the yellow (magenta) ellipse the 95$\%$ error region of 3FGL (2FGL) catalogue. The XRT sources detected in this work are displayed as green circles with radius equal to the XRT error radius. The cyan ellipses show the error circles of the NVSS radio sources and the white crosses are the\textit{Swift}/XRT objects of the 1SXPS catalogue. \texttt{Upper panel right}: Close up of XRT J033829+130216 sky map. The white cross is the position of the XRT source in the 1SXPS catalogue. The blue and magenta crosses are the positions of WISE and USNO objects and the green diamonds correspond to 2MASS sources.
\newline
\texttt{Bottom panel}: Broadband SED of 2FGL J0338.2+1306 created using the SED Builder tool of the ASI ASDC Data Centre. We combine NVSS radio data (blue point), WISE and 2MASS IR data (violet and green points), USNO B1.0 optical data (light blue points) with the HE $\gamma$-ray data (green points) from the 3FGL catalog. The X-ray flux points are obtained from  1SXPS catalogue (blue points).}
  \label{fig:0338_ass}
\end{figure*}

This $\gamma$-ray emitter is a \textit{Fermi} source with detection significance of 11.90 (5.8) $\sigma$ and an error-box of 1.8' (6.6') in the 3FGL (2FGL) catalogue. 
In the 3FGL catalogue, this source is classified as an active galaxy of uncertain type (BCU-II).
It was observed by \textit{Swift}/XRT on 4th July 2012 with an exposure time of 3344 sec. 
The resulting XRT sky map is shown in Fig. \ref{fig:0338_ass}-(upper-left panel). 
We found only one X-ray source, XRT J033829+130216, within the 3FGL error-box (yellow ellipse). 
Therefore we decide to propose it as the most likely X-ray counterpart, according to the 3FGL association. 
From the image analysis, the XRT positional error for this source is 2.1'' and its count rate is (7.160$\times$10$^{-2}$ $\pm$ 4.653$\times$10$^{-3}$) cts/s with 242 total counts found. 
Using an appropriate absorbed model, the integral flux in the energy range 0.3--10 keV is 4.1319$\times$10$^{-12}$ ergs cm$^{-2}$ s$^{-1}$.
Looking at the close-up image (Fig. \ref{fig:0338_ass} upper-right panel), we can see that the radio source NVSS J033829+130215, the infrared sources WISE J033829+130215 and 2MASS 03382926+1302151, and the optical USNOB 1030-0045117 are spatially coincident with the X-ray object. 
The multi-frequency SED (bottom panel) is obtained by combining the data-points of these objects.

\subsection{2FGL J1129.5+3758} 
\label{1129_ass}

\begin{figure*}
\raisebox{-.5\height}  {\includegraphics[width=.7\textwidth]{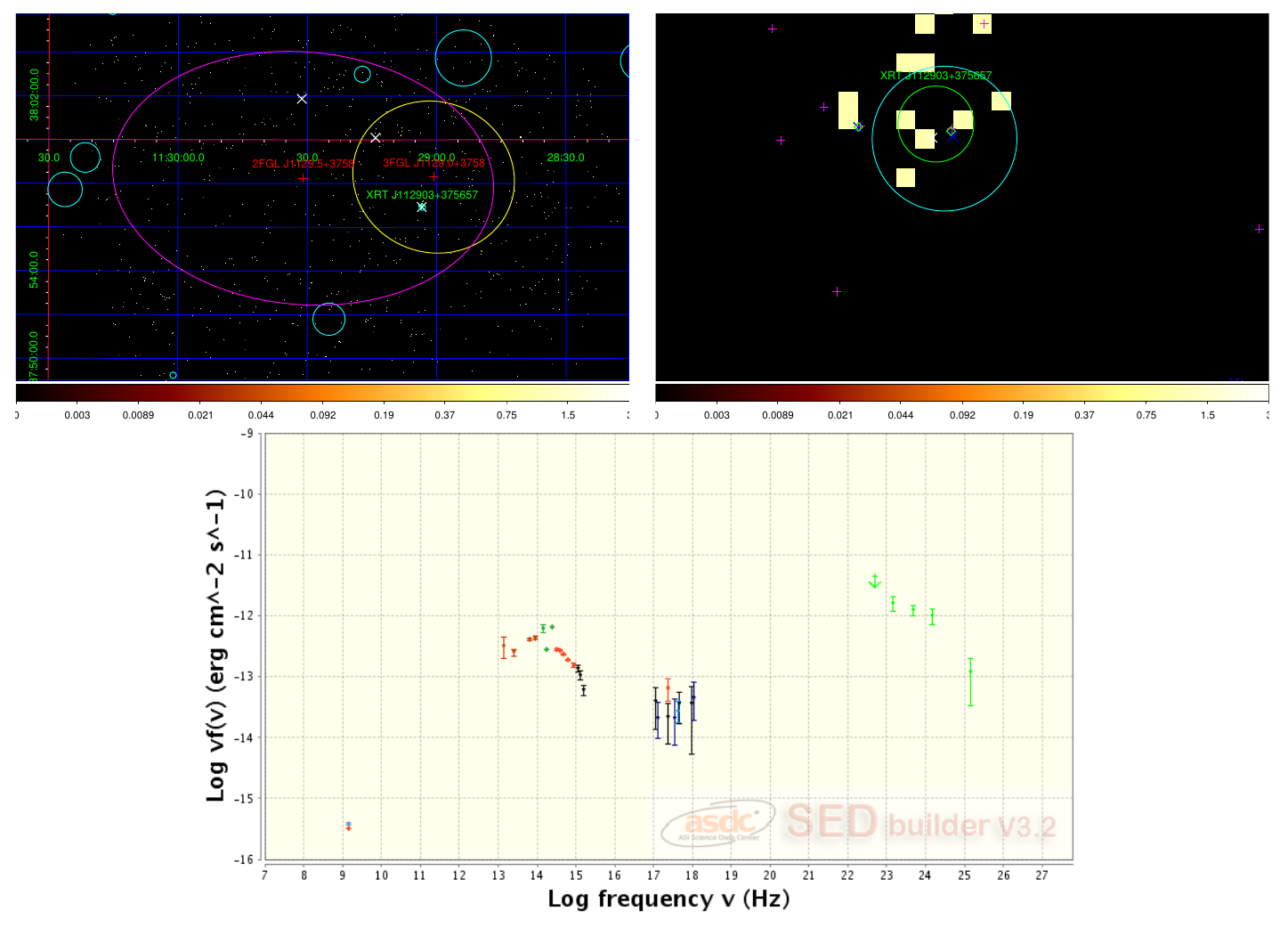} } 
  \caption{\texttt{Upper panel left}:  \textit{Swift}/XRT images of the UGS 2FGL J1129.5+3758 (3FGL J1129.0+3758) created using the online data analysis tool of UK Swift Science Data Centre. The red crosses are the 2FGL J1129.5+3758 and 3FGL J1129.0+3758 position and the yellow (magenta) ellipse the 95$\%$ error region of 3FGL (2FGL) catalogue. The XRT sources detected in this work are displayed as green circles with radius equal to the XRT error radius. The cyan ellipses show the error circles of the NVSS radio sources and the white crosses are the \textit{Swift}/XRT objects of the 1SXPS catalogue. \texttt{Upper panel right}:  Close-up of  XRT J112903-375857 sky map. The white cross is the position of the XRT source in the 1SXPS catalogue. The blue and magenta crosses are the positions of WISE and USNO objects and the green diamonds correspond to 2MASS sources. \newline
\texttt{Bottom panel}:  Broadband SED of the 2FGL J1129.5+3758 - 3FGL J1129.0+3758  created using the SED Builder tool of the ASI ASDC Data Centre. We combine WISE and 2MASS IR data (brown points and dark green points respectively) with the HE $\gamma$-ray data (green points) from the 3FGL catalogue. The X-ray flux estimate by 1SXPS catalogue are in blue and the black data points are taken from \citep{takeuchi2013}.}
  \label{fig:1129_ass}
\end{figure*}

In the 3FGL catalogue 2FGL J1129.5+3758 is still an unidentified object with a detection significance = 10.25 $\sigma$ and a 95\% semi major axis of 3.6'.
In 2014 the \textit{Swift} satellite provided about 4700 seconds of data. 
Through the X-ray image analysis, we found that within the reduced 3FGL error box of this source there is one X-ray source (XRT J112903+375857 with (RA,DEC)=21 15 22.08, 12 18 01.8) detected (Fig. \ref{fig:1129_ass}, upper panel left).
We propose it as the likely X-ray counterpart of this UGS and the close-up image shows its multi-frequency counterparts within the X-ray error circle of radius 4.7'': the radio source NVSS J112903+375655, the IR objects WISE J112903+375655 and 2MASS 11290325+3756564, and the optical object SDSS7 587739099132657672. 
In the bottom panel the corresponding MWL SED is reported.

\subsection{2FGL J1410.4+7411}
\label{1410_ass}
\begin{figure*}
\raisebox{-.5\height}  {\includegraphics[width=.7\textwidth]{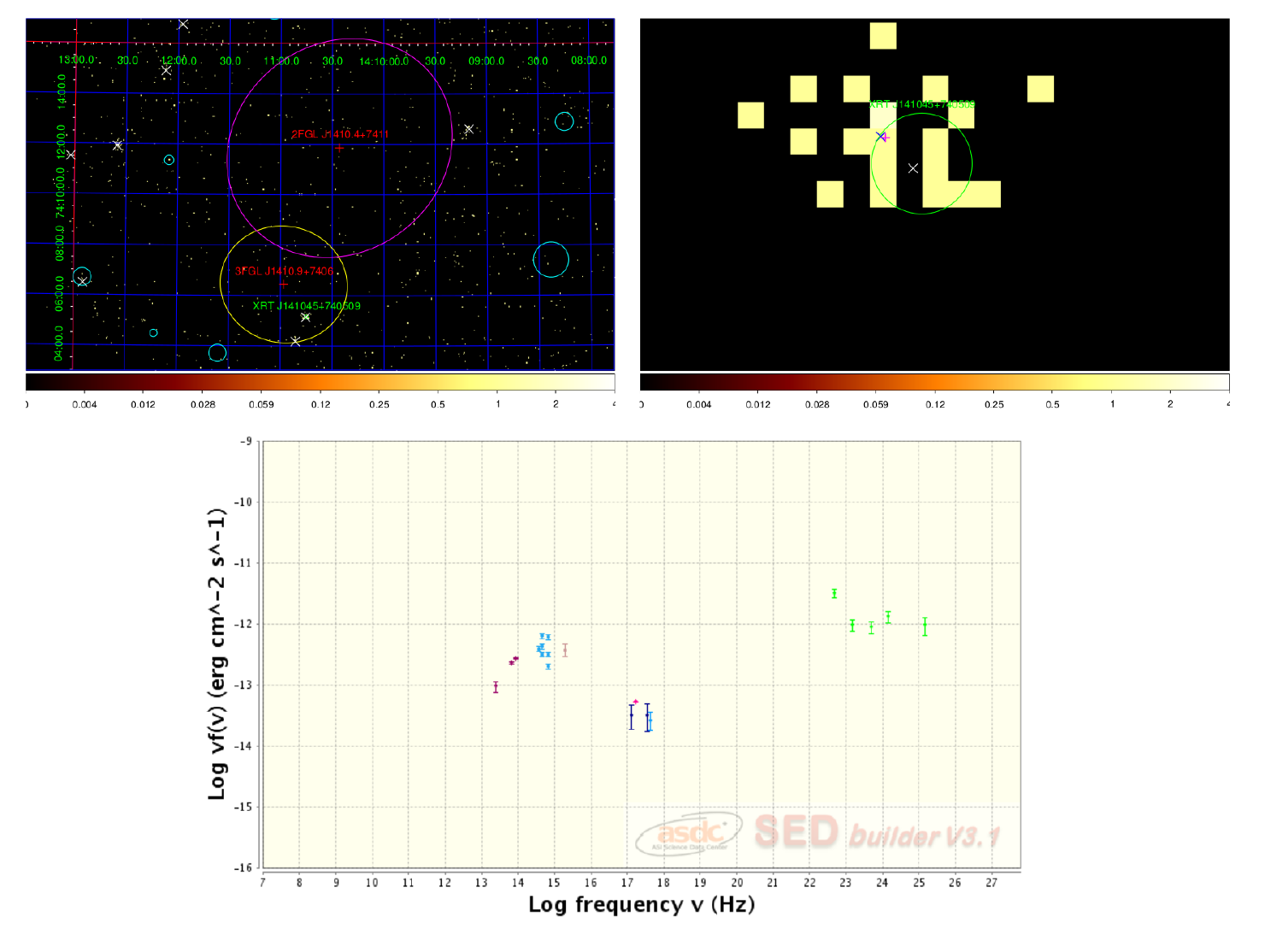} } 
  \caption{\texttt{Upper panel left}:  \textit{Swift}/XRT images of the UGS 2FGL J1410.4+7411 (3FGL J1410.9+7406) created using the online data analysis tool of UK Swift Science Data Centre. The red crosses are the 2FGL J1410.4+7411 and 3FGL J1410.9+7406 position and the yellow (magenta) ellipse the 95$\%$ error region of 3FGL (2FGL) catalogue. The XRT sources detected in this work are displayed as green circles with radius equal to the XRT error radius. The cyan ellipses show the error circles of the NVSS radio sources and the white crosses are the\textit{Swift}/XRT objects of the 1SXPS catalogue. \texttt{Upper panel right}:  Close-up of  XRT J141045+740609 sky map. The white cross is the position of the XRT source in the 1SXPS catalogue. The blue and magenta crosses are the positions of WISE and USNO objects and the green diamonds correspond to 2MASS sources. \newline
\texttt{Bottom panel}:  Broadband SED of the UGS 2FGL J1410.4+7411 - 3FGL J1410.9+7406  created using the SED Builder tool of the ASI ASDC Data Centre. We combine WISE IR data (violet points), USNO B1.0 optical data (light blue points) with the HE $\gamma$-ray data (green points) from the 3FGL catalogue. The X-ray flux data points are taken from the 1SXPS catalogue (blue points).}
  \label{fig:1410_ass}
\end{figure*}

In the 2FGL catalogue 2FGL J1410.4+7411 has a 9.8$\sigma$ significance and a \textit{semi major axis} of 4.8’. 
In the 3FGL catalogue the unassociated source 3FGL J1410.9+7406 is reported with a detection significance of 15.76$\sigma$ and a \textit{semi major axis} of 2.4'. 
We suggest that 2FGL J1410.4+7411 and 3FGL J1410.9+7406 are the same $\gamma$-ray emitter.  
There are several short \textit{Swift}/XRT observations provided between 2011 and 2014, and from the resulting XRT sky map, shown in Fig. \ref{fig:1410_ass}-(upper-left panel), we can found two X-ray 1SXPS sources (white crosses). We suggest the brightest one, the source XRT J141045+740609, as likely X-ray counterpart for 3FGL J1410.9+7406. 
From the image analysis, the XRT positional error for this source is 4.5''.
Looking at the close-up image (upper-right panel), we can see that the infrared sources WISE J141046+740511 and the optical USNOB 1640-0083647  are spatially coincident with the X-ray object.
 The multi-frequency SED of 2FGL J1410.4+7411 - 3FGL J1410.9+7406 is obtained by combining the data-points of these objects and the X-ray data points provided by the 1SXPS catalogue (blue points).

\subsection{2FGL J1502.1+5548} 
\label{1502_ass}

\begin{figure*}
\raisebox{-.5\height}  {\includegraphics[width=.7\textwidth]{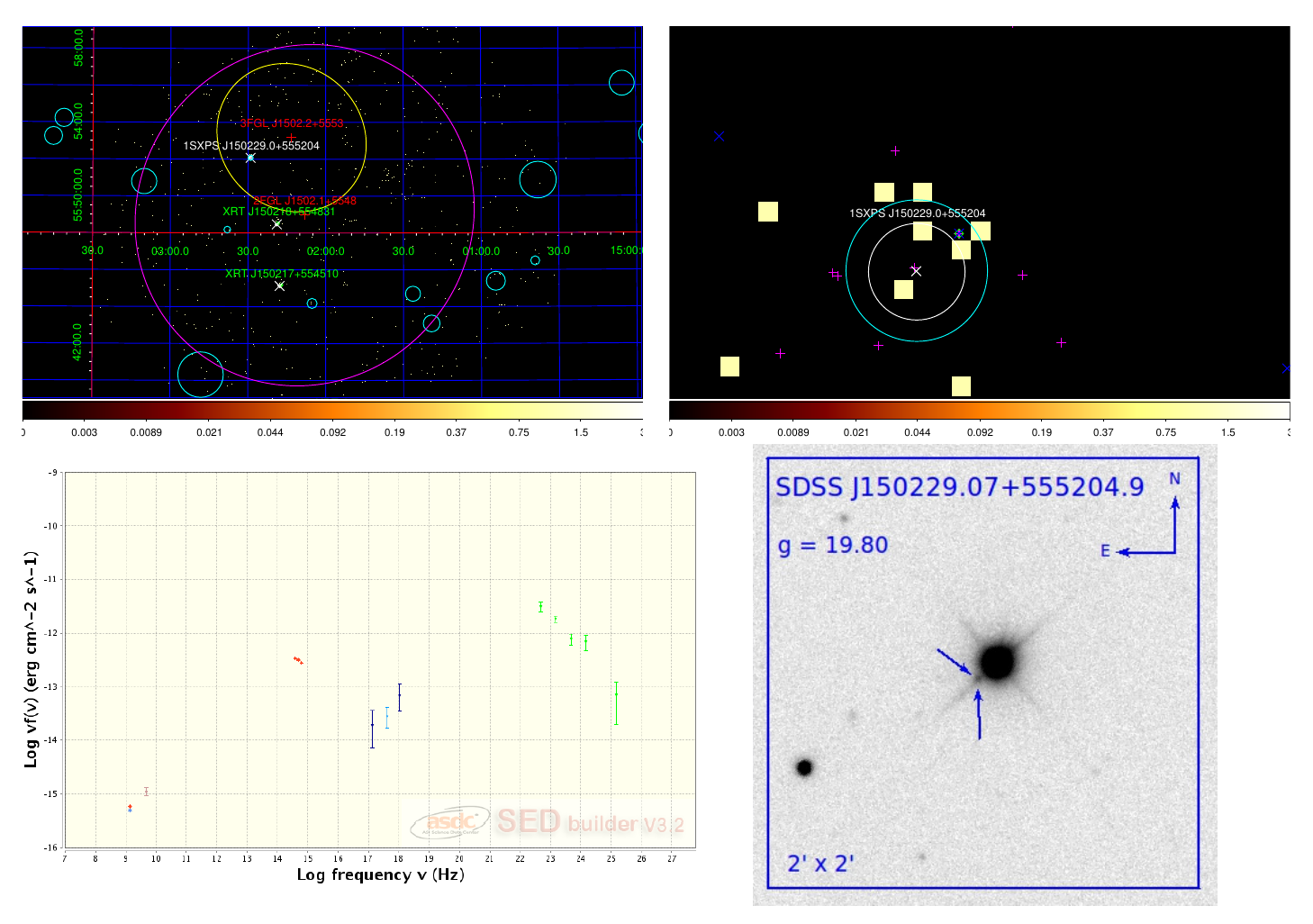} } 
  \caption{\texttt{Upper panel left}:  \textit{Swift}/XRT image of the 2FGL J1502.1+5548 created using the online data analysis tool of UK Swift Science Data Centre. The red cross is the 3FGL position and the yellow (magenta) ellipse the 95$\%$ error region of 3FGL (2FGL) catalogue. The XRT sources detected in this work are displayed as green circles. The cyan ellipses show the error circles of the NVSS radio sources and the white crosses are \textit{Swift}/XRT objects of the 1SXPS Swift XRT Point Source catalogue \citep{evans2014_catalog}. 
\texttt{Upper panel right}:   Close-up of 1SXPS J150229.0+555204 sky map. The white cross is the position of the XRT source in the 1SXPS catalogue. The blue and magenta crosses are the positions of WISE and SDSS objects and the green diamonds corresponds to 2MASS sources. \newline
\texttt{Bottom panel left}:  Broadband SED for the source 2FGL J1502.1+5548 created using the SED Builder tool of the ASI ASDC Data Centre. We combine the SDSS optical data (red points) with the HE $\gamma$-ray data (green points) from the 3FGL catalogue. The X-ray flux points are from the 1SXPS catalogue (blue points). \newline 
\texttt{Bottom panel right}: Optical finding chart centered on the optical source SDSS J150229.07+555204.9, the optical counterpart proposed for the UGS 2FGL J1502.1+5548. \newline 
}
  \label{fig:1502_ass}
\end{figure*}

In the 3FGL catalogue, this UGS is reported with a detection significance of 12.64$\sigma$ and a 95\% semi major axis of 4.2'. 2FGL J1502.1+5548 has been observed by Swift/XRT and $\sim$4000 seconds of good exposure time was collected. 
The XRT sky map is shown in Fig. \ref{fig:1502_ass}(upper-left panel), and we were able to detect two X-ray sources that are located outside the 3FGL error-box, but within the 3FGL error-box. 
From the high quality 1SXSP catalogue \citep{evans2014_catalog}, we note that the source 1SXSP J150229.0+555204 (with a positional error radius of 5.9''), coincident with the radio source NVSS J150229+555204, can be considered the most likely counterpart for 2FGL J1502.1+5548. 
In the close-up image of 1SXSP J150229.0+555204 (upper-right panel) we find that, besides the radio source, only the optical source SDSS J150229.07+555204.9 is spatially coincident with the X-ray counterpart. 
There are not IR sources detected and the closest one is located outside the X-ray error-box and associated to the star very close to SDSS J150229.07+555204.9 (see the finding chart in Fig. \ref{fig:1502_ass}-bottom-right panel).  
In the same figure (bottom-left panel) the corresponding broad-band SED built is shown.

\subsection{2FGL J1511.8-0513}	
\label{1511_ass}

\begin{figure*}
\raisebox{-.5\height}  {\includegraphics[width=.7\textwidth]{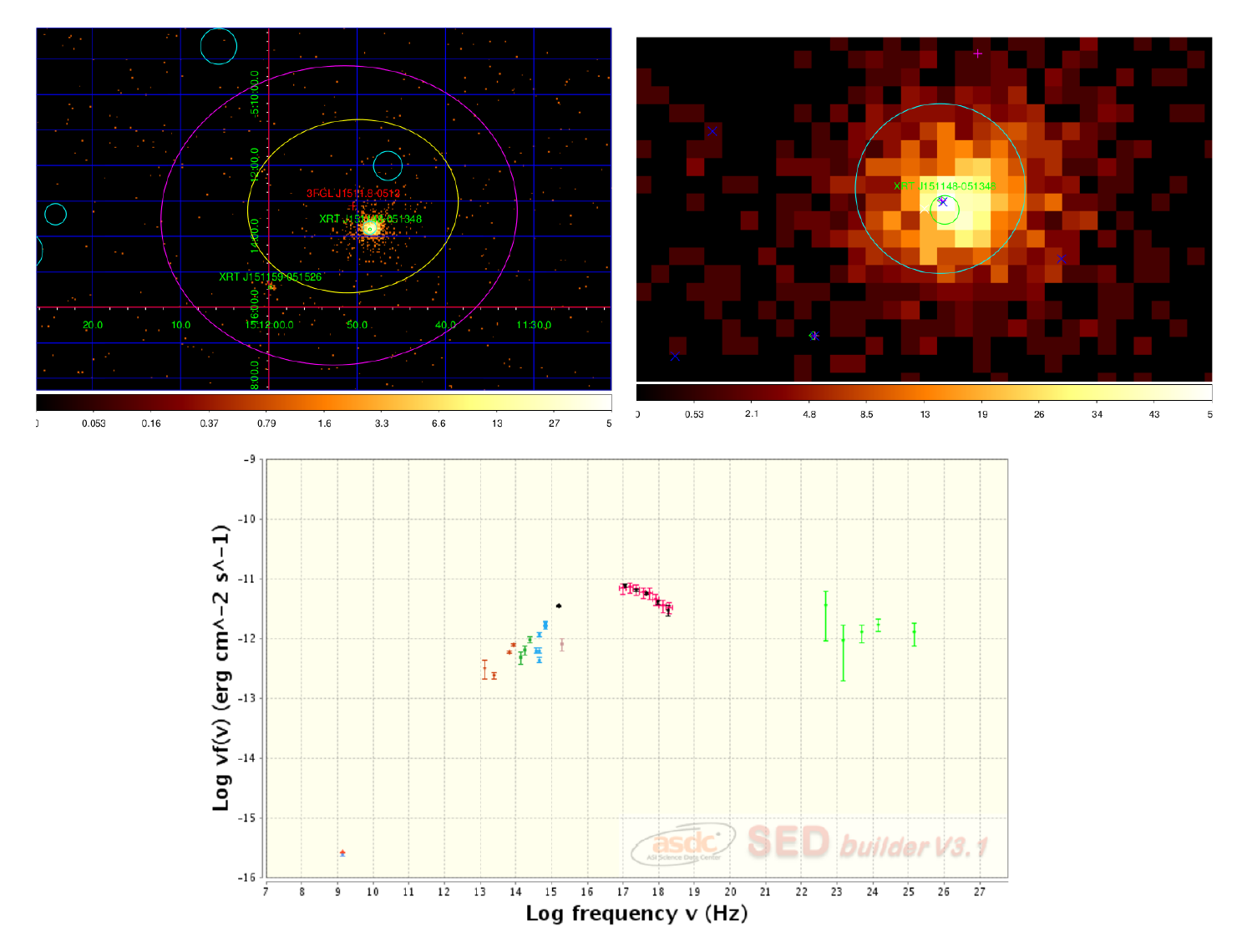} } 
  \caption{\texttt{Upper panel left}:  \textit{Swift}/XRT image of the 2FGL J1511.8-0513 created using the online data analysis tool of UK Swift Science Data Centre. The red cross is the 3FGL position and the yellow (magenta) ellipse the 95$\%$ error region of 3FGL (2FGL) catalogue. The XRT sources detected in this work are displayed as green circles. The cyan ellipses show the error circles of the NVSS radio sources and the white crosses are \textit{Swift}/XRT objects of the 1SXPS Swift XRT Point Source catalogue \citep{evans2014_catalog}. 
\texttt{Upper panel right}:   Close-up of XRT J151148-051348 sky map. The white cross is the position of the XRT source in the 1SXPS catalogue. The blue and magenta crosses are the positions of WISE and USNOB1.0 objects and the green diamonds corresponds to 2MASS sources. \newline
\texttt{Bottom panel}:  Broadband SED for the source 2FGL J1511.8-0513 created using the SED Builder tool of the ASI ASDC Data Centre. We combine the WISE IR data (brown points), the 2MASS IR data (green points), the USNOB1.0 optical data (light blue points) with the HE $\gamma$-ray data (green points) from the 2FGL catalogue. The X-ray flux points  are from the \textit{Swift}-XRT data analysis of performed this work (magenta points), from the 1SXPS catalogue (blue points) and from \citep{takeuchi2013} (black points). \newline }
  \label{fig:1511_ass}
\end{figure*}

In the 3FGL (2FGL) catalogue this source shows a detection significance of 10.59 (7.8) $\sigma$ and a \textit{semi major axis} of 3' (4.8'). 
In the 3FGL this source is not unassociated but it is classified as an active galaxy with uncertain type. 
By pointing at the position of the 1FGL source, \textit{Swift}/XRT observed the 2FGL J1511.8-0513 sky region in 2010 for a total time of 4160 sec. 
The XRT data analysis was performed by the UK XRT analysis tool and the resulting sky map is shown in Fig. \ref{fig:1511_ass} (upper-left panel). 
Within the 3FGL error-box we found only one X-ray source, with (RA,DEC) = (15 11 48.55, -05 13 48.00) and a 90$\%$ error radius of 1.9''. 
We propose it as X-ray counterpart in agreement with the 3FGL association.
From the close-up image in Fig. \ref{fig:1511_ass} (upper-right panel), the radio source NVSS 151148-051345, the IR sources WISE J151148-051346 and 2MASS 15114857-0513467, and the optical source USNOB U0825.08626045 are spatially coincident with the X-ray position of XRT J151148-051348. 
The multi-wavelength SED (bottom panel) is built by combining all available flux data of this set of counterparts. 
The X-ray flux is derived by our dedicated \textit{Swift}-XRT analysis (magenta points) and in addition we plot the X-ray data points taken from the 1SXPS catalogue (blue points) and \citep{takeuchi2013} (black points).

\subsection{2FGL J1544.5-1126} %
\label{1544_ass}

2FGL J1544.5-1126 shows a detection significance of 10.85 (5.79) $\sigma$ and a \textit{95$\%$ semi major axis} of 4.8' (8.4') in the 3FGL (2FGL) catalogue. 
\textit{Swift}/XRT did not observe it directly, but pointed to the ROSAT source 1RXS J154439.4-112820, from 2006 to 2012, for an exposure time of 13350 seconds. 
This object is the brightest X-ray source within the 3FGL error-box (yellow ellipse) of 2FGL J1544.5-1126 (Fig. \ref{1544_ass}, upper-left panel). 
We suggest it as the likely X-ray counterpart. 
From the XRT data analysis, we find that this X-ray counterpart has an error-box of 1.7''. 
In the close-up image (upper-right panel) the IR object WISE J154439-112804 and the optical source USNOB1.0 0785-0287377 are positionally coincident and hence we consider them as associated to 2FGL J1544.5-1126. 
The estimated XRT count rate is (7.003$\times$10$^{-2}$ $\pm$ 2.311$\times$10$^{-3}$) cts/s and the integrated 0.3-10 keV flux is 4.4394$\times$10$^{-12}$ ergs cm$^{-2}$ s$^{-1}$ (935 total counts). 
The X-ray differential spectrum (magenta points) is plotted in the multi-wavelength SED (bottom panel) together with the X-ray spectrum taken from \citet{takeuchi2013} (black points) and the data-points provided by the 1SXPS catalogue (blue points).

\begin{figure*}
\raisebox{-.5\height}  {\includegraphics[width=.7\textwidth]{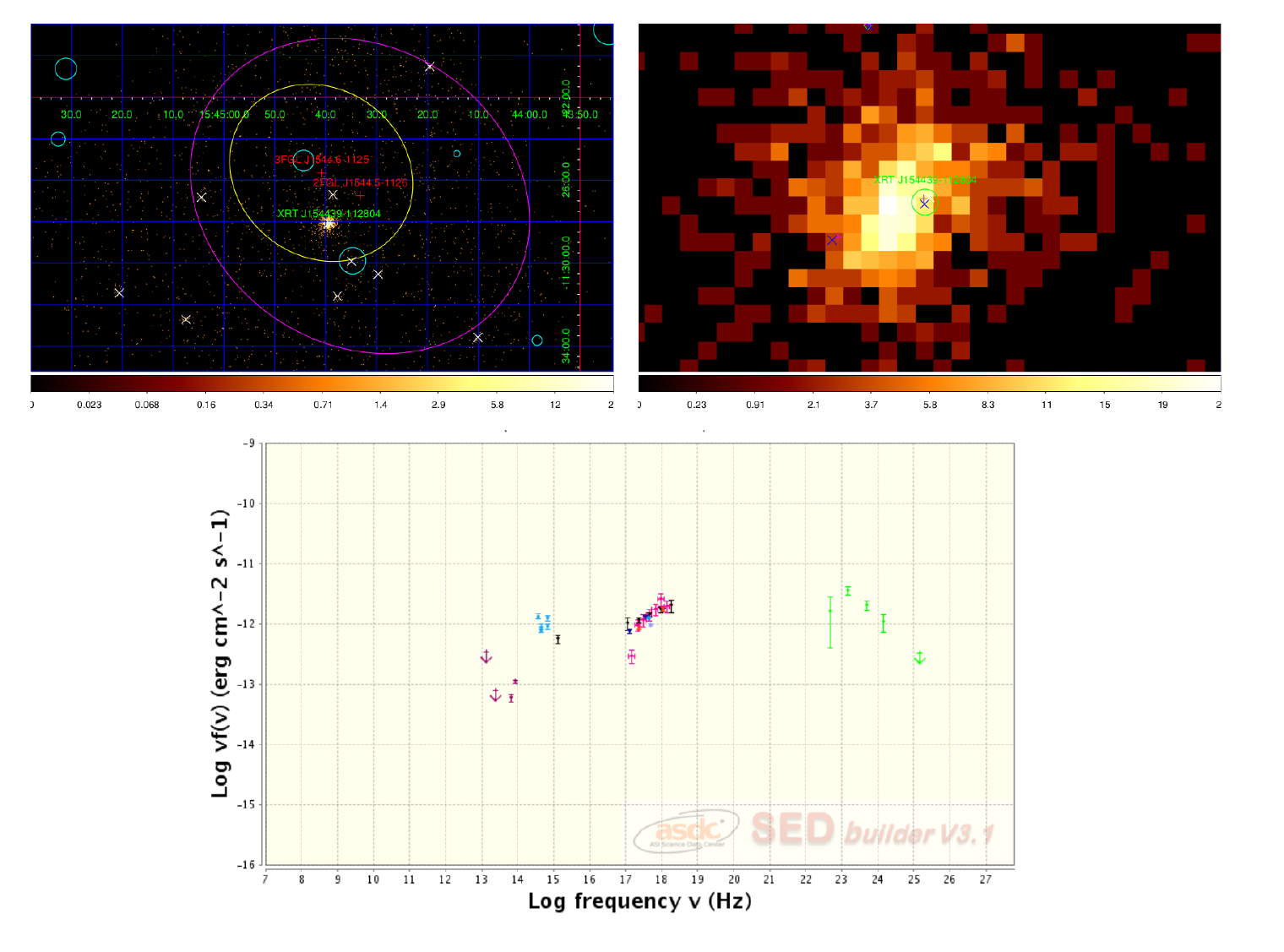} } 
  \caption{\texttt{Upper panel left}:  \textit{Swift}/XRT images of the 2FGL J1544.5-1126 created using the online data analysis tool of UK Swift Science Data Centre. The red crosses are the 2FGL and 3FGL position and the yellow (magenta) ellipse the 95$\%$ error region of 3FGL (2FGL) catalogue. The XRT sources detected in this work are displayed as green circles with radius equal to the XRT error radius. The cyan ellipses show the error circles of the NVSS radio sources and the white crosses are the\textit{Swift}/XRT objects of the 1SXPS catalogue. \texttt{Upper panel right}:  Close-up of  XRT J154439-112804 sky map. The white cross is the position of the XRT source in the 1SXPS catalogue. The blue and magenta crosses are the positions of WISE and USNO objects and the green diamonds correspond to 2MASS sources. \newline
\texttt{Bottom panel}:  Broadband SED of the UGS 2FGL J1544.5-1126 created using the SED Builder tool of the ASI ASDC Data Centre. We combine WISE IR data (violet points), USNO B1.0 optical data (light blue points) with the HE $\gamma$-ray data (green points) from the 3FGL catalogue. The X-ray data points given from our \textit{Swift}/XRT analysis are in magenta, while the data taken from \citet{takeuchi2013} in black and from the 1SXPS catalogue in blue.}
  \label{fig:1544_ass}
\end{figure*}


\subsection{2FGL J1614.8+4703} 
\label{1614_ass}

\begin{figure*}
\raisebox{-.5\height}  {\includegraphics[width=.7\textwidth]{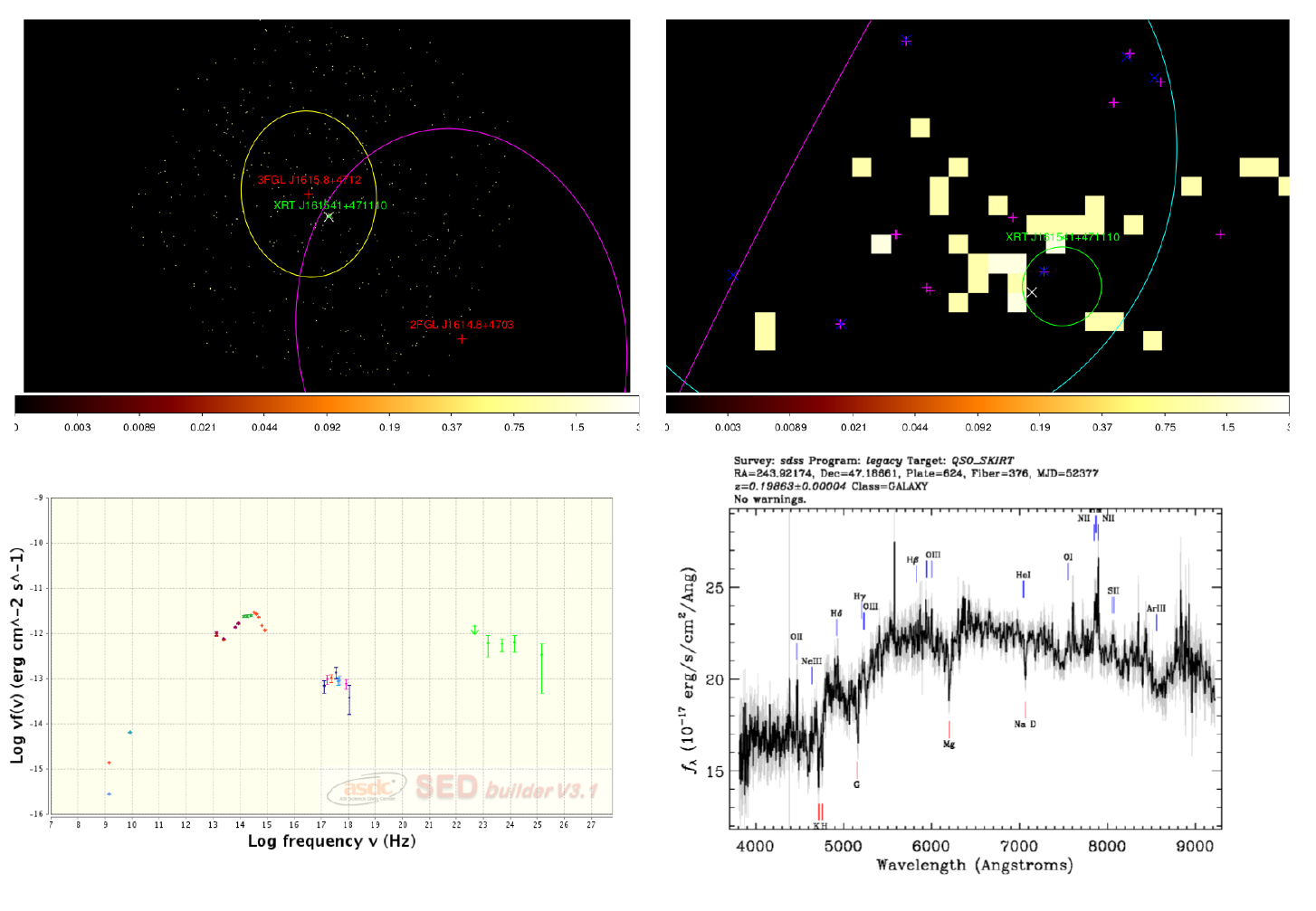} } 
  \caption{\texttt{Upper panel left}:  \textit{Swift}/XRT image of the 2FGL J1614.8+4703 created using the online data analysis tool of UK Swift Science Data Centre. The red crosses are the 2FGL J1614.8+4703 position and the yellow (magenta) ellipse is the 95$\%$ error region of 3FGL (2FGL) catalogue. The XRT sources detected in this work are displayed as green circles with radius equal to the XRT error radius. The cyan circles show the error circles of the NVSS radio sources and the white crosses are the\textit{Swift}/XRT objects of the 1SXPS catalogue. \texttt{Upper panel right}:  Close-up of XRT J161541+471110  sky-map. The white cross is the position of the XRT source in the 1SXPS catalogue. The blue and magenta crosses are the positions of WISE and SDSS objects and the green diamonds correspond to 2MASS sources. \newline
\texttt{Bottom panel left}:  Broadband SED for the UGS 2FGL J1614.8+4703 created using the SED Builder tool of the ASI ASDC Data Centre. We combine radio data,  WISE IR data (brown points), the 2MASS IR data (green points), the SDSS10 optical data (red points) with the HE $\gamma$-ray data (green points) from the 3FGL catalog. The X-ray flux points (magenta points) are performed from the \textit{Swift}-XRT data analysis of this work, and the blue points are the X-ray data taken from 1SXPS catalog. \texttt{Upper panel right}:  Optical spectrum for the optical counterpart provided by the SDSS survey.}
  \label{fig:1614_ass}
\end{figure*}

2FGL J1614.8+4703 is a very faint $\gamma$-ray emitter with a detection significance of 4.59$\sigma$ and a rather large \textit{Fermi 95$\%$ semi major axis}  of 13.8'. 
In the 3FGL and 3LAC catalogues, this object is a 6.30 $\sigma$ source with a \textit{95$\%$ semi major axis} of 5.4' and it is associated to the source TXS 1614+473, classified as LSP \textit{blazar}.  
The \textit{Swift}/XRT pointings for this source were targeted to the IR source 2MASX J16154117+47111 for 4990 sec (see Fig. \ref{fig:1614_ass}-upper panel left). 
Only the source XRT J161541+471110 is detected and we suggest it as likely X-ray counterpart, in agreement with the 3FGL association.
The close up image around XRT J161541+47111 position (upper panel right) shows that within the XRT positional error of 4.8'', and we can find the IR objects WISE J161541+471111 and 2MASS 16154121+4711118, and the optical object SDSS10 588007004192637004 spatially coincident. 
For the latter, the SDSS survey\footnote{http://skyserver.sdss3.org/dr10/en/tools/chart/navi.aspx} identifies the source with an elliptical galaxy at redshift of 0.19 (bottom panel right).
The multi-wavelength SED for 2FGL J1614.8+4703 is displayed in the bottom-left panel by combining all flux data of the proposed counterparts. The magenta points indicate the X-ray spectrum estimated through our UK online analysis of the XRT J161541+471110 data, while the blue points are the X-ray data flux taken from 1SXPS catalogue. 

\subsection{\textbf{2FGL J1704.3+1235}} 
\label{1704_12_ass}

\begin{figure*}
\raisebox{-.5\height}  {\includegraphics[width=.7\textwidth]{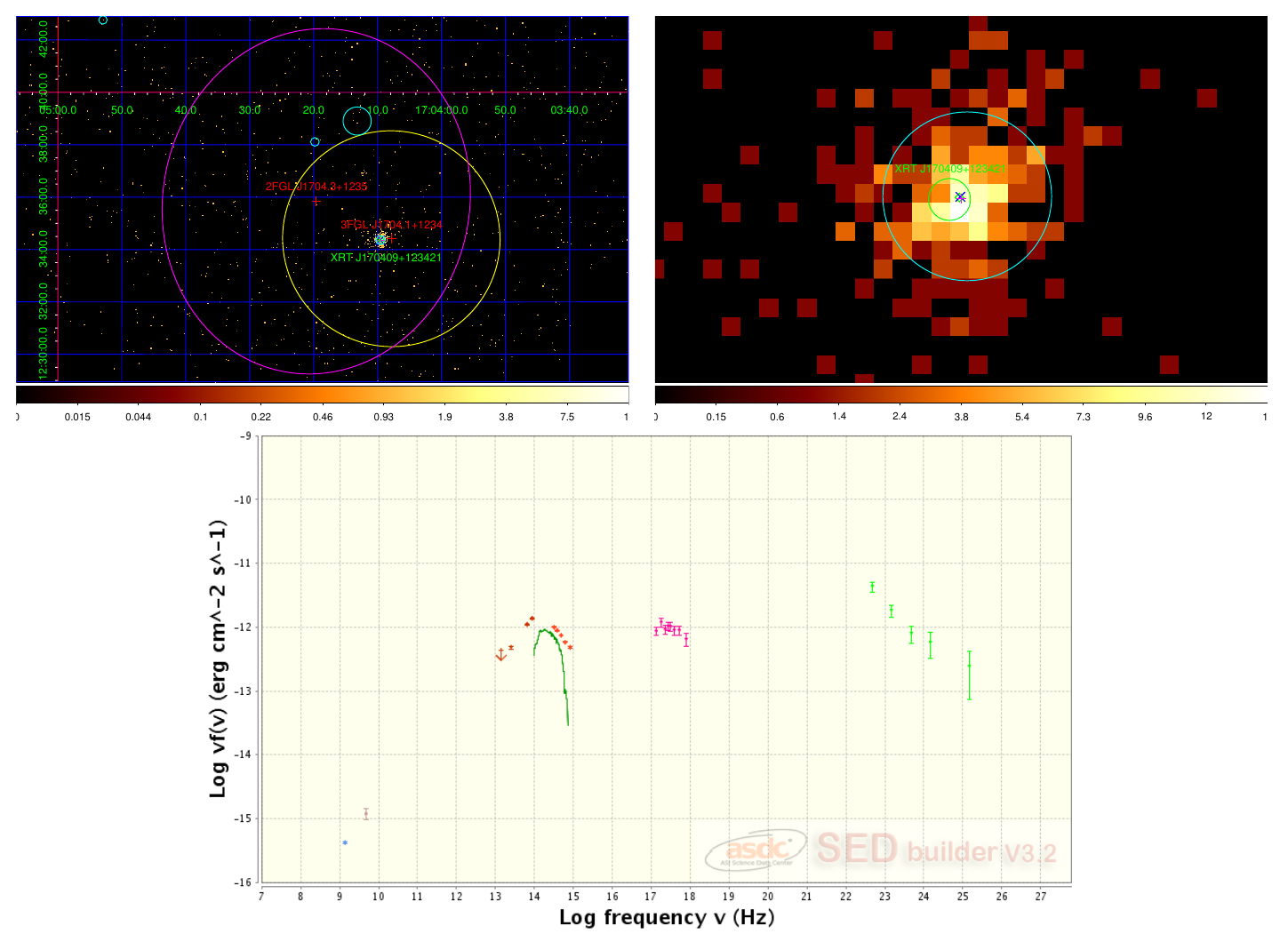} } 
  \caption{\texttt{Upper panel left}:  \textit{Swift}/XRT image of the 2FGL J1704.3+1235 created using the online data analysis tool of UK Swift Science Data Centre. The red crosses are the 2FGL and 3FGL position and the yellow ellipse is the 95$\%$ error region of 2FGL catalogue. The XRT sources detected in this work are displayed as green circles with radius equal to the XRT error radius. The cyan circles show the error circles of the NVSS radio sources and the white crosses are the\textit{Swift}/XRT objects of the 1SXPS catalog. \texttt{Upper panel right}:  Close-up of XRT J170409+123421 sky map. The blue and magenta crosses are the positions of WISE and SDSS objects and the green diamonds correspond to 2MASS sources. \newline
\texttt{Bottom panel}:  Broadband SED for the 2FGL J1704.3+1235 created using the SED Builder tool of the ASI ASDC Data Centre. We combine the WISE IR data (brown points), the 2MASS IR data (green points), the SDSS10 optical data (red points) with the HE $\gamma$-ray data (green points) from the 3FGL catalogue. The X-ray flux points (magenta points) are performed from the \textit{Swift}-XRT data analysis of this work.}
  \label{fig:1704_12_ass}
\end{figure*}

Through the UK online analysis of the 2013 XRT data ($\sim$ 4800 seconds), covering the 2FGL J1704.3+1235 sky region, within the 3FGL (2FGL) \textit{95$\%$ semi major axis} error box of 4.2' (6.6'), we found only one bright X-ray source, XRT J170409+123421 (Fig. \ref{1704_12_ass}, upper-left panel), with a count rate of 6.418$\times$10$^{-2}$ $\pm$3.68$\times$10$^{-3}$. 
We consider it as the X-ray counterpart for this UGS. 
Looking at the close-up image (upper-right panel) this X-ray source, with positional error of 2.6'',  is spatially coincident with the radio source NVSS J170409+123421, the infrared source WISE J170409+123421 and the optical object SDSS10 1237665106510021484. 
The broad-band SED is built by combining all the corresponding flux data and shown in the bottom panel: the magenta points are the X-ray differential spectrum obtained by the UK online analysis. 
We find evidence for the possible contribution of a host galaxy assumed it at z~=~0.3, as illustrated by the green spectrum in Fig. \ref{fig:1704_12_ass}.

\subsection{\textbf{2FGL J2115.4+1213}} 
\label{2115_ass}

\begin{figure*}
\raisebox{-.5\height}  {\includegraphics[width=.7\textwidth]{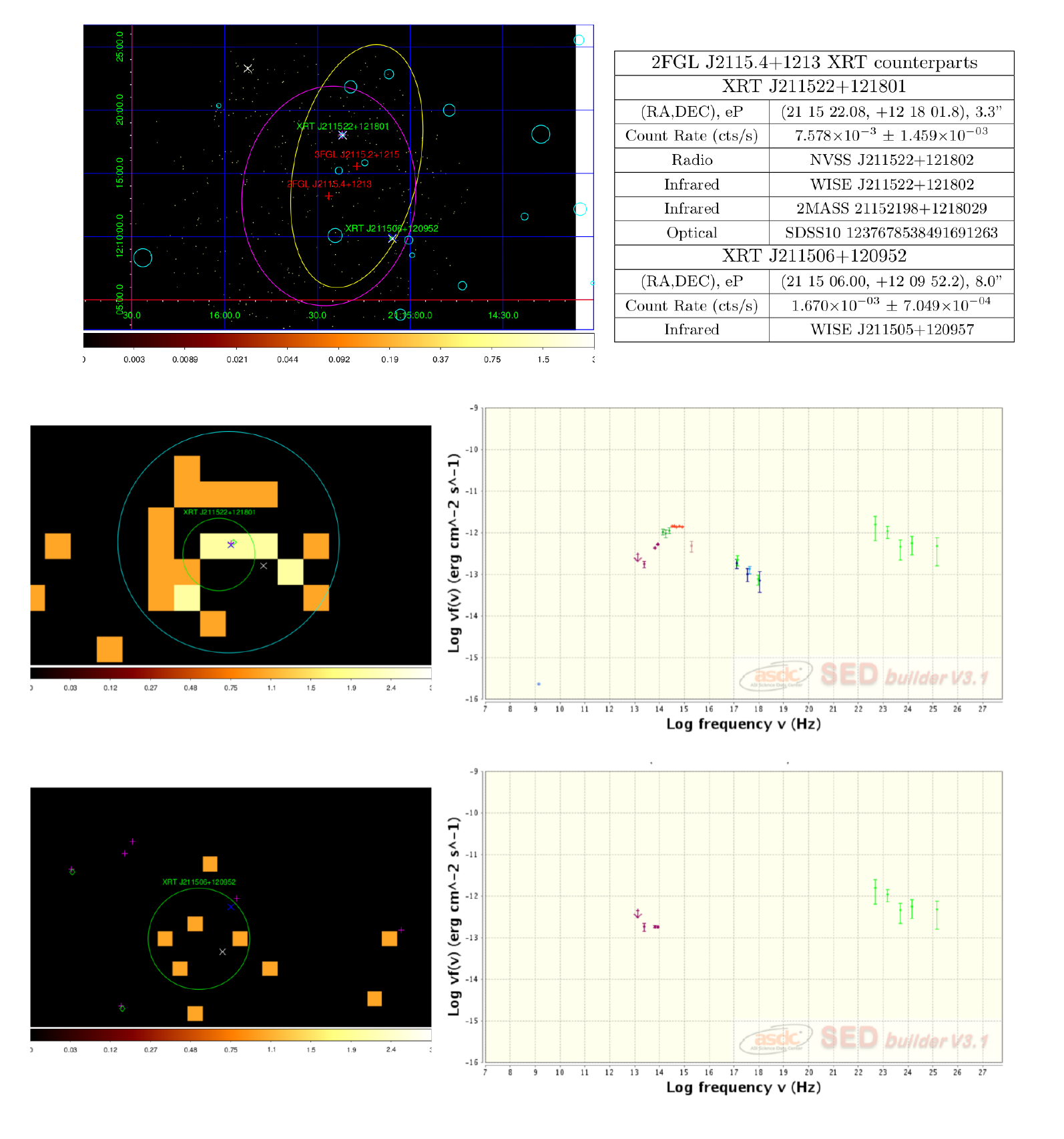} } 
  \caption{\texttt{Upper panel}:  \textit{Swift}/XRT image of the UGS 2FGL J2115.4+1213 created using the online data analysis tool of UK Swift Science Data Centre. The red crosses are the 3FGL (2FGL) position and the yellow (magenta) ellipse the 95$\%$ error region of 3FGL (2FGL) catalogue. The XRT sources detected in this work are displayed as green circles. The cyan ellipses show the error circles of the NVSS radio sources and the white crosses are \textit{Swift}/XRT objects of the 1SXPS Swift XRT Point Source catalogue \citep{evans2014_catalog}. \newline
\texttt{Middle panel left}:   Close-up of XRT J211522+121801 sky map. The white cross is the position of the XRT source in the 1SXPS catalogue. The blue and magenta crosses are the positions of WISE and USNOB1.0 objects and the green diamonds corresponds to 2MASS sources. \texttt{Middle panel right}:  Broadband SED for XRT J211522+121801  created using the SED Builder tool of the ASI ASDC Data Centre. We combine the WISE IR data (brown points), the 2MASS IR data (green points), the USNOB1.0 optical data (light blue points) with the HE $\gamma$-ray data (green points) from the 2FGL catalogue. The X-ray flux points (green points) are from the \textit{Swift}-XRT data analysis of performed this work and from the 1SXPS catalogue (blue ponts). \newline
\texttt{Bottom panel left}:  \textit{Left}: Close-up of XRT J211506+120952 sky map. The white cross is the position of the XRT source in the 1SXPS catalogue. The blue and magenta crosses are the positions of WISE and USNOB1.0 objects and the green diamonds corresponds to 2MASS sources. \texttt{Bottom panel right}:  Broadband SED for the XRT J211506+120952 created using the SED Builder tool of the ASI ASDC Data Centre. We combine the WISE (violet points) with the HE $\gamma$-ray data (green points) from the 2FGL catalogue.}
  \label{fig:2115_ass}
\end{figure*}

In the 3FGL (2FGL) catalogue 2FGL J2115.4+1213 is an unidentified object with a detection significance = 6.15 (5.11) $\sigma$ and a \textit{95$\%$ semi major axis} of 9.6' (8.4'). 
In 2012 the \textit{Swift} satellite provided about 3800 seconds of X-ray data. 
Through the image analysis, we found that within the \textit{Fermi} error box there are two X-ray sources detected (Fig. \ref{fig:2115_ass}, upper panel). 
We propose as the likely X-ray counterpart of this UGS the brightest X-ray object (details in the corresponding table) with 30 counts and (RA,DEC)=(21 15 22.08, 12 18 01.8). 
In the middle panel the multi-wavelength SED of XRT J211522+121801 with the close-up image that shows the multi-frequency counterparts for XRT J211522+121801 within the X-ray error circle of radius 3.3'': the radio source NVSS J211522+121802, the IR objects WISE J211522+121802 and 2MASS 21152198+1218029, and the optical object SDSS10 1237678538491691263.

\subsection{2FGL J2246.3+1549} 
\label{2246_ass}
\begin{figure*}
\raisebox{-.5\height}  {\includegraphics[width=.7\textwidth]{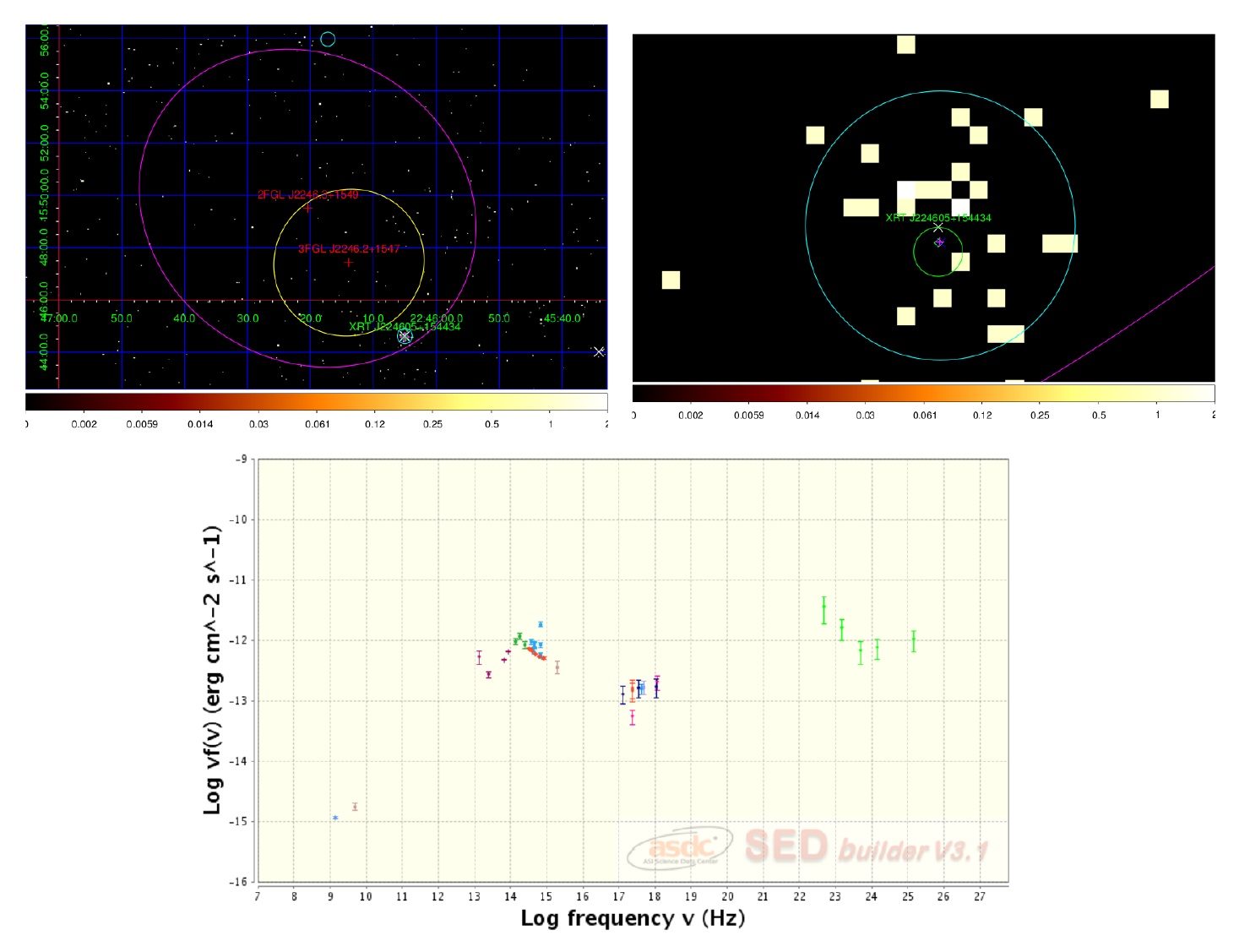} } 
  \caption{\texttt{Upper panel left}:  \textit{Swift}/XRT image of the  2FGL J2246.3+1549 created using the online data analysis tool of UK Swift Science Data Centre. The red crosses are the 2FGL J2246.3+1549 position and the yellow (magenta) ellipses is the 95$\%$ error region of 3FGL (2FGL) catalogue. The XRT sources detected in this work are displayed as green circles with radius equal to the XRT error radius. The cyan circles show the error circles of the NVSS radio sources and the white crosses are the\textit{Swift}/XRT objects of the 1SXPS catalogue. \texttt{Upper panel right}:  Close-up of XRT J224605+154434 sky map. The blue and magenta crosses are the positions of WISE and SDSS objects and the green diamonds correspond to 2MASS sources. \newline
\texttt{Bottom panel left}:  Broadband SED for the  2FGL J2246.3+1549 created using the SED Builder tool of the ASI ASDC Data Centre. We combine radio data, the WISE IR data (brown points), the 2MASS IR data (green points), the SDSS10 optical data (red points) with the HE $\gamma$-ray data (magenta points) from the 2FGL catalogue. The X-ray flux points are provided by the our \textit{Swift}-XRT data analysis (magenta points) and by the 1SXPS catalogue (blue points).}
  \label{fig:2246_ass}
\end{figure*}

In the 3FGL (2FGL) catalogue, this \textit{Fermi} object has a detection significance of 9.47 (8.21) $\sigma$ and a \textit{95$\%$ semi major axis} of 3' (6.6'). 
This object is associated and classified as an active galaxy of uncertain type. 
\textit{Swift}/XRT observed 2FGL J2246.3+1549 in 2010 for a total of 3381 seconds. 
The XRT sky map of the 2FGL J2246.3+1549 region is shown in Fig. \ref{fig:2246_ass}-(upper-left panel) and only one X-ray source, with (RA,DEC)=(22 46 05.1, +15 44 34.07) is detected within the 2FGL error ellipse, with a count rate of 8.718$\times$10$^{-3}$ $\pm$ 1.648$\times$10$^{-3}$ cts/s and the integrated 0.3-10 keV flux is 4.1156$\times$10$^{-13}$ ergs cm$^{-2}$ s$^{-1}$. 
However this X-ray source is not inside the 3FGL error region, but we decide to consider it as the likely X-ray counterpart for 2FGL J2246.3+1549 because it is the only X-ray source detected in the larger 2FGL error region and moreover this choice is in agreement with the association provide from the 3FGL catalog. 
From the close-up image (upper-right panel), within the X-ray positional error of 3.2'' we see that the radio source NVSS J224605+154437, the IR sources WISE J224604+154435 and 2MASS 22460500+1544352, and the optical object SDSS10 237680091651178703 can be associated to XRT J224605+154434. In the bottom panel the broad-band SED built by combining the flux data-points of the multi frequency counterparts.

\subsection{2FGL J2347.2+0707} 
\label{2347_ass}

\begin{figure*}
\raisebox{-.5\height}  {\includegraphics[width=.9\textwidth]{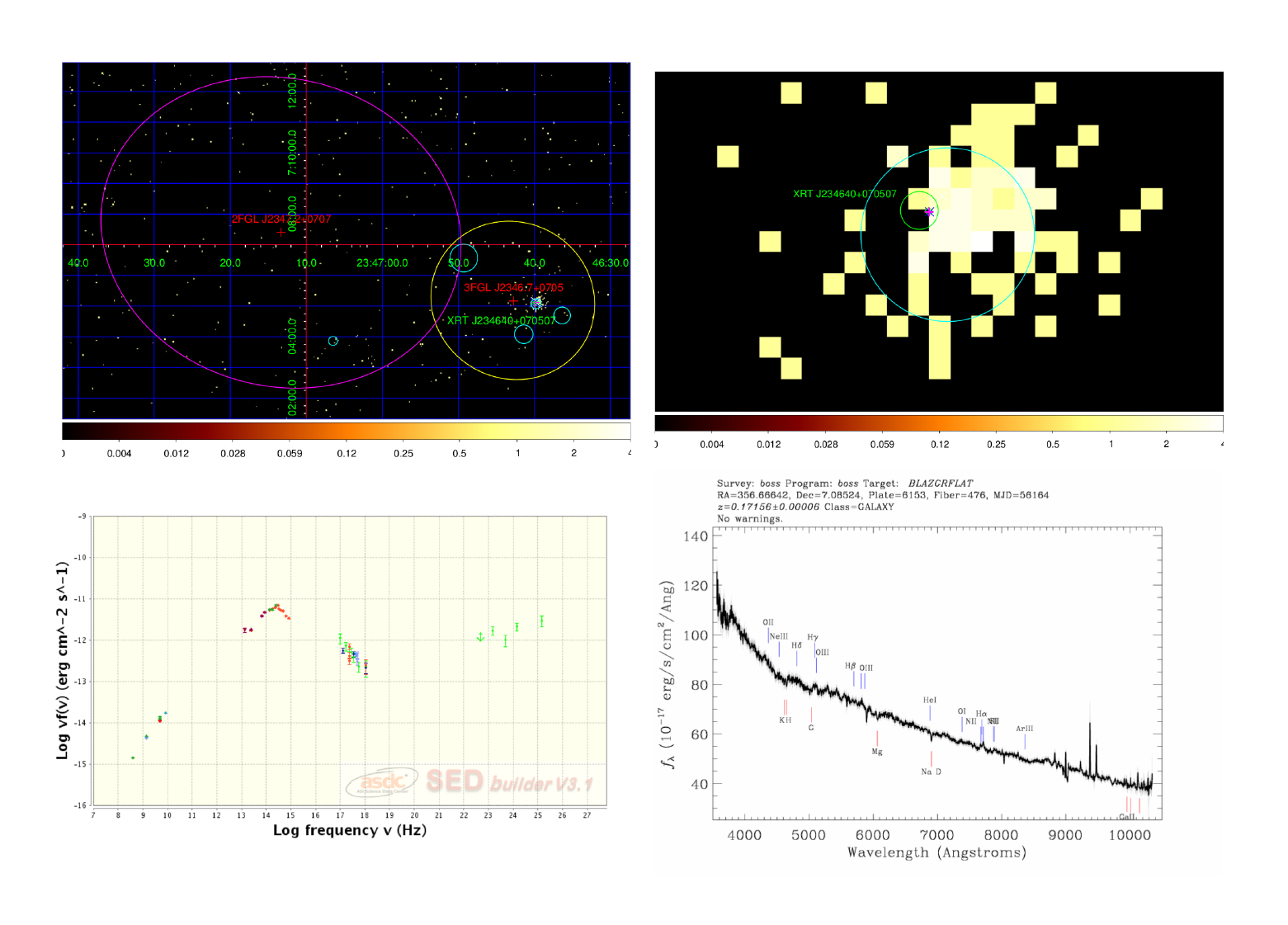} } 
  \caption{\texttt{Upper panel left}:  \textit{Swift}/XRT image of the 2FGL J2347.2+0707 created using the online data analysis tool of UK Swift Science Data Centre. The red crosses are the 3FGL and 2FGL position and the yellow (magenta) ellipse is the 95$\%$ error region of 3FGL (2FGL) catalogue. The XRT sources detected in this work are displayed as green circles with radius equal to the XRT error radius. The cyan circles show the error circles of the NVSS radio sources and the white crosses are the\textit{Swift}/XRT objects of the 1SXPS catalogue. \texttt{Upper panel right}:  Close-up of XRT J234640+070507 sky map. The blue and magenta crosses are the positions of WISE and SDSS objects and the green diamonds correspond to 2MASS sources. \newline
\texttt{Bottom panel left}:  Broadband SED for the 2FGL J2347.2+0707 created using the SED Builder tool of the ASI ASDC Data Centre. We combine the WISE IR data (brown points), the 2MASS IR data (green points), the SDSS10 optical data (red points) with the HE $\gamma$-ray data (green points) from the 2FGL catalogue. The X-ray flux points are performed from the \textit{Swift}-XRT data analysis of this work (green points) and taken from 1SXPS catalogue (blue points).
\texttt{Bottom panel right}:  the SDSS optical spectrum of the optical counterpart  2FGL J2347.2+0707, which shows a power law shape and indicates a \textit{blazar} origin. 
}
  \label{fig:2347_ass}
\end{figure*}

2FGL J2347.2+0707 is an object of the 3FGL (2FGL) catalogue with detection significance = 13.83 (7.2)$\sigma$ and an \textit{95$\%$ semi axis error} of 3' (6.0'). 
In the 3FGL catalogue the source is classified as an active galaxy of uncertain type and associated to the object TXS 2344+068. From the UK online analysis of the 2011 XRT data ($\sim$5000 seconds) we obtain the X-ray count map of the 2FGL J2347.2+0707 sky region (Fig. \ref{fig:2347_ass}, upper-left panel). 
Inside the 3FGL error ellipse, we have only one X-ray source detected from \textit{Swift} with (RA,DEC)=(23 46 40.01, +07 05 07.0) with a count rate of 2.090$\times$10$^{-2}$ $\pm$ 2.055$\times$10$^{-3}$ cts/s. 
We propose this object as the most likely X-ray counterpart for 2FGL J2347.2+0707 in agreement with the 3FGL association.
From the close-up image focused on XRT J234640+070507 (upper-right panel) within the X-ray error circle of 2.1''  we have the radio source NVSS J234639+070504, the optical source SDSS10 1237669517440385146, and the IR objects WISE J234639+070506 and 2MASS 23463993+0705068. In the bottom panel the multi-wavelength SED of 2FGL J2347.2+0707, is shown with the X-ray differential spectrum obtained by our dedicated X-ray analysis (green points) and the 1SXPS data points (blue points).

\label{lastpage}

\end{document}